\pdfoutput=1

\documentclass[11pt,twoside,a4paper,cmspaper,final,collab]{cms-tdr}
\def\svnVersion{334497}\def\cmsCernNoTag{CERN-PH-EP/2013-037}\def\cmsCernDate{\today}\def\cmsMessage{Published in the European Physical Journal C as \href{http://dx.doi.org/10.1140/epjc/s10052-016-3988-x}{\doi{10.1140/epjc/s10052-016-3988-x.}}}

\begin{document}\cmsNoteHeader{GEN-14-001}

\hyphenation{had-ron-i-za-tion}
\hyphenation{cal-or-i-me-ter}
\hyphenation{de-vices}
\RCS$HeadURL: svn+ssh://svn.cern.ch/reps/tdr2/notes/GEN-14-001/trunk/GEN-14-001.tex $
\RCS$Id: GEN-14-001.tex 234706 2014-04-01 13:24:56Z pgunnell $
\newlength\cmsFigWidth
\ifthenelse{\boolean{cms@external}}{\setlength\cmsFigWidth{0.85\columnwidth}}{\setlength\cmsFigWidth{0.4\textwidth}}
\ifthenelse{\boolean{cms@external}}{\providecommand{\cmsLeft}{top}}{\providecommand{\cmsLeft}{left}}
\ifthenelse{\boolean{cms@external}}{\providecommand{\cmsRight}{bottom}}{\providecommand{\cmsRight}{right}}
\ifthenelse{\boolean{cms@external}}{\providecommand{\cmsTable}[1]{#1}}{\providecommand{\cmsTable}[1]{\resizebox{\textwidth}{!}{#1}}}
\providecommand{\NA}{---\xspace}

\newcommand*{\pthat}{$\hat{p}_{\rm T}$}
\newcommand*{\ptzero}{$p_{\rm T0}$}
\newcommand*{\UE}{underlying event}
\newcommand*{\TR}{transverse}
\newcommand*{\tmax}{TransMAX}
\newcommand*{\tmin}{TransMIN}
\newcommand*{\tdif}{TransDIF}
\newcommand*{\tave}{TransAVE}
\newcommand*{\LJ}{leading jet}
\newcommand*{\BB}{back-to-back}
\newcommand*{\ptmax}{$p_{\rm T}^{\rm max}$}
\newcommand*{\ptsum}{$p_{\rm T}^{\rm sum}$}
\newcommand*{\etsum}{$ET{\rm sum}$}
\newcommand*{\ptcut}{$p_{\rm T}\!>\!0.5\GeV$}
\newcommand*{\etacut}{$|\eta|\!<\!0.8$}
\newcommand*{\etabig}{$|\eta|\!<\!2.0$}
\newcommand*{\aveN}{$\langle\!N_{\rm chg}\!\rangle$}
\newcommand*{\avept}{$\langle\!p_{\rm T}\!\rangle$}
\newcommand*{\etaphi}{$\eta$-$\phi$}
\newcommand*{\delphi}{$\Delta\phi$}
\newcommand*{\absdelphi}{$|\Delta\phi|$}
\newcommand*{\delphicut}{$|\Delta\phi|>150^\circ$}
\newcommand*{\pyold}{\textsc{pythia6}}
\newcommand*{\pyoldhyphen}{\textsc{pythia6}}
\newcommand*{\pynew}{\textsc{pythia8}}
\newcommand*{\pynewhyphen}{\textsc{pythia8}}
\newcommand*{\hwpp}{\textsc{herwig++}}
\newcommand*{\hera}{HERAPDF$1.5$LO}
\newcommand*{\tunee}{UE-EE-$5$C}
\newcommand*{\tunec}{Tune $4$C}
\newcommand*{\cuePA}{CUETP$6$S$1$-CTEQ$6$L1}
\newcommand*{\cuePAH}{CUETP$6$S$1$-HERAPDF1.5LO}
\newcommand*{\cuePB}{CUETP$8$S$1$-CTEQ$6$L1}
\newcommand*{\cuePH}{CUETP$8$S$1$-HERAPDF1.5LO}
\newcommand*{\cuePM}{CUETP$8$M$1$}
\newcommand*{\cueHW}{CUETHppS$1$}
\newcommand*{\cdpWA}{CDPSTP$8$S$1$-Wj}
\newcommand*{\cdpWB}{CDPSTP$8$S$2$-Wj}
\newcommand*{\cdpJA}{CDPSTP$8$S$1$-$4$j}
\newcommand*{\cdpJB}{CDPSTP$8$S$2$-$4$j}
\newcommand*{\tunelep}{Tune Z$2^*$lep}
\newcommand*{\tuneZ}{Tune Z$2^*$}
\newcommand*{\eff}{$\sigma_{\rm eff}$}
\newcommand*{\relpt}{$\Delta^{\rm rel}p_{\rm T}$}
\newcommand*{\RIVET}{\textsc{rivet}}
\newcommand*{\PROFESSOR}{\textsc{professor}}

\cmsNoteHeader{GEN-14-001}
\title{Event generator tunes obtained from underlying event and multiparton scattering measurements}

\abstract{New sets of parameters (``tunes'') for the underlying-event (UE) modelling of the \textsc{pythia}8, \textsc{pythia6} and 
\textsc{herwig++}  Monte Carlo event generators are constructed using different parton distribution functions. 
 Combined fits to CMS UE proton-proton ($\Pp \Pp$) data at $\sqrt{s} = 7\TeV$ and to UE proton-antiproton ($\Pp \PAp$) data from the CDF experiment at lower $\sqrt{s}$, are used to study the UE models and constrain their parameters, providing thereby improved predictions for proton-proton collisions at 13\TeV. In addition, it is investigated whether the values of the parameters obtained from fits to UE observables are consistent with the values determined from fitting observables sensitive to double-parton scattering processes. Finally, comparisons are presented of the UE tunes to ``minimum bias'' (MB) events, multijet, and Drell--Yan ($ \PQq \PAQq \rightarrow \PZ / \gamma^* \rightarrow$ lepton-antilepton+jets) observables at 7 and 8\TeV, as well as predictions for MB and UE observables at 13\TeV.}

\hypersetup{%
pdfauthor={CMS Collaboration},%
pdftitle={Event generator tunes obtained from underlying event and multiparton scattering measurements at CMS},%
pdfsubject={CMS},%
pdfkeywords={CMS, Underlying Event, Double Parton Scattering, Generator tuning}}

\maketitle

\section{Introduction}
\label{section-1}

Monte Carlo (MC) event generators of hadron-hadron collisions based on perturbative quantum chromodynamics (QCD) contain several components. The ``hard-scattering'' part of the event consists of particles resulting from the hadronization of the two partons (jets) produced in the hardest scattering, and in their associated hard initial- and final-state radiation (ISR and FSR).  The \UE\ (UE) consists of particles from the hadronization of beam-beam remnants (BBR), of multiple-parton interactions (MPI), and their associated ISR and FSR.  The BBR include hadrons from the fragmentation of spectator partons that do not exchange any appreciable transverse momentum (\pt) in the collision. The MPI are additional 2-to-2 parton-parton scatterings that occur within the same hadron-hadron collision, and are softer in transverse momentum ($\pt \lesssim 3\GeV$) than the hard scattering.

{\tolerance=1200
The perturbative 2-to-2 parton-parton differential cross section diverges like $1/\hat p_{\rm T}^4$, where $\hat p_{\rm T}$ is the transverse momentum of the outgoing partons in the parton-parton center-of-mass (c.m.) frame.  Usually, QCD MC models such as \PYTHIA~\cite{Sjostrand:2006za,Sjostrand:2001yu,Sjostrand:1987su,Bengtsson:1986gz,oai:arXiv.org:0710.3820} regulate this divergence by including a smooth phenomenological cutoff $p_{\rm T0}$ as follows:

\begin{equation}\label{pTzero}
1/\hat p_{\rm T}^4\rightarrow 1/(\hat p_{\rm T}^2+p_{\rm T0}^2)^2.
\end{equation}
\par}

This formula approaches the perturbative result for large scales and is finite as $\hat p_{\rm T}\rightarrow0$.  The divergence of the strong coupling $\alpha_{\rm s}$ at low $\hat p_{\rm T}$ is also regulated through Eq.~(\ref{pTzero}). The primary hard $2$-to-$2$ parton-parton scattering process and the MPI are regulated in the same way through a single $p_{\rm T0}$ parameter. However, this cutoff is expected to have a dependence on the center-of-mass energy of the hadron-hadron collision $\sqrt{s}$. In the \PYTHIA MC event generator this energy dependence is parametrized with a power-law function with exponent $\epsilon$: 

\begin{equation}\label{Ezero}
p_{\rm T0}(\sqrt{s})= p_{\rm T0}^{\rm ref} \,  (\sqrt{s}/\sqrt{s_0})^\epsilon,                        
\end{equation}

where $\sqrt{s_0}$ is a given reference energy and $p_{\rm T0}^{\rm ref}$ is the value of $p_{\rm T0}$ at $\sqrt{s_0}$. At a given $\sqrt{s}$, the amount of MPI depends on $p_{\rm T0}$, the parton distribution functions (PDF), and the overlap of the matter distributions (or centrality) of the two colliding hadrons. Smaller values of $p_{\rm T0}$ provide more MPI due to a larger MPI cross section.  Table~\ref{table1} shows the parameters in \pyold~\cite{Sjostrand:2006za} and \pynew~\cite{oai:arXiv.org:0710.3820} that, together with the selected PDF, determine the energy dependence of MPI. Recently, in \hwpp~\cite{Bahr:2008pv,Bellm:2013lba} the same formula has been adopted to provide an energy dependence to their MPI cutoff, which is also shown in Table~\ref{table1}. The QCD MC generators have other parameters that can be adjusted to control the modelling of the properties of the events, and a specified set of such parameters adjusted to fit certain prescribed aspects of the data is referred to as a ``tune''~\cite{Albrow:2006rt,Field:2009zz,Skands:2009zm}. 

\begin{table*}[htbp]
\renewcommand{\arraystretch}{1.2}
\begin{center}
\topcaption{Parameters in \pyold~\cite{Sjostrand:2006za}, \pynew~\cite{oai:arXiv.org:0710.3820}, and \hwpp~\cite{Bahr:2008pv,Bellm:2013lba} MC event generators that, together with some chosen PDF, determine the energy dependence of MPI. }
\label{table1}
\cmsTable{ 
\begin{tabular}{l c c c} \hline 
 Parameter					&  \pyold  &  \pynew  &  \hwpp \\ \hline
   MPI cutoff, $p_{\rm T0}^{\rm ref}$, at $\sqrt{s} = \sqrt{s_0}$	& PARP($82$)    & MultipartonInteractions:pT0Ref & MPIHandler:pTmin0 \\ 
  Reference energy, $\sqrt{s_0}$	& PARP($89$)    & MultipartonInteractions:ecmRef & MPIHandler:ReferenceScale \\ 
  Exponent of $\sqrt{s}$ dependence, $\epsilon$	& PARP($90$) & MultipartonInteractions:ecmPow & MPIHandler:Power \\ \hline
\end{tabular}
}
\end{center}
\end{table*}

{\tolerance=5000
In addition to hard-scattering processes, other processes contribute to the inelastic cross section in hadron-hadron collisions: single-diffraction dissociation (SD), double-diffraction dissociation (DD), and central-diffraction (CD). In SD and DD  events, one or both beam particles are excited into high-mass color-singlet states (\ie~into some resonant $\mathrm{N}^*$), which then decay. The SD and DD processes correspond to color-singlet exchanges between the beam hadrons, while CD corresponds to double color-singlet exchange with a diffractive system produced centrally. For non-diffractive processes (ND), color is exchanged, the outgoing remnants are no longer color singlets, and this separation of color generates a multitude of quark-antiquark pairs that are created via vacuum polarization. The sum of all components except SD corresponds to non single-diffraction (NSD) processes.
\par}

Minimum bias (MB) is a generic term that refers to events selected by requiring minimal activity within the detector. This selection accepts a large fraction of the overall inelastic cross section. Studies of the UE are often based on MB data, but it should be noted that the dominant particle production mechanisms in MB collisions and in the UE are not exactly the same. On the one hand, the UE is studied in collisions in which a hard 2-to-2 parton-parton scattering has occurred, by analyzing the hadronic activity in different regions of the event relative to the back-to-back azimuthal structure of the hardest particles emitted~\cite{Aaltonen:2015aoa}. On the other hand, MB collisions are often softer and include diffractive interactions that, in the case of \textsc{pythia}, are modelled via a Regge-based approach~\cite{Schuler:1993wr}. 
 
The MPI are usually much softer than primary hard scatters, however, occasionally two hard $2$-to-$2$ parton scatters can take place within the same hadron-hadron collision.  This is referred to as double-parton scattering (DPS)~\cite{Manohar:2012pe,DelFabbro:2001rs,Blok:2013bpa,Bartalini:2011jp}, and is typically described in terms of an effective cross section parameter, \eff, defined as:

\begin{equation}\label{sigeff}
\sigma_{\rm AB} = \frac{\sigma_{\rm A} \sigma_{\rm B}}{\sigma_{\rm eff}},                        
\end{equation}

where $\sigma_{\rm A}$ and $\sigma_{\rm B}$ are the inclusive cross sections for individual hard scattering processes of generic type A and B, respectively, and $\sigma_{\rm AB}$ is the cross section for producing both scatters in the same hadron-hadron collision.  If A and B are indistinguishable, as in four-jet production, a statistical factor of $1/2$ must be inserted on the right-hand side of Eq.~(\ref{sigeff}). Furthermore, \eff\ is assumed to be independent of A and B. However, \eff\ is  not a directly observed quantity, but can be calculated from the overlap function of the two transverse profile distributions of the colliding hadrons, as implemented in any given MPI model. 

The UE tunes have impact in both soft and hard particle production in a given pp collision. First, about half of the particles produced in a MB collision originate from the hadronization of partons scattered in MPI, and have their differential cross sections in \pt\ regulated via Eq.(\ref{pTzero}), using the same \ptzero\ cutoff used to tame the hardest 2-to-2 parton-parton scattering in the event. The tuning of the cross-section regularization affects therefore all (soft and hard) parton-parton scatterings and provides a prediction for the behavior of the ND cross section. Second, the UE tunes parametrize the distribution in the transverse overlap of the colliding protons and thereby the probability of two hard parton-parton scatters that is then used to estimate DPS-sensitive observables. 

{\tolerance=5000
In this paper, we study the $\sqrt{s}$ dependence of the UE using recent CDF proton-antiproton data from the Fermilab Tevatron at $0.3$, $0.9$, and $1.96\TeV$~\cite{Aaltonen:2015aoa}, together with CMS pp data from the CERN LHC at $\sqrt{s} = 7\TeV$~\cite{CMS:2012kca}. The $0.3$ and $0.9\TeV$ data are from the ``Tevatron energy scan'' performed just before the Tevatron was shut down.  
Using the \RIVET\ (version 1.9.0) and \PROFESSOR\ (version 1.3.3) frameworks~\cite{Rivet,Buckley:2009bj}, we construct: (i)  new \pynewhyphen\ (version 8.185) UE tunes  using several PDF sets (CTEQ$6$L1~\cite{Pumplin:2002vw},  \hera ~\cite{Sarkar:2014zua}, and NNPDF$2.3$LO~\cite{Ball:2013hta,Ball:2011uy}), (ii) new \pyoldhyphen\ (version 6.327) UE tunes (using CTEQ$6$L1 and \hera ), and (iii) a new \hwpp\ (version 2.7.0) UE tune for CTEQ$6$L1. The \RIVET\ software is a tool for producing predictions of physics quantities obtained from MC event generators. It is used for generating sets of MC predictions with a different choice of parameters related to the UE simulation. The predictions are then included in the \PROFESSOR\ framework, which parametrizes the generator response and returns the set of tuned parameters that best fits the input measurements.
\par}

{\tolerance=1200
In addition, we construct several new CMS ``DPS tunes''  and investigate whether the values of the UE parameters determined from fitting the UE observables in a hard-scattering process are consistent with the values determined from fitting DPS-sensitive observables. The \PROFESSOR\ software also offers the possibility of extracting ``eigentunes'', which provide an estimate of the uncertainties in the fitted parameters. The eigentunes consist of a collection of additional tunes, obtained through the covariance matrix of the data-theory fitting procedure, to determine independent directions in parameter space that provide a specific modification in the goodness of the fit, $\chi^2$ (Section~\ref{section-2}). All of the CMS UE and DPS tunes are provided with
eigentunes. In Section~\ref{section-4}, predictions using the CMS UE tunes are compared to other UE measurements not used in determining the tunes, and we examine how well Drell--Yan, MB, and multijet observables can be predicted using the UE tunes.  In Section~\ref{section-5}, predictions of the new tunes are shown for UE observables at $13\TeV$, together with a comparison to the first MB distribution measured. Section~\ref{section-6} has a brief summary and conclusions. The appendices contain additional comparisons between the \textsc{pythia6} and \textsc{herwig++} UE tunes and the data, information about the tune uncertainties, and predictions for some MB and DPS observables at 13\TeV. 
\par}

\section{The CMS UE tunes}
\label{section-2}

Previous UE studies have used the charged-particle jet with largest \pt\ \cite{Chatrchyan:2011id,Chatrchyan:2013gfi} or a $\PZ$ boson~\cite{Chatrchyan:2012tb,Aaltonen:2015aoa} as the leading (\ie highest \pt) objects in the event.  The CDF and CMS data, used for the tunes, select the charged particle with largest \pt\ in the event (\ptmax) as the ``leading object'', and use just the charged particles with \ptcut\ and $|\eta| < 0.8$ to characterize the UE. 

\begin{figure*}[htbp]
\begin{center}
\includegraphics[scale=0.8]{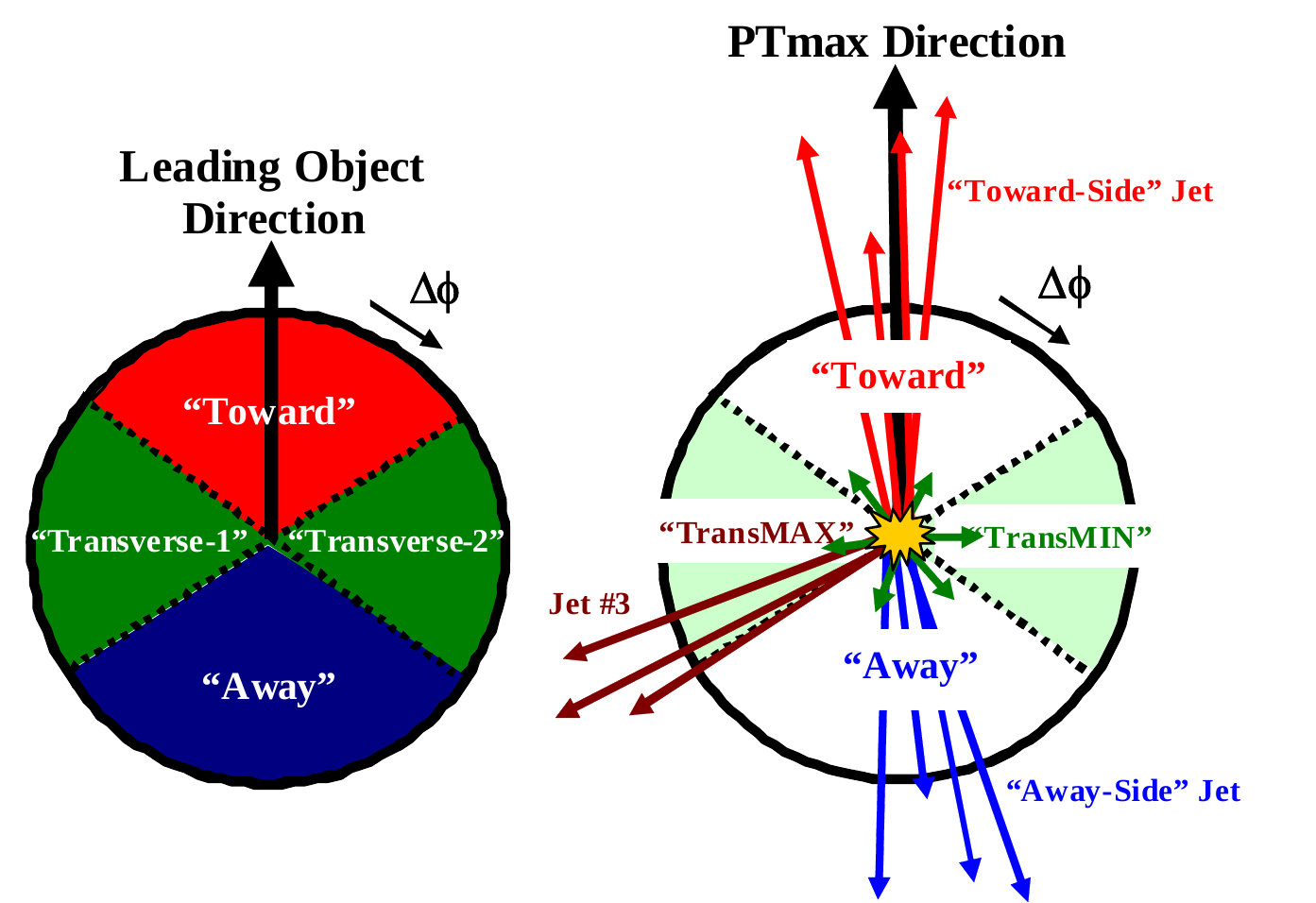}
\caption{Left: Illustration of the azimuthal regions in an event defined by the \delphi\ angle relative to the direction of the leading object~\cite{Aaltonen:2015aoa}. Right: Illustration of the topology of a hadron-hadron collision in which a hard parton-parton collision has occurred, and the leading object is taken to be the charged particle of largest \pt\ in the event, \ptmax.}
\label{PUB_Newfig1}
\end{center}
\end{figure*}

On an event-by-event basis, the leading object is used to define regions of pseudorapidity-azimuth (\etaphi) space. The ``toward'' region relative to this direction, as indicated in Fig.~\ref{PUB_Newfig1}, is defined by $|\Delta\phi|<\pi/3$ and $|\eta| < 0.8$, and the ``away'' region by $|\Delta\phi|>2\pi/3$ and $|\eta| < 0.8$. The charged-particle and the scalar-\pt\ sum densities in the \TR\ region are calculated as the sum of the contribution in the two regions: ``Transverse-1'' ($\pi/3<\Delta\phi<2\pi/3$, $|\eta| < 0.8$) and ``Transverse-2'' ($\pi/3<-\Delta\phi<2\pi/3$, $|\eta| < 0.8$), divided by the area in \etaphi\ space, $\Delta\eta\Delta\phi = 1.6\times 2\pi/3$. The \TR\ region is further separated into the ``\tmax'' and ``\tmin'' regions, also shown in Fig.~1. This defines on an event-by-event basis the regions with more (\tmax) and fewer (\tmin) charged particles ($\mathrm{N}_{\rm ch}$), or greater (\tmax) or smaller (\tmin) scalar-\pt\ sums (\ptsum).
The UE particle and \pt\ densities are constructed by dividing by the area in \etaphi\ space, where the \tmax\ and \tmin\ regions each have an area of $\Delta\eta\Delta\phi = 1.6\times 2\pi/6$. The \TR\ density (also referred to as ``\tave'') is the average of the \tmax\ and the \tmin\ densities. For events with hard initial- or final-state radiation, the \tmax\ region often contains a third jet, but both the \tmax\ and \tmin\ regions receive contributions from the MPI and beam-beam remnant components. The \tmin\ region is very sensitive to the MPI and beam-beam remnant components of the UE, while ``\tdif'' (the difference between \tmax\ and \tmin\ densities) is very sensitive to ISR and FSR~\cite{Pumplin:1997ix}. 

The new UE tunes are determined by fitting UE observables, and using only those parameters that are most sensitive to the UE data.  Since it is not possible to tune all parameters of a MC event generator at once, the parameters that affect, for example, the parton shower, the fragmentation, and the intrinsic-parton \pt\ are fixed to the values given by an initially established reference tune. The initial reference tunes used for \textsc{pythia8} are Tune 4C~\cite{Corke:2010yf} and the Monash Tune~\cite{Skands:2014pea}. For \textsc{pythia6}, the reference tune is Tune Z2*lep~\cite{Chatrchyan:2013gfi}, and for \textsc{herwig++} it is Tune UE-EE-5C~\cite{Seymour:2013qka}. 

\ifthenelse{\boolean{cms@external}}{\subsection{The PYTHIA8 UE tunes}}{\subsection{The \textsc{pythia8} UE tunes}}

{\tolerance=5000
Taking as the reference tune the set of parameters of \textsc{pythia8} \tunec\ \cite{Corke:2010yf}, we construct two new UE tunes, one using CTEQ$6$L1 (\cuePB) and one using \hera\ (\cuePH). \verb+CUET+ (read as ``cute'') stands for ``CMS UE tune'', and \verb+P8S1+ stands for \pynew\ ``Set 1''. 
\par}

{\tolerance=1200
The tunes are extracted by varying the four parameters in Table~\ref{table4} in fits to the \tmax\ and \tmin\ charged-particle and \ptsum\ densities at three energies, for $ \Pp \PAp$ collisions at $\sqrt{s} = 0.9$ and $1.96$, and $ \Pp \Pp$ collisions at $7\TeV$. The measurements of \tave\ and \tdif\ densities are not included in the fit, since they can be constructed from \tmax\ and \tmin.  The new tunes use an exponentially-falling matter-overlap function between the two colliding protons of the form $\exp(-b^{\verb+expPow+}$), with $b$ being the impact parameter of the collision. The parameters that are varied are  \verb+expPow+, the MPI energy-dependence parameters (Table~\ref{table1}) and the range, \ie the probability, of color reconnection (CR). A small (large) value of the final-state CR parameter tends to increase (reduce) the final particle multiplicities. In \pynew, unlike in \pyold, only one parameter determines the amount of CR, which includes a \pt\ dependence, as defined in Ref.~\cite{oai:arXiv.org:0710.3820}. 
\par}

The generated inelastic events include ND and diffractive (DD$+$SD$+$CD) contributions, although the UE observables used to determine the tunes are sensitive to single-diffraction dissociation, central-diffraction, and double-diffraction dissociation only at very small \ptmax\ values (\eg $p_{\rm T}^{\rm max}<1.5\GeV$). The ND component dominates for $p_{\rm T}^{\rm max}$ values greater than ${\approx} 2.0\GeV$, since the cross section of the diffractive components rapidly decreases as a function of \pthat. The fit is performed by minimizing the $\chi^2$ function:

\begin{equation}\label{chi}
\chi^2(p)=\sum_{i}\frac{(f^{i}(p)-R_{i})^2}{\Delta_{i}^2},
\end{equation}

where the sum runs over each bin $i$ of every observable. The $f^{i}(p)$ functions correspond to the interpolated MC response for the simulated observables as a function of the parameter vector $p$, $R_i$ is the value of the measured observable in bin $i$, and $\Delta_i$ is the total experimental uncertainty of $R_i$. We do not use the Tevatron data at $\sqrt{s}=300\GeV$, as we are unable to obtain an acceptable $\chi^2$ in a fit of the four parameters in Table~\ref{table4}. The $\chi^2$ per degree of freedom (dof) listed in Table \ref{table4} refers to the quantity $\chi^2(p)$ in Eq.~(\ref{chi}), divided by the number of dof in the fit. The eigentunes (Appendix~A) correspond to the tunes in which the changes in the $\chi^2$ ($\Delta\chi^2$) of the fit relative to the best-fit value equals the $\chi^2$ value obtained in the tune, \ie $\Delta\chi^2$ = $\chi^2$. For both tunes in Table~\ref{table4}, the fit quality is very good, with $\chi^2$/dof values very close to 1. 

The contribution from CR changes in the two new tunes; it is large for the HERAPDF1.5LO and small for the CTEQ6L1 PDF. This is a result of the shape of the parton densities at small fractional momenta $x$, which is different for the two PDF sets. While the parameter $p_{\rm T0}^{\rm ref}$ in Eq.~(\ref{Ezero}) stays relatively constant between \tunec\ and the new tunes, the energy dependence $\epsilon$ tends to increase in the new tunes, as do the matter-overlap profile functions.

\begin{table*}[htbp]
\renewcommand{\arraystretch}{1.2}
\begin{center}
\topcaption{The \pynew\ parameters, tuning range, \tunec\ values~\cite{Corke:2010yf}, and best-fit values for \cuePB\ and \cuePH , obtained from fits to the \tmax\ and \tmin\ charged-particle and \ptsum\ densities, as defined by the leading charged-particle \ptmax at $\sqrt{s} = 0.9$, $1.96$, and $7\TeV$. The $\sqrt{s}=300\GeV$ data are excluded from the fit.}
\label{table4}
\cmsTable{
\begin{tabular}{l c c c c}  \hline
\pynew\ Parameter & Tuning Range  & \tunec &  CUETP8S1 & CUETP8S1 \\ \hline
   PDF                 &    \NA               &  CTEQ6L1       &  CTEQ6L1         & HERAPDF$1.5$LO \\ \hline
   MultipartonInteractions:pT0Ref [GeV]  & $1.0$--$3.0\phantom{0}$  & $2.085$ & $2.101$ & $2.000$ \\ 
   MultipartonInteractions:ecmPow & $0.0$--$0.4\phantom{0}$  & $0.19\phantom{0}$ & $0.211$ & $0.250$ \\ 
   MultipartonInteractions:expPow & $0.4$--$10.0$ & $2.0\phantom{00}$  & $1.609$ & $1.691$ \\ 
   ColourReconnection:range    & $0.0$--$9.0\phantom{0}$  & $1.5\phantom{00}$ & $3.313$ & $6.096$ \\ \hline
   MultipartonInteractions:ecmRef [GeV]  & \NA      & $1800$       & $1800^*$       & $1800^*$  \\ \hline 
   $\chi^2$/dof                & \NA      & \NA            & $0.952$        & $1.13\phantom{00}$  \\ \hline
   \multicolumn{5}{r}{\footnotesize $^*$ Fixed at \tunec\ value.}
\end{tabular}
}
\end{center}
\end{table*}

{\tolerance=1200
The \textsc{pythia8} Monash Tune~\cite{Skands:2014pea} combines updated fragmentation parameters with the NNPDF$2.3$LO PDF.
\par}

{\tolerance=5000
 The NNPDF$2.3$LO PDF has a gluon distribution at small $x$ that is different compared to CTEQ$6$L1 and \hera, and this affects predictions in the forward region of hadron-hadron collisions. Tunes using the NNPDF2.3LO PDF provide a more consistent description of the UE and MB observables in both the central and forward regions, than tunes using other PDF. 
 \par}
 
 {\tolerance=1200
A new \textsc{pythia8} tune \cuePM\ (labeled with \verb+M+ for Monash) is constructed using the parameters of the Monash Tune and fitting the two MPI energy-dependence parameters
of Table~\ref{table1} to UE data
at $\sqrt{s} = 0.9$, $1.96$, and $7\TeV$.
Varying the CR range and the exponential slope of the matter-overlap function freely in the minimization of the $\chi^2$ leads to suboptimal best-fit values. The CR range is therefore fixed to the value of the the Monash Tune, and the exponential slope of the matter-overlap function \verb+expPow+ is set to 1.6, which is similar to the value determined in CUETP8S1-CTEQ6L1.
The best-fit values of the two tuned parameters are shown in Table~\ref{table5}. Again, we exclude the $300\GeV$ data, since we are unable to get a good $\chi^2$ in the fit.
The parameters obtained for \cuePM\ differ slightly from the ones of the Monash Tune. The  obtained energy-dependence parameter $\epsilon$ is larger, while a very similar value is obtained for $p_{\rm T0}^{\rm ref}$.
\par}

\begin{table*}[htbp]
\renewcommand{\arraystretch}{1.2}
\begin{center}
\topcaption{The \pynew\ parameters, tuning range, Monash values~\cite{Skands:2014pea}, and best-fit values for \cuePM, obtained from fits to the \tmax\ and \tmin\ charged-particle and \ptsum\ densities, as defined by the leading charged-particle \ptmax at $\sqrt{s} =  0.9$, $1.96$, and $7\TeV$. The $\sqrt{s}=300\GeV$ data are excluded from the fit.}
\label{table5}
\cmsTable{
\begin{tabular}{l c c c}  \hline
 \pynew\ Parameter  &  Tuning Range & Monash   & CUETP8M1\\ \hline
   PDF                 &      \NA              & NNPDF$2.3$LO    & NNPDF$2.3$LO \\ \hline
   MultipartonInteractions:pT0Ref [GeV]  & $1.0$--$3.0$  & $2.280$ & $2.402$ \\ 
   MultipartonInteractions:ecmPow & $0.0$--$0.4$  & $0.215$      & $0.252$ \\ \hline
   MultipartonInteractions:expPow & \NA            & $1.85\phantom{0}$          & $1.6^{*}\phantom{0}$   \\ 
   ColourReconnection:range    & \NA            & $1.80\phantom{0}$          & $1.80^{**}$ \\
   MultipartonInteractions:ecmRef [GeV]  & \NA      & $7000$          & $7000^{**}$  \\ \hline 
   $\chi^2$/dof           & \NA            & \NA               & $1.54\phantom{0}$  \\ \hline
   \multicolumn{4}{r}{\footnotesize $^{*}$ Fixed at \cuePB\ value.} \\ [-3pt]
   \multicolumn{4}{r}{\footnotesize $^{**}$ Fixed at Monash Tune value.} 
\end{tabular}
}
\end{center}
\end{table*}

\begin{figure*}[htbp]
\begin{center}
\includegraphics[scale=0.65]{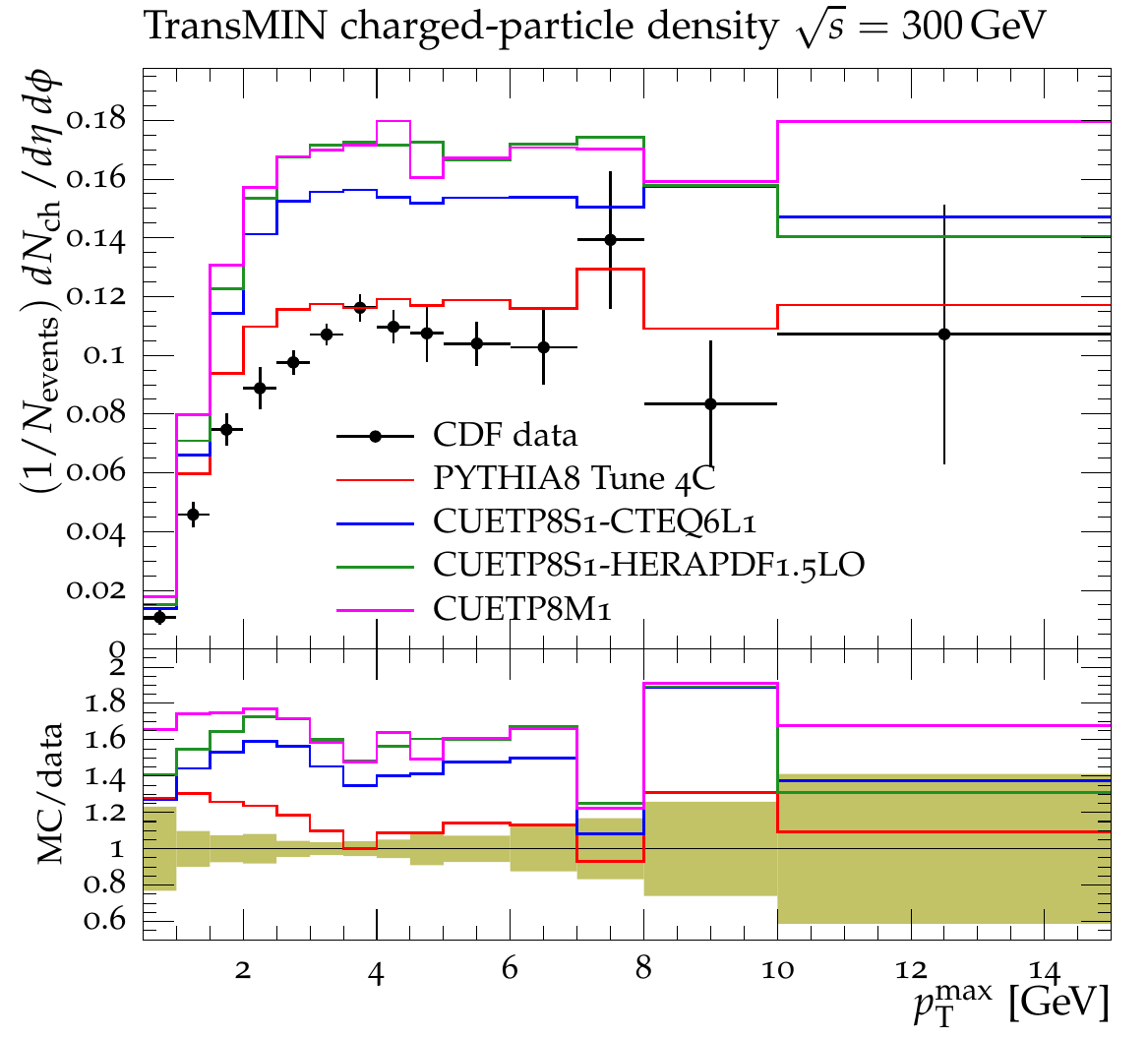}
\includegraphics[scale=0.65]{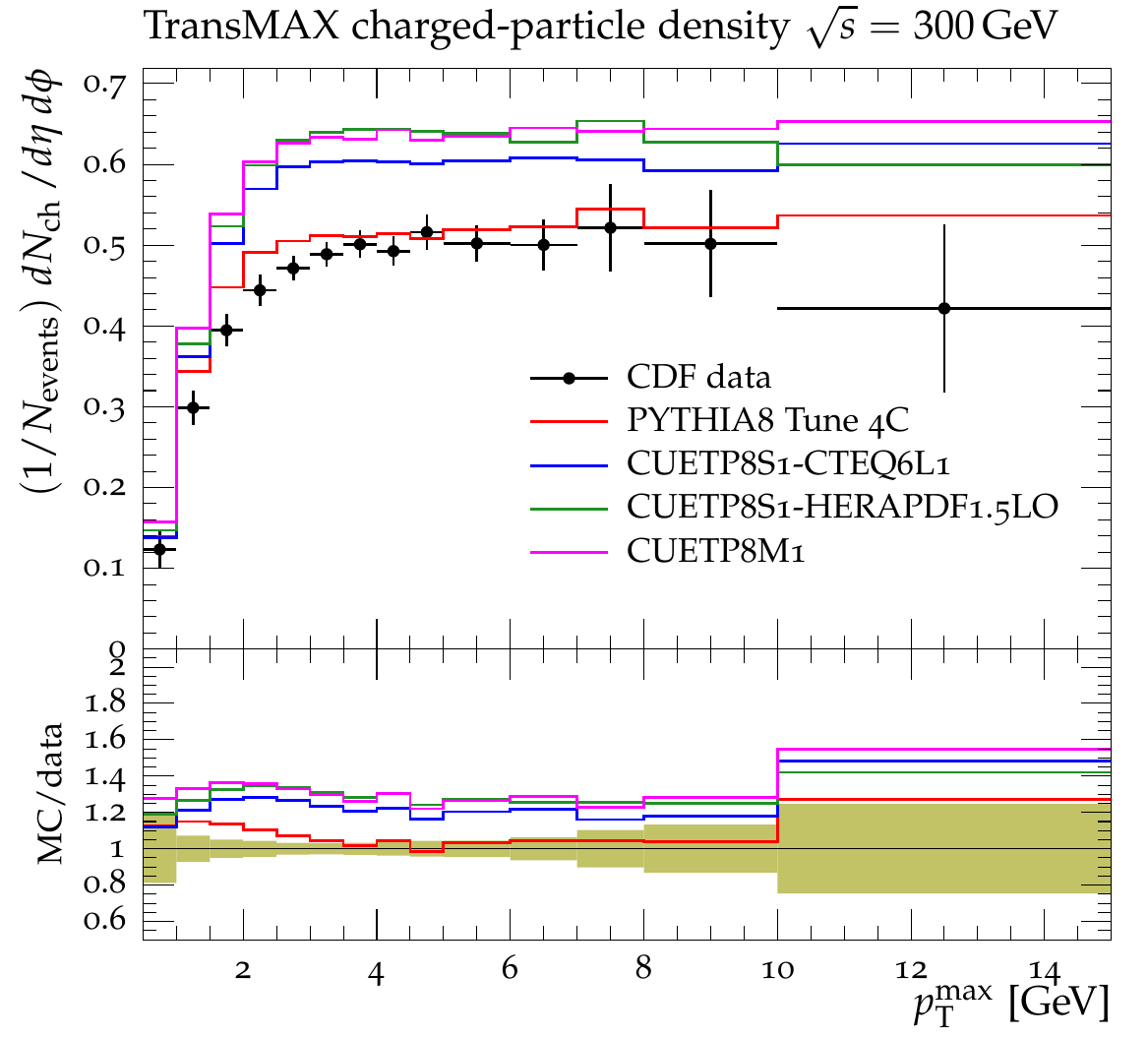}\\
\includegraphics[scale=0.65]{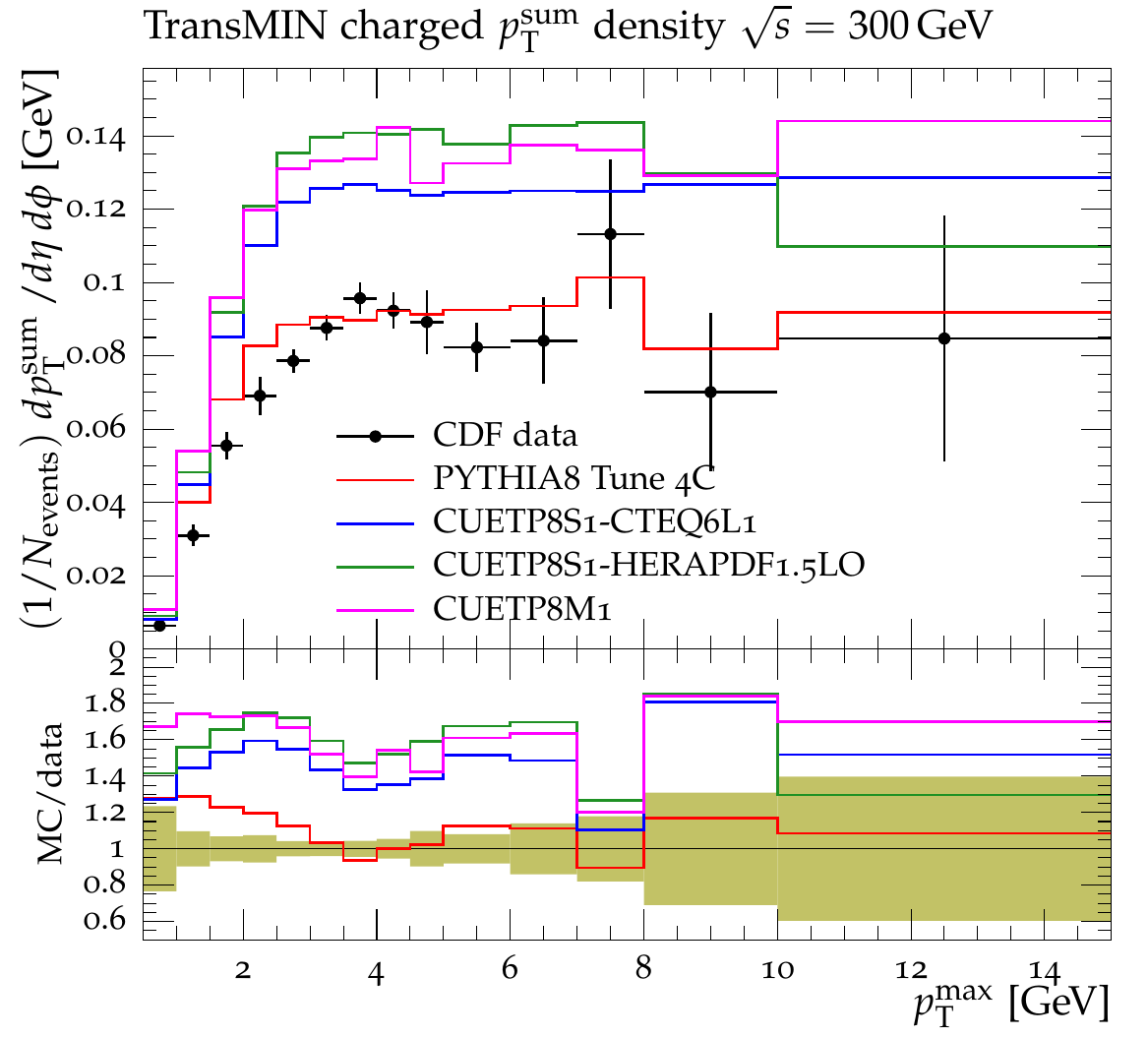}
\includegraphics[scale=0.65]{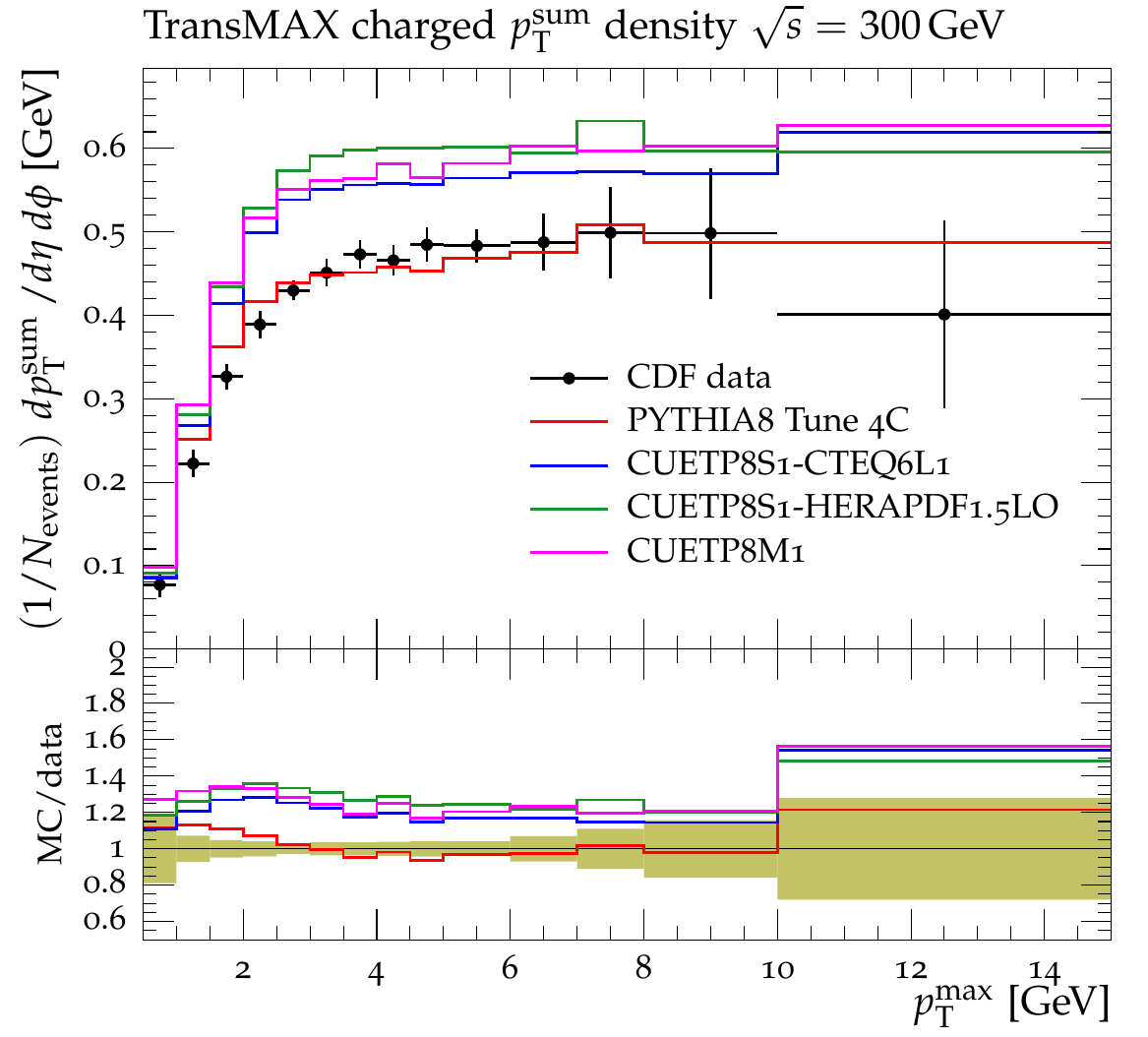}
\caption{CDF data at $\sqrt{s}=300\GeV$~\cite{Aaltonen:2015aoa} on particle (top) and \ptsum\ densities (bottom) for charged particles with \ptcut\ and \etacut\ in the \tmin\ (left) and \tmax\ (right) regions as defined by the leading charged particle, as a function of the transverse momentum of the leading charged-particle \ptmax. The data are compared to \pynewhyphen\ \tunec,  \cuePB, \cuePH, and \cuePM. The ratios of MC events to data are given below each panel. The data at $\sqrt{s}=300\GeV$ are not used in determining these tunes. The green bands in the ratios represent the total experimental uncertainties.}
\label{PUB_fig7}
\end{center}
\end{figure*}

\begin{figure*}[htbp]
\begin{center}
\includegraphics[scale=0.65]{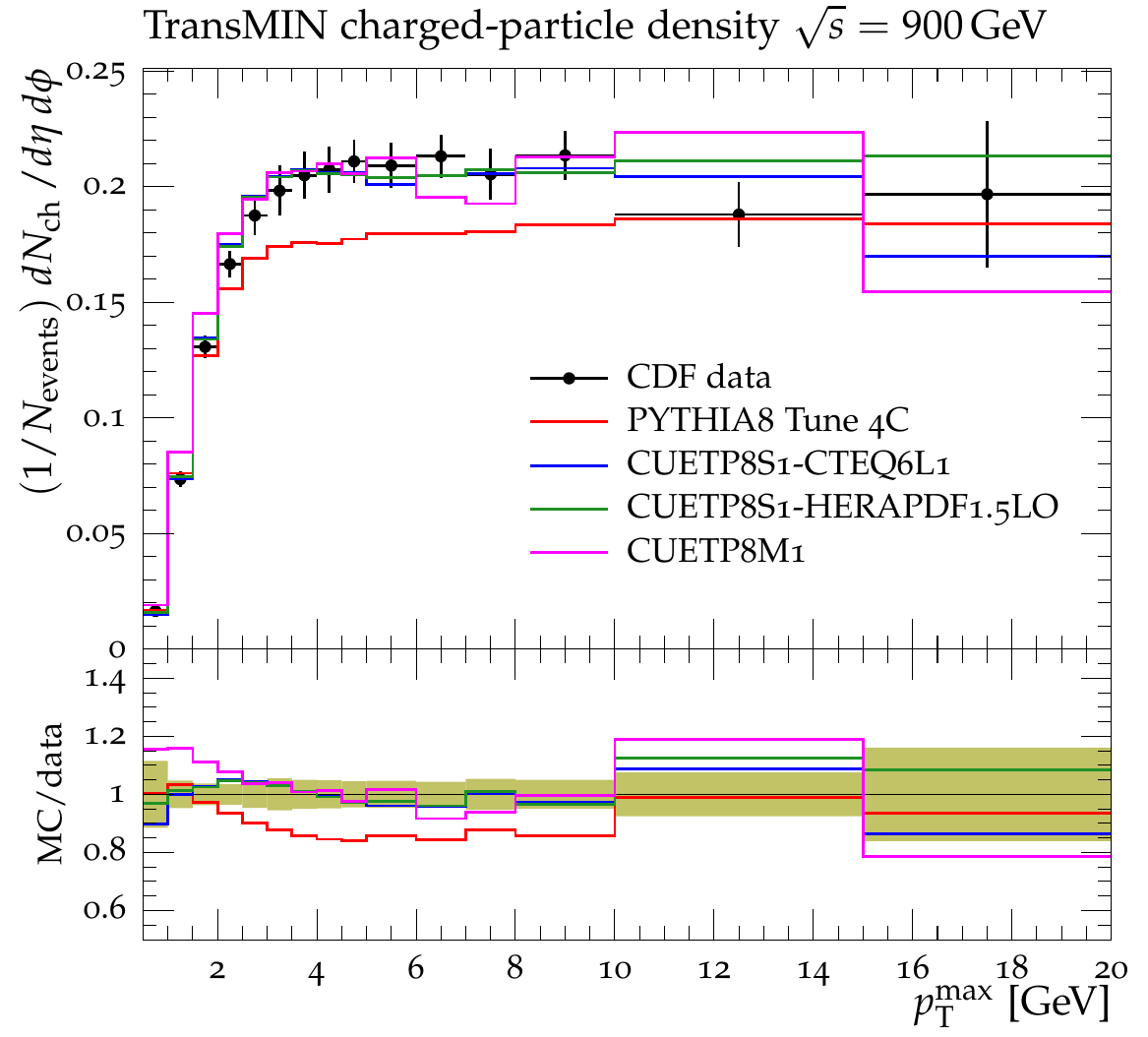}
\includegraphics[scale=0.65]{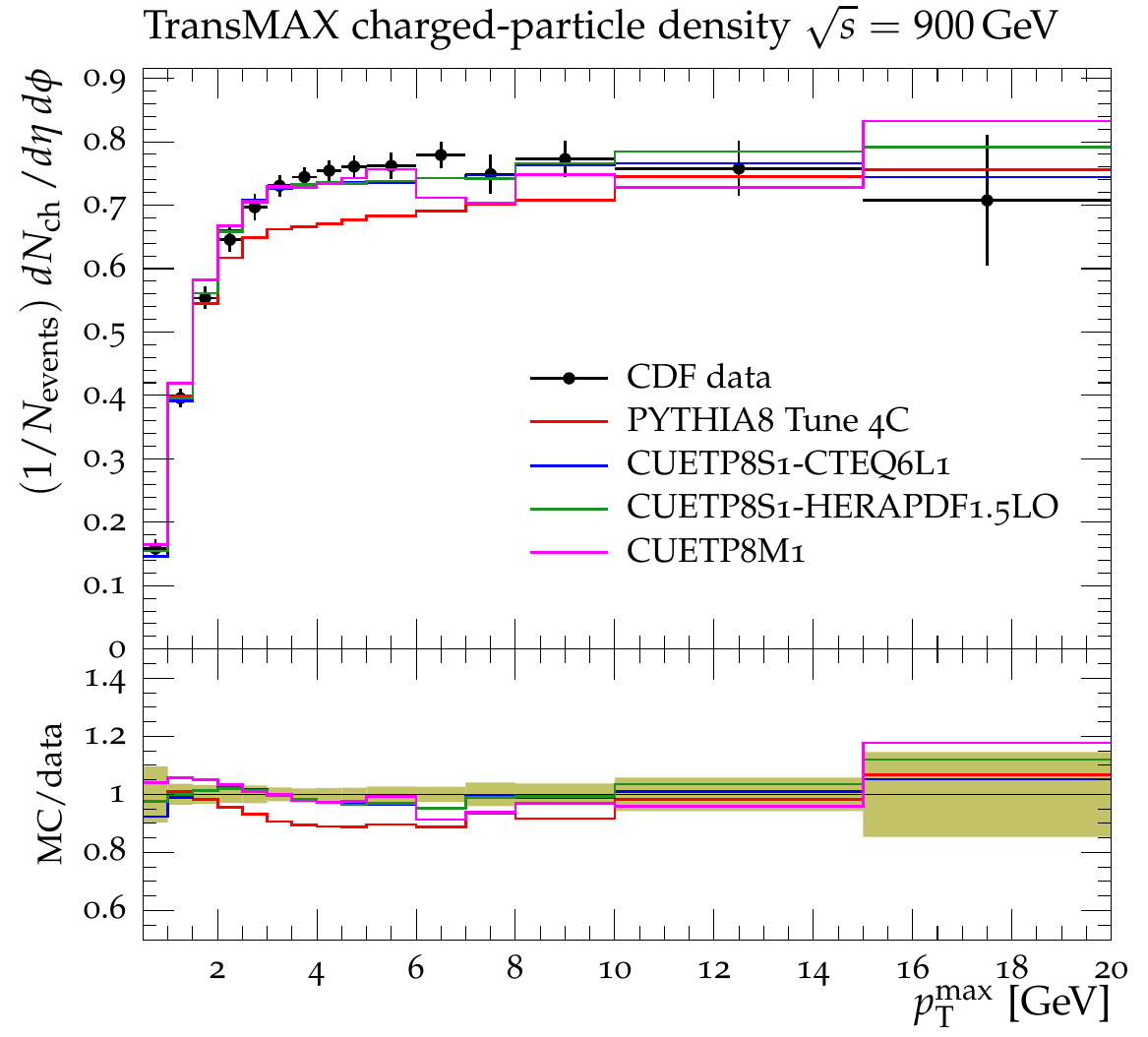}\\
\includegraphics[scale=0.65]{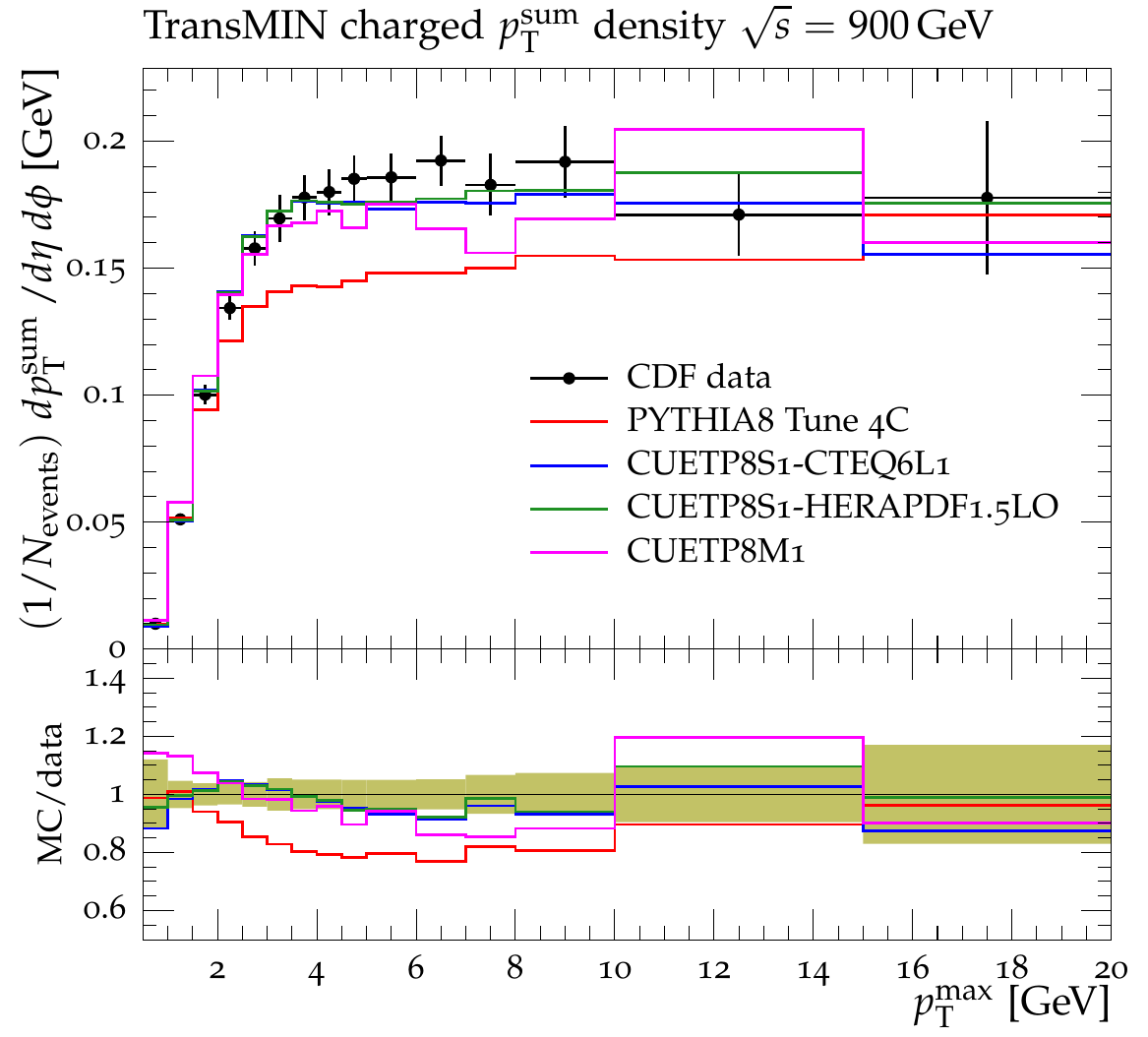}
\includegraphics[scale=0.65]{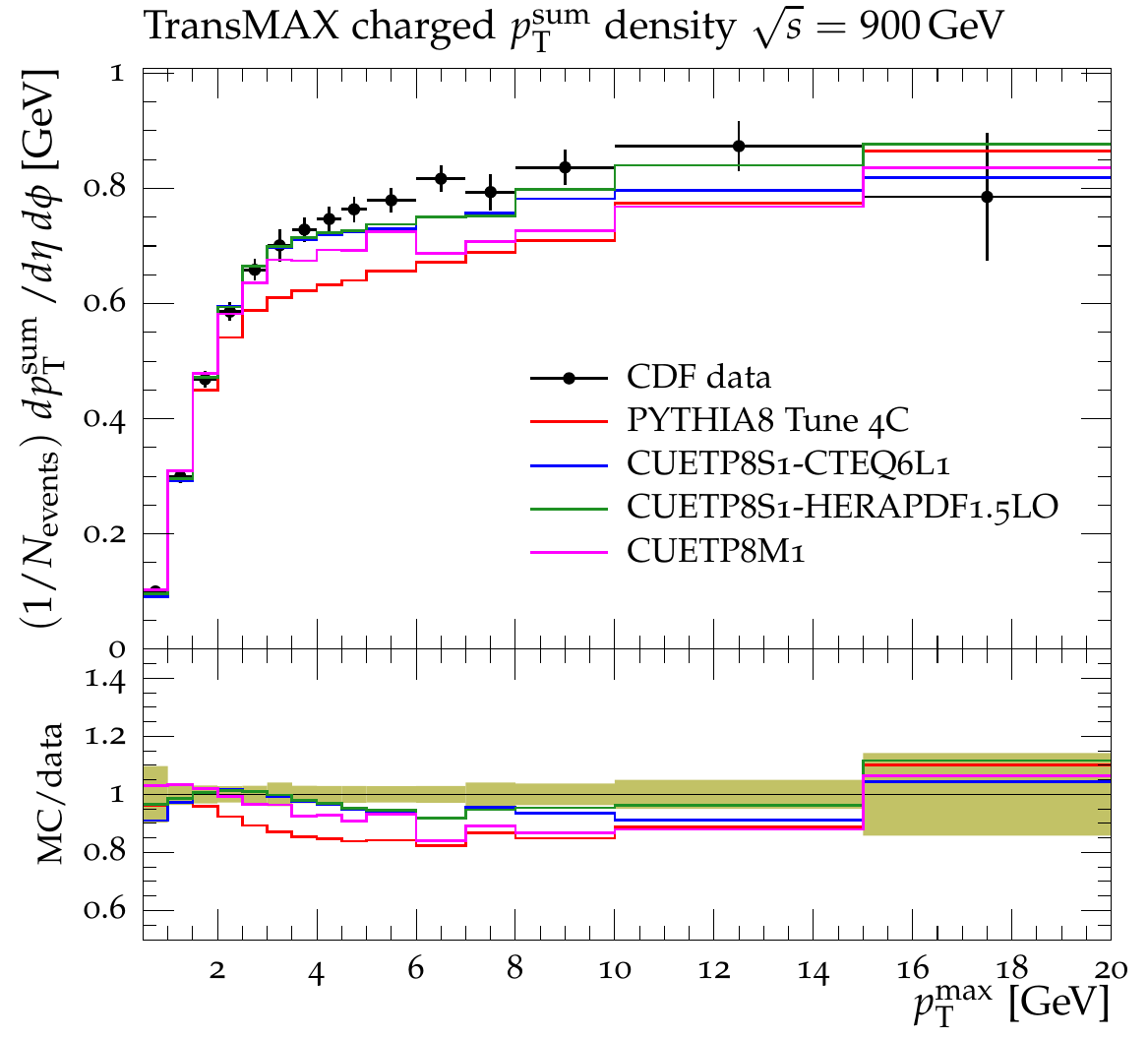}
\caption{CDF data at $\sqrt{s}=900\GeV$~\cite{Aaltonen:2015aoa} on particle (top) and \ptsum\ densities (bottom) for charged particles with \ptcut\ and \etacut\ in the \tmin\ (left) and \tmax\ (right) regions as defined by the leading charged particle, as a function of the transverse momentum of the leading charged-particle \ptmax. The data are compared to \pynewhyphen\ \tunec, \cuePB, \cuePH, and \cuePM. The ratios of MC events to data are given below each panel. The green bands in the ratios represent the total experimental uncertainties.}
\label{PUB_fig8}
\end{center}
\end{figure*}

\begin{figure*}[htbp]
\begin{center}
\includegraphics[scale=0.65]{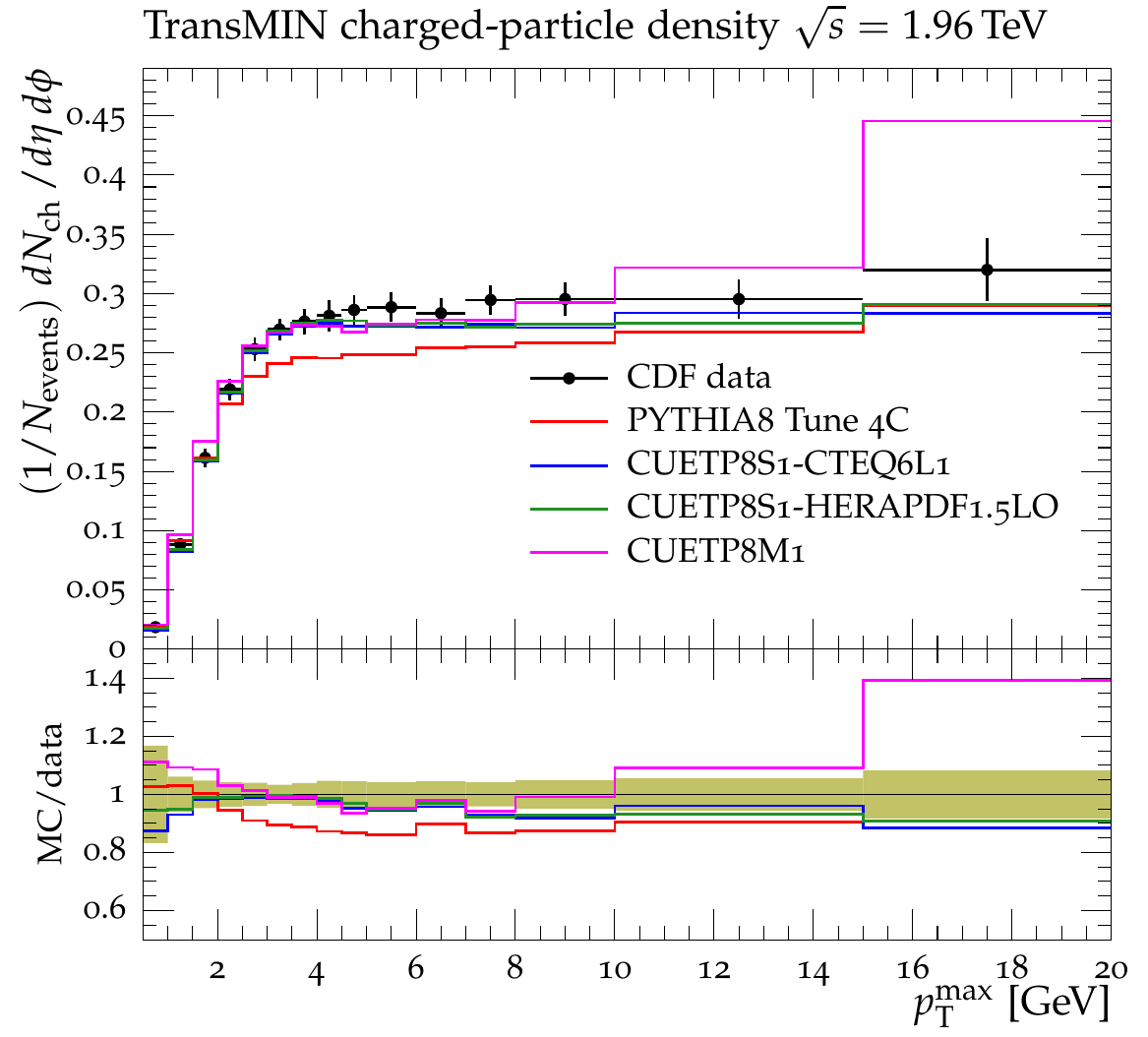}
\includegraphics[scale=0.65]{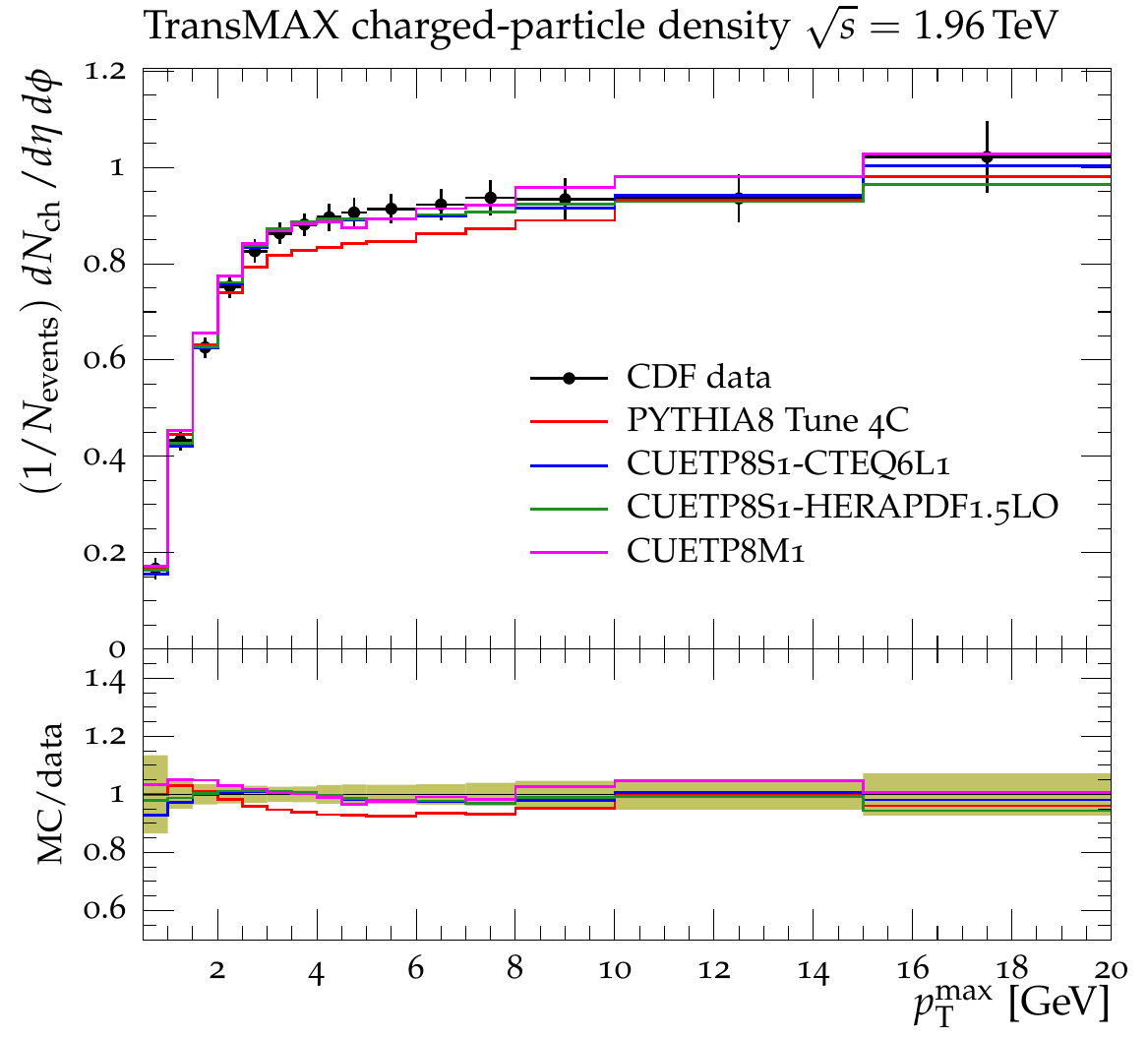}\\
\includegraphics[scale=0.65]{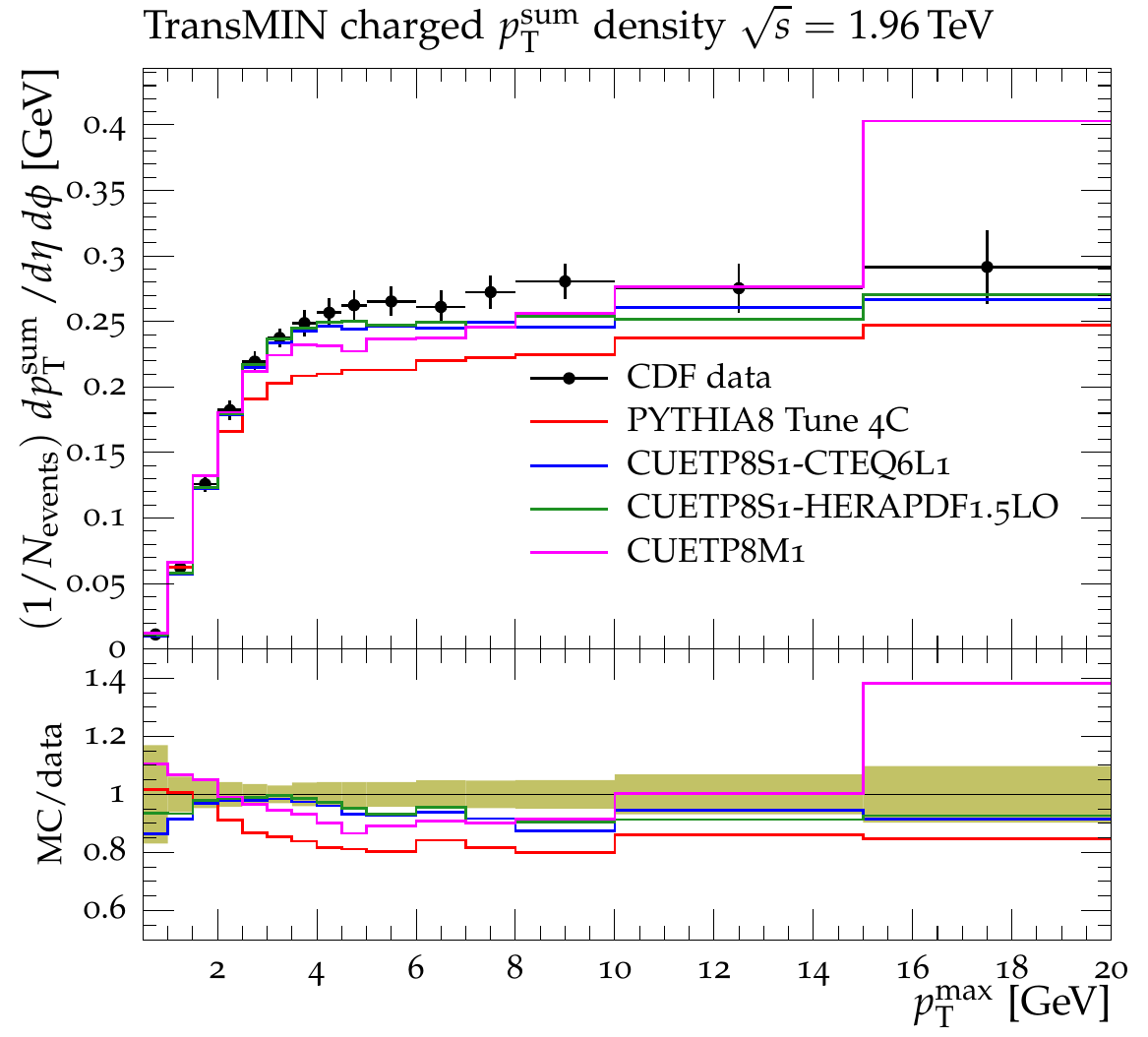}
\includegraphics[scale=0.65]{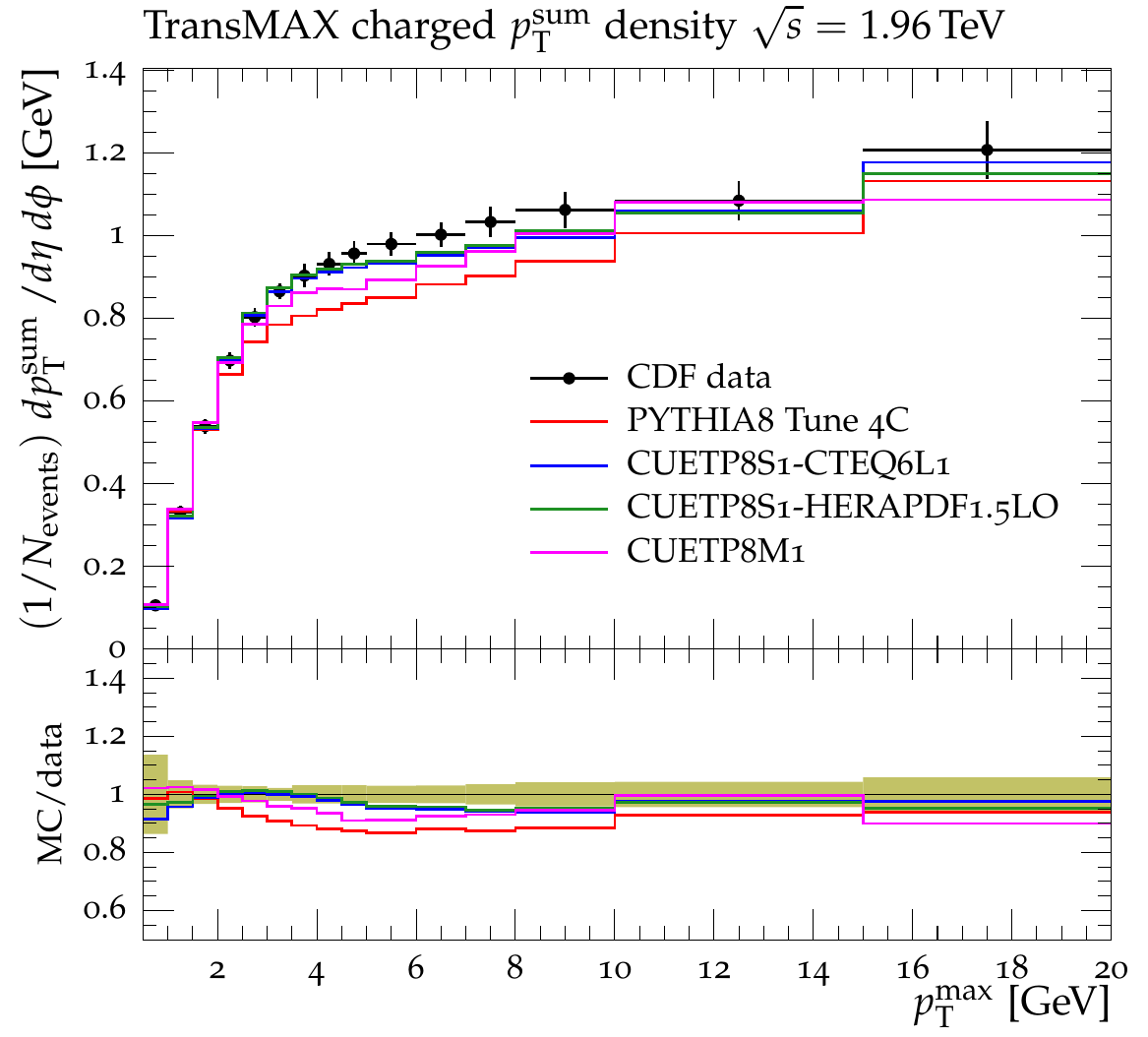}
\caption{CDF data at $\sqrt{s}=1.96\TeV$~\cite{Aaltonen:2015aoa} on particle (top) and \ptsum\ densities (bottom) for charged particles with \ptcut\ and \etacut\ in the \tmin\ (left) and \tmax\ (right) regions as defined by the leading charged particle, as a function of the transverse momentum of the leading charged-particle \ptmax. The data are compared to \pynewhyphen\ \tunec, \cuePB, \cuePH, and \cuePM. The ratios of MC events to data are given below each panel. The green bands in the ratios represent the total experimental uncertainties.}
\label{PUB_fig9}
\end{center}
\end{figure*}

\begin{figure*}[htbp]
\begin{center}
\includegraphics[scale=0.65]{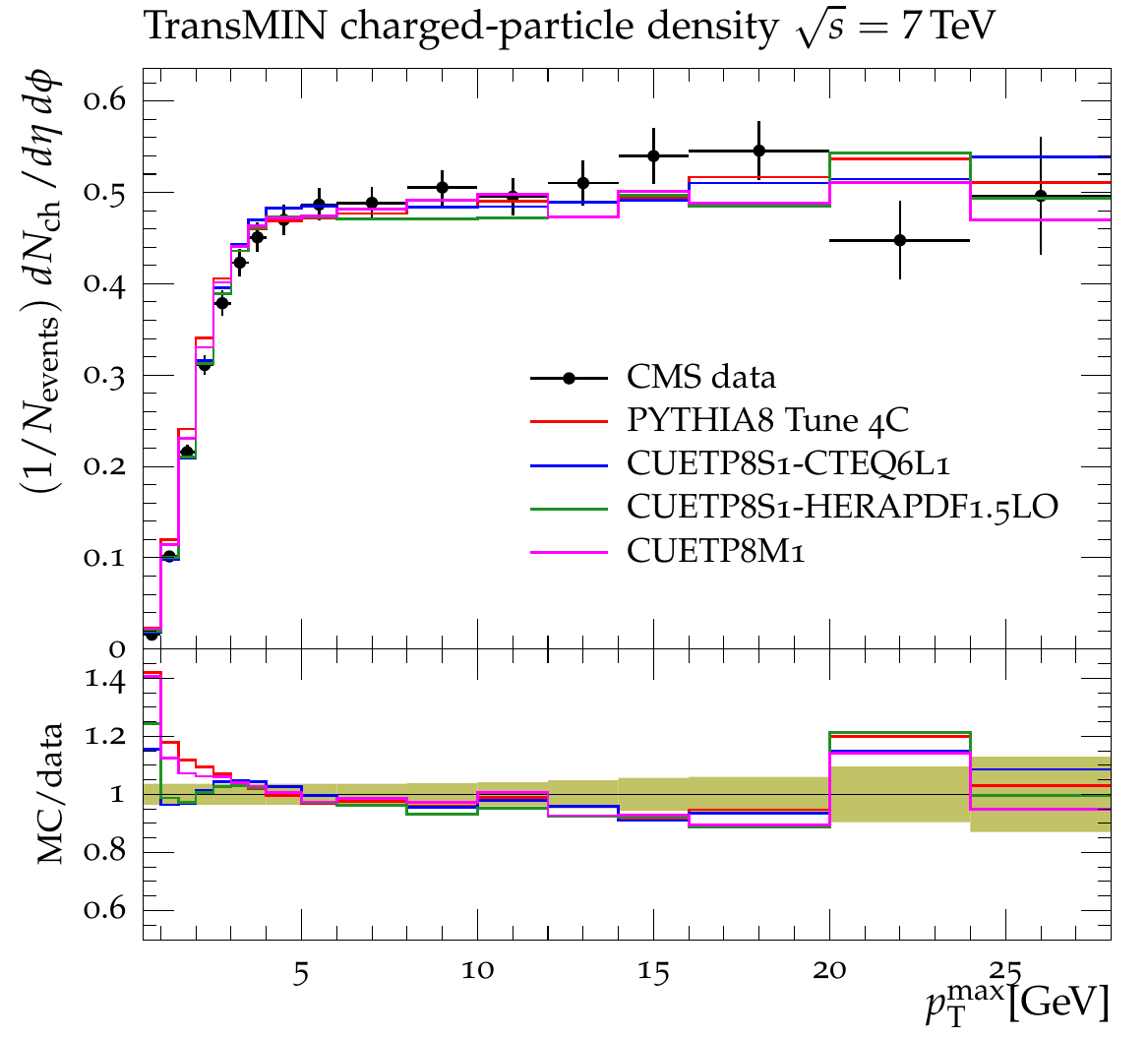}
\includegraphics[scale=0.65]{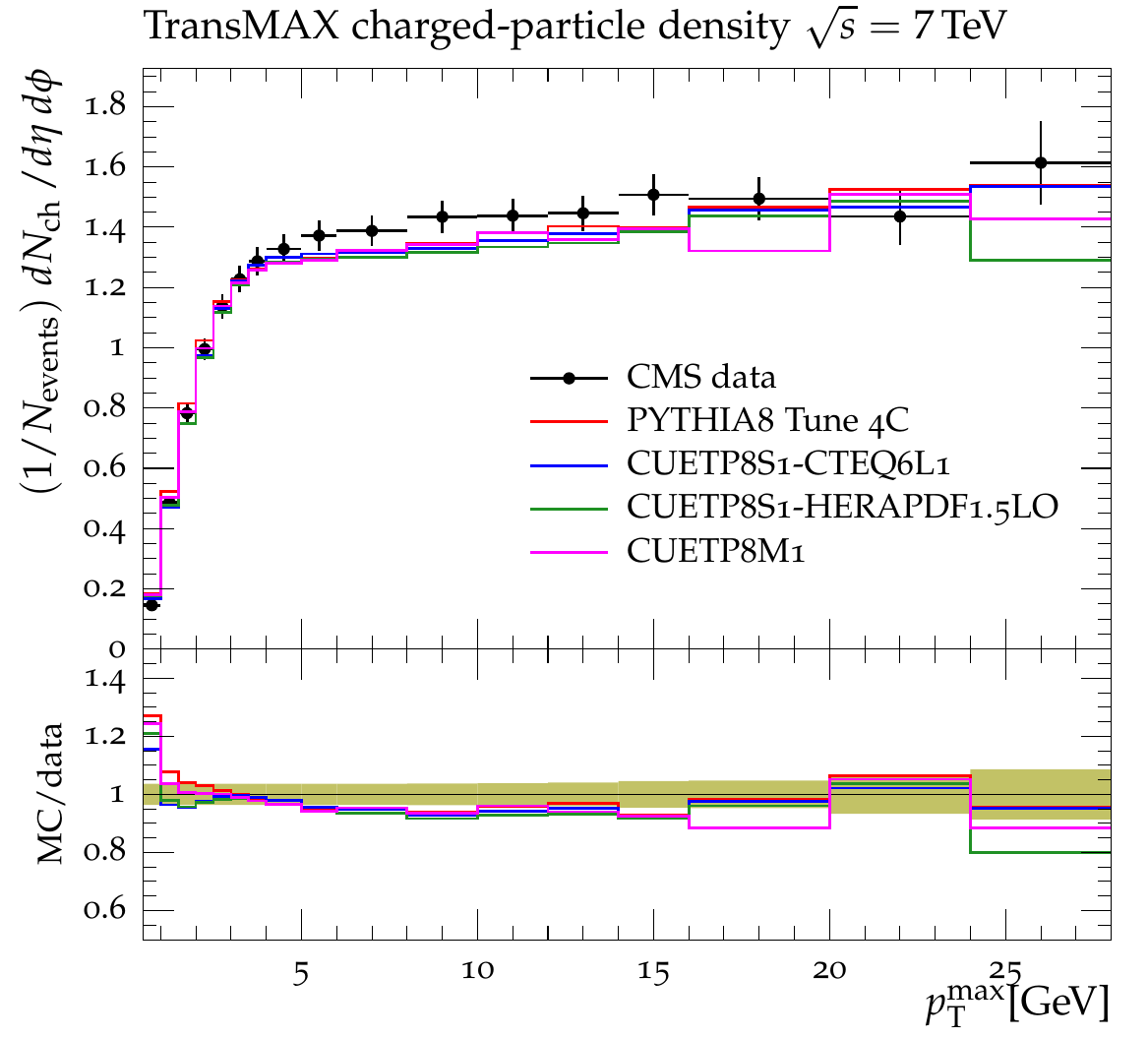}\\
\includegraphics[scale=0.65]{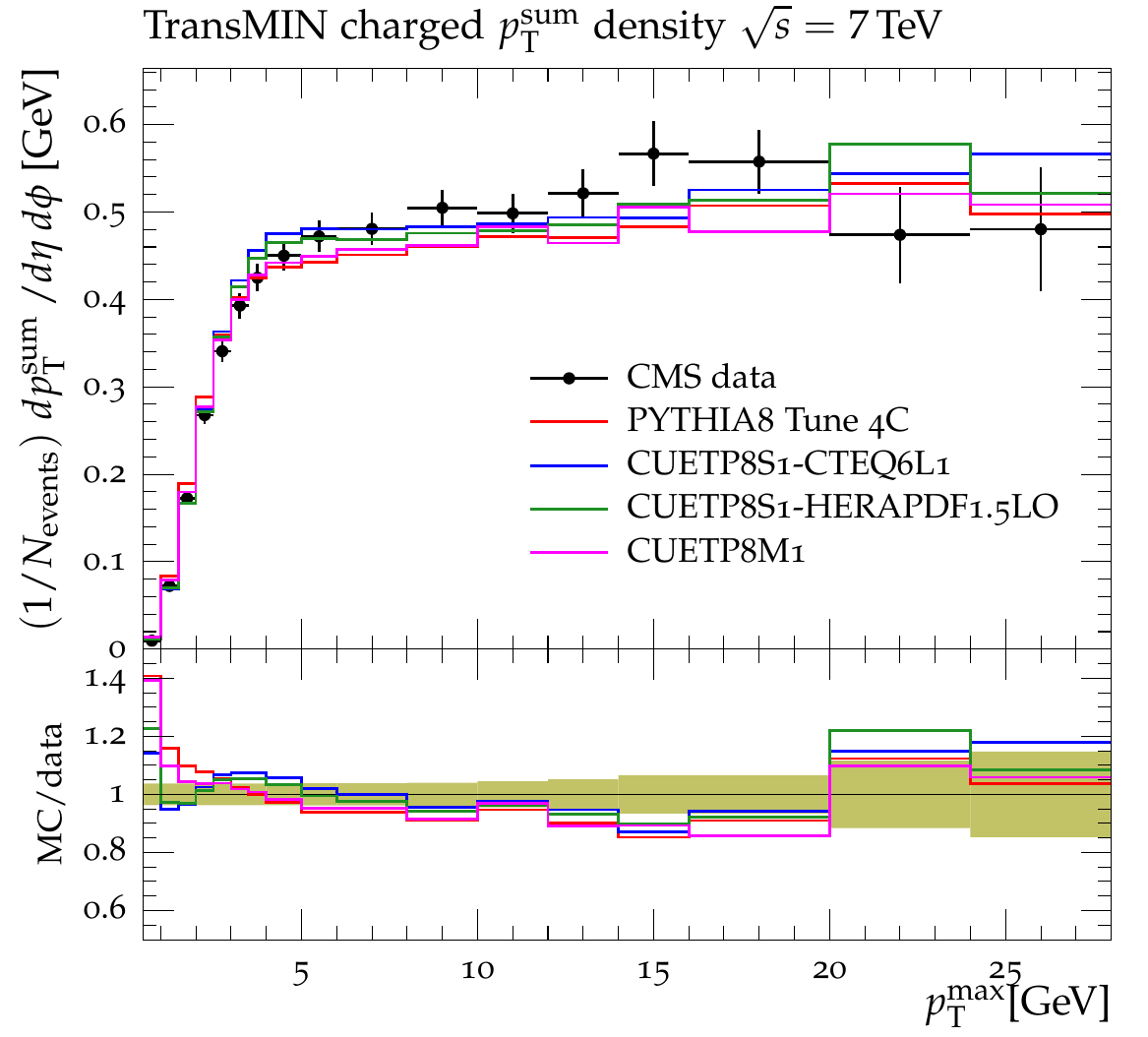}
\includegraphics[scale=0.65]{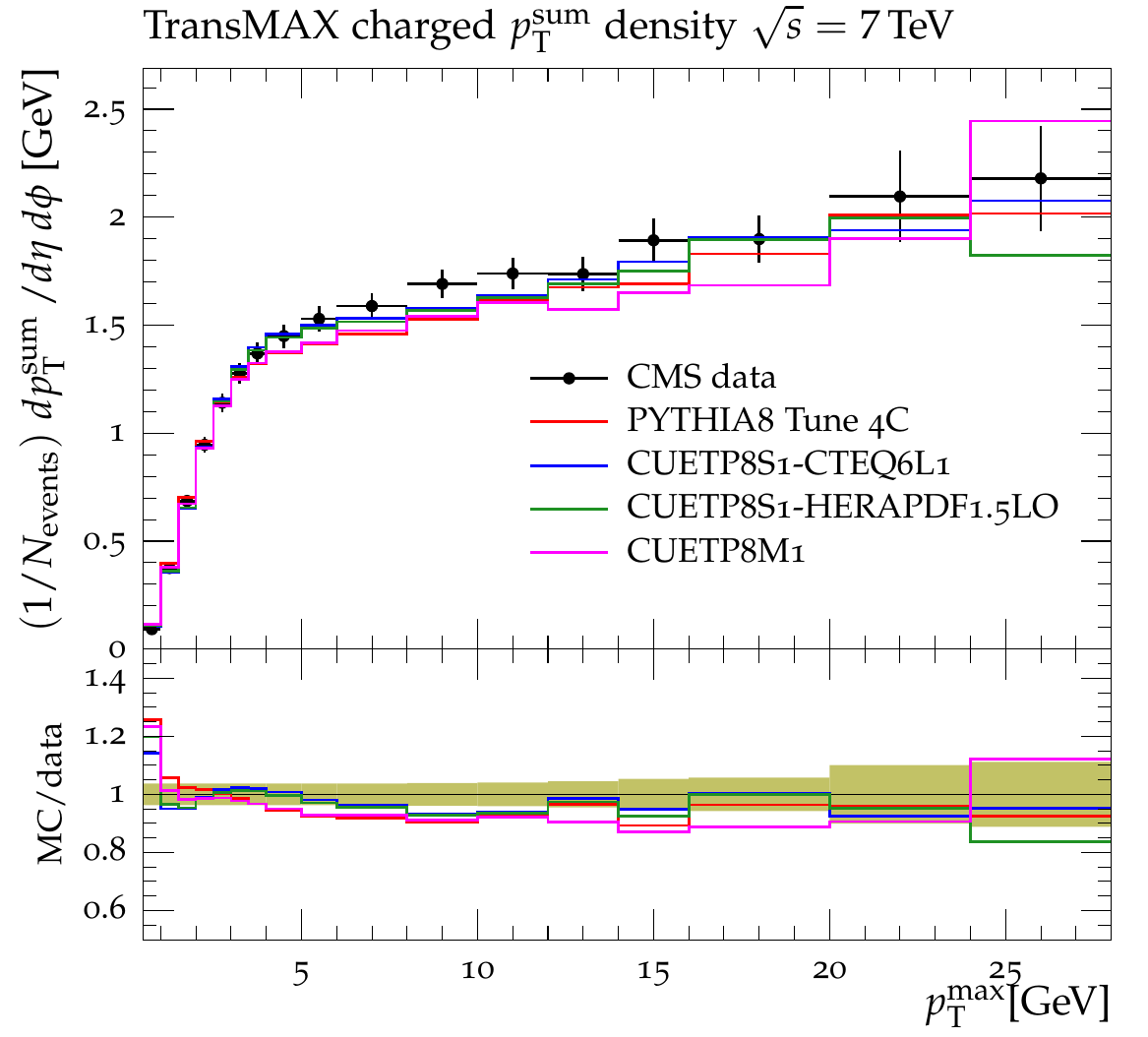}
\caption{CMS data at $\sqrt{s}=7\TeV$~\cite{CMS:2012kca} on particle (top) and \ptsum\ densities (bottom) for charged particles with \ptcut\ and \etacut\ in the \tmin\ (left) and \tmax\ (right) regions as defined by the leading charged particle, as a function of the transverse momentum of the leading charged-particle \ptmax. The data are compared to \pynewhyphen\ \tunec, and \cuePB, \cuePH, and \cuePM. The ratios of MC events to data are given below each panel. The green bands in the ratios represent the total experimental uncertainties.}
\label{PUB_fig10}
\end{center}
\end{figure*}

{\tolerance=1200
Figures~\ref{PUB_fig7}--\ref{PUB_fig10} show the CDF data at $0.3$, $0.9$, and $1.96\TeV$, and the CMS data at $7\TeV$ for charged-particle and \ptsum\ densities in the \tmin\ and \tmax\ regions as a function of \ptmax, compared to predictions obtained with the \pynewhyphen\ \tunec\ and with the new CMS tunes: \cuePB, \cuePH, and \cuePM. Predictions from the new tunes cannot reproduce the $\sqrt{s}=300\GeV$ data, but describe very well the data at the higher $\sqrt{s} = 0.9$, $1.96$, and $7\TeV$. In particular, the description provided by the new tunes significantly improves relative to the old \tunec, which is likely due to the better choice of parameters used in the MPI energy dependence and the extraction of the CR in the retuning.
\par}

\ifthenelse{\boolean{cms@external}}{\subsection{The PYTHIA6 UE tunes}}{\subsection{The \textsc{pythia6} UE tunes}}

The \pyoldhyphen\ \tunelep~\cite{Chatrchyan:2013gfi} uses the improved fragmentation parameters from fits to the LEP e$^+$e$^-$ data~\cite{Skands:2010ak}, and a double-Gaussian matter profile for the colliding protons but corresponds to an outdated CMS UE tune.  It was constructed by fitting the CMS charged-particle jet UE data at $0.9$ and $7\TeV$~\cite{Chatrchyan:2011id} using data on the \tave\ charged-particle and \ptsum\ densities, since data on \tmax, \tmin, and \tdif\ were not available at that time.  

{\tolerance=5000
Starting with \tunelep\ parameters, two new \pyoldhyphen\ UE tunes are constructed, one using CTEQ$6$L1 (\cuePA) and one using \hera\ (\cuePAH),  with  \verb+P6S1+ standing for \pyold\ ``Set $1$''. The tunes are constructed by fitting the five parameters shown in Table~\ref{table3} to the \tmax\ and \tmin\ charged-particle and \ptsum\ densities at $\sqrt{s} = 0.3$, $0.9$, $1.96$, and $7\TeV$. In addition to varying the MPI energy-dependence parameters (Table~\ref{table1}), we also vary the core-matter fraction PARP($83$), which parametrizes the amount of matter contained within the radius of the proton core, the CR strength PARP($78$), and the CR suppression PARP($77$). The PARP($78$) parameter reflects the probability for a given string to retain its color history, and therefore does not change the color and other string pieces, while the PARP($77$) parameter introduces a \pt\ dependence on the CR probability~\cite{Sjostrand:2006za}. 
\par}

Inelastic events (ND$+$DD$+$SD$+$CD) are generated with \pyold . The best-fit values of the five parameters are  shown in Table~\ref{table3}.  The matter-core fraction is quite different in the two new \pyold\ tunes. This is due to the fact that this parameter is very sensitive to the behaviour of the PDF at small $x$. Predictions obtained with \pyoldhyphen\ \tunelep\ , CUETP6S1-CTEQ6L1 and CUETP6S1-HERAPDF$1.5$LO  are compared in Appendix B to the UE data. The new \pyoldhyphen\ tunes significantly improve the description of the UE data relative to \pyoldhyphen\ \tunelep\ at all considered energies, due to the better choice of parameters governing the MPI energy dependence.

\begin{table*}[htbp]
\renewcommand{\arraystretch}{1.2}
\begin{center}
\topcaption{The \pyoldhyphen\ parameters, tuning range, \tunelep\ values~\cite{Skands:2010ak}, and best-fit values for \cuePA\ and \cuePAH  , obtained from fits to the \tmax\ and \tmin\ charged-particle and \ptsum\ densities as defined by the \ptmax\ of the leading charged particle at $\sqrt{s} = 0.3$ , $0.9$, $1.96$, and $7\TeV$.}
\label{table3}
\cmsTable{
\begin{tabular}{l c c c c} \hline 
  \pyold\ Parameter                &   Tuning Range &   \tunelep &   CUETP6S1 &   CUETP6S1 \\ \hline
   PDF	                             & \NA                  &  CTEQ6L1         &  CTEQ6L1      &  HERAPDF$1.5$LO \\ \hline
   PARP(82) - MPI cutoff [GeV]	     & $1.6$--$2.2$      & $1.921$        & $1.910$    & $1.946$       \\ 
   PARP(90) - Exponent of $\sqrt{s}$ dependence& $0.18$--$0.28$    & $0.227$        & $0.248$    & $0.250$       \\ 
   PARP(77) - CR suppression          & $0.25$--$1.15$    & $1.016$        & $0.665$    & $0.667$       \\ 
   PARP(78) - CR strength	     & $0.2$--$0.8$      & $0.538$        & $0.545$    & $0.537$       \\ 
   PARP(83) - Matter fraction in core & $0.1$--$1.0$      & $0.356$        & $0.822$    & $0.490$       \\ \hline
   PARP(89) - Reference energy [GeV]  & \NA                  & $1800$         & $1800^*$    & $1800^*$       \\ \hline
   $\chi^2$/dof                   & \NA                  & \NA              & $0.915$     & $1.004$        \\ \hline
   \multicolumn{5}{r}{\footnotesize $^*$ Fixed at \tunelep\ value.}
\end{tabular}
}
\end{center}
\end{table*}

\ifthenelse{\boolean{cms@external}}{\subsection{The HERWIG++ UE tunes}} {\subsection{The \textsc{herwig++} UE tunes}}

Starting with the parameters of \hwpp\ Tune UE-EE-5C~\cite{Seymour:2013qka}, we construct a new \hwpp\ UE tune, \cueHW, where \verb+Hpp+ stands for \hwpp. This tune is obtained by varying the four parameters shown in Table~\ref{table6} in the fit to \tmax\ and \tmin\ charged-particle and \ptsum\ densities at the four $\sqrt{s} = 0.3$, $0.9$, $1.96$, and $7\TeV$.  We set the MPI cutoff \ptzero\ and the reference energy $\sqrt{s_0}$ to the Tune UE-EE-5C values, and vary the MPI c.m. energy extrapolation parameter in Table~\ref{table1}. We also vary the inverse radius that determines the matter overlap and the range of CR. The CR model in \hwpp\ is defined by two parameters, one (\verb+colourDisrupt+) ruling the color structure of soft interactions (\pt\ $<$ \ptzero), and one (\verb+ReconnectionProbability+) giving the probability of CR without a \pt\ dependence for color strings. We include all four center-of-mass energies, although at each energy we exclude the first two \ptmax\ bins. These first bins, \eg for $p_{\rm T}^{\rm max}<1.5\GeV$, are sensitive to single-diffraction dissociation, central-diffraction, and double-diffraction dissociation, but \hwpp\ contains only the ND component. 

In Table~\ref{table6}, the parameters of the new CUETHppS1 are listed and compared to those from Tune UE-EE-5C. The parameters of the two tunes are very similar. The $\chi^2$/dof, also indicated in Table~\ref{table6}, is found to be ${\approx} 0.46$, which is smaller than the value obtained for other CMS UE tunes. This is due to the fact that the first two bins as a function of \ptmax, which have much smaller statistical uncertainties than the higher-\ptmax\ bins, are excluded from the fit because they cannot be described by any reasonable fit-values. In Appendix~C, predictions obtained with \hwpp\ Tune UE-EE-5C and CUETHppS1 are compared to the UE data. The two tunes are both able to reproduce the UE data at all energies. With  the new CUETHppS1 tune,  uncertainties can be estimated using the eigentunes (Appendix~A).

\begin{table*}[htbp]
\renewcommand{\arraystretch}{1.2}
\begin{center}
\topcaption{The \hwpp\ parameters, tuning range, Tune UE-EE-5C values~\cite{Seymour:2013qka}, and best-fit values for \cueHW, obtained from a fit to the \tmax\ and \tmin\ charged-particle and \ptsum\ densities as a function of the leading charged-particle \ptmax at $\sqrt{s} = 0.3$ , $0.9$, $1.96$, and $7\TeV$.}
\label{table6}
\cmsTable{
\begin{tabular}{l c c c}  \hline
  \hwpp\ Parameter  &   Tuning Range  &   UE-EE-5C    &   CUETHppS1 \\ \hline
   PDF                  &    \NA             &  CTEQ6L1    &  CTEQ6L1 \\ \hline
   MPIHandler:Power & $0.1$--$0.5$           & $0.33$    & $0.371$ \\ 
   RemnantDecayer:colourDisrupt & $0.1$--$0.9$ & $0.8\phantom{0}$   & $0.628$       \\ 
   MPIHandler:InvRadius [GeV$^2$] & $0.5$--$2.7$ & $2.30$ & $2.255$ \\ 
   ColourReconnector:ReconnectionProbability & $0.1$--$0.9$ & $0.49$ & $0.528$ \\ \hline
   MPIHandler:pTmin0 [GeV]  &\NA             & $3.91$    & $3.91^*$ \\ 
   MPIHandler:ReferenceScale [GeV]  &\NA     & $7000$    & $7000^*$      \\ \hline
   $\chi^2$/dof            &\NA          &\NA             & $0.463$  \\ \hline
   \multicolumn{4}{r}{\footnotesize $^*$ Fixed at Tune UE-EE-5C value.}
\end{tabular}
}
\end{center}
\end{table*}

In conclusion, both \hwpp\ tunes, as well as the new CMS \pyoldhyphen\ UE tunes reproduce the UE data at all four $\sqrt{s}$. The \pynewhyphen\ UE tunes, however, do not describe well the data at $\sqrt{s}=300\GeV$,
which may be related to the modelling of the proton-proton overlap function. The \pyoldhyphen\ \tunelep, and the new CMS UE tunes use a double-Gaussian matter distribution, while all the \pynewhyphen\ UE tunes use a single exponential matter overlap. The \hwpp\ tune, on the other hand, uses a matter-overlap function that is related to the Fourier transform of the electromagnetic form factor with $\mu^2$~\cite{Bellm:2013lba} playing the role of an effective inverse proton radius (\ie the \verb+InvRadius+ parameter in Table~\ref{table6}). 
However, predictions from a tune performed with \pynew\ using a double-Gaussian matter distribution were not able to improve the quality of the fit
as a fit obtained without interleaved FSR in the simulation of the UE (as it is implemented in \pyold) did not show any improvement. Further investigations are needed to resolve this issue.

\section{The CMS DPS tunes}
\label{section-3}

{\tolerance=1200
Traditionally, \eff\ is determined by fitting the DPS-sensitive observables with two templates~\cite{afsdps,cdfdps,uadps,Aad:2013bjm,Chatrchyan:2013xxa} that are often based on distributions obtained from QCD MC models.  One template is constructed with no DPS, \ie just single parton scattering (SPS), while the other represents DPS production.  This determines \eff\ from the relative amounts of SPS and DPS contributions needed to fit the data.  
Here we use an alternative method that does not require construction of templates from MC samples. Instead, we fit the DPS-sensitive observables directly and then calculate the resulting \eff\ from the model. For example, in \pynew, the value of \eff\ is calculated by multiplying the ND cross section by an enhancement or a depletion factor, which expresses the dependence of DPS events on the collision impact parameter. As expected, more central collisions have a higher probability of a second hard scattering than peripheral collisions. The enhancement/depletion factors depend on the UE parameters, namely, on the parameters that characterize the matter-overlap function of the two protons,  which for \verb+bProfile+ $=3$ is determined by the exponential parameter  \verb+expPow+, on the MPI regulator $p_{\rm T0}$ in Eq.~(\ref{Ezero}), and the range of the CR. 
\pynewhyphen\ \tunec\ gives \eff\ $\approx$ $30.3$ mb at $\sqrt{s}=7\TeV$.  
\par}

In Section~\ref{section-2}, we determined the MPI parameters by fitting UE data.  Here we determine the MPI parameters by fitting to observables which involve correlations among produced objects in hadron-hadron collisions that are sensitive to DPS. Two such observables used in the fit, $\Delta S$ and \relpt, are defined as follows:

\begin{equation}\label{dels}
\Delta {\rm S}=\arccos\left(\frac{\vec{p}_{\textrm T}(\textrm{object}_1)\cdot \vec{p}_{\textrm T} (\textrm{object}_2)}{|\vec{p}_T(\textrm{object}_1)| \times|\vec{p}_{\textrm T}(\textrm{object}_2)|}\right),\\
\end{equation}
\begin{equation}\label{deltarel}
\Delta^{\rm rel}p_{\rm T} = \frac{|\vec{p}_{\rm T}^{\ \rm jet_1}+\vec{p}_{\rm T}^{\ \rm jet_2}|}{|\vec{p}_{\rm T}^{\ \rm jet_1}|+|\vec{p}_{\rm T}^{\ \rm jet_2}|},\\
\end{equation}
where, for $\PW$+dijet production, object$_1$ is the $\PW$ boson and object$_2$ is the dijet system.  For four-jet production, object$_1$ is the hard-jet pair and object$_2$ is the soft-jet pair. For \relpt\ in $\PW$+dijet production, jet$_1$ and jet$_2$ are the two jets of the dijet system, while in four-jet production, jet$_1$ and jet$_2$ refer to the two softer jets.

The \pynewhyphen\ UE parameters are fitted to the DPS-sensitive observables measured by CMS in $\PW$+dijet~\cite{Chatrchyan:2013xxa} and in four-jet production~\cite{Chatrchyan:2013qza}. After extracting the MPI parameters, the value of \eff\ in Eq.~(\ref{sigeff}) can be calculated from the underlying MPI model. In \pynew, \eff\ depends primarily on the matter-overlap function and, to a lesser extent, on the value of \ptzero\ in Eq.~(\ref{Ezero}), and the range of the CR. We obtain two separate tunes for each channel: in the first one, we vary just the matter-overlap parameter \verb+expPow+, to which the \eff\ value is most sensitive, and in the second one, the whole set of parameters is varied. These two tunes allow to check whether the value of \eff\ is stable relative to the choice of parameters.

The $\PW$+dijet and the four-jet channels are fitted separately. The fit to DPS-sensitive observables in the $\PW$+dijet channel gives a new determination of \eff\ which can be compared to the value measured through the template method in the same final state~\cite{Chatrchyan:2013xxa}. Fitting the same way to the observables in the four-jet final state provides an estimate of \eff\ for this channel. 

\subsection{Double-parton scattering in W+dijet production}

\begin{figure*}[htbp]
\begin{center}
\includegraphics[scale=0.65]{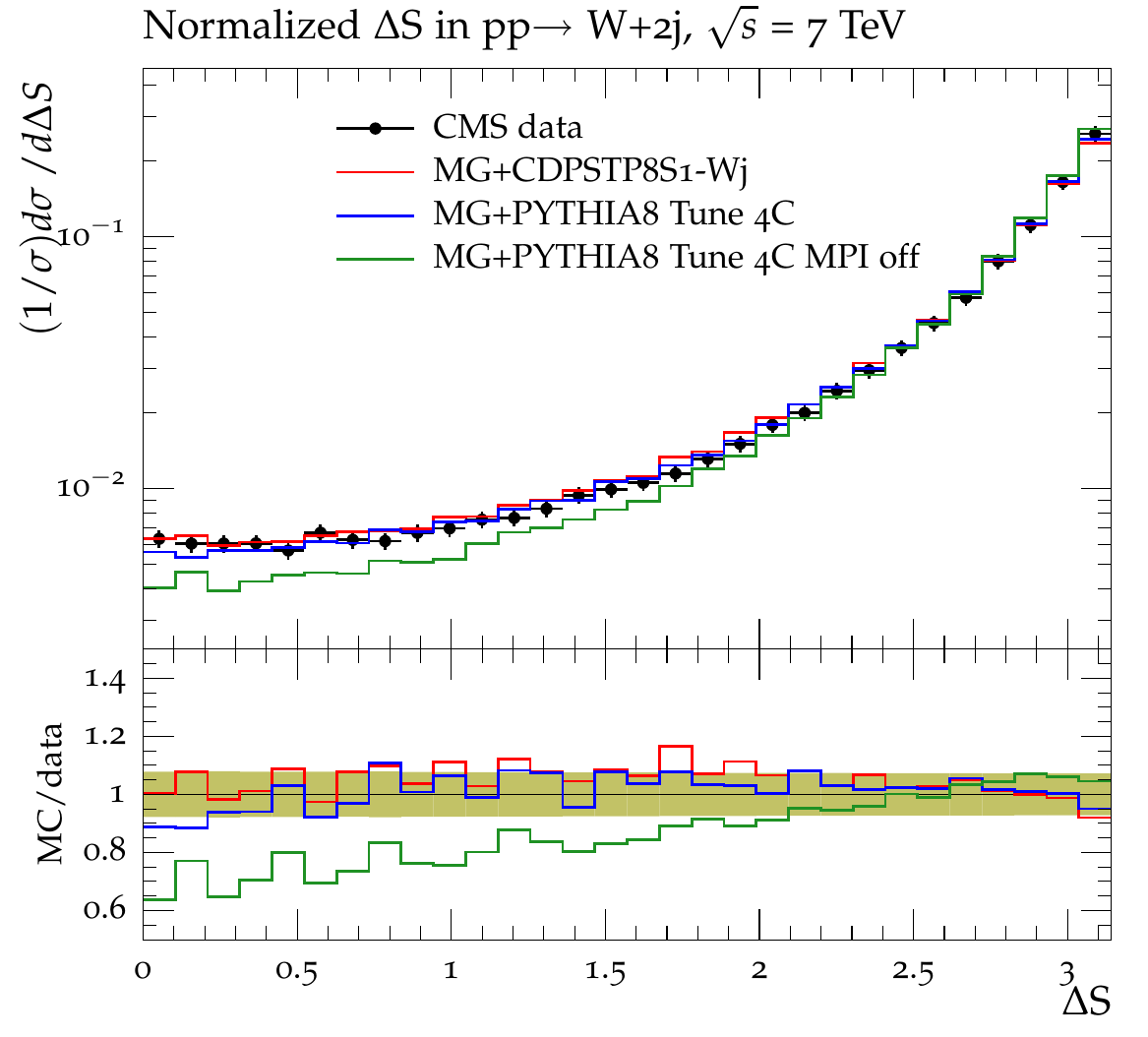}
\includegraphics[scale=0.65]{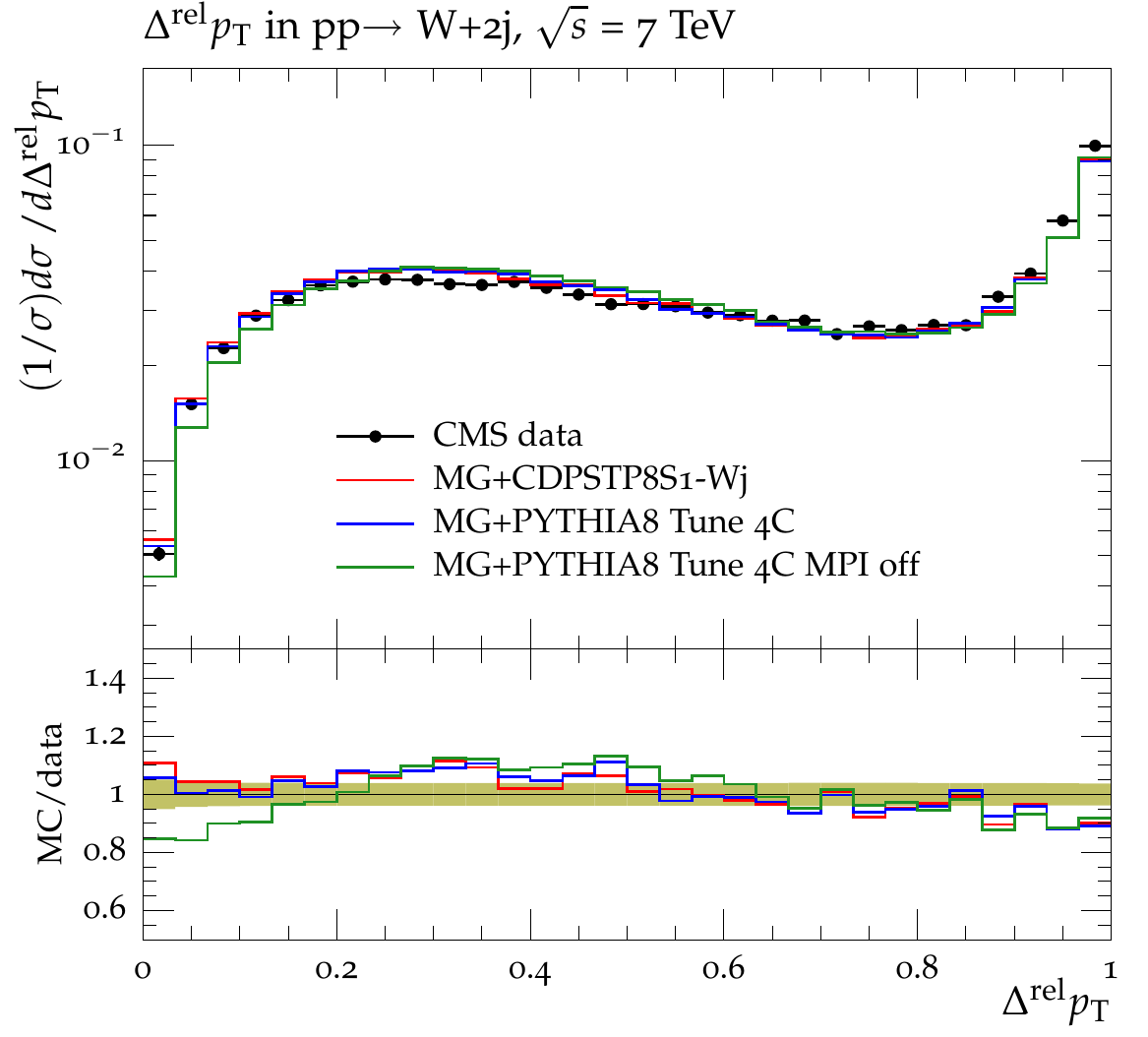}\\
\includegraphics[scale=0.65]{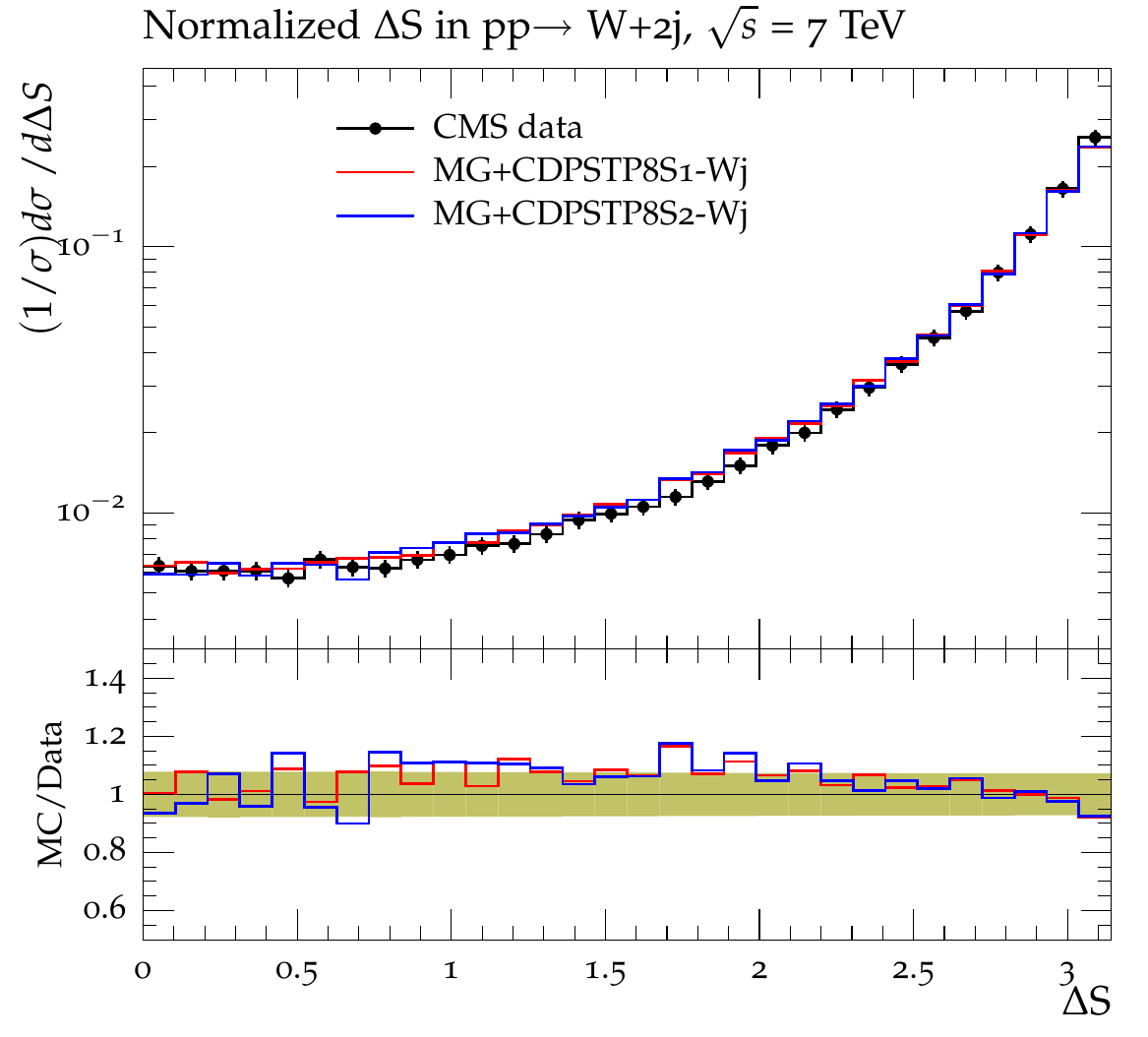}
\includegraphics[scale=0.65]{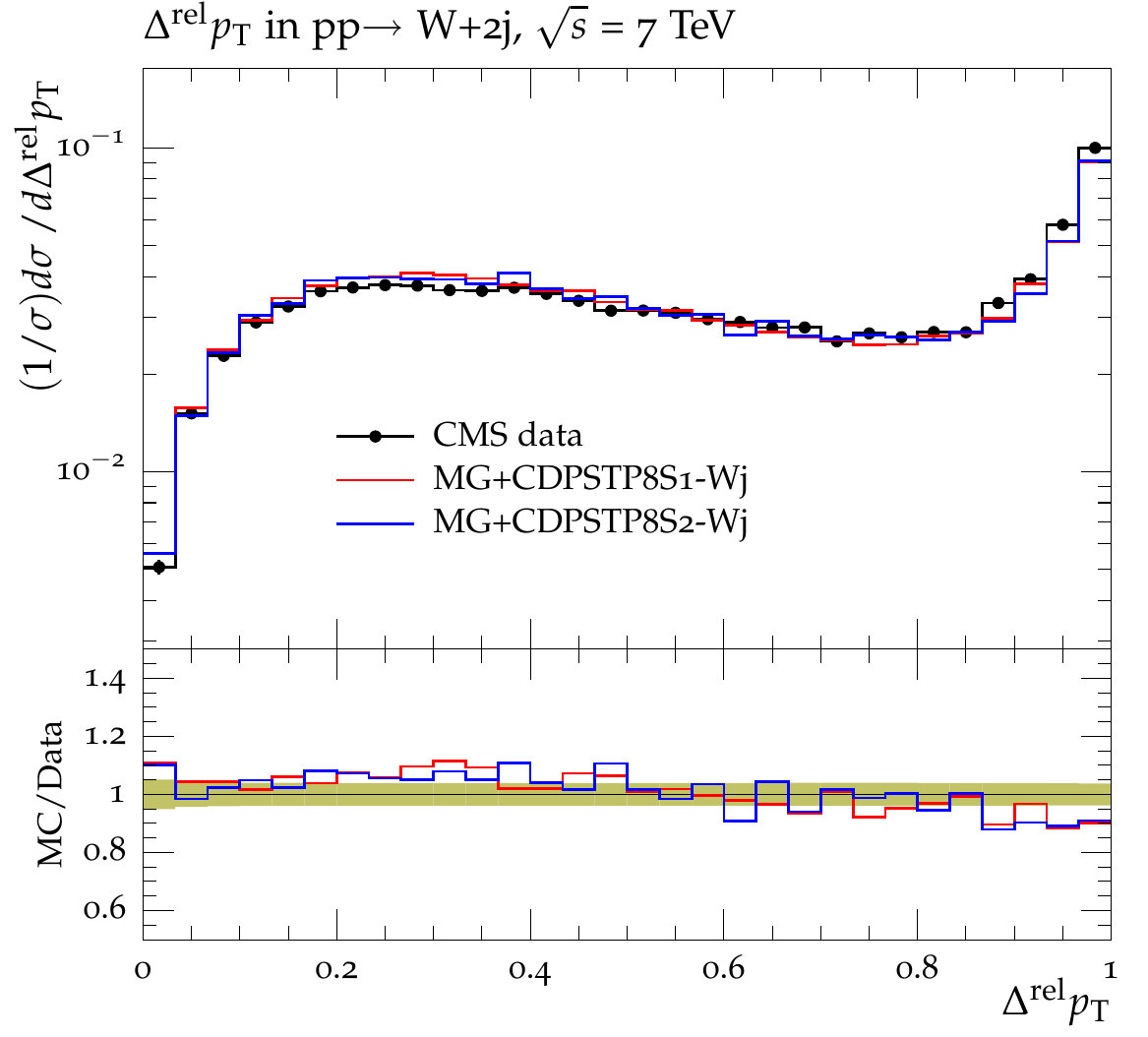}
\caption{CMS data at $\sqrt{s}=7\TeV$~\cite{Chatrchyan:2013xxa} for the normalized distributions of the correlation observables $\Delta$S (left), and \relpt\ (right) in the $\PW$+dijet channel, compared to \MADGRAPH (MG) interfaced to: \pynewhyphen\ \tunec, \tunec\ with no MPI, and the CMS \pynewhyphen\ DPS partial \cdpWA\ (top); and \cdpWA, and \cdpWB\ (bottom). The bottom panels of each plot show the ratios of these tunes to the data, and the green bands around unity represent the total experimental uncertainty.}
\label{PUB_fig18}
\end{center}
\end{figure*}

To study the dependence of the DPS-sensitive observables on MPI parameters, we construct two $\PW$+dijet DPS tunes, starting from the parameters of \pynewhyphen\ \tunec.  In a partial tune only the parameter of the exponential distribution \verb+expPow+ is varied, and in a full tune all four parameters in Table~\ref{table7} are varied. In a comparison of models with $\PW$+dijet events~\cite{Chatrchyan:2013xxa}, it was shown that higher-order SPS contributions (not present in \PYTHIA) fill a similar region of phase-space as the DPS signal. When such higher-order SPS diagrams are neglected, the measured DPS contribution to the $\PW$+dijet channel can be overestimated (\ie \eff\ underestimated). We therefore interface the LO matrix elements (ME) generated by \MADGRAPH5~(version 1.5.14)~\cite{Alwall:2011uj} with \pynew, and tune to the normalized distributions of the correlation observables in Eqs.~(\ref{dels}) and (\ref{deltarel}). For this study, we produce \MADGRAPH parton-level events with a $\PW$ boson and up to four partons in the final state. The cross section is calculated using the CTEQ6L1 PDF with a matching scale for ME and parton shower (PS) jets set to 20 GeV. (In Section~\ref{section-4}, we show that the CMS UE tunes can be interfaced to higher-order ME generators without additional tuning of MPI parameters). Figure~\ref{PUB_fig18} shows the CMS data~\cite{Chatrchyan:2013xxa} for the observables $\Delta$S and \relpt\ measured in $\PW$+dijet production, compared to predictions from \MADGRAPH interfaced to \pynewhyphen\ \tunec, to \tunec\ with no MPI, to the  partial \cdpWA, as well as to the full \cdpWB\ (\verb+CDPST+ stands for ``CMS DPS tune'').  Table~\ref{table7} gives the best-fit parameters and the resulting \eff\ values at $\sqrt{s}=7\TeV$. The uncertainties quoted for \eff\ are computed from the uncertainties of the fitted parameters given by the eigentunes. For \tunec, the uncertainty in \eff\ is not provided since no eigentunes are available for that tune.
The resulting values of \eff\ are compatible with the value measured by CMS using the template method of $\sigma_{\rm eff} = 20.6\pm  0.8 \stat \pm  6.6 \syst$\unit{mb}~\cite{Chatrchyan:2013xxa}.

\begin{table*}[htbp]
\renewcommand{\arraystretch}{1.2}
\begin{center}
\topcaption{The \pynewhyphen\ parameters, tuning ranges, \tunec\  values~\cite{Corke:2010yf} and best-fit values of \cdpWA\ and \cdpWB, obtained from fits to DPS observables in $\PW$+dijet production with the \MADGRAPH event generator interfaced to \pynew.  Also shown are the predicted values of \eff\ at $\sqrt{s}=7\TeV$, and the uncertainties obtained from the eigentunes.}
\label{table7}
\cmsTable{
\begin{tabular}{l c c c c}  \hline
  \pynew\ Parameter & Tuning Range  & \tunec & \cdpWA & \cdpWB \\ \hline
   PDF                  &                     &  CTEQ6L1       &  CTEQ6L1      &  CTEQ6L1\\ \hline
   MultipartonInteractions:pT0Ref [GeV] & $1.0$--$3.0$  & $2.085$    & $2.085^*$   & $2.501$ \\ 
   MultipartonInteractions:ecmPow & $0.0$--$0.4$  & $0.19\phantom{0}$     & $0.19^*\phantom{0}$    & $0.179$ \\ 
   MultipartonInteractions:expPow & $0.4$--$10.0$ & $2.0\phantom{00}$      & $1.523$  & $1.120$ \\ 
   ColourReconnection:range & $0.0$--$9.0$  & $1.5\phantom{00}$      & $1.5^*\phantom{0}$     & $2.586$ \\ \hline
   MultipartonInteractions:ecmRef [GeV]  & \NA        & $1800$     & $1800^*$    & $1800^*$  \\ 
     $\chi^2$/dof                   & \NA  &  \NA & 0.118 & 0.09 \\
   Predicted \eff\ (in mb) &  \NA & $30.3$ & $25.9^{+2.4}_{-2.9}$ & $25.8^{+8.2}_{-4.2}$ \\ \hline
   \multicolumn{5}{r}{\footnotesize $^*$ Fixed at \tunec\ value.}
\end{tabular}
}
\end{center}
\end{table*}

\subsection{Double-parton scattering in four-jet production}

\begin{figure*}[htbp]
\begin{center}
\includegraphics[scale=0.65]{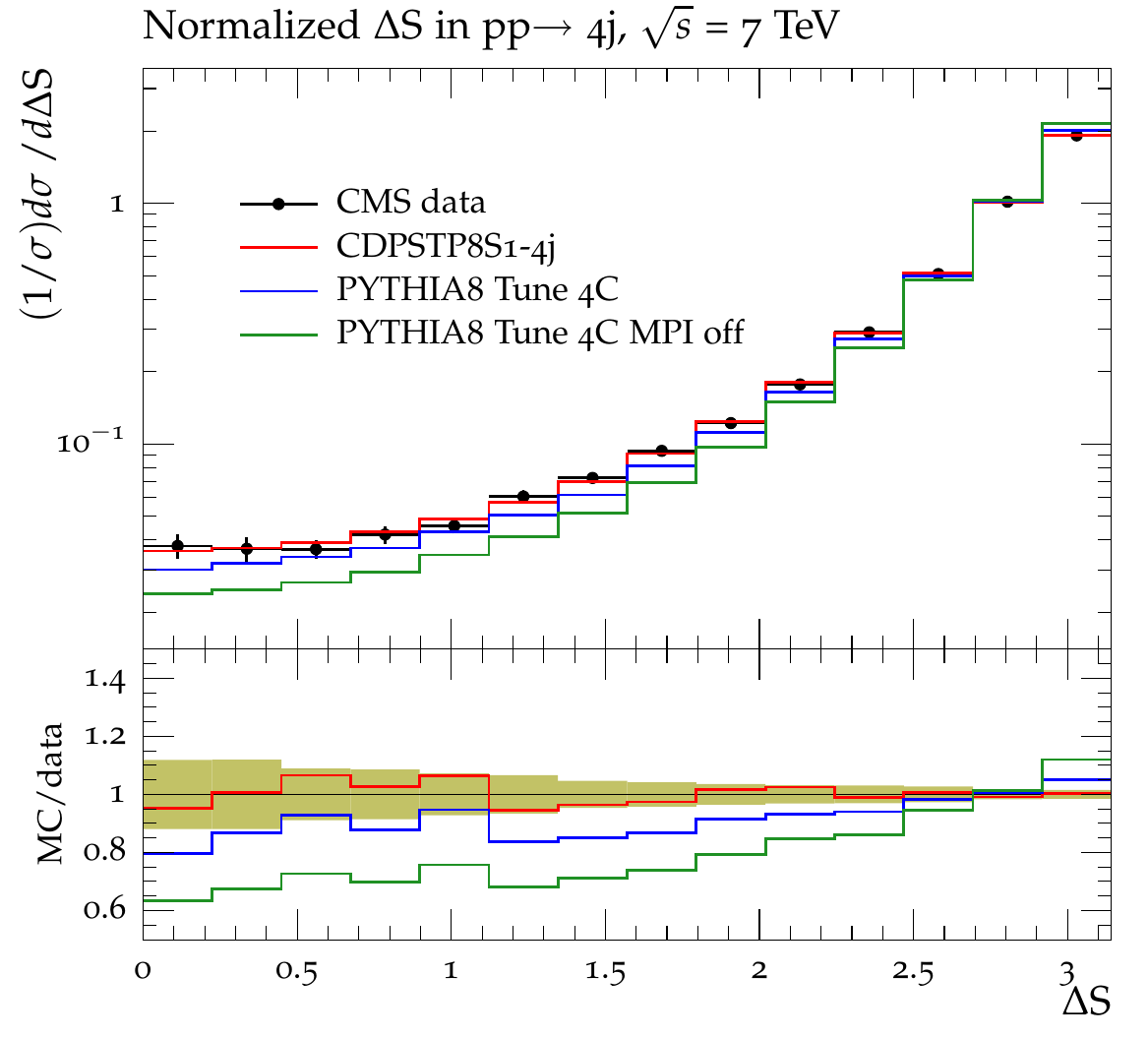}
\includegraphics[scale=0.65]{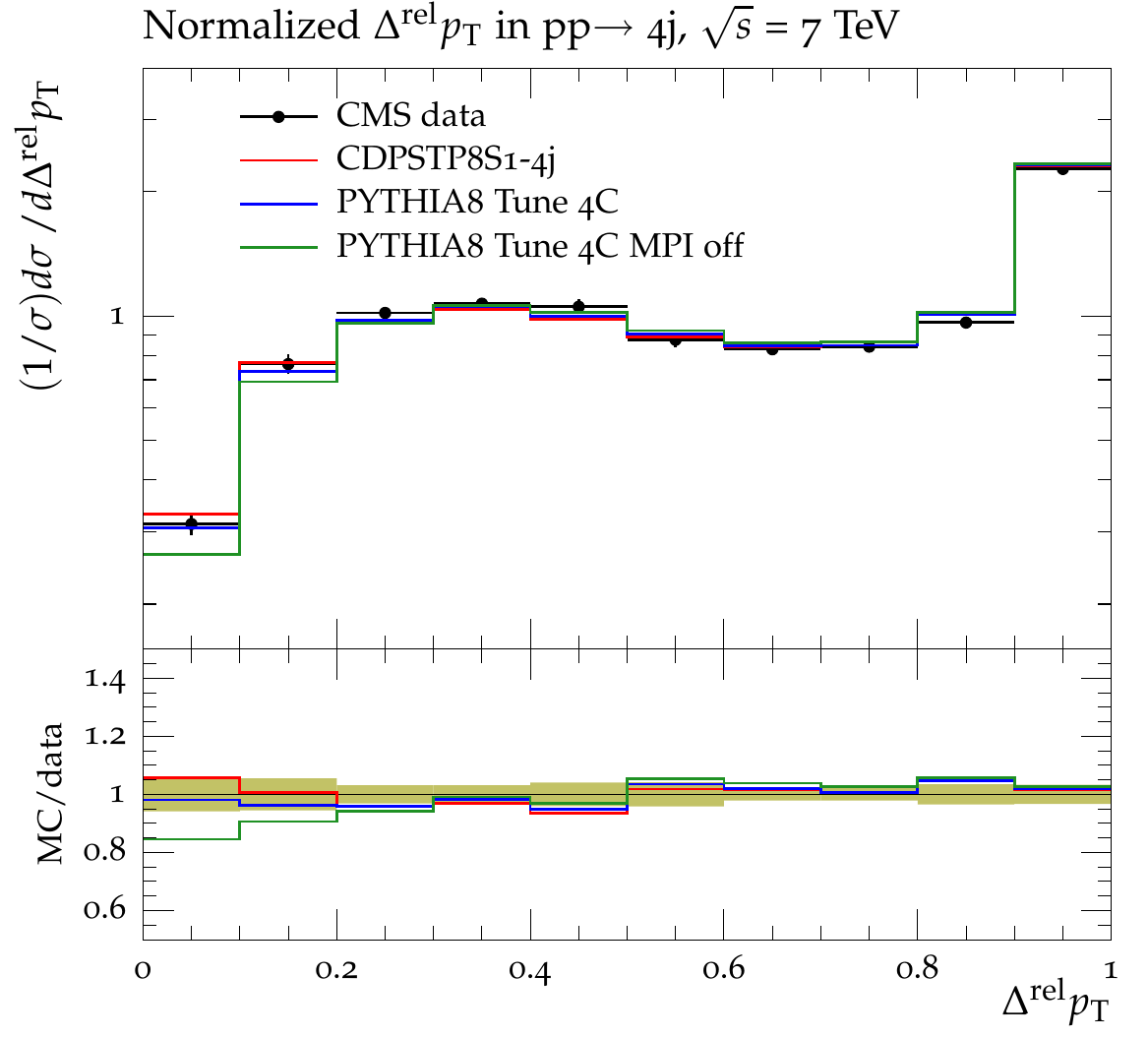}\\
\caption{Distributions of the correlation observables $\Delta$S (left) and \relpt\ (right) measured in four-jet production at $\sqrt{s}=7\TeV$~\cite{Chatrchyan:2013qza} compared to \textsc{pythia8} Tune 4C, Tune 4C with no MPI, and CDPSTP8S1-4j. The bottom panels of each plot show the ratios of these predictions to the data, and the green bands around unity represent the total experimental uncertainty.}
\label{PUB_fig19bis}
\end{center}
\end{figure*}

\begin{figure*}[htbp]
\begin{center}
\includegraphics[scale=0.65]{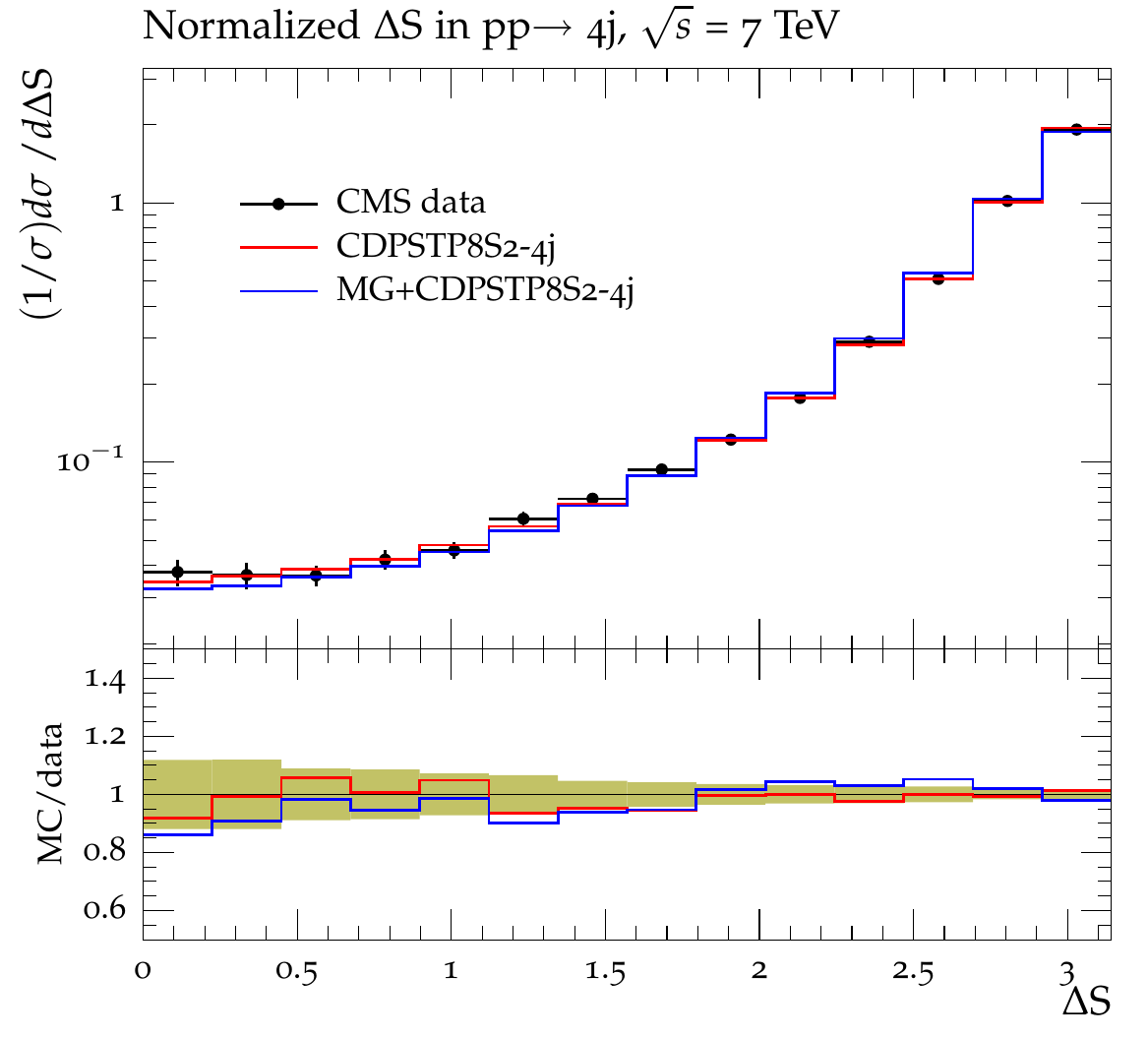}
\includegraphics[scale=0.65]{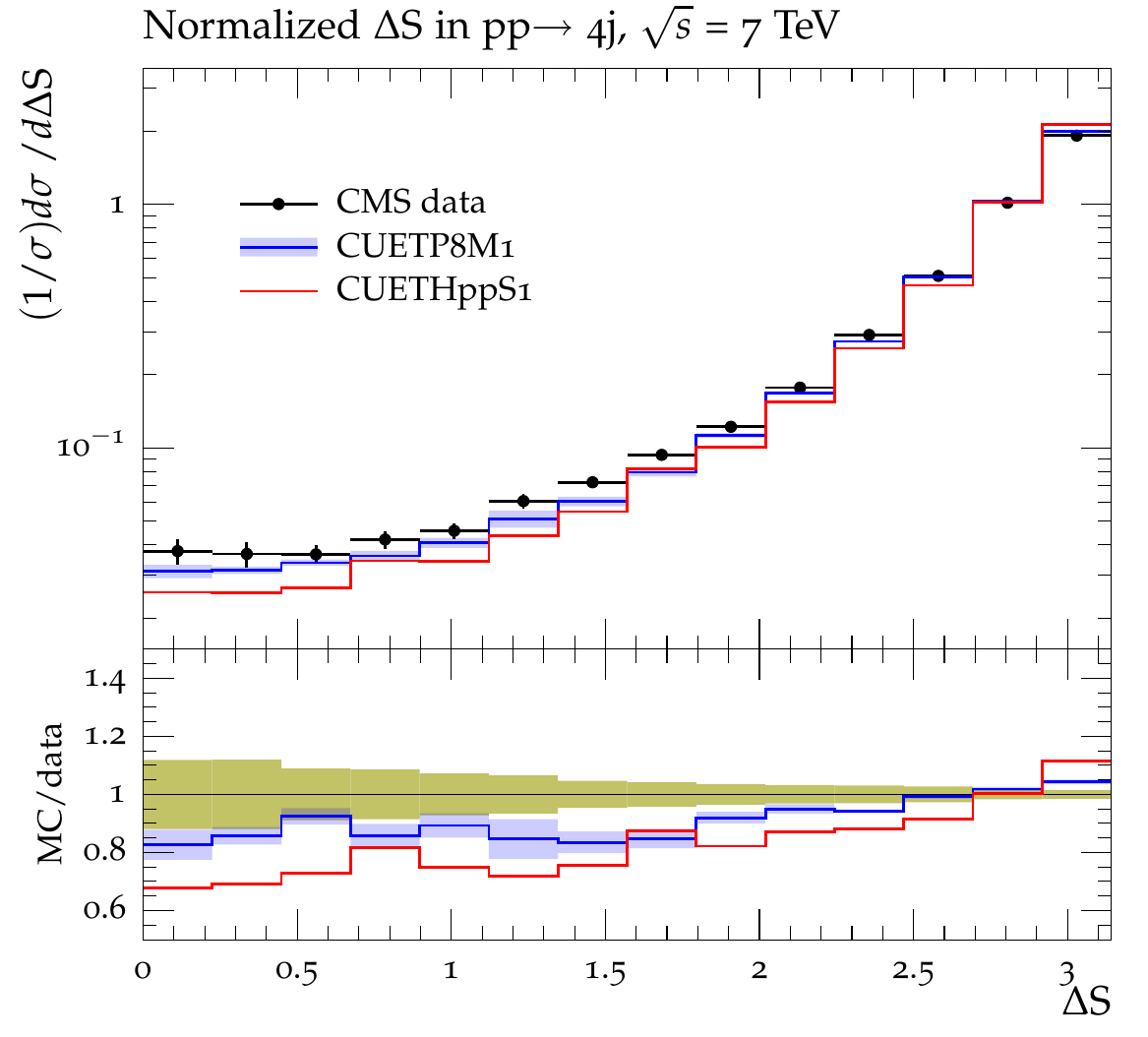}\\
\includegraphics[scale=0.65]{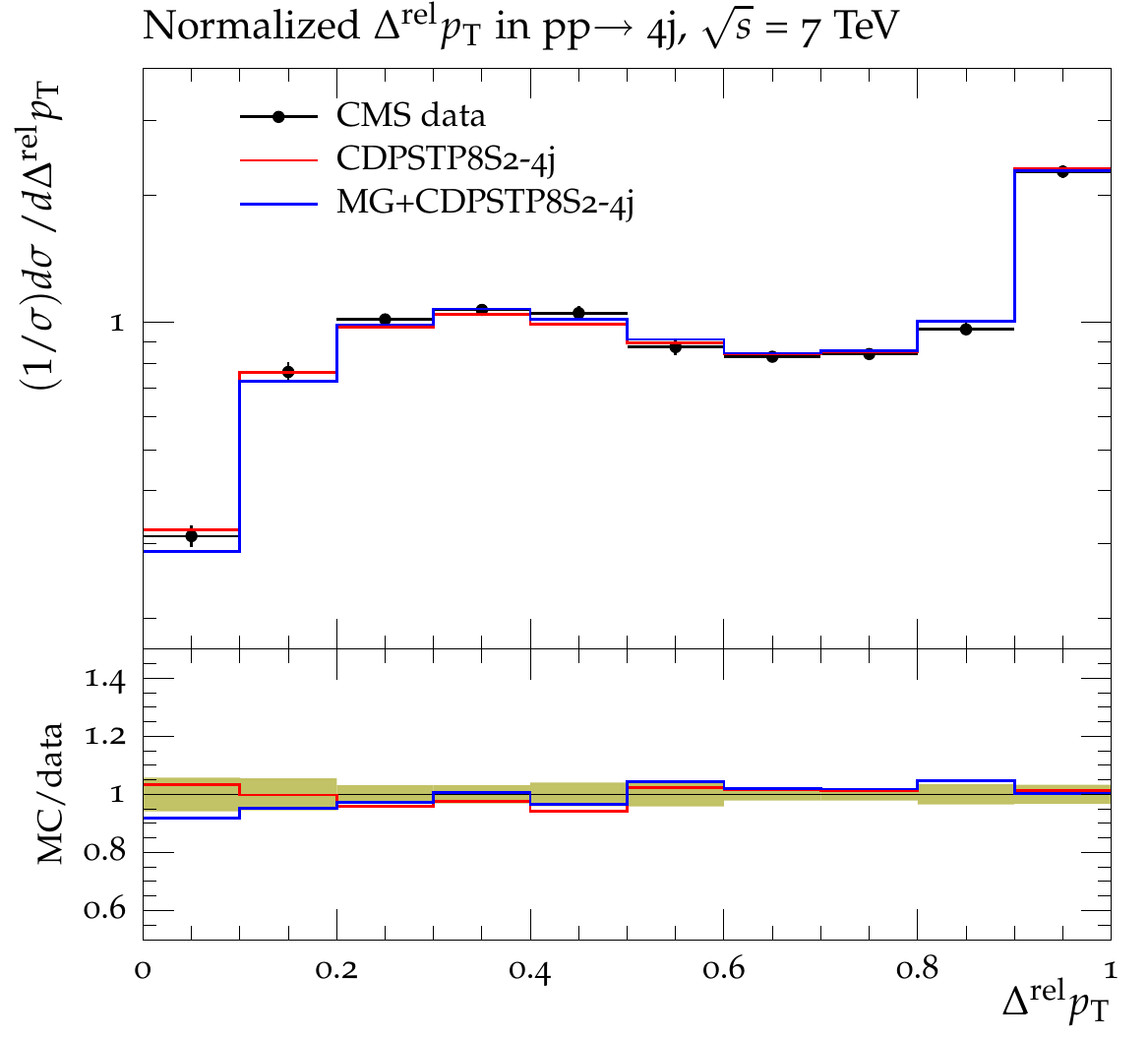}
\includegraphics[scale=0.65]{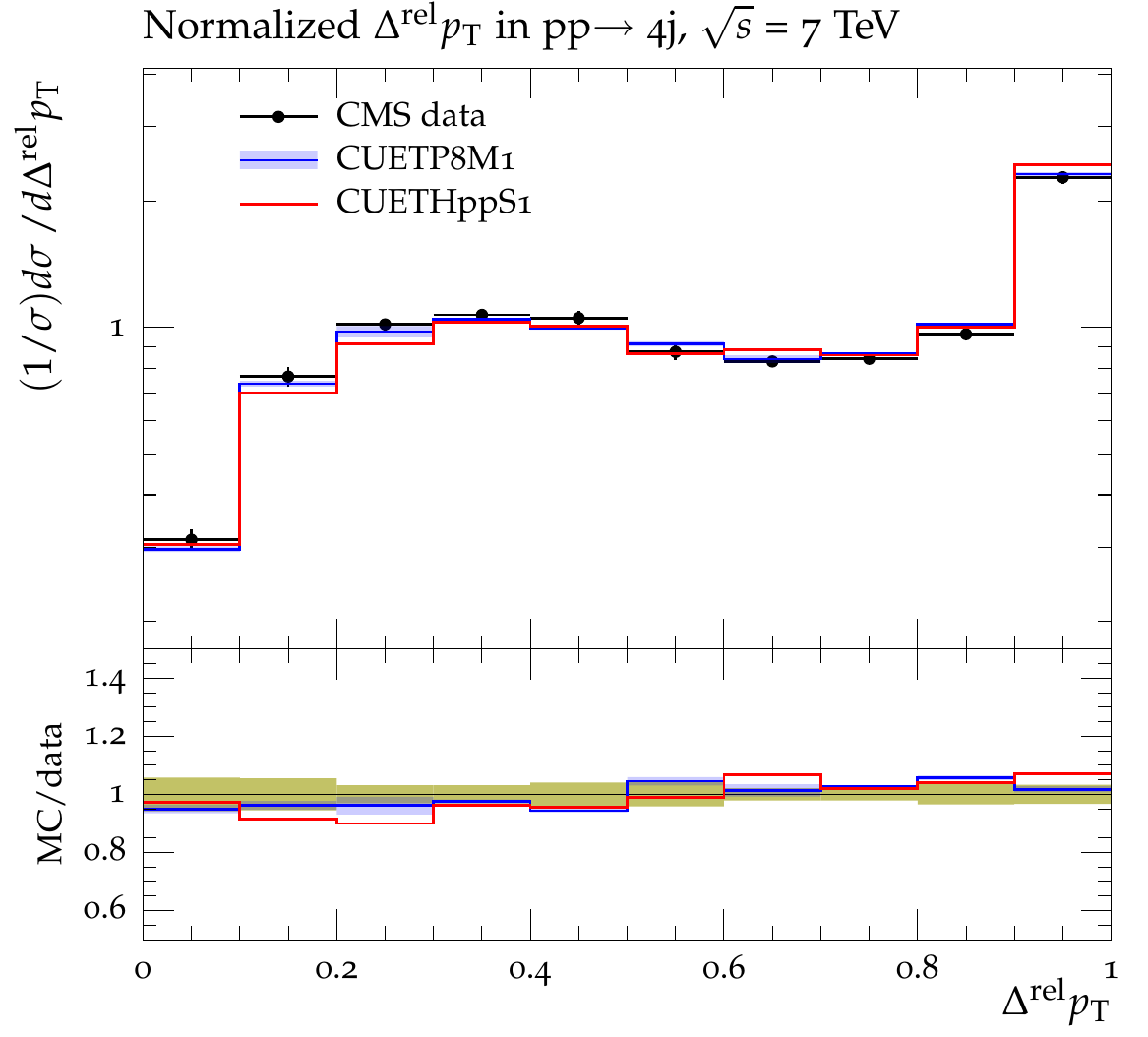}
\caption{
Distributions in the correlation observables $\Delta$S (top) and \relpt\ (bottom) measured in four-jet production at $\sqrt{s}=7\TeV$~\cite{Chatrchyan:2013qza}, compared to predictions of \pynewhyphen\ using \cdpJB\ and of \MADGRAPH (MG) interfaced to  \pynewhyphen\ using \cdpJB\ (left) and  \pynewhyphen\ using \cuePM\ and \hwpp\ with \cueHW\ (right). Also shown are the ratios of the predictions to the data. Predictions for \cuePM\ (right) are shown with an error band corresponding to the total uncertainty obtained from the eigentunes (Appendix~A). The green bands around unity represent the total experimental uncertainty.}
\label{PUB_fig19}
\end{center}
\end{figure*}

{\tolerance=1200
Starting from the parameters of \pynewhyphen\ \tunec, we construct two different four-jet DPS tunes.  As in the $\PW$+dijet channel, in the partial tune just the exponential-dependence parameter, \verb+expPow+, while in the full tune all four parameters of Table~\ref{table8} are varied.  We obtain a good fit to the four-jet data without including higher-order ME contributions.  However, we also obtain a good fit when higher-order (real) ME terms are generated with \MADGRAPH. 
In Fig.~\ref{PUB_fig19bis} and \ref{PUB_fig19}  the correlation observables $\Delta$S and \relpt\ in four-jet production~\cite{Chatrchyan:2013qza} are compared to predictions obtained with \pynewhyphen\ \tunec, \tunec\ without MPI,  \cdpJA,  \cdpJB, and \MADGRAPH interfaced to \cdpJB.  Table~\ref{table8} gives the best-fit parameters and the resulting \eff\ values. The values of \eff\ extracted from the CMS \pynewhyphen\ DPS tunes give the first determination of \eff\ in four-jet production at $\sqrt{s}=7\TeV$.  
The uncertainties quoted for \eff\  are obtained from the eigentunes.
\par}

\begin{table*}[htbp]
\renewcommand{\arraystretch}{1.2}
\begin{center}
\topcaption{The \pynewhyphen\ parameters, tuning ranges, \tunec\  values~\cite{Corke:2010yf} and best-fit values of \cdpJA\  and  \cdpJB, obtained from fits to DPS observables in four-jet production. Also shown are the predicted values of \eff\ at $\sqrt{s}=7\TeV$, and the uncertainties obtained from the eigentunes. }
\label{table8}
\cmsTable{
\begin{tabular}{l c c c c} \hline 
  \pynew\ Parameter  &   Tuning Range  &   \tunec &   \cdpJA &   \cdpJB \\ \hline
   PDF                  &                     &  CTEQ6L1       &  CTEQ6L1      &  CTEQ6L1\\ \hline
   MultipartonInteractions:pT0Ref [GeV] & $1.0$--$3.0$  & $2.085$    & $2.085^*$   & $2.125$ \\ 
   MultipartonInteractions:ecmPow & $0.0$--$0.4$  & $0.19\phantom{0}$     & $0.19^*\phantom{0}$    & $0.179$ \\ 
   MultipartonInteractions:expPow & $\phantom{0}0.4$--$10.0$ & $2.0\phantom{00}$      & $1.160\phantom{0}$     & $0.692$ \\ 
   ColourReconnection:range & $0.0$--$9.0$  & $1.5\phantom{00}$      & $1.5^*\phantom{00}$     & $6.526$ \\ \hline
   MultipartonInteractions:ecmRef [GeV]  &\NA       & $1800$     & $1800^*$    & $1800^*$  \\ 
   $\chi^2$/dof                   &\NA & \NA&  0.751 & 0.428 \\
   Predicted \eff\ (in mb) & \NA  & $30.3$ & $21.3^{+1.2}_{-1.6}$ & $19.0^{+4.7}_{-3.0}$ \\ \hline
  \multicolumn{5}{r}{\footnotesize $^*$ Fixed at \tunec\ value.}
\end{tabular}
}
\end{center}
\end{table*}

\section{Validation of CMS tunes}
\label{section-4}

Here we discuss the compatibility of the UE and DPS tunes.  In addition, we compare the CMS UE tunes with UE data that have not been used in the fits, and we examine how well Drell--Yan and MB observables can be predicted from MC simulations using the UE tunes.  We also show that the CMS UE tunes can be interfaced to higher-order ME generators without additional tuning of the MPI parameters.

\subsection{Compatibility of UE and DPS tunes}

\begin{figure*}[htbp]
\begin{center}
\includegraphics[scale=0.65]{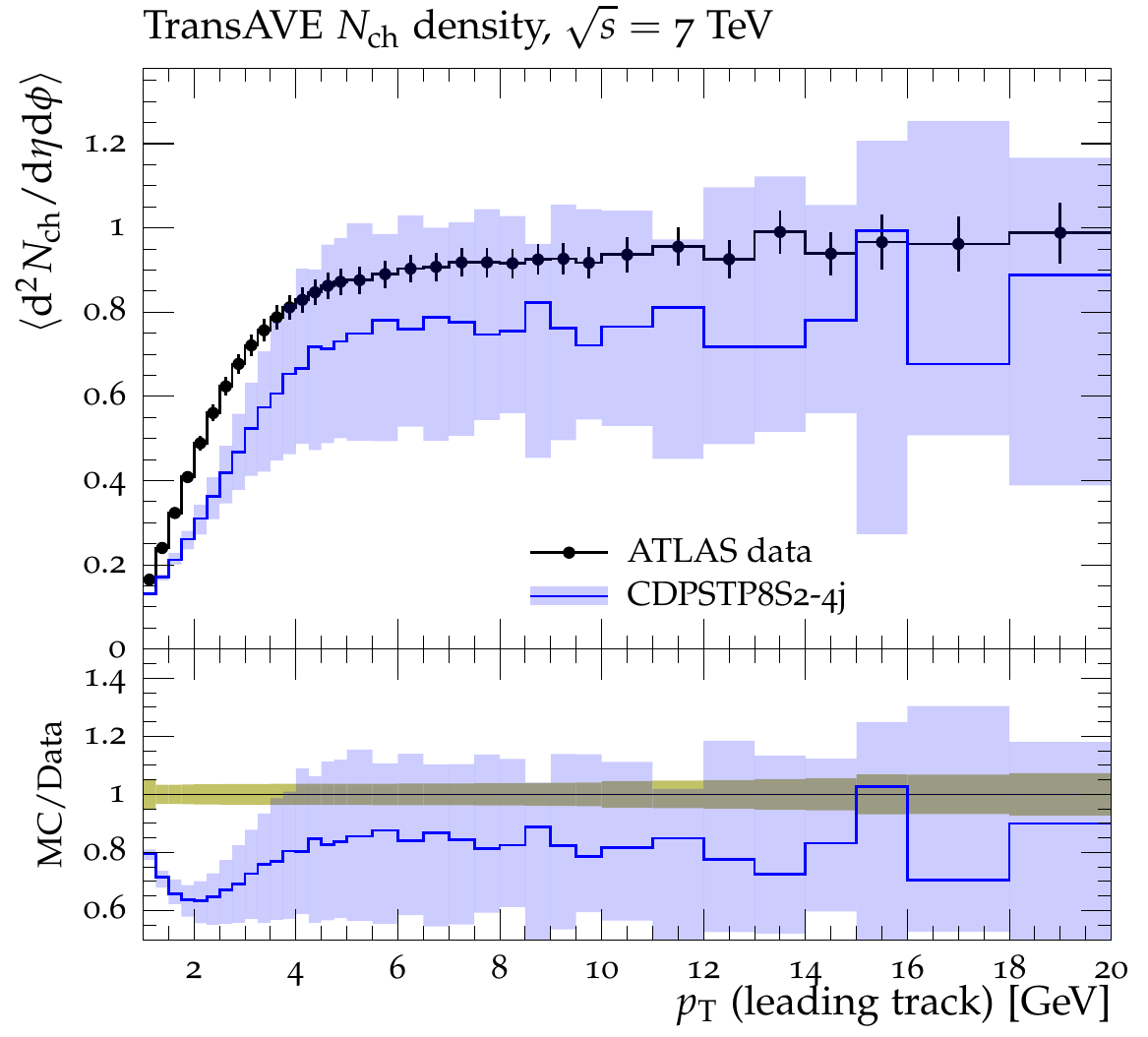}    
\includegraphics[scale=0.65]{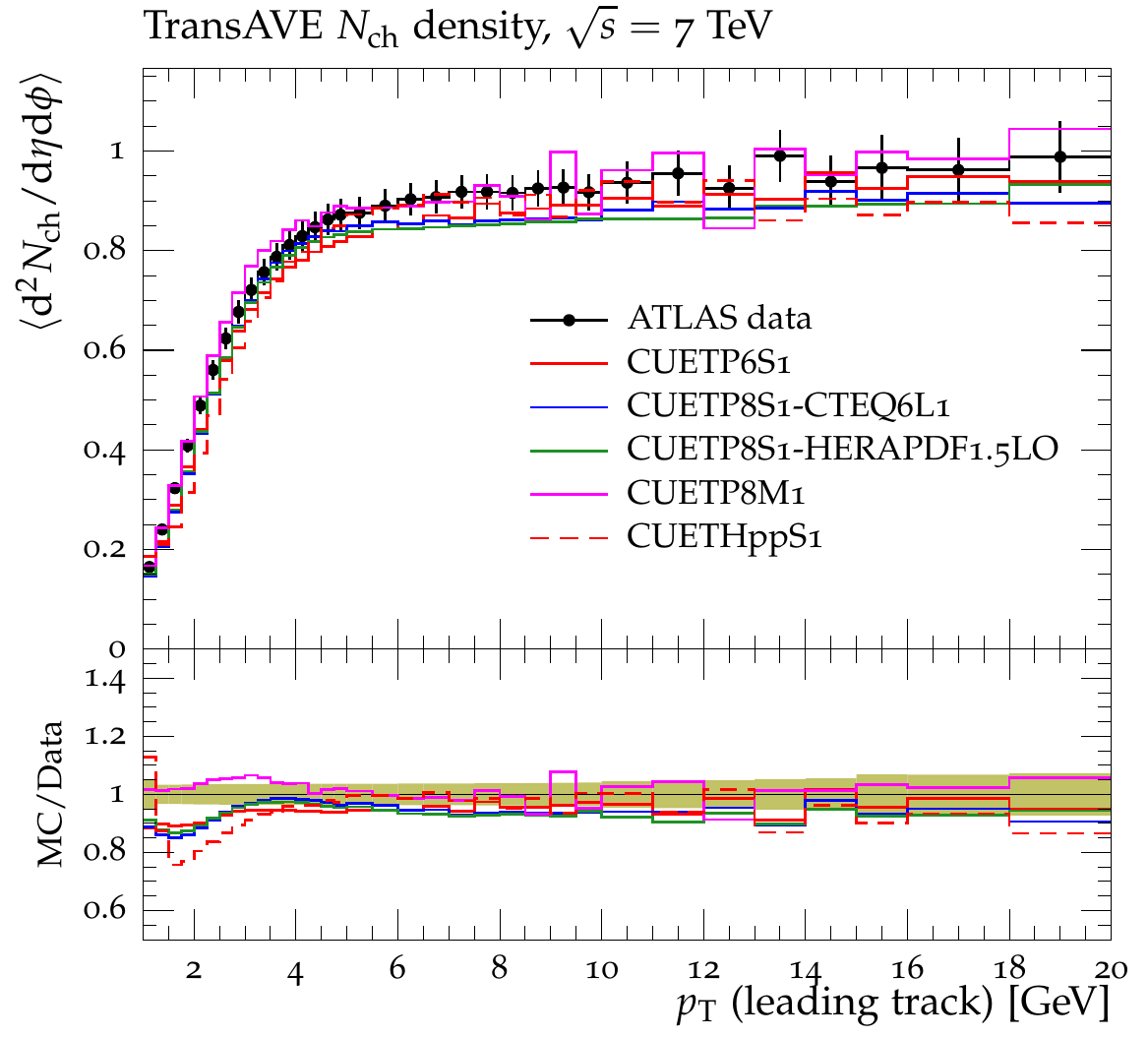}\\                                                                                
\includegraphics[scale=0.65]{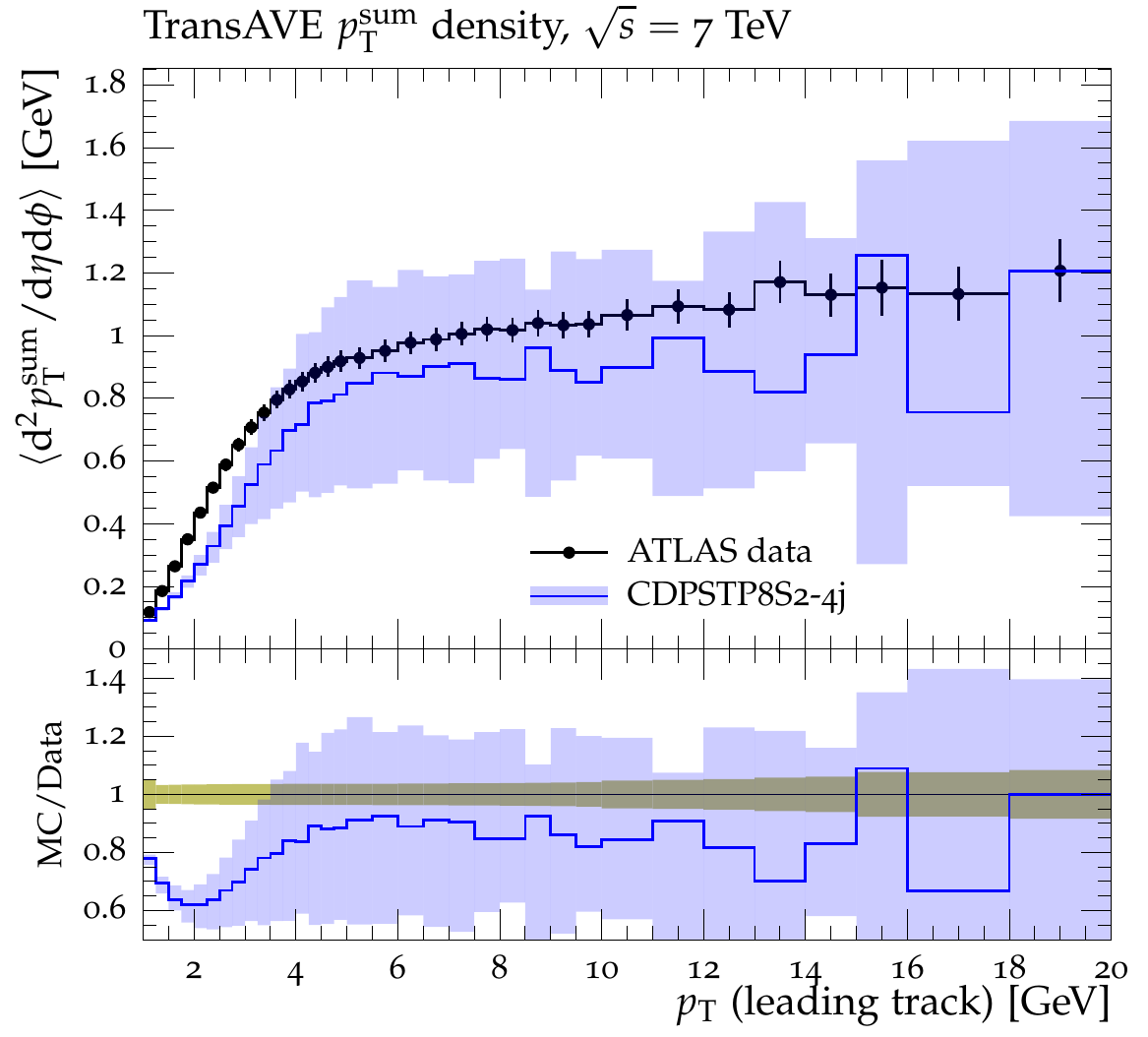}
\includegraphics[scale=0.65]{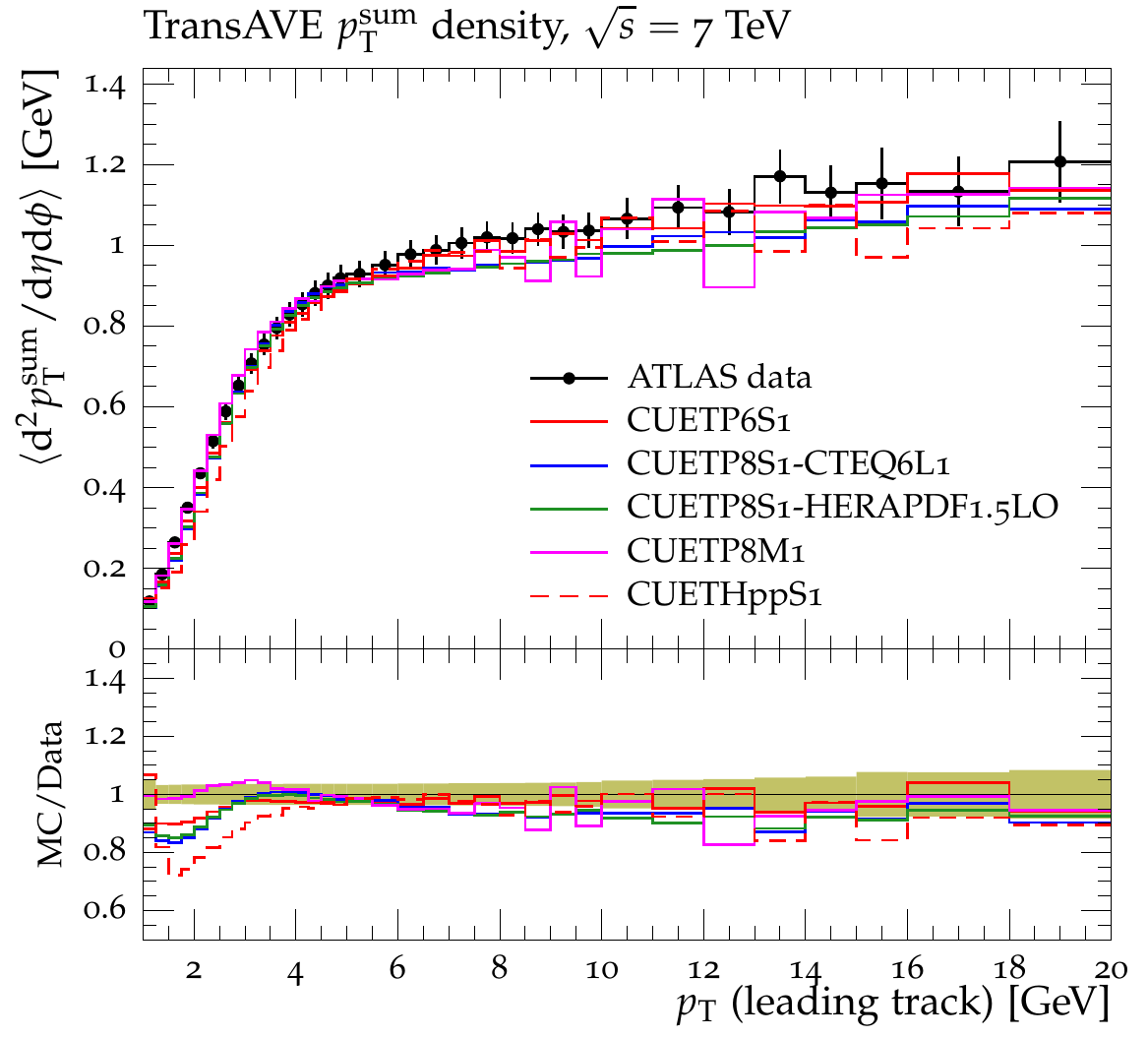}
\caption{ATLAS data at $\sqrt{s}=7\TeV$~\cite{Aad:2010fh} for charged-particle (left) and \ptsum\ densities (right) with \ptcut\ and \etabig\ in the transverse (\tave) region
 compared to predictions of \pynewhyphen\  using \cdpJB\ (left) and \cuePB, \cuePH, and \cuePM, plus \hwpp\ using \cueHW\ (right). The predictions of \cdpJB\ are shown with an error band corresponding to the total uncertainty obtained from the  eigentunes  (Appendix~A). The bottom panels of each plot show the ratios of these predictions to the data, and the green bands around unity represent the total experimental uncertainty.}
\label{PUB_fig20}
\end{center}
\end{figure*}

{\tolerance=5000
The values of \eff\ obtained from simulations applying the CMS \pynewhyphen\ UE and DPS tunes at $\sqrt{s}=7\TeV$ and $\sqrt{s}=13\TeV$ are listed in Table~\ref{table9}. The uncertainties, obtained from eigentunes are also quoted in Table~\ref{table9}.  At $\sqrt{s}=7\TeV$, the CMS DPS tunes give values of \eff\ $\approx$ 20\unit{mb}, while the CMS \pynewhyphen\ UE tunes give slightly higher values  in the range $26$--$29$ mb as shown in Figs.~\ref{PUB_fig19} and ~\ref{PUB_fig20}. Figure~\ref{PUB_fig19} shows the CMS DPS-sensitive data for four-jet production at $\sqrt{s}=7\TeV$ compared to predictions using \cdpJB, \cuePM, and \cueHW. Figure~\ref{PUB_fig20} shows ATLAS UE data at $\sqrt{s}=7\TeV$~\cite{Aad:2010fh} compared to predictions obtained with various tunes: \cdpJB\ with uncertainty bands, \cuePA, \cuePB, \cuePH, \cuePM, and \cueHW.  Predictions from \pynewhyphen\ using \cuePM\ describe reasonably well the DPS observables, but do not fit them as well as predictions using the DPS tunes. On the other hand, predictions using  \cdpJB\ do not fit the UE data as well as the UE tunes do. 
\par}

\begin{table*}[htbp]
\renewcommand{\arraystretch}{1.2}
\begin{center}
\topcaption{Values of \eff\ at $\sqrt{s}=7\TeV$ and $13\TeV$ for  \cuePB, \cuePH, and \cuePM,  \cueHW, and for   \cdpJA\ and \cdpJB. At $\sqrt{s}=7\TeV$, also shown are the uncertainties in \eff\ obtained from the eigentunes.}
\label{table9}
\begin{tabular}{l c c}  \hline
 CMS Tune &  \eff (mb) at $7\TeV$  &  \eff (mb) at $13\TeV$ \\ \hline
   \cuePB      & $27.8^{+1.2}_{-1.3}$       & $29.9^{+1.6}_{-2.8}$ \\ 
   \cuePH    & $29.1^{+2.2}_{-2.0}$       & $31.0^{+3.8}_{-2.6}$ \\ 
   \cuePM      & $26.0^{+0.6}_{-0.2}$       & $27.9^{+0.7}_{-0.4}$ \\ 
   \cueHW      & $15.2^{+0.5}_{-0.6}$       & $15.2^{+0.5}_{-0.6}$ \\ 
   \cdpJA      & $21.3^{+1.2}_{-1.6}$       & $21.8^{+1.0}_{-0.7}$ \\ 
   \cdpJB      & $19.0^{+4.7}_{-3.0}$       & $22.7^{+10.0}_{-5.2}$ \\ \hline
\end{tabular}
\end{center}
\end{table*}

As discussed previously, the \pynew\ tunes use a single exponential matter-overlap function, while the \hwpp\ tune uses a matter-overlap function that is related to the Fourier transform of the electromagnetic form factor.
The \cueHW\ gives a value of \eff\ $\approx$ 15\unit{mb}, while UE and DPS tunes give higher values of \eff.  It should be noted that \eff\ is a parton-level observable and  its importance is not in the modelled value of \eff, but in what is learned about the transverse proton profile (and its energy evolution), and how well the models describe the DPS-sensitive observables.  As can be seen in Fig.~\ref{PUB_fig19}, predictions using  \cuePM\ describe the DPS-sensitive observables better than \cueHW, but not quite as well as the DPS tunes. We performed a simultaneous \pynewhyphen\ tune that included both the UE data and DPS-sensitive observables, however, the quality of the resulting fit was poor. This confirms the difficulty of describing soft and hard MPI within the current \PYTHIA and \hwpp\ frameworks. Recent studies~\cite{Blok:2015rka,Diehl:2014vaa} suggest the need for introducing parton correlation effects in the MPI framework in order to achieve a consistent description of both the UE and DPS observables.

\subsection{Comparisons with other UE measurements}

\begin{figure*}[htbp]
\begin{center}
\includegraphics[scale=0.6]{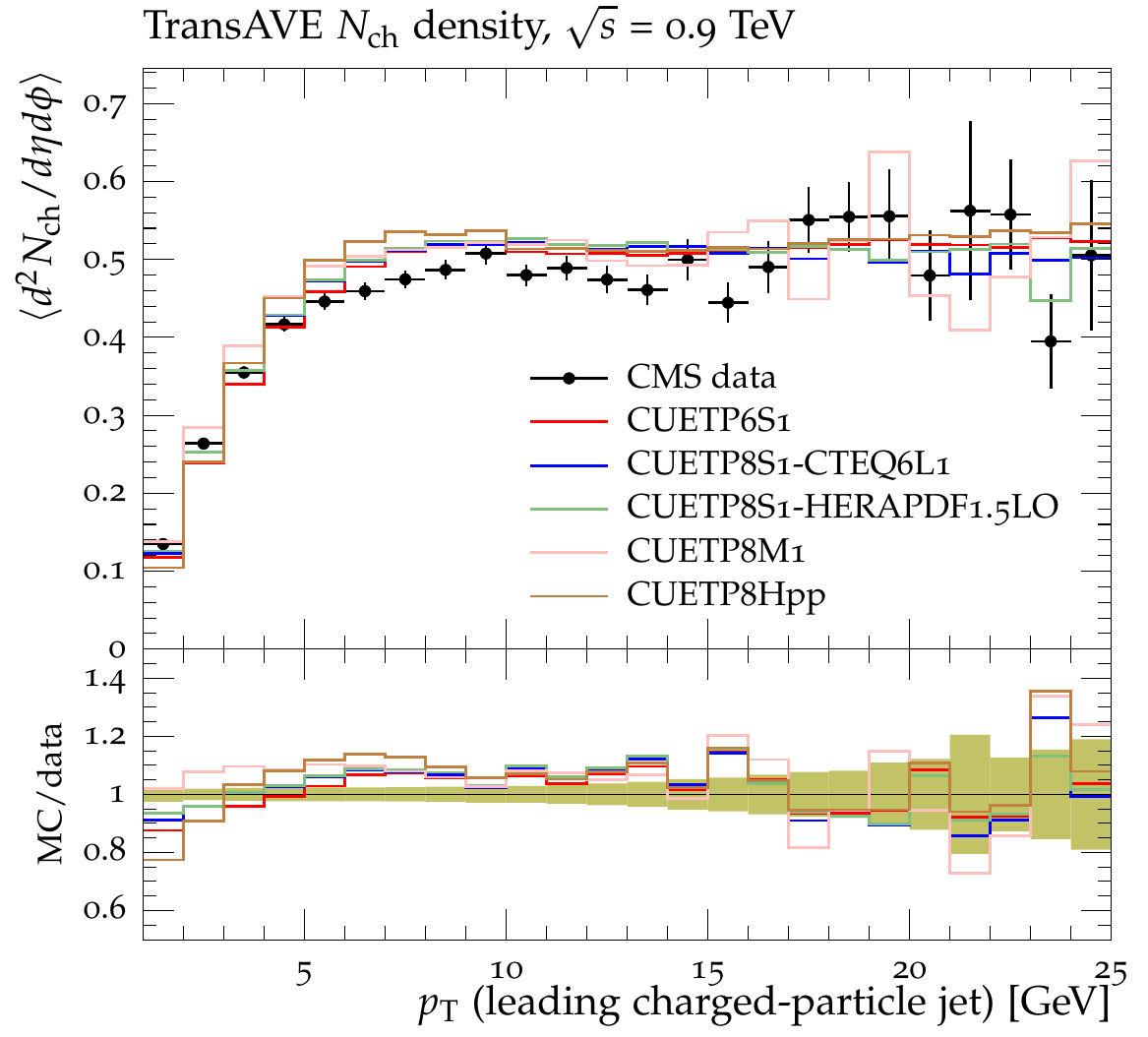}                                                                               
\includegraphics[scale=0.6]{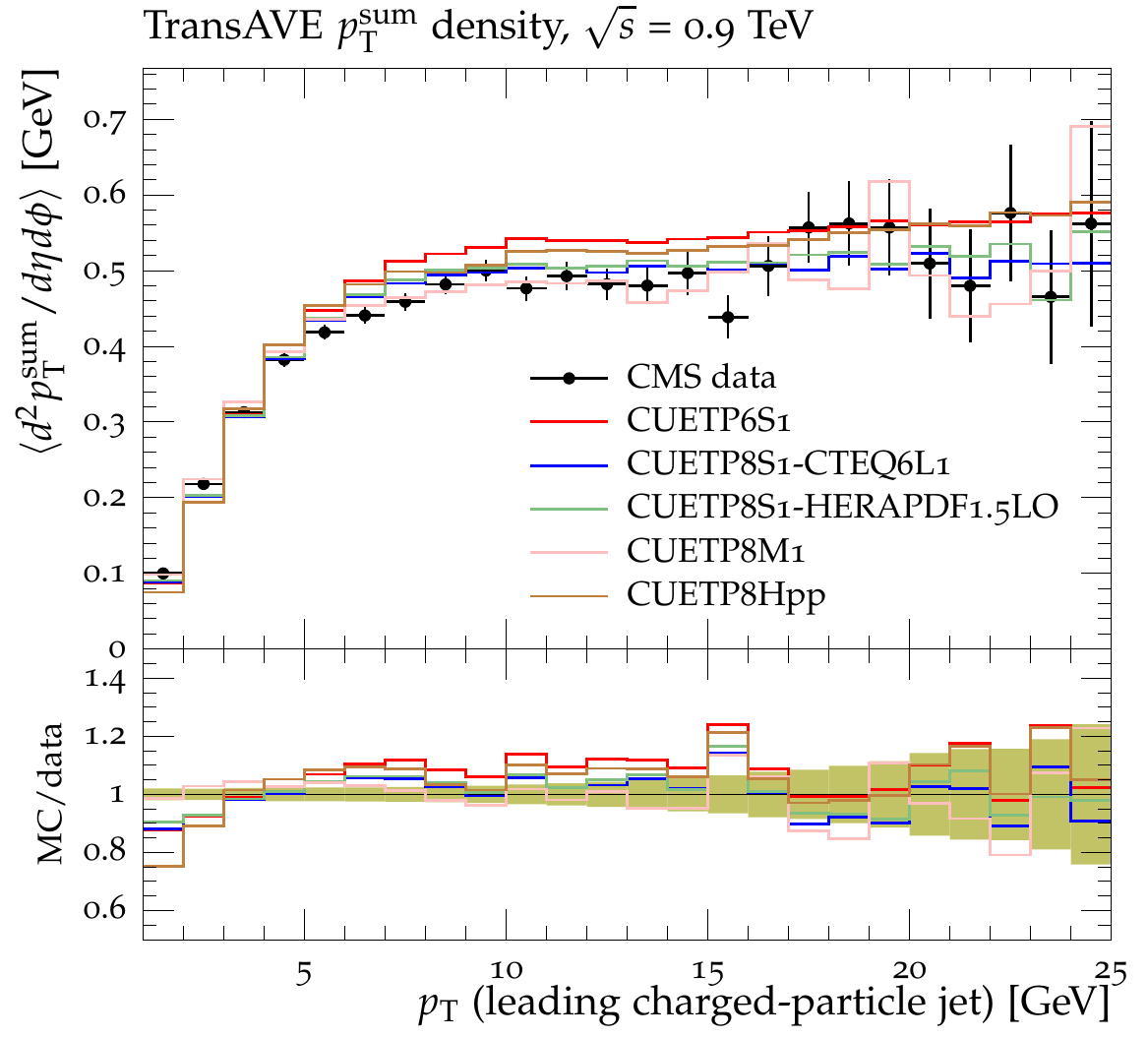}\\
\includegraphics[scale=0.6]{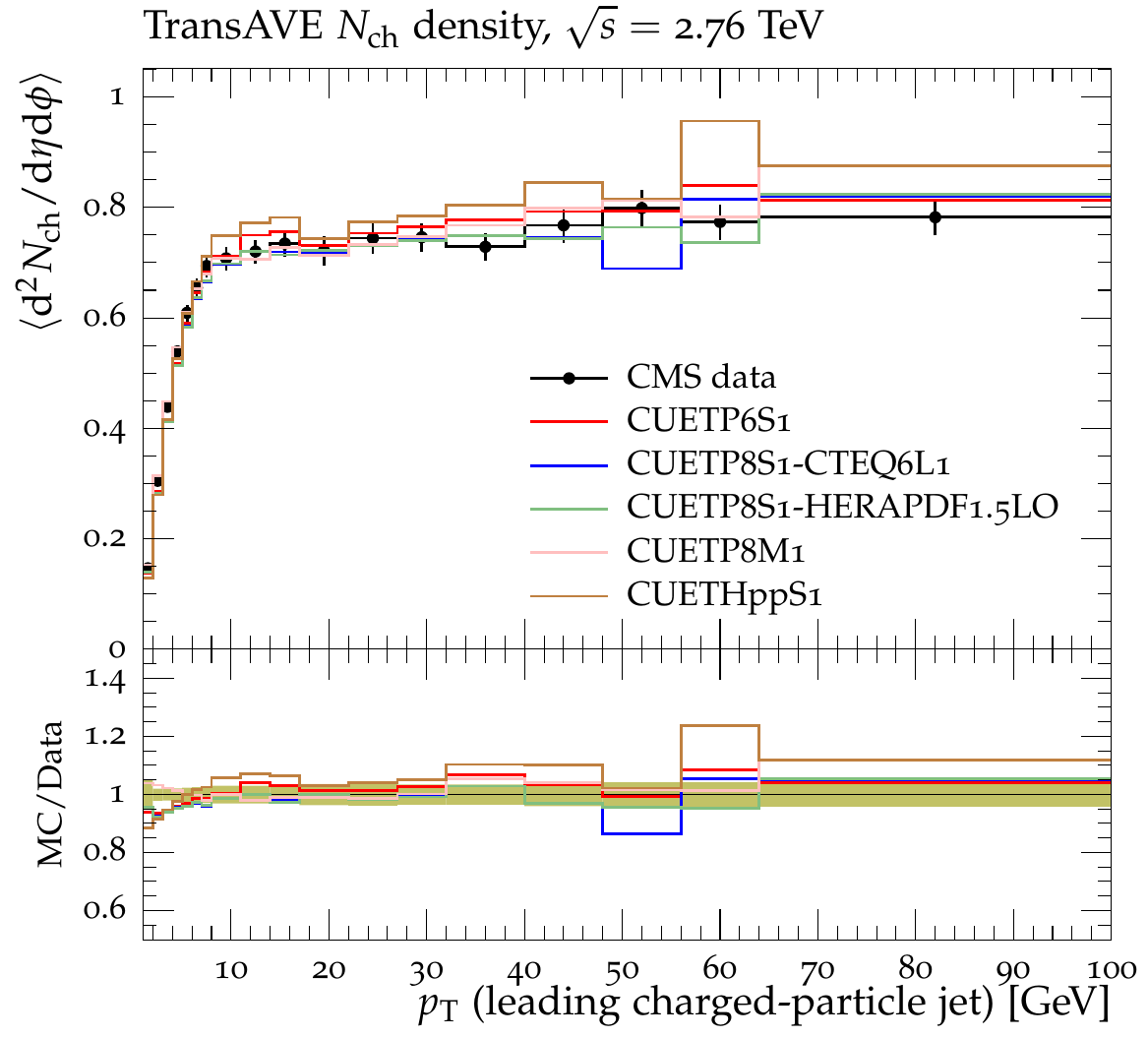}                                                                               
\includegraphics[scale=0.6]{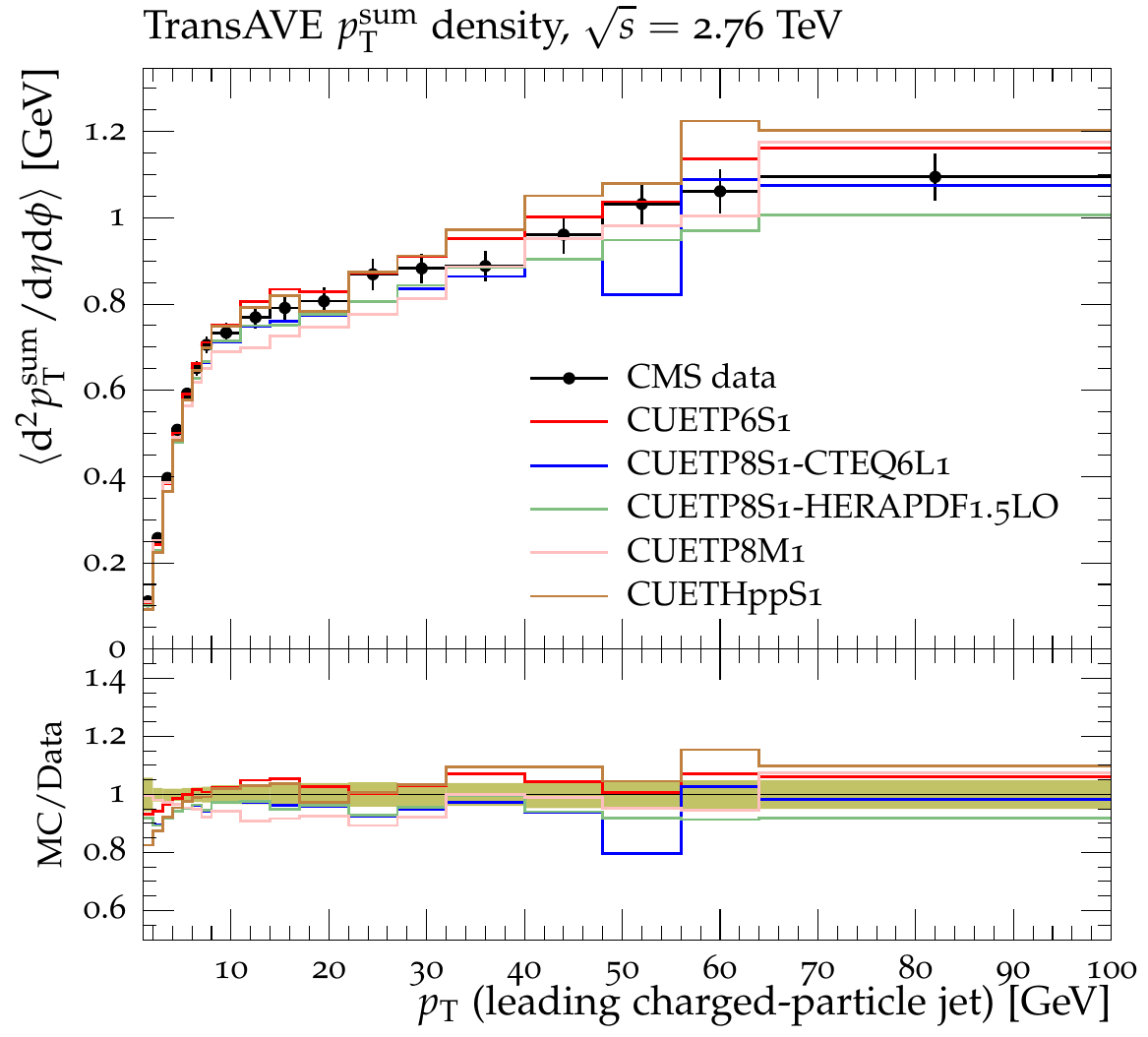}\\
\includegraphics[scale=0.6]{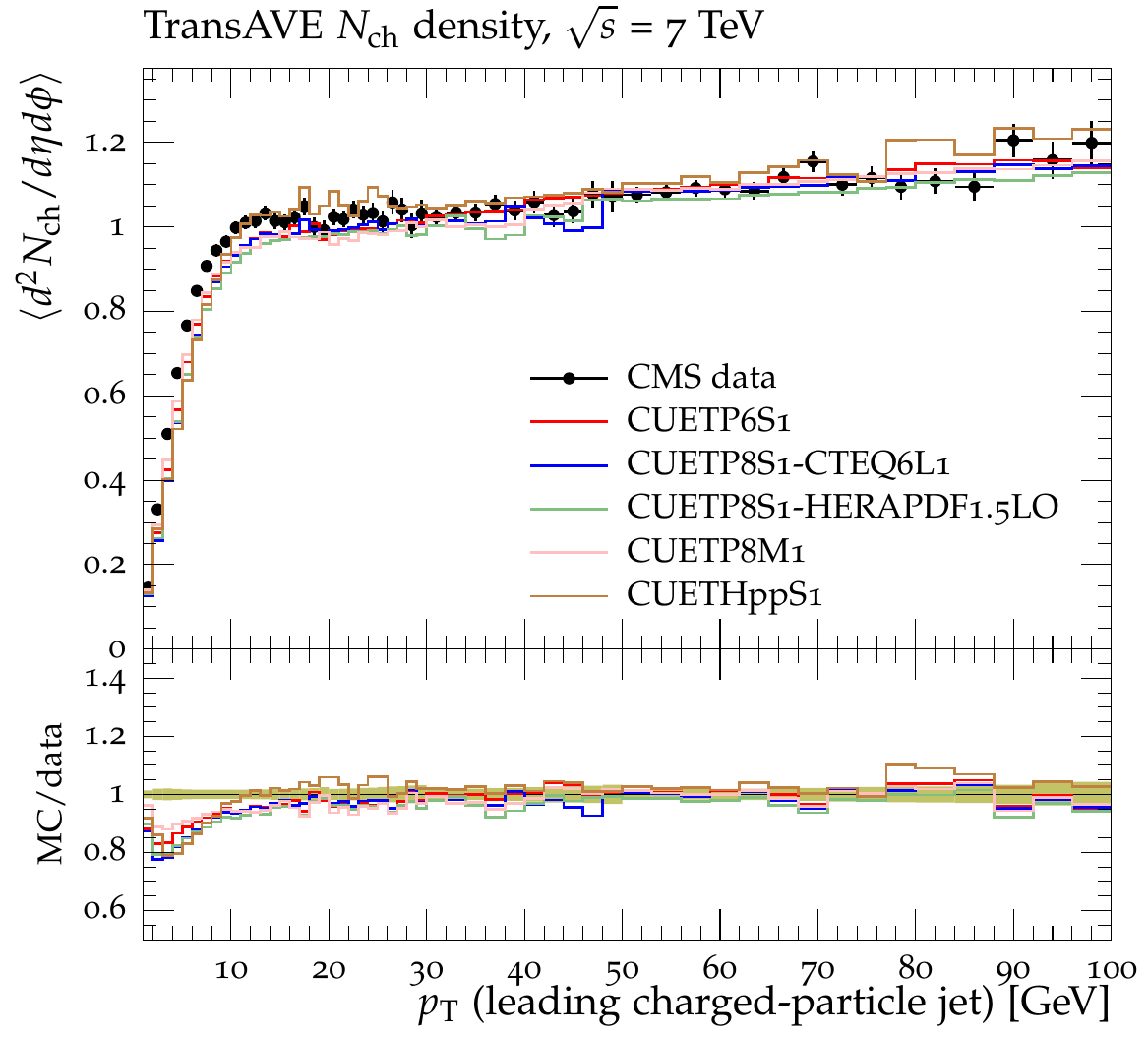}
\includegraphics[scale=0.6]{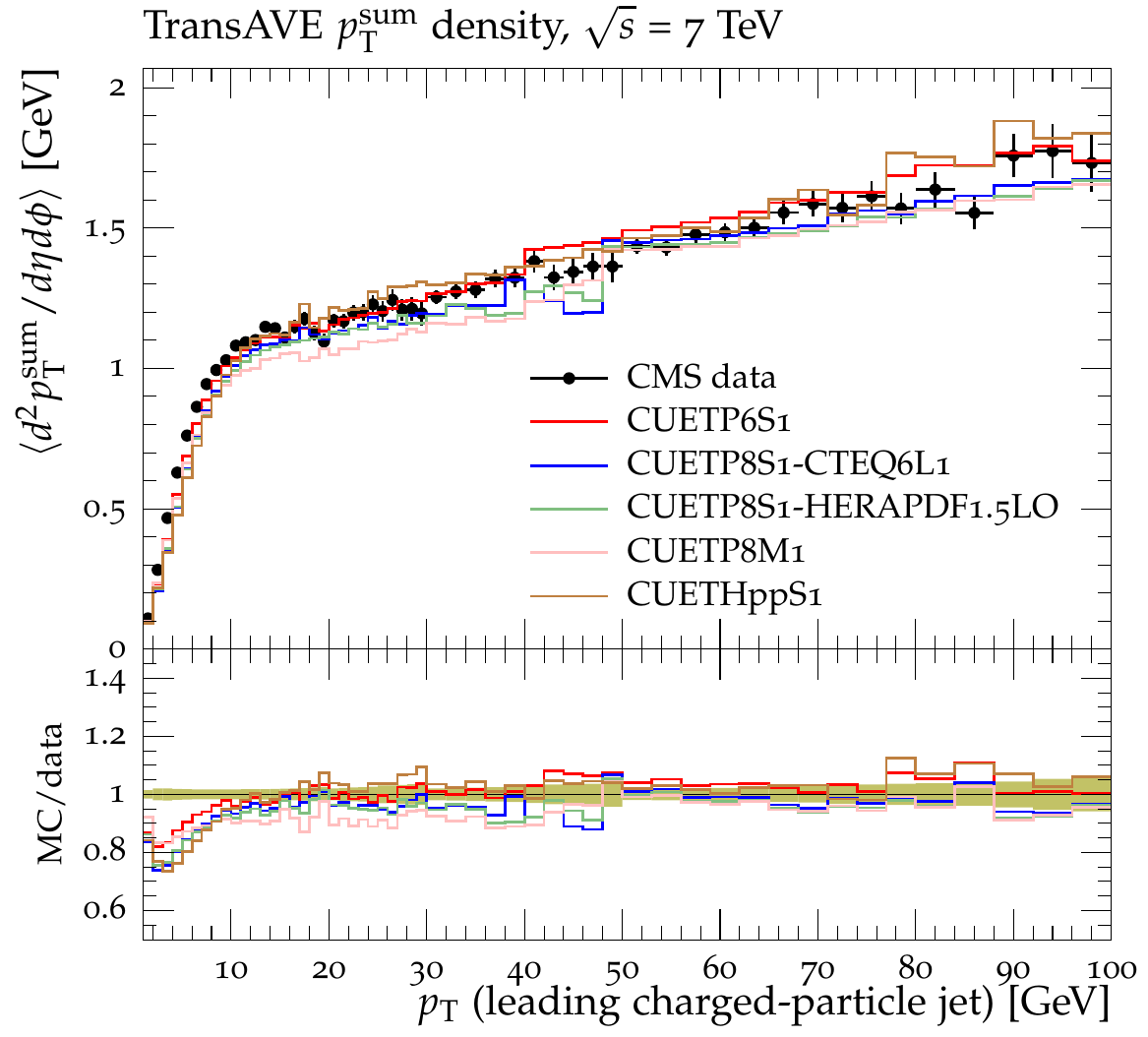}
\caption{CMS data on charged-particle (left) and \ptsum\ (right) densities at $\sqrt{s}$ = $0.9$~\cite{Chatrchyan:2011id} (top), $2.76$~\cite{Khachatryan:2015jza} (middle), and $7\TeV$~\cite{Chatrchyan:2011id} (bottom) with \ptcut\ and \etabig\ in the \TR\ (\tave) region as defined by the leading charged-particle jet, as a function of the transverse momentum of the leading charged-particle jet. The data are compared to predictions of  \pyoldhyphen\ using \cuePA,  \pynewhyphen\ using \cuePB, \cuePH, and \cuePM, and \hwpp\ using \cueHW. The bottom panels of each plot show the ratios of these predictions to the data, and the green bands around unity represent the total experimental uncertainty.}
\label{PUB_fig22}
\end{center}
\end{figure*}

Figure~\ref{PUB_fig22} shows charged particle and \ptsum\ densities~\cite{Chatrchyan:2011id,Khachatryan:2015jza} at $\sqrt{s} = 0.9$, $2.76$, and $7\TeV$ with \ptcut\ and \etabig\ in the \tave\ region, as defined by the leading jet reconstructed by using just the charged particles (also called ``leading track-jet'') compared to predictions using the CMS UE tunes.
The CMS UE tunes describe quite well the UE measured using the leading charged particle as well as the leading charged-particle jet.

\begin{figure*}[htbp]
\begin{center}
\includegraphics[scale=0.65]{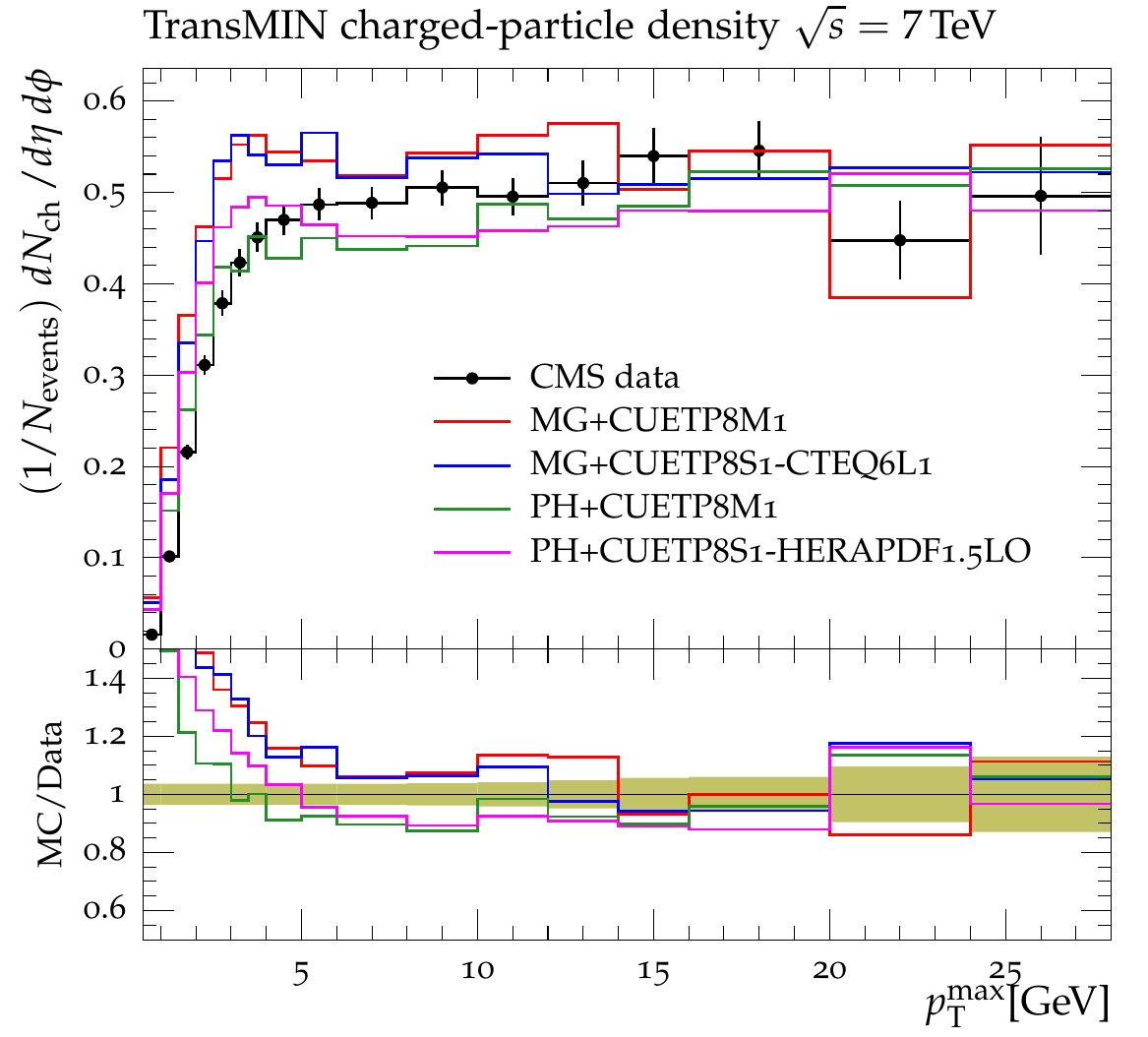}                                                                               
\includegraphics[scale=0.65]{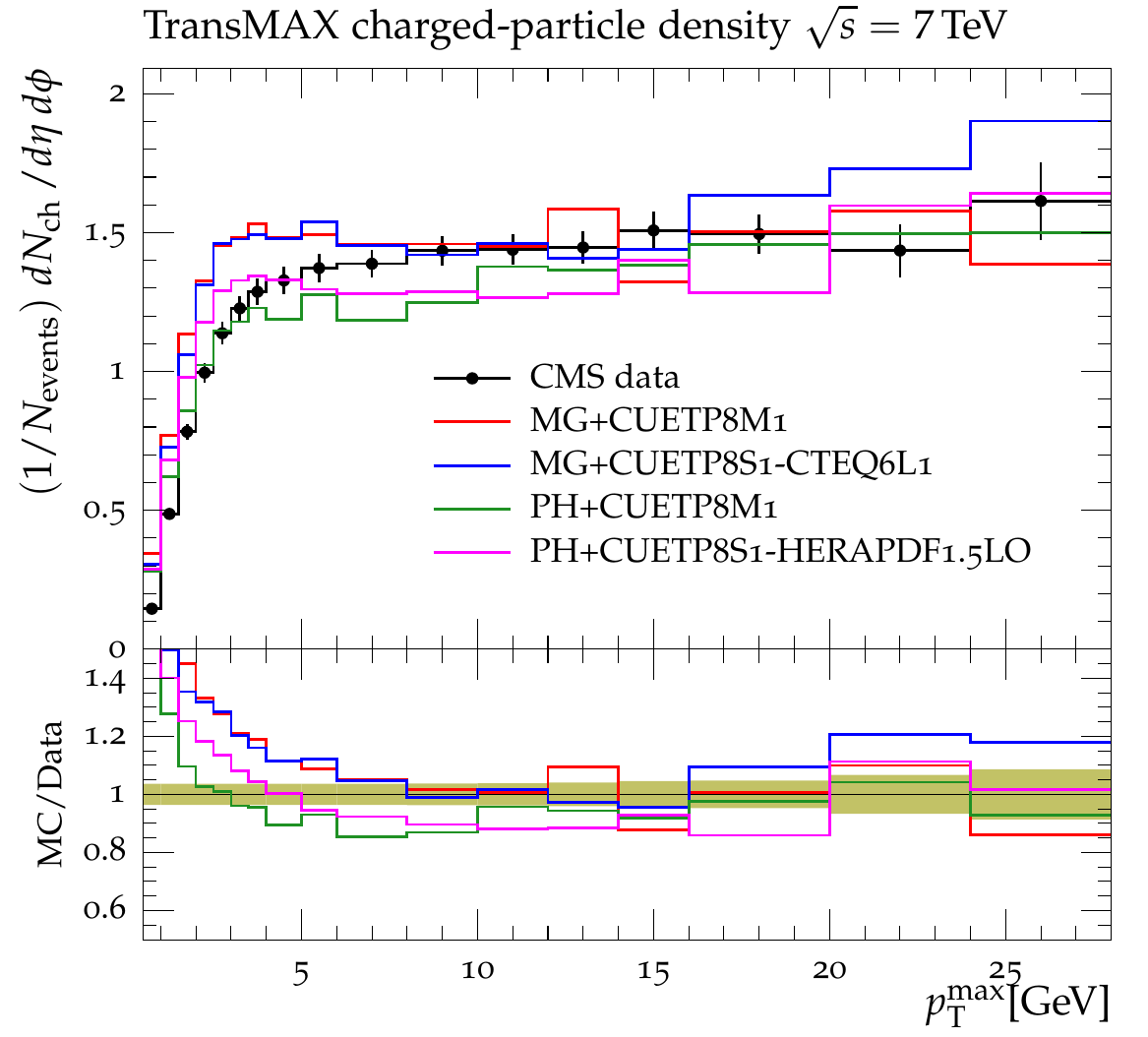}\\                                                                               
\includegraphics[scale=0.65]{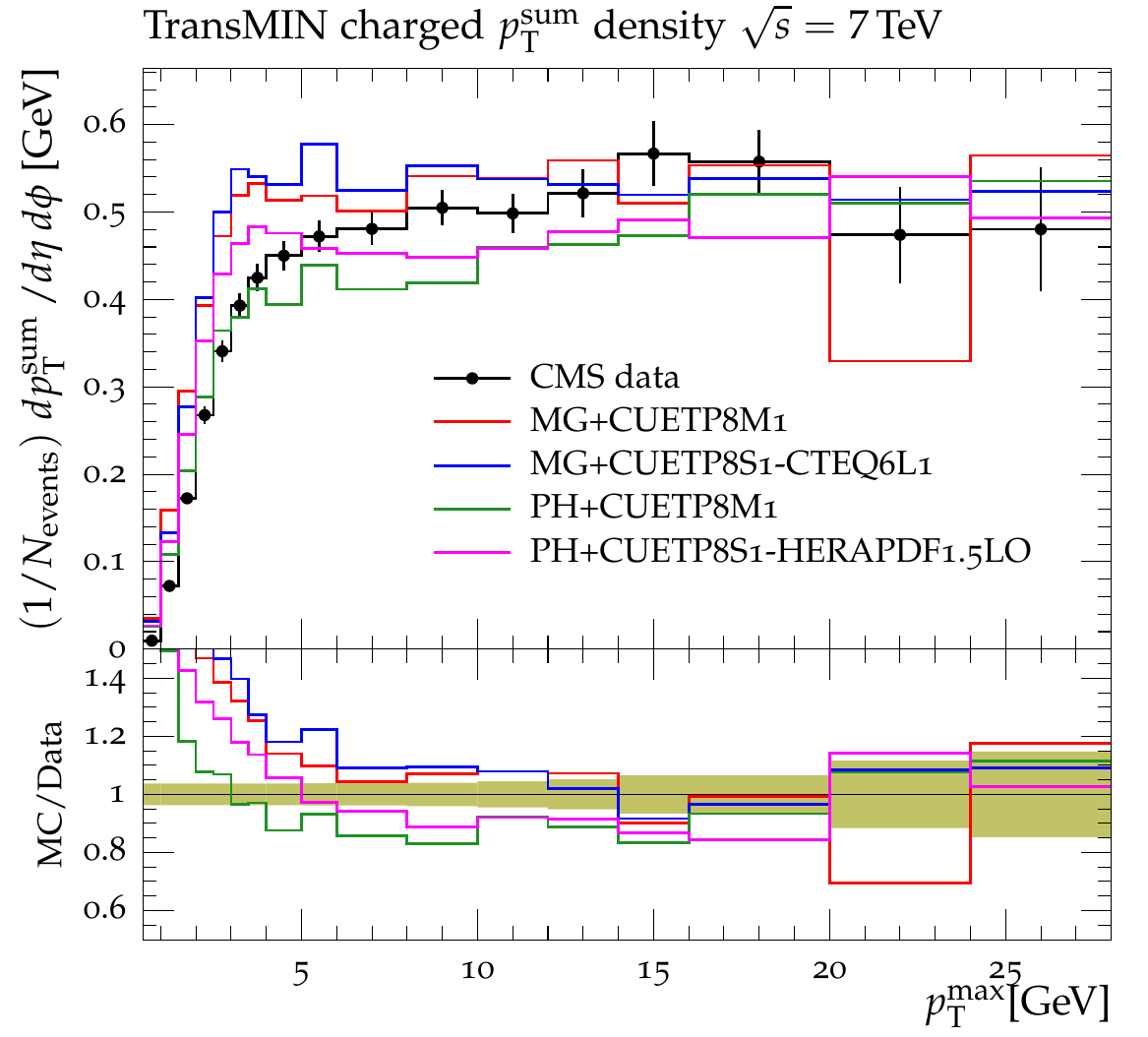}
\includegraphics[scale=0.65]{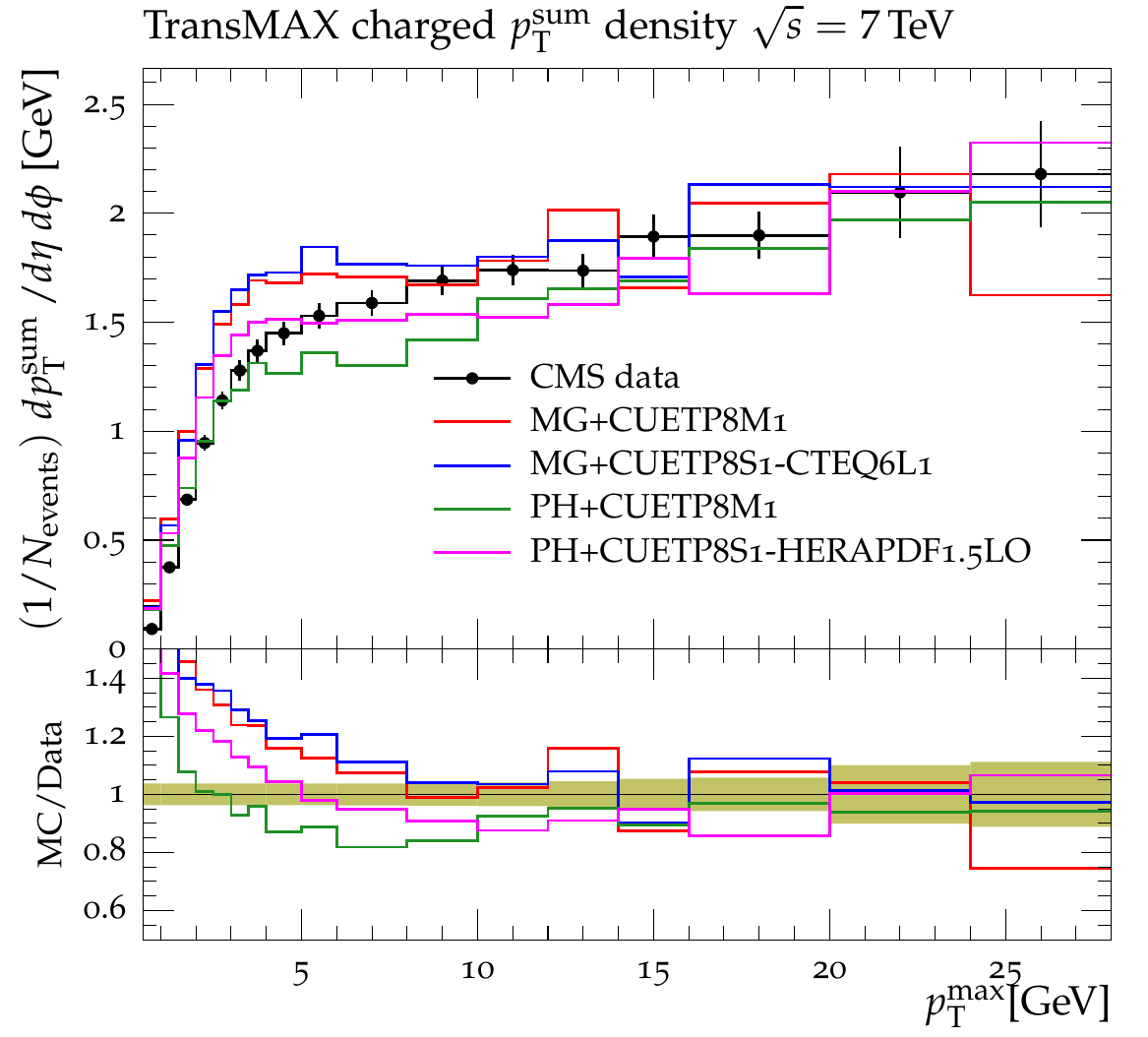}
\caption{CMS data  at $\sqrt{s}=7\TeV$ ~\cite{CMS:2012kca} for particle (top) and \ptsum\ densities (bottom) for charged particles  with \ptcut\ and \etacut\ in the \tmin\ (left) and \tmax\ (right) regions, as defined by the leading charged particle, as a function of the transverse momentum of the leading charged-particle \ptmax. The data are compared to \MADGRAPH (MG), interfaced to \pynewhyphen\ using \cuePB\ and \cuePM, and to \POWHEG (PH), interfaced to \pynewhyphen\ using \cuePH\ and \cuePM. The bottom panels of each plot show the ratios of these predictions to the data, and the green bands around unity represent the total experimental uncertainty.}
\label{PUB_fig24bis}
\end{center}
\end{figure*}

Tunes obtained from fits to UE data and combined with higher-order ME calculations~\cite{Cooper:2011gk} can also be cross-checked against the data. The CMS UE tunes can be interfaced to higher-order ME generators without spoiling their good description of the UE. In Fig.~\ref{PUB_fig24bis}, the charged-particle and \ptsum\ densities in the \tmin\ and \tmax\ regions as a function of \ptmax, are compared to predictions obtained with \MADGRAPH and \POWHEG~\cite{Nason:2010ap,Alioli:2010xd} interfaced to \pynewhyphen\ using \cuePB\ and \cuePM. In \MADGRAPH , up to four partons are simulated in the final state. The cross section is calculated with the CTEQ6L1 PDF. The ME/PS matching scale is taken to be 10\GeV. The \POWHEG predictions are based on next-to-leading-order (NLO) dijet using the CT10nlo PDF~\cite{Lai:2010vv} interfaced to  \pynewhyphen\ based on \cuePM, and HERAPDF1.5NLO~\cite{Sarkar:2014zua} interfaced to the \pynewhyphen\ using \cuePH . 

The poor agreement below \ptmax $=5\GeV$ in Fig.~\ref{PUB_fig24bis} is not relevant as the minimum \pthat\ for \MADGRAPH and \POWHEG is $5\GeV$.  The agreement with the UE data in the plateau region of \ptmax $> 5\GeV$ is good. All these figures show that CMS UE tunes interfaced to higher-order ME generators do not spoil their good description of the UE data.   

\subsection{Predicting MB observables}

\begin{figure*}[htbp]
\begin{center}
\includegraphics[scale=0.65]{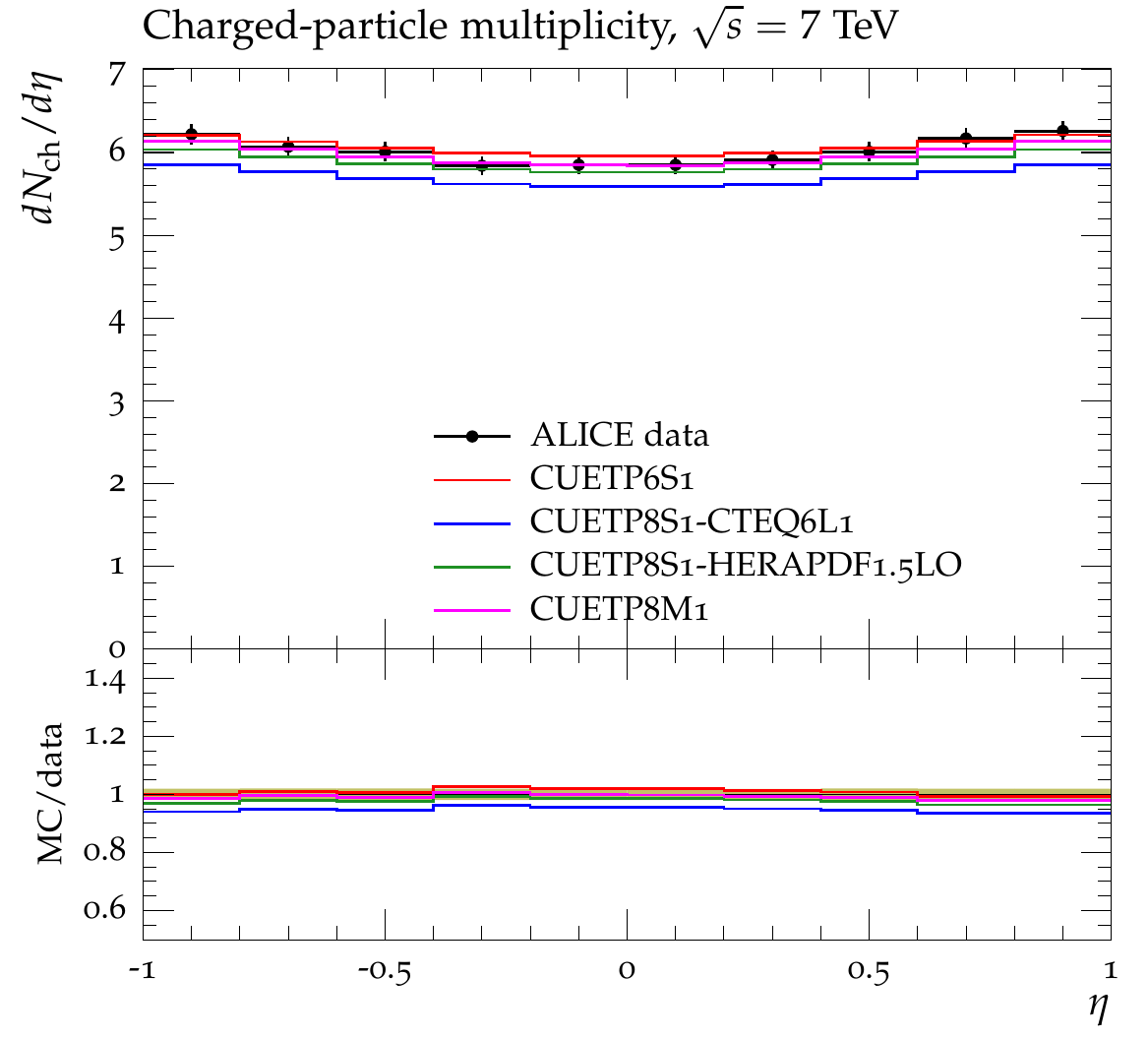}
\includegraphics[scale=0.65]{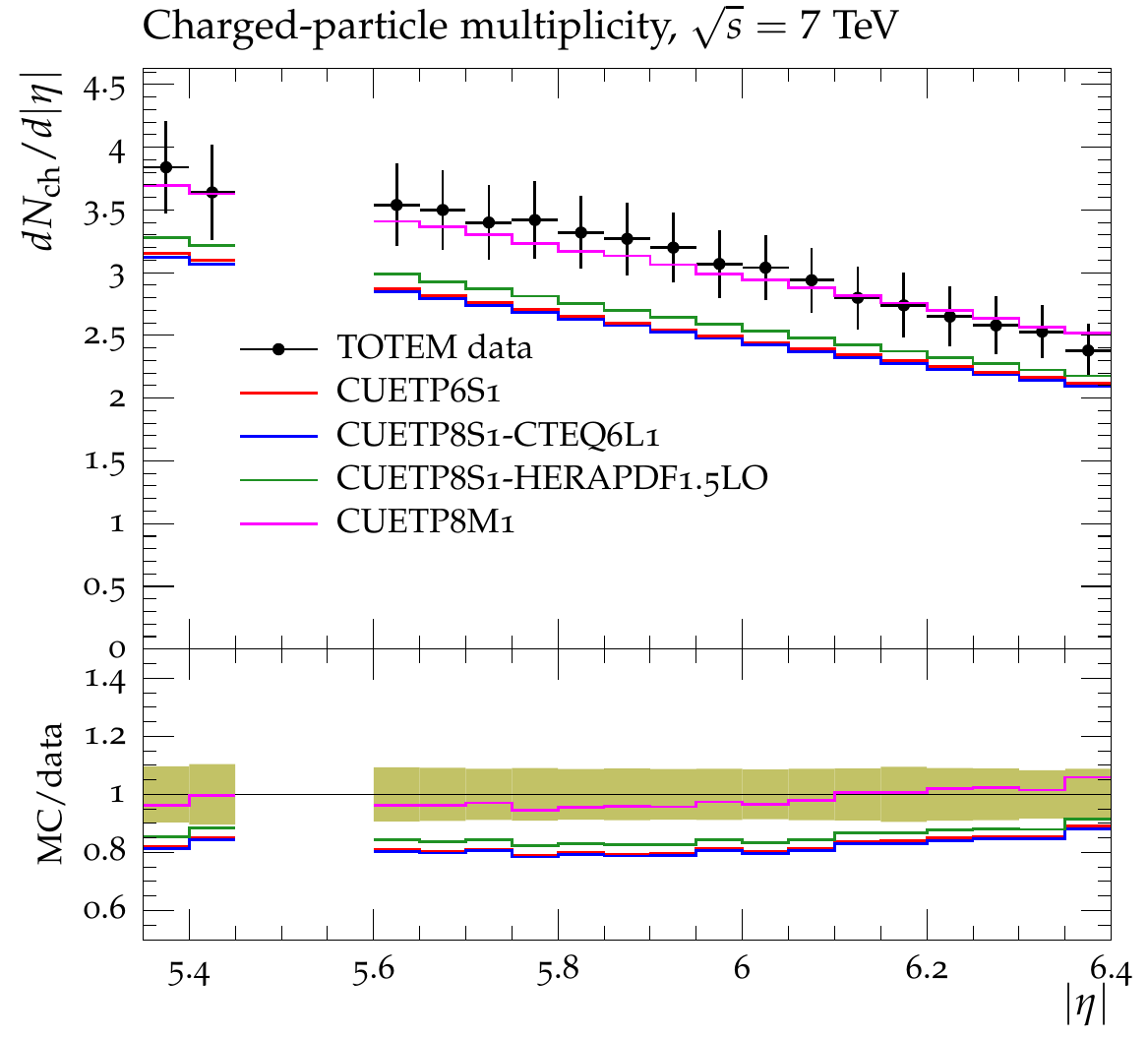}\\
\includegraphics[scale=0.65]{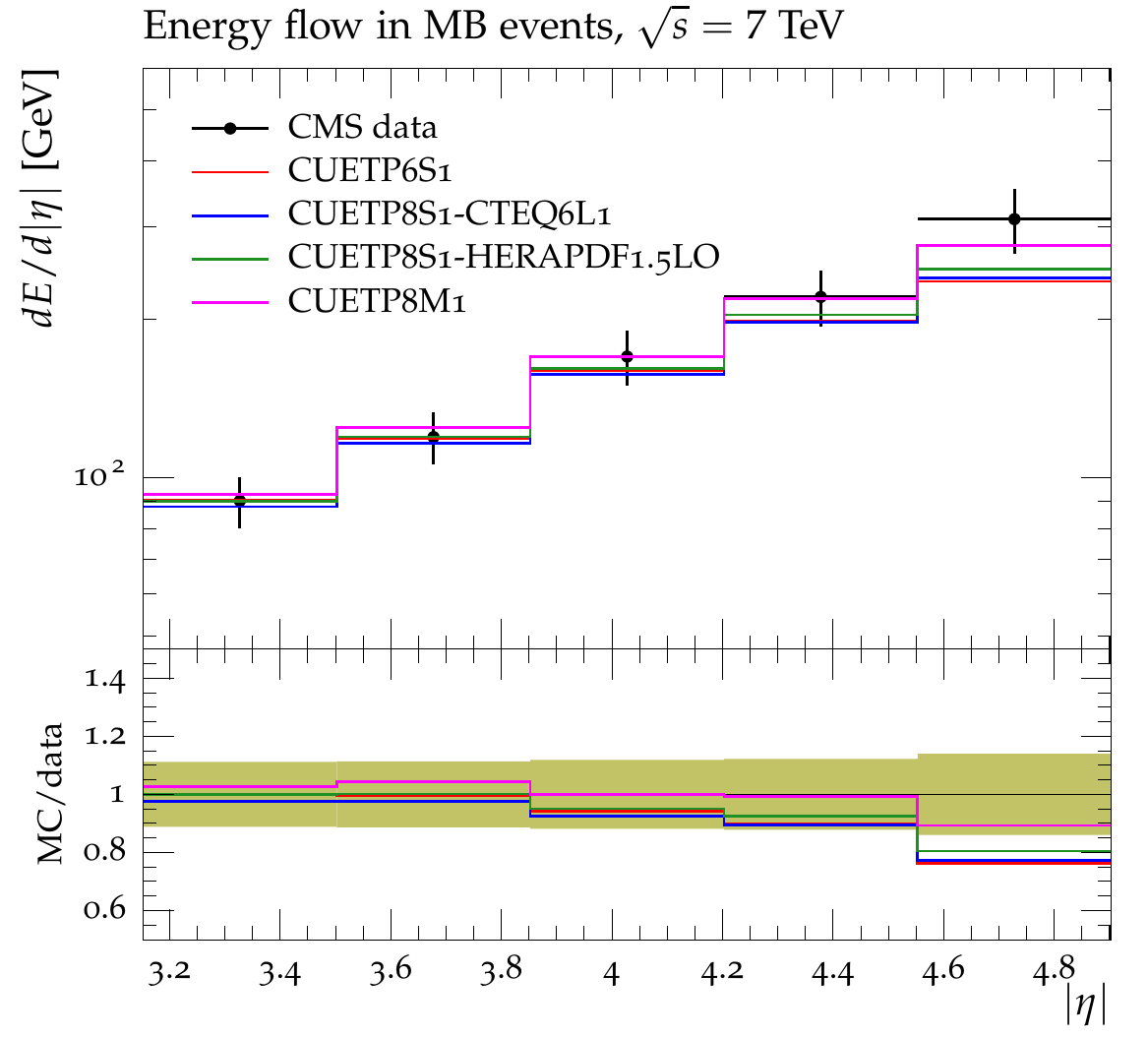}
\caption{ALICE data at $\sqrt{s}=7\TeV$~\cite{Aamodt:2010pp} for the charged-particle pseudorapidity distribution, $\rd \mathrm{N}_{\text{ch}}/ \rd \eta$, in inclusive inelastic $\Pp \Pp$ collisions (top left). TOTEM data at $\sqrt{s}=7\TeV$~\cite{Antchev:2011vs} for the charged-particle pseudorapidity distribution, $\rd \mathrm{N}_{\text{ch}}/ \rd \eta$, in inclusive inelastic pp collisions ($p_{\rm T}>40\MeV$, $\mathrm{N}_{\rm chg}\ge1$) (top right). CMS data at $\sqrt{s}=7\TeV$~\cite{Chatrchyan:2014qka} for the energy flow $\rd E/ \rd \eta$, in MB $\Pp \Pp$ collisions.  The data are compared to  \pyoldhyphen\ using \cuePA, and to \pynewhyphen\ using \cuePB, \cuePH, and \cuePM. The bottom panels of each plot show the ratios of these predictions to the data, and the green bands around unity represent the total experimental uncertainty.}
\label{PUB_fig24}
\end{center}
\end{figure*}

\begin{figure*}[htbp]
\begin{center}
\includegraphics[scale=0.65]{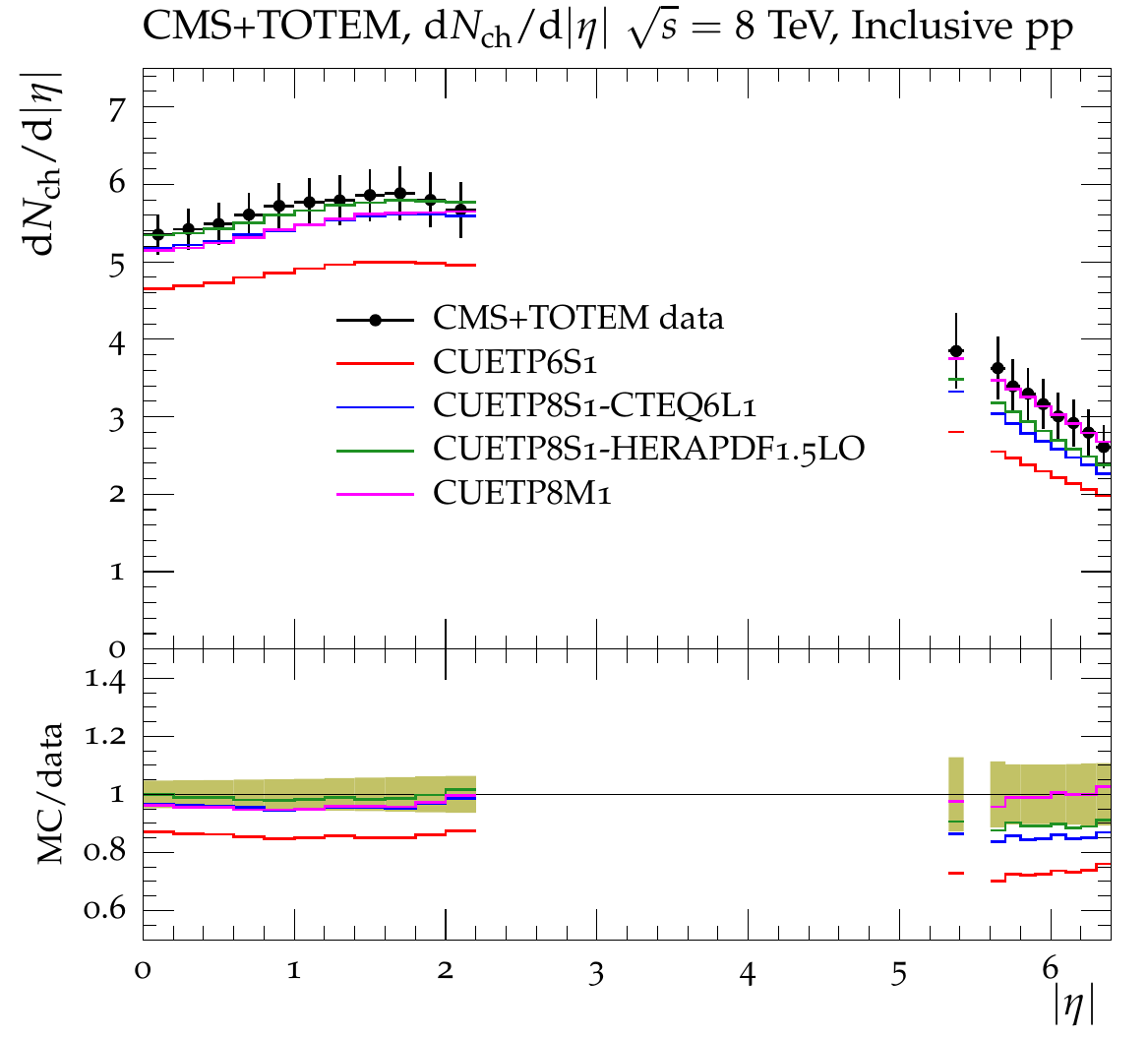}
\includegraphics[scale=0.65]{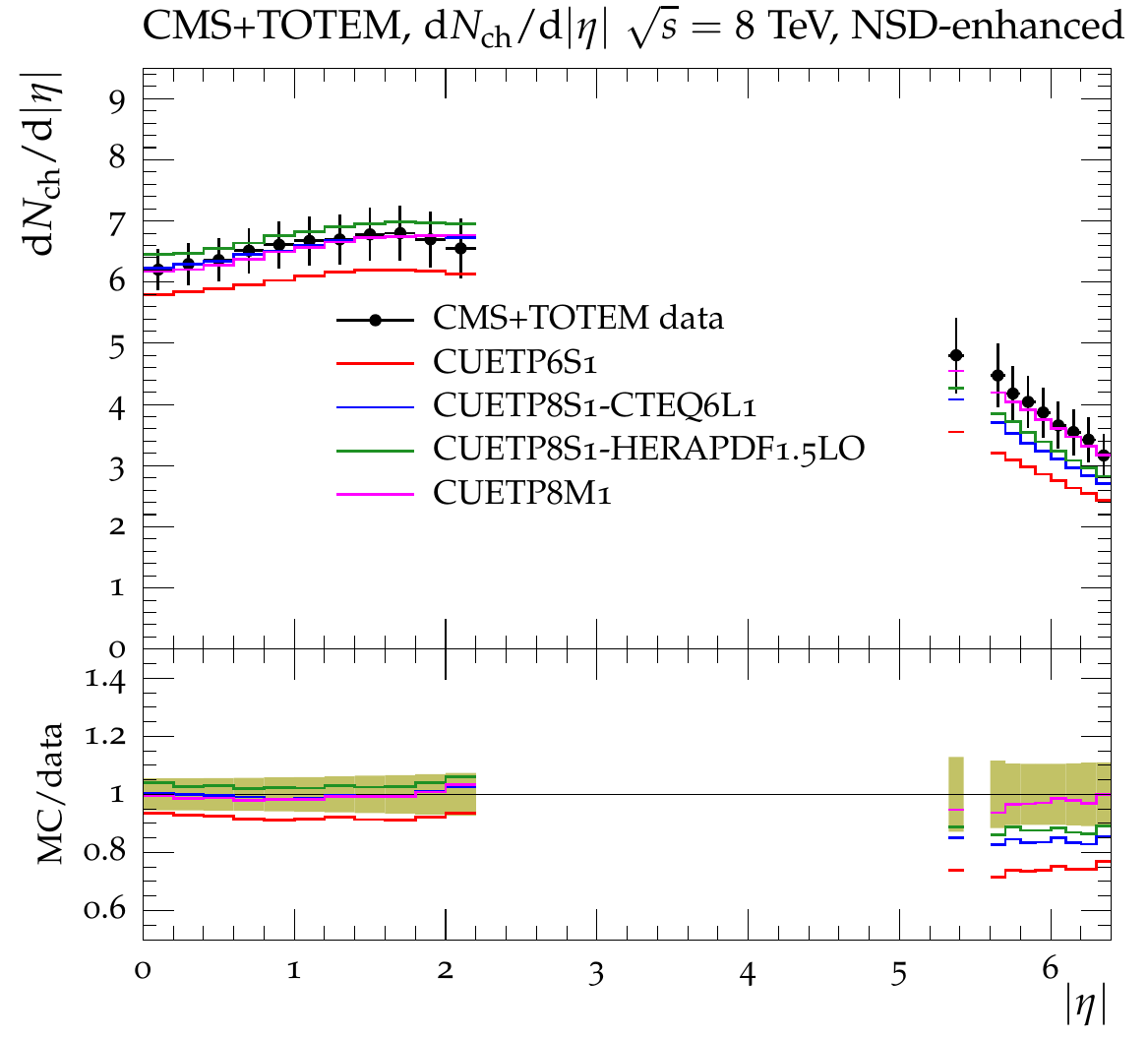}\\
\includegraphics[scale=0.65]{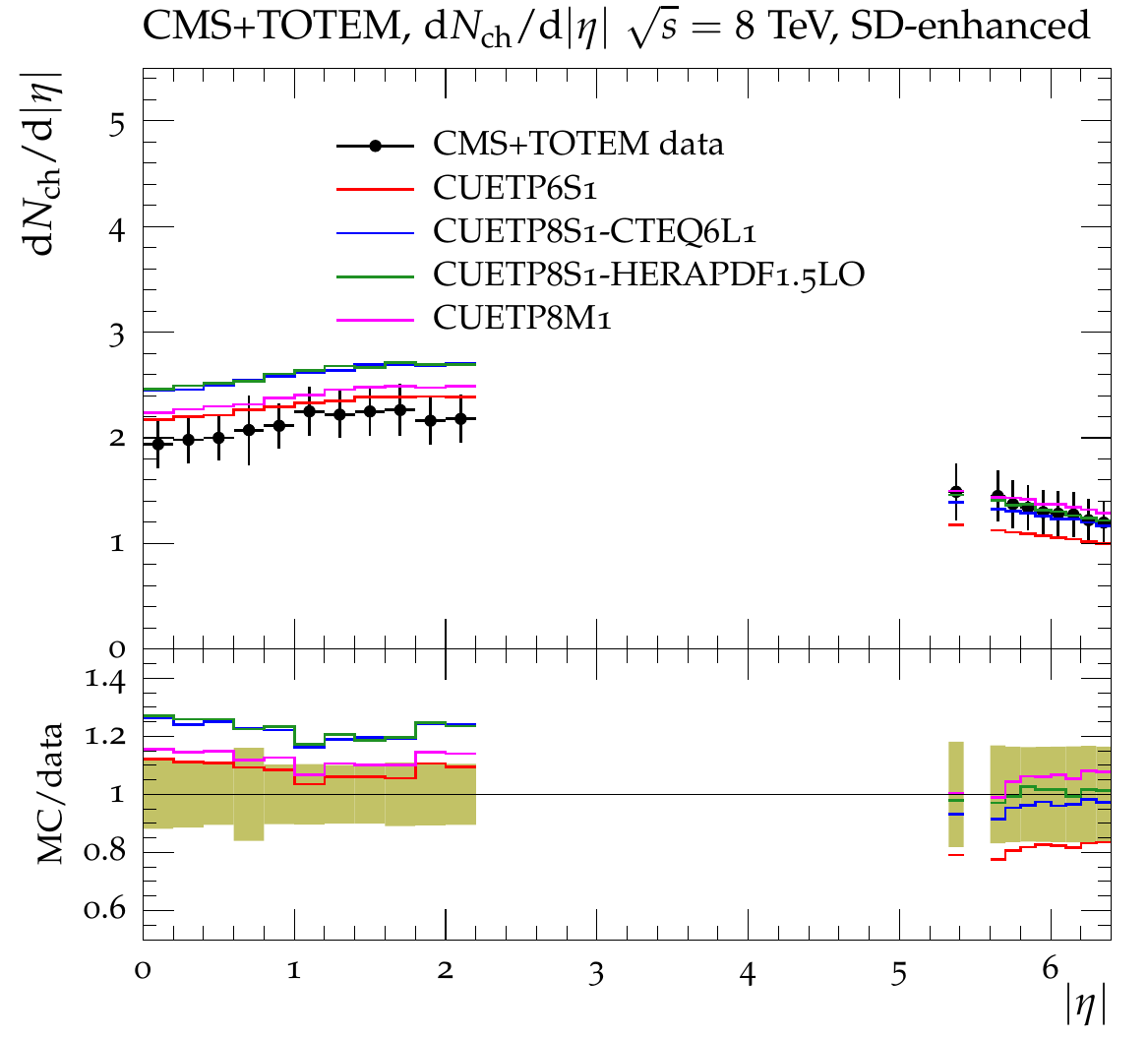}
\caption{Combined CMS and TOTEM data at $\sqrt{s}=8\TeV$~\cite{Chatrchyan:2014qka} for the charged-particle distribution $\rd \mathrm{N}_{\text{ch}}/ \rd \eta$, in inclusive inelastic (top left), NSD-enhanced (top right), and SD-enhanced (bottom) pp collisions. The data are compared to  \pyoldhyphen\ using \cuePA, and to \pynewhyphen\ using \cuePB, \cuePH, and \cuePM. The bottom panels of each plot show the ratios of these predictions to the data, and the green bands around unity represent the total experimental uncertainty.}
\label{PUB_fig25}
\end{center}
\end{figure*}

{\tolerance=5000
The UE is studied in events containing a hard scatter, whereas most of the MB collisions are softer and can include diffractive scatterings. It is however interesting to see how well predictions based on the CMS UE tunes can describe the properties of MB distributions. Figure~\ref{PUB_fig24} shows predictions using CMS UE tunes for the ALICE~\cite{Aamodt:2010pp} and TOTEM data~\cite{Antchev:2011vs} at $\sqrt{s}=7\TeV$ for the charged-particle pseudorapidity distribution, $\rd \mathrm{N}_{\text{ch}}/\rd \eta$, and for $\rd E/\rd \eta$~\cite{Chatrchyan:2011wm} at $\sqrt{s}=7\TeV$.  These observables are sensitive to single-diffraction dissociation, central-diffraction, and double-diffraction dissociation, which are modelled in \PYTHIA.  Since \hwpp\ does not include a model for single-diffraction dissociation, central-diffraction, and double-diffraction dissociation, we do not show it here. Figure~\ref{PUB_fig25} shows predictions using the CMS UE tunes for the combined CMS$+$TOTEM data at $\sqrt{s}=8\TeV$~\cite{Chatrchyan:2014qka} for the charged-particle pseudorapidity distribution, $\rd \mathrm{N}_{\text{ch}}/\rd \eta$, for inelastic, non single-diffraction-enhanced, and single-diffraction-enhanced proton-proton collisions. 
\par}

\begin{figure*}[htbp]
\begin{center}
\includegraphics[trim=0cm 13.5cm 0cm 0cm,scale=0.58]{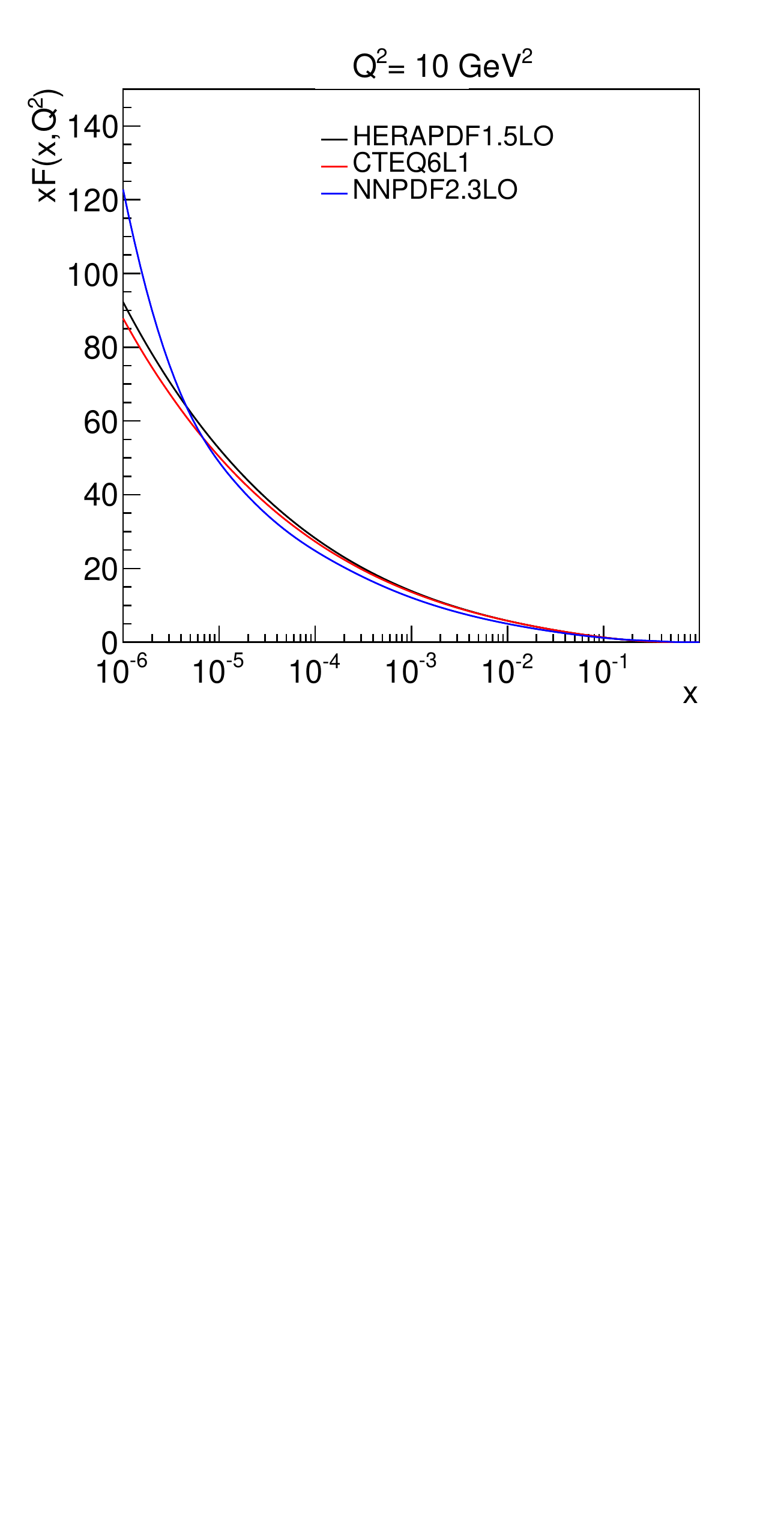}
\includegraphics[trim=0cm 13.5cm 0cm 0cm,scale=0.58]{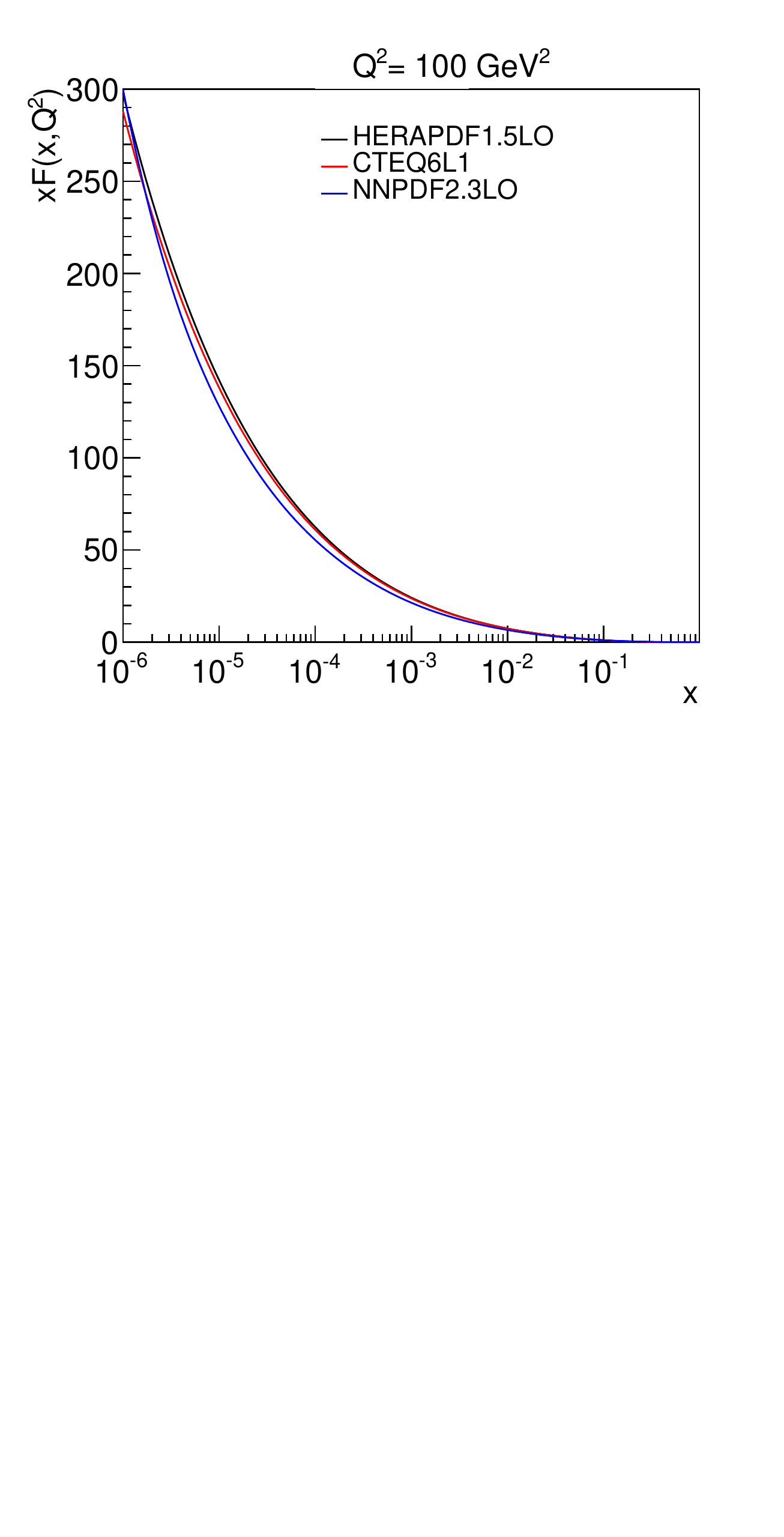}
\caption{Comparison of gluon distributions in the proton for the CTEQ6L1, HERAPDF1.5LO, and NNPDF2.3LO PDF sets, at the $Q^2$ = 10 GeV$^2$ (left) and 100 GeV$^2$ (right).}
\label{PUB_fig26gluon}
\end{center}
\end{figure*}

The \pynewhyphen\ event generator using the UE tunes describes the MB data better than \pyold\ with the UE tune, which is likely due to the improved modelling of single-diffraction dissociation, central-diffraction, and double-diffraction dissociation in \pynew.  Predictions with all the UE tunes describe fairly well MB observables in the central region ($|\eta|<2$), however, only predictions obtained with \cuePM\ describe the data in the forward region ($|\eta|>4$).  This is due to the PDF used in \cuePM.  As can be seen in Fig.~\ref{PUB_fig26gluon}, the NNPDF$2.3$LO PDF at scales $Q^2$ = 10 GeV$^2$ (corresponding to hard scatterings with \pthat\ $\sim$ 3 GeV) and small $x$, features a larger gluon density than in CTEQ$6$L1 and \hera , thereby contributing to more particles (and more energy) produced in the forward region. We have checked that increasing the gluon distribution in \hera\ at values below 10$^{-5}$ improved the description of the charged-particle multiplicity measurements in the forward region. 

\subsection{Comparisons with inclusive jet production}

\begin{figure*}[htbp]
\begin{center}
\includegraphics[scale=0.6]{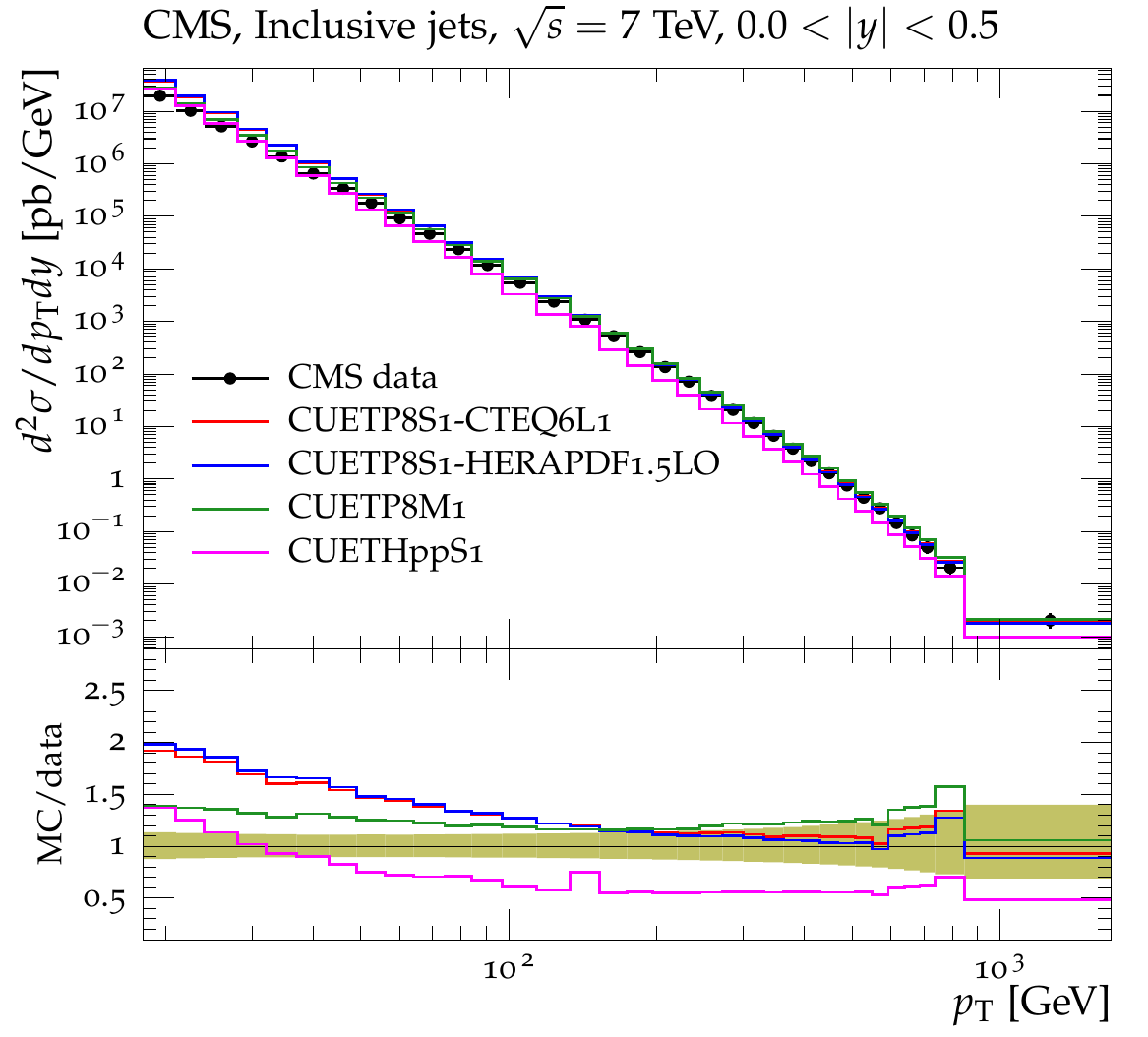}
\includegraphics[scale=0.6]{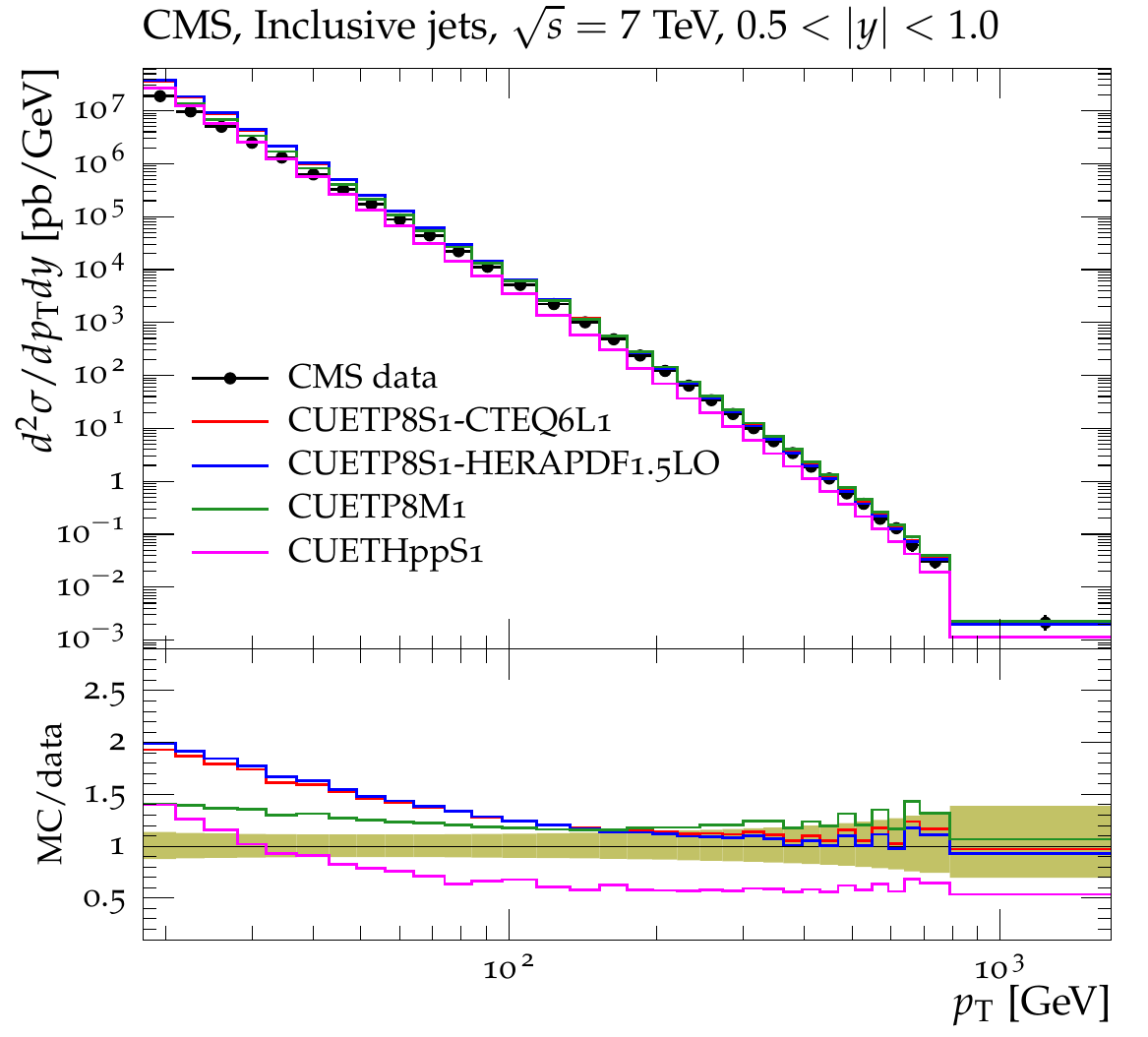}\\
\includegraphics[scale=0.6]{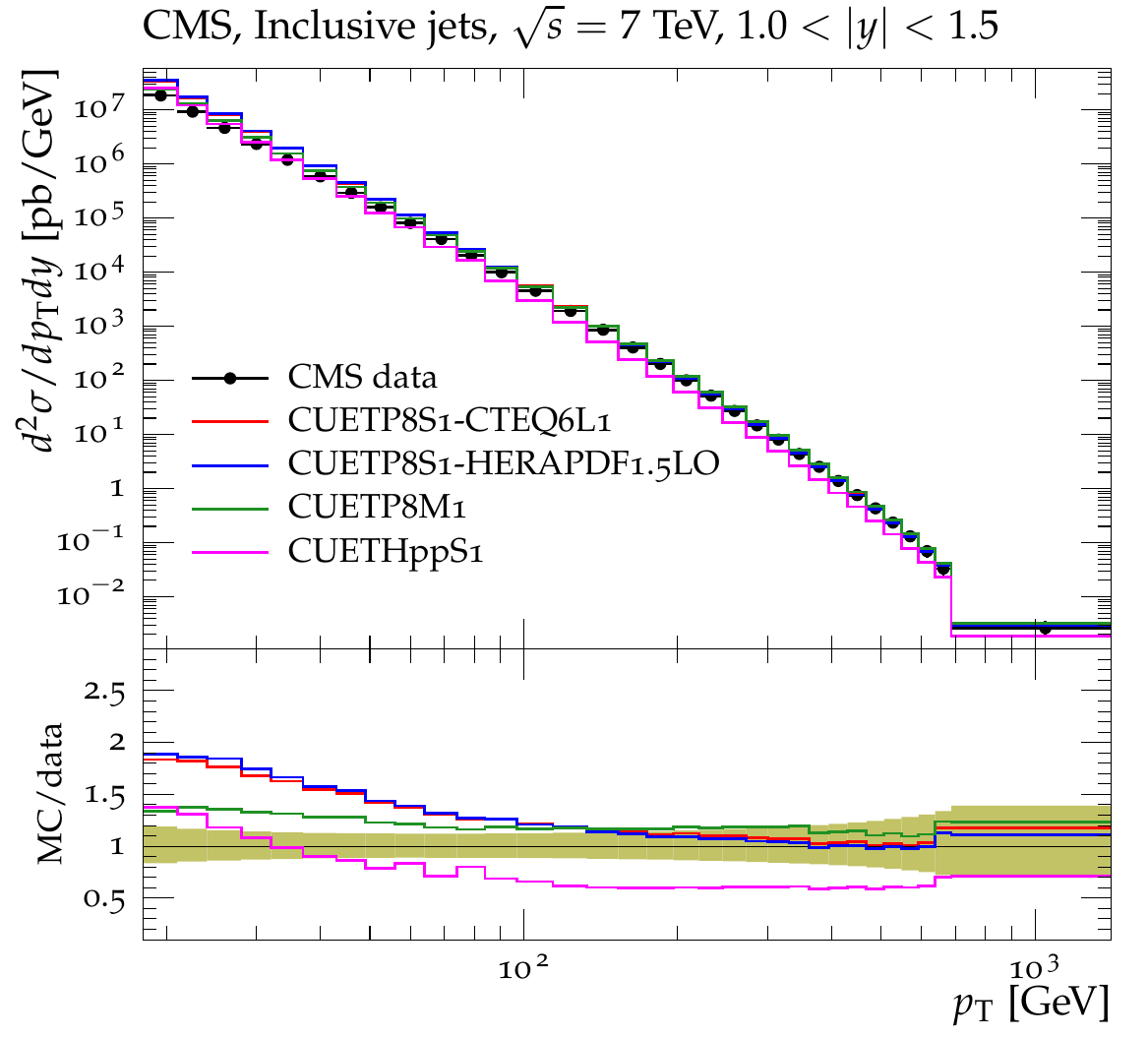}
\includegraphics[scale=0.6]{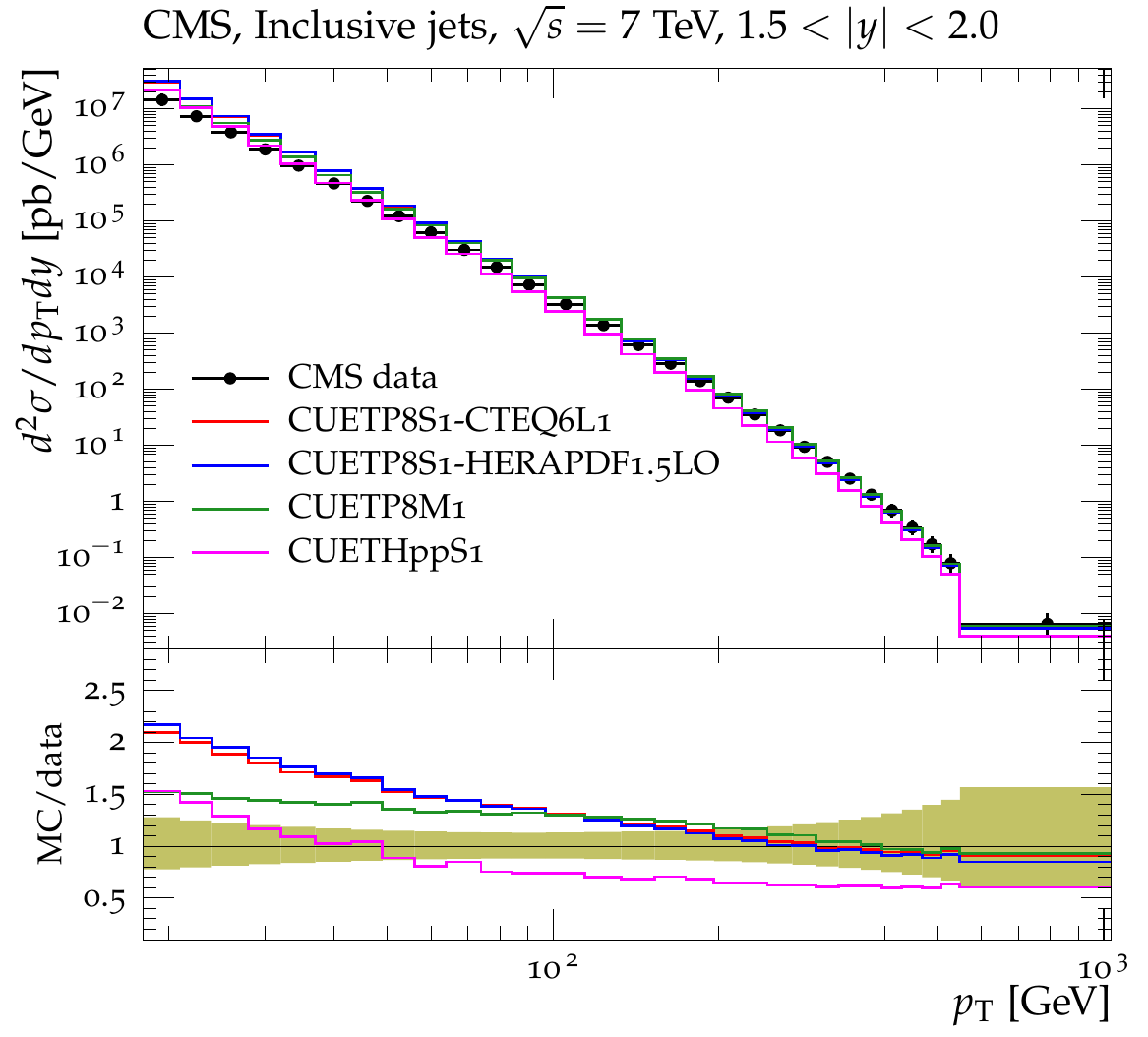}\\
\includegraphics[scale=0.6]{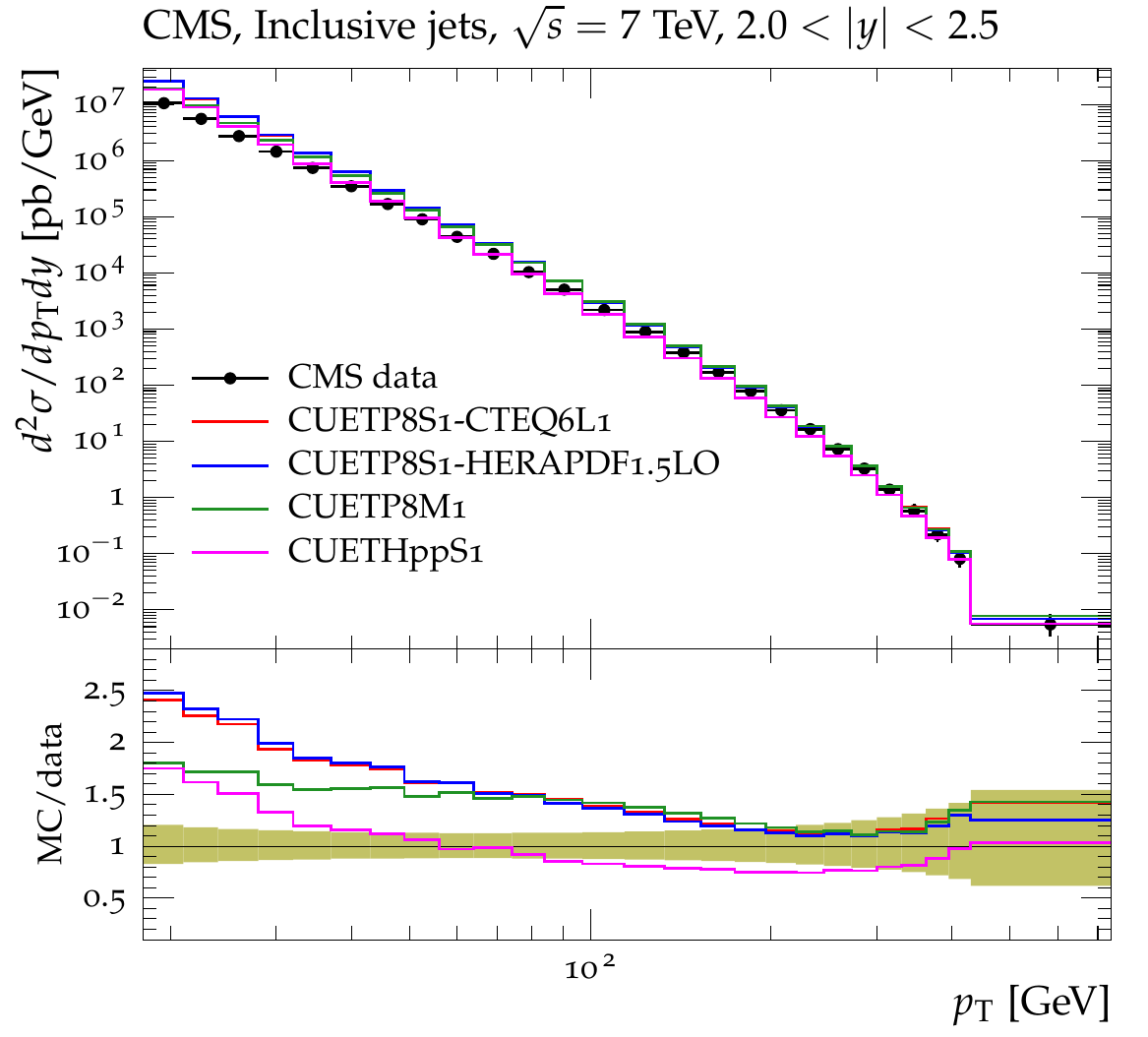}
\includegraphics[scale=0.6]{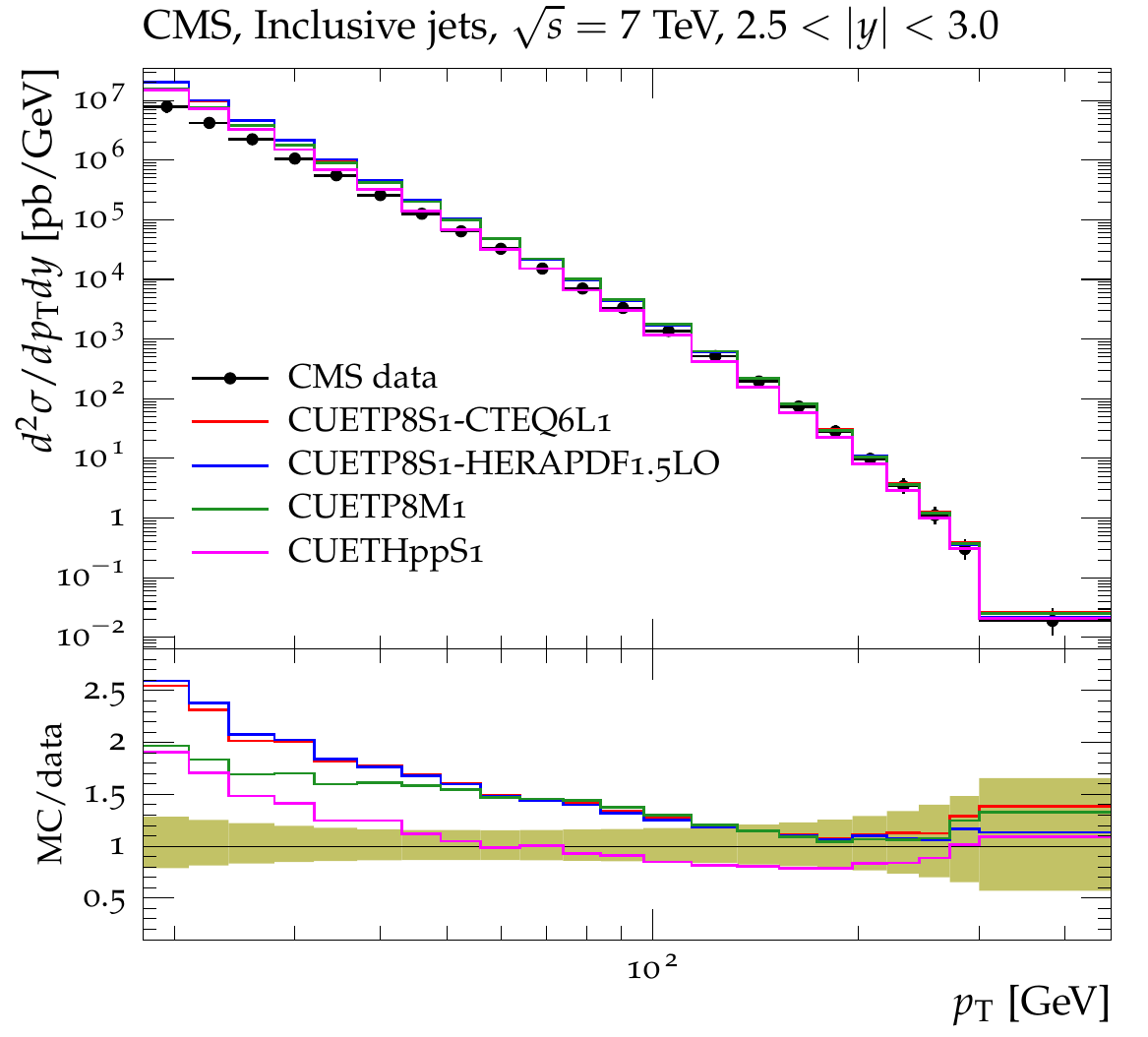}
\caption{CMS data at $\sqrt{s}=7\TeV$~\cite{CMS:2011ab} for the inclusive jet cross section as a function of \PT in different rapidity ranges  compared to predictions of \textsc{pythia8} using CUETP8S1-CTEQ6L1, CUETP8S1-HERAPDF, and CUETP8M1, and of \textsc{herwig++} using CUETHppS1. The bottom panels of each plot show the ratios of these predictions to the data, and the green bands around unity represent the total experimental uncertainty.}
\label{PUB_fig27}
\end{center}
\end{figure*}

\begin{figure*}[htbp]
\begin{center}
\includegraphics[scale=0.6]{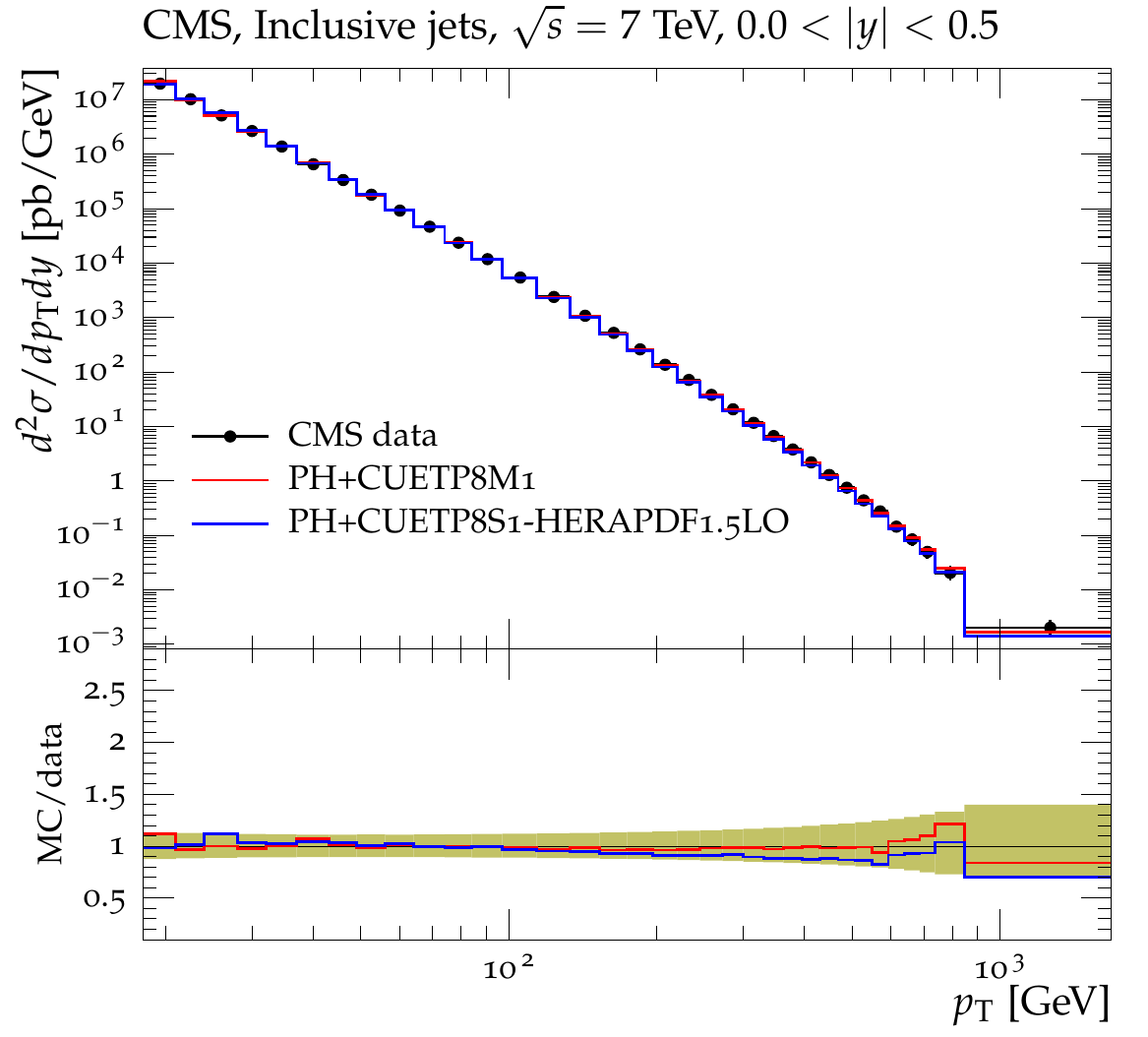}
\includegraphics[scale=0.6]{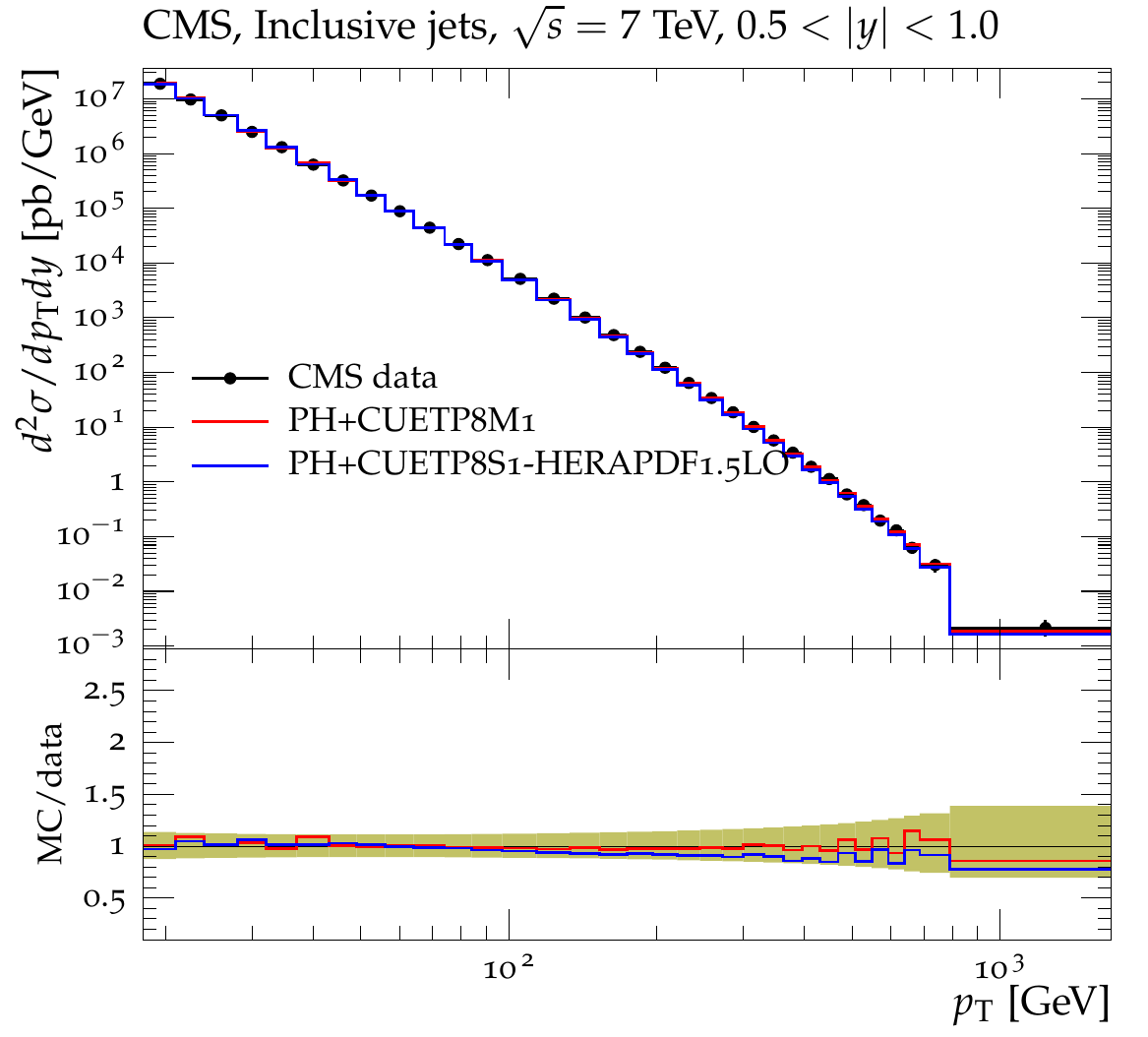}\\
\includegraphics[scale=0.6]{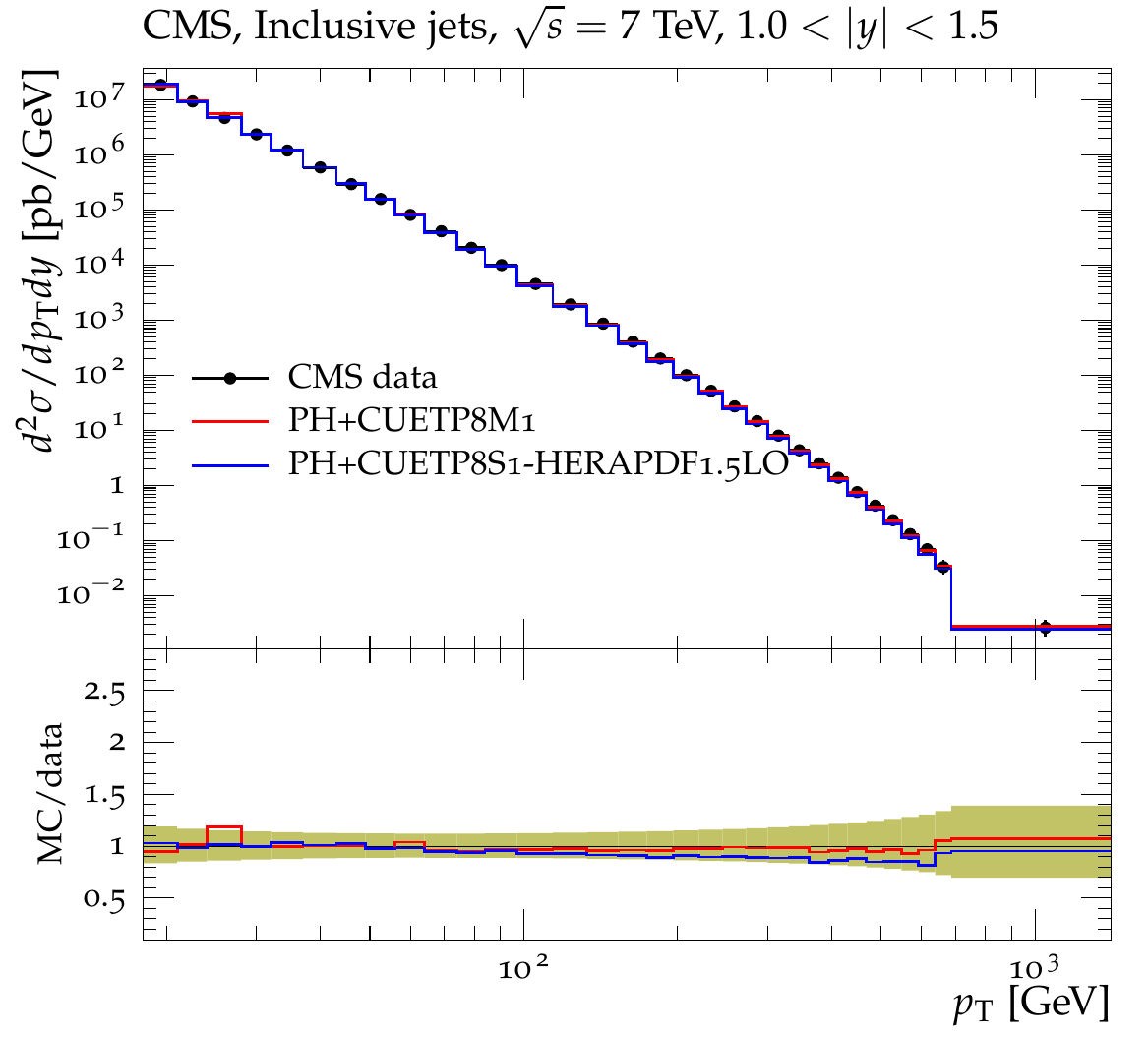}
\includegraphics[scale=0.6]{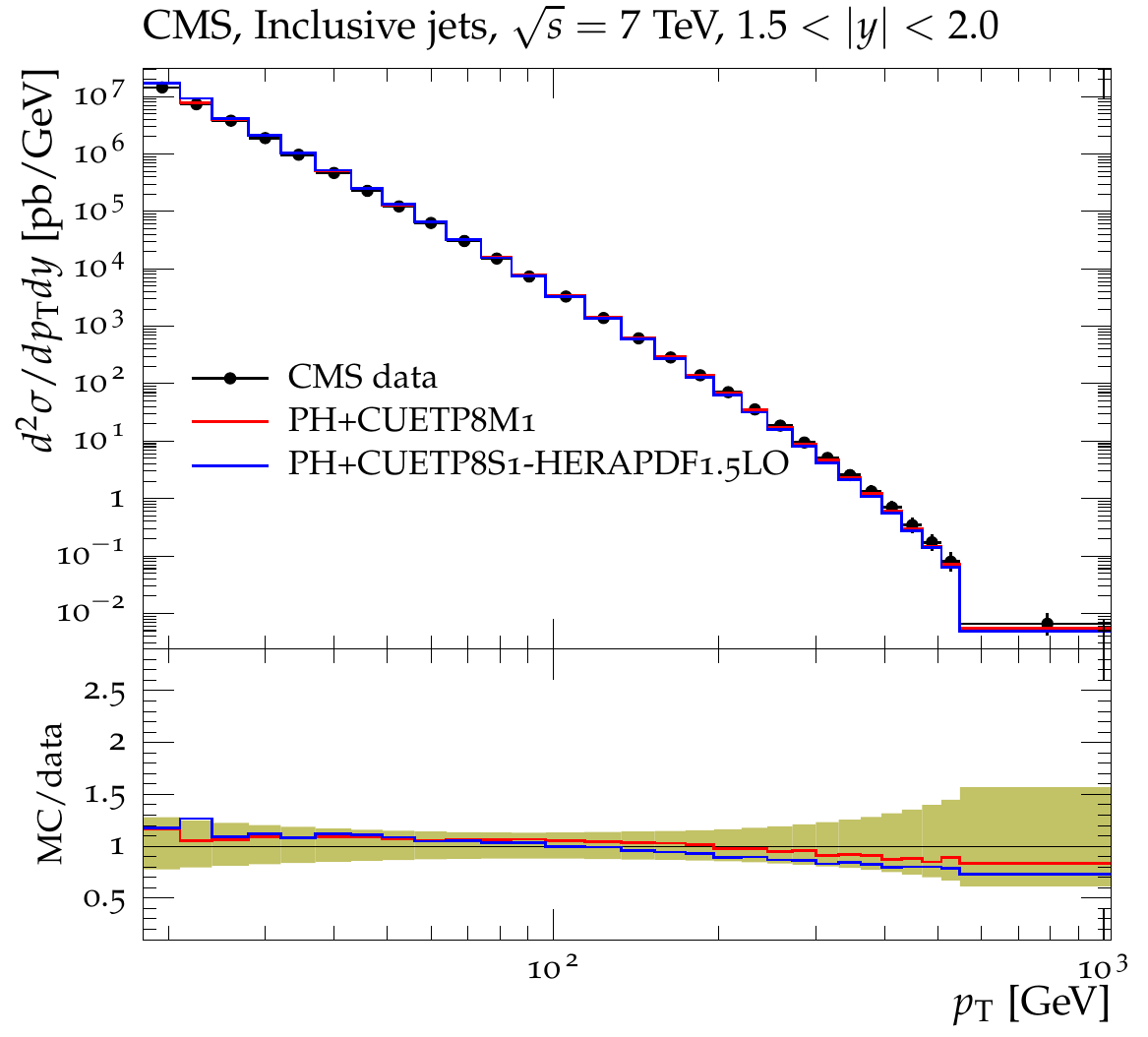}\\
\includegraphics[scale=0.6]{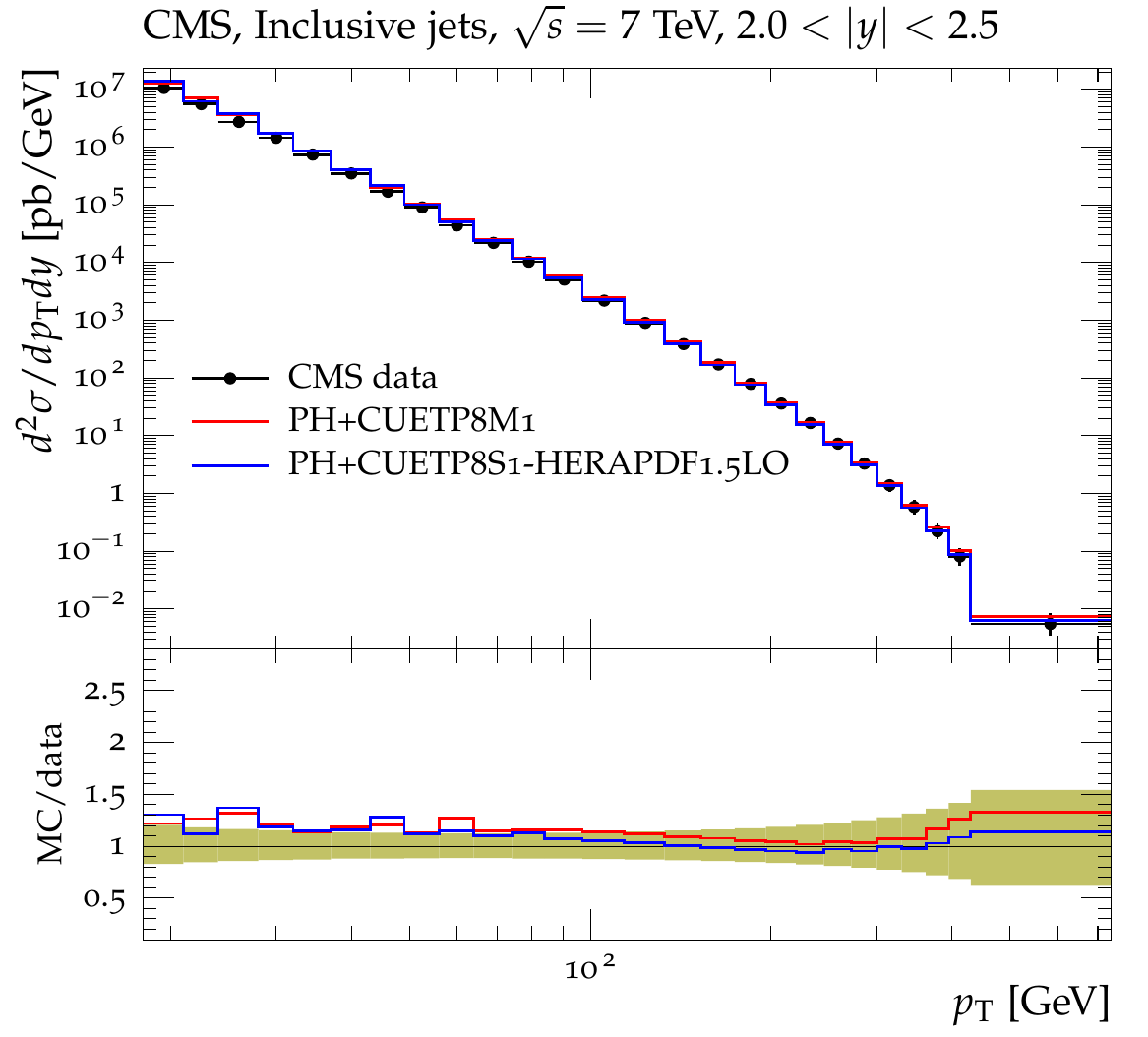}
\includegraphics[scale=0.6]{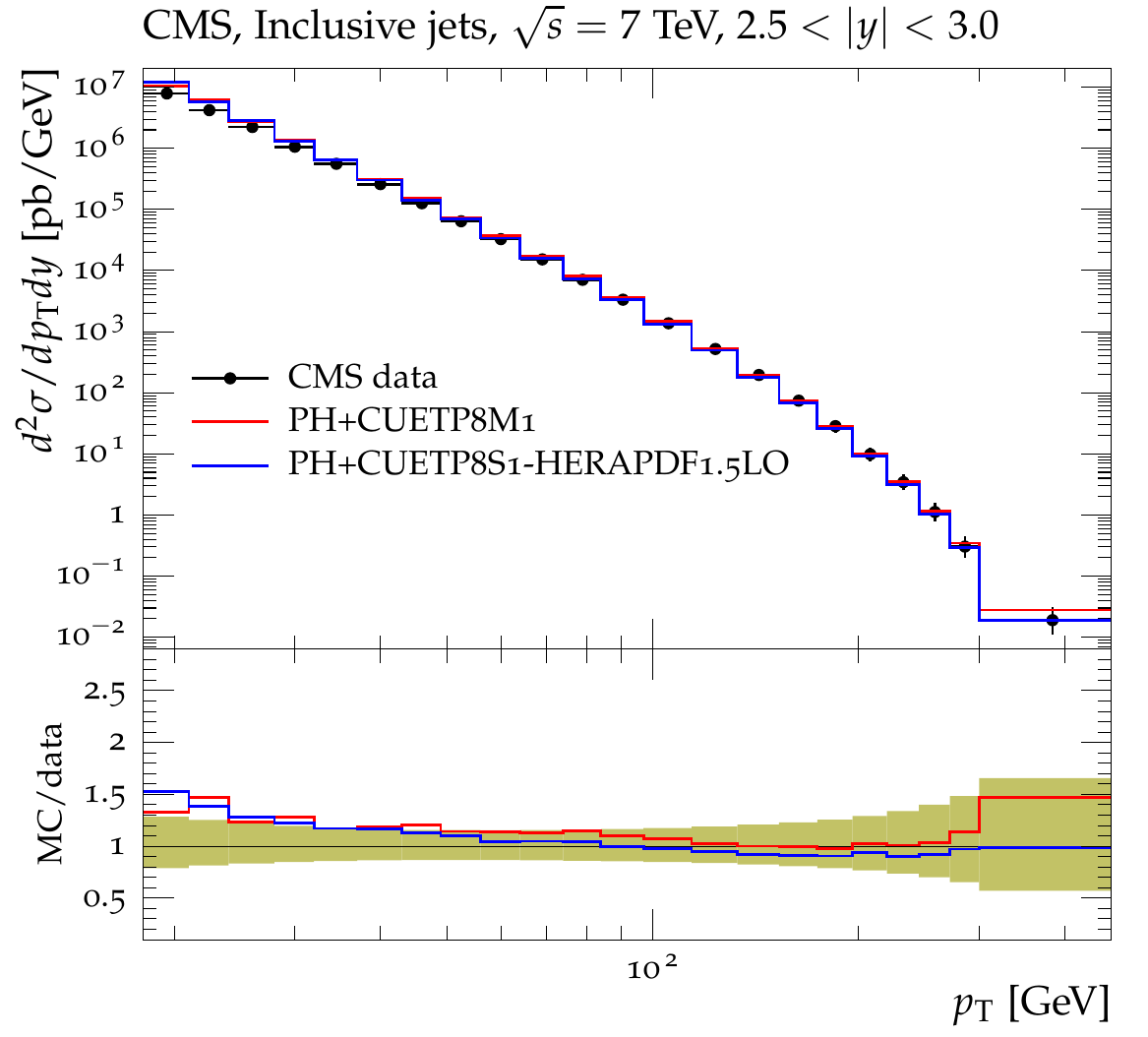}
\caption{CMS data at $\sqrt{s}=7\TeV$~\cite{CMS:2011ab} for the inclusive jet cross section as a function of \PT in different rapidity ranges  compared to predictions of \POWHEG interfaced to \pynewhyphen\ using \cuePH\ and \cuePM. The bottom panels of each plot show the ratios of these predictions to the data, and the green bands around unity represent the total experimental uncertainty.}
\label{PUB_fig27bis}
\end{center}
\end{figure*}

{\tolerance=5000
In Fig.~\ref{PUB_fig27} predictions using \cuePB, \cuePH, and \cuePM, and CUETHppS1 are compared to inclusive jet cross section at $\sqrt{s}=7\TeV$~\cite{CMS:2011ab} in several rapidity ranges. Predictions using \cuePM\ describe the data best, however, all the tunes overshoot the jet spectra at small \pt. Predictions from the CUETHppS1 underestimate the high $p_{\rm T}$ region at central rapidity ($|y|$ $<$ 2.0). In Fig.~\ref{PUB_fig27bis}, the inclusive jet cross sections are compared to predictions from \POWHEG interfaced to \pynewhyphen\ using \cuePH\ and \cuePM. A very good description of the measurement is obtained.
\par}

\subsection{Comparisons with Z boson production}

\begin{figure*}[htbp]
\begin{center}
\includegraphics[scale=0.65]{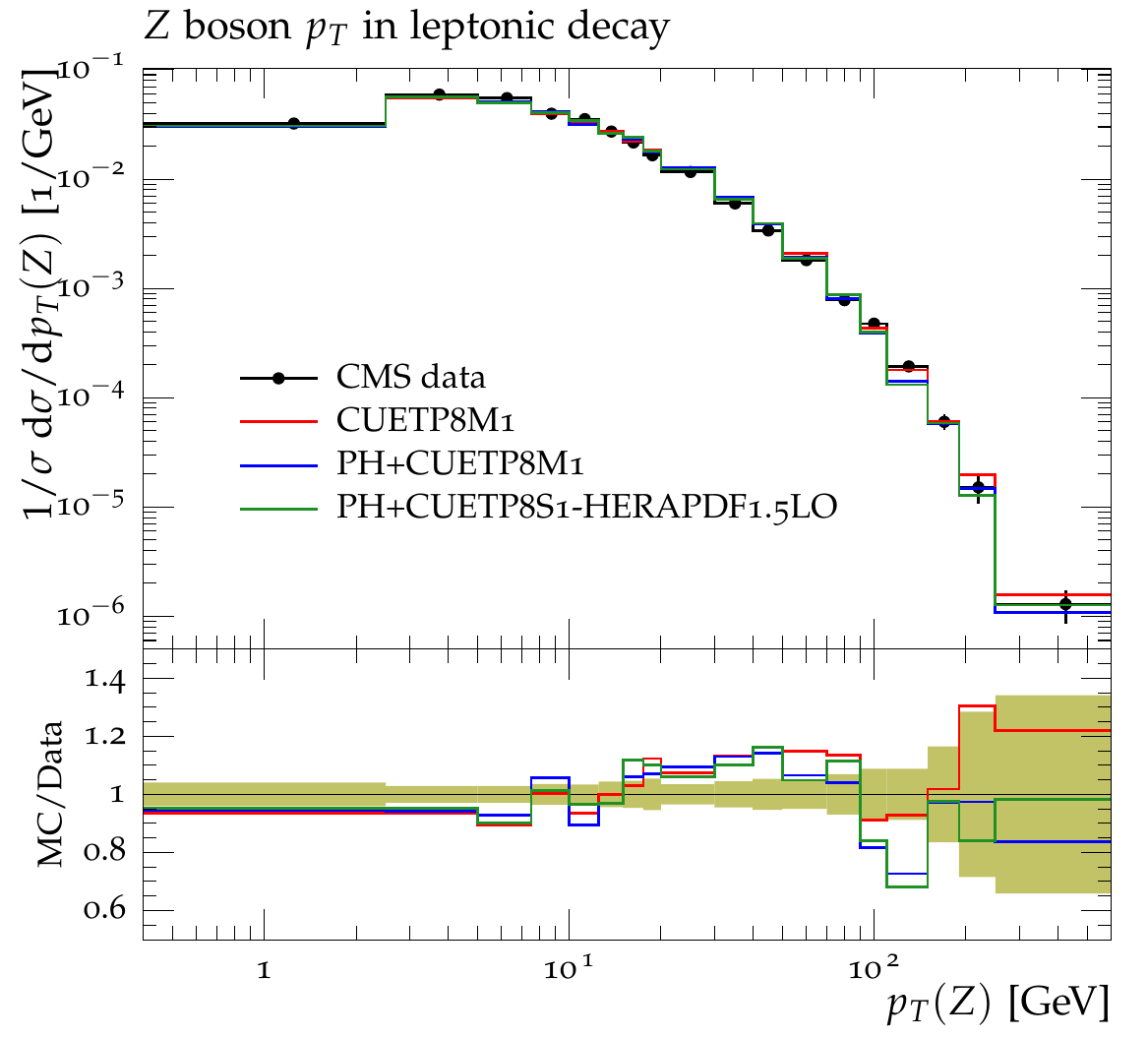}
\includegraphics[scale=0.65]{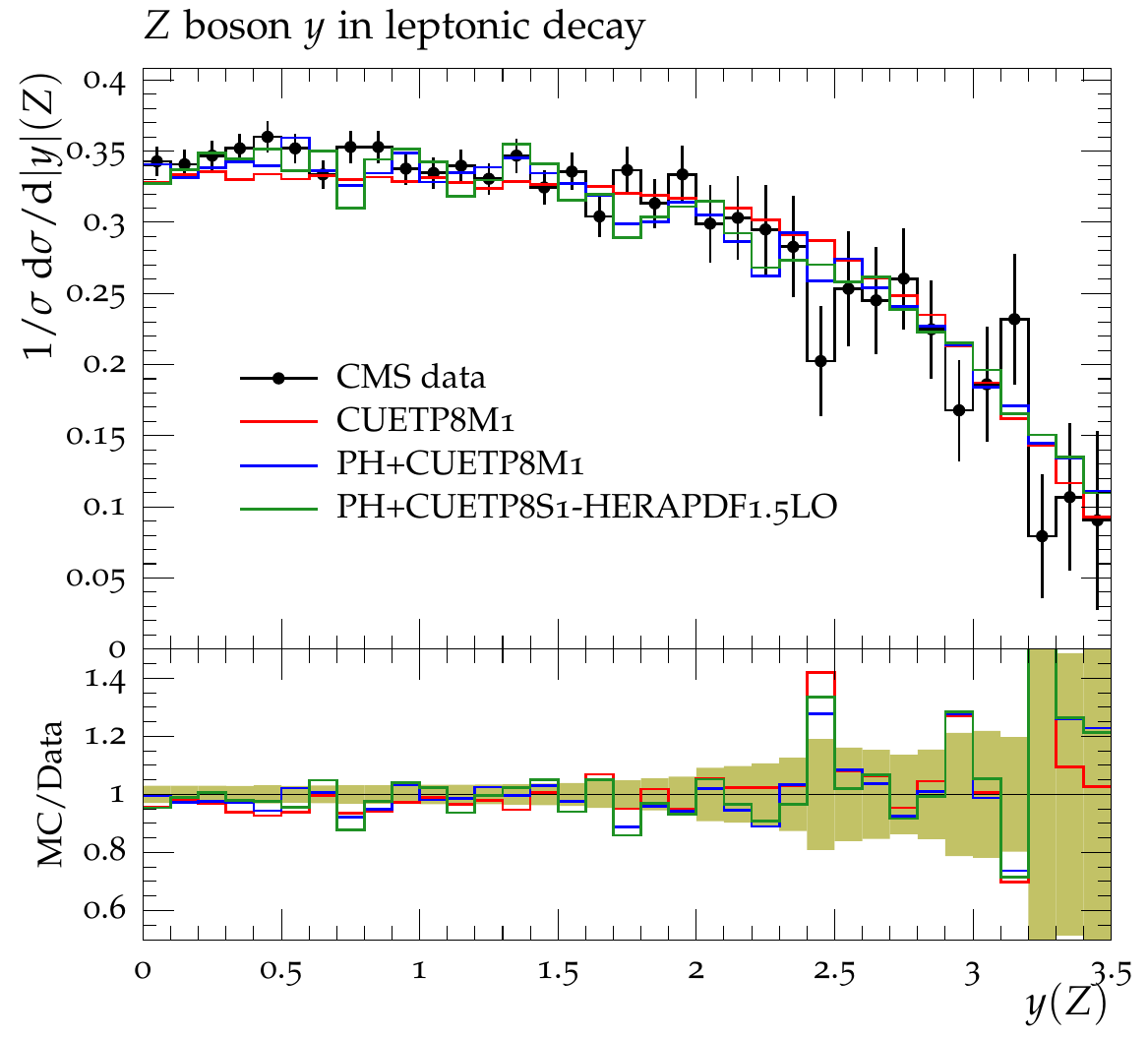}\\
\caption{Transverse momentum \pt\ (left) and rapidity distributions (right) of $\PZ$ boson production in pp collisions at $\sqrt{s}=7\TeV$~\cite{Chatrchyan:2011wt}. The data are compared to \pynewhyphen\ using \cuePM, and to \POWHEG interfaced to  \pynewhyphen\ using \cuePB\ and \cuePM. The green bands in the ratios represent the total experimental uncertainty.}
\label{PUB_fig28}
\end{center}
\end{figure*}

{\tolerance=5000
In Fig.~\ref{PUB_fig28} 
the \pt\ and rapidity distributions of the $\PZ$ boson in pp collisions at $\sqrt{s}=7\TeV$~\cite{Chatrchyan:2011wt} are shown and compared to  \pynewhyphen\ using \cuePM, and to \POWHEG interfaced to  \pynewhyphen\ using \cuePB\ and \cuePM.  The prediction using \pynewhyphen\ with \cuePM\ (without \POWHEG ) agrees reasonably well with the distribution of the $\PZ$ boson at small \pt\ values. Also, when interfaced to \POWHEG, which implements an inclusive $\PZ$ boson NLO calculation, the agreement is good over the whole spectrum.
\par}

\begin{figure*}[htbp]
\begin{center}
\includegraphics[scale=0.6]{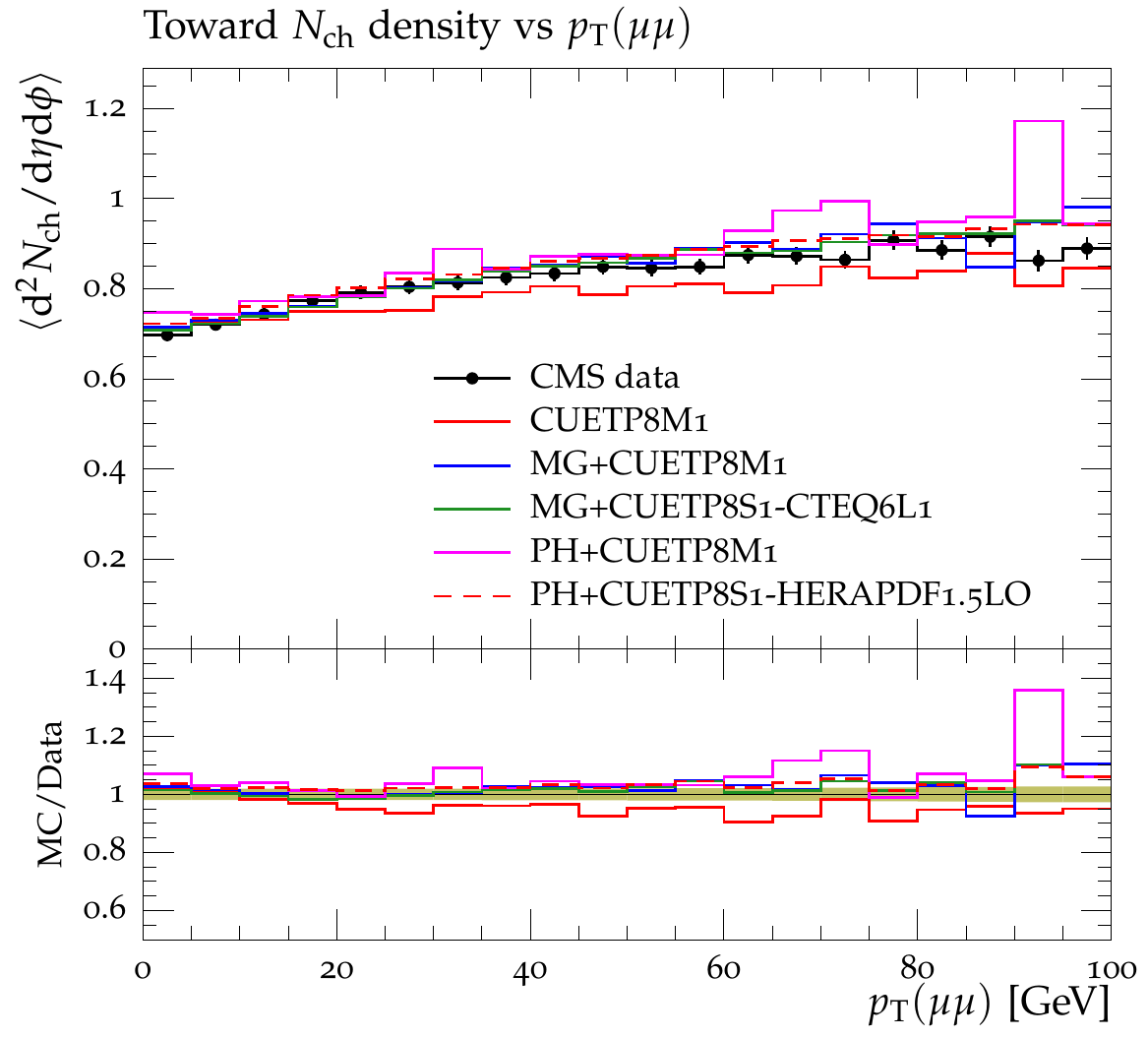}
\includegraphics[scale=0.6]{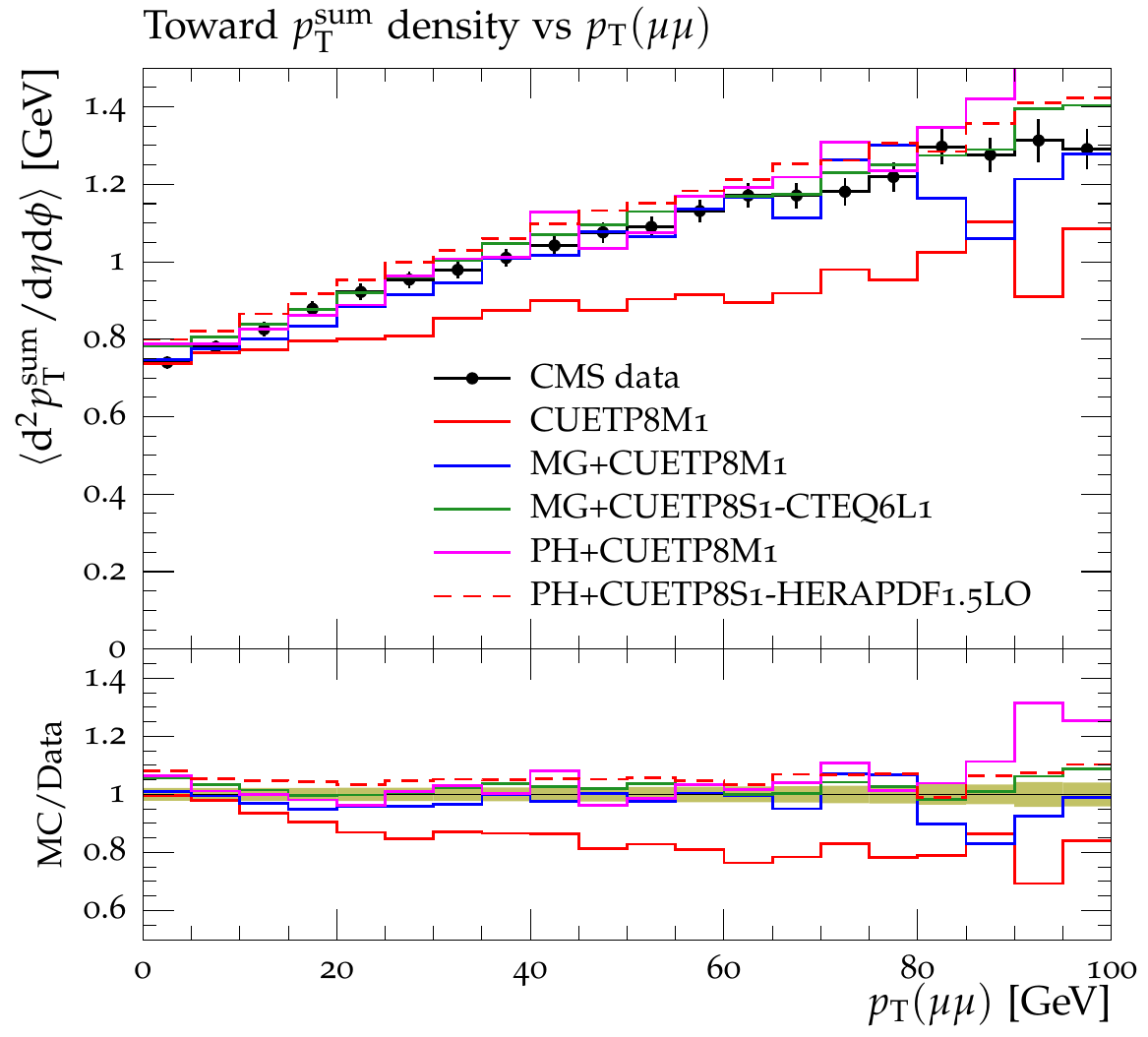}\\
\includegraphics[scale=0.6]{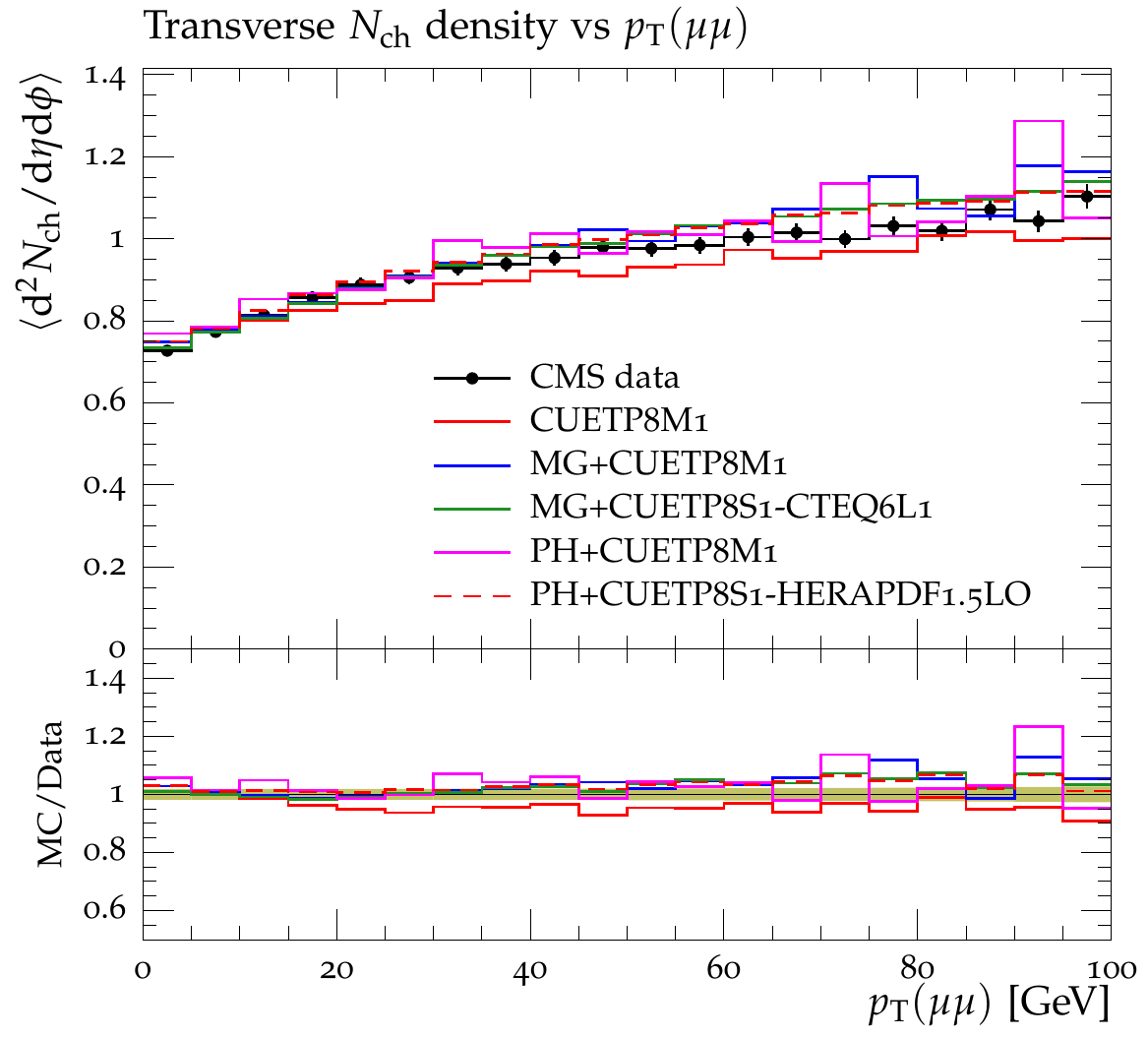}
\includegraphics[scale=0.6]{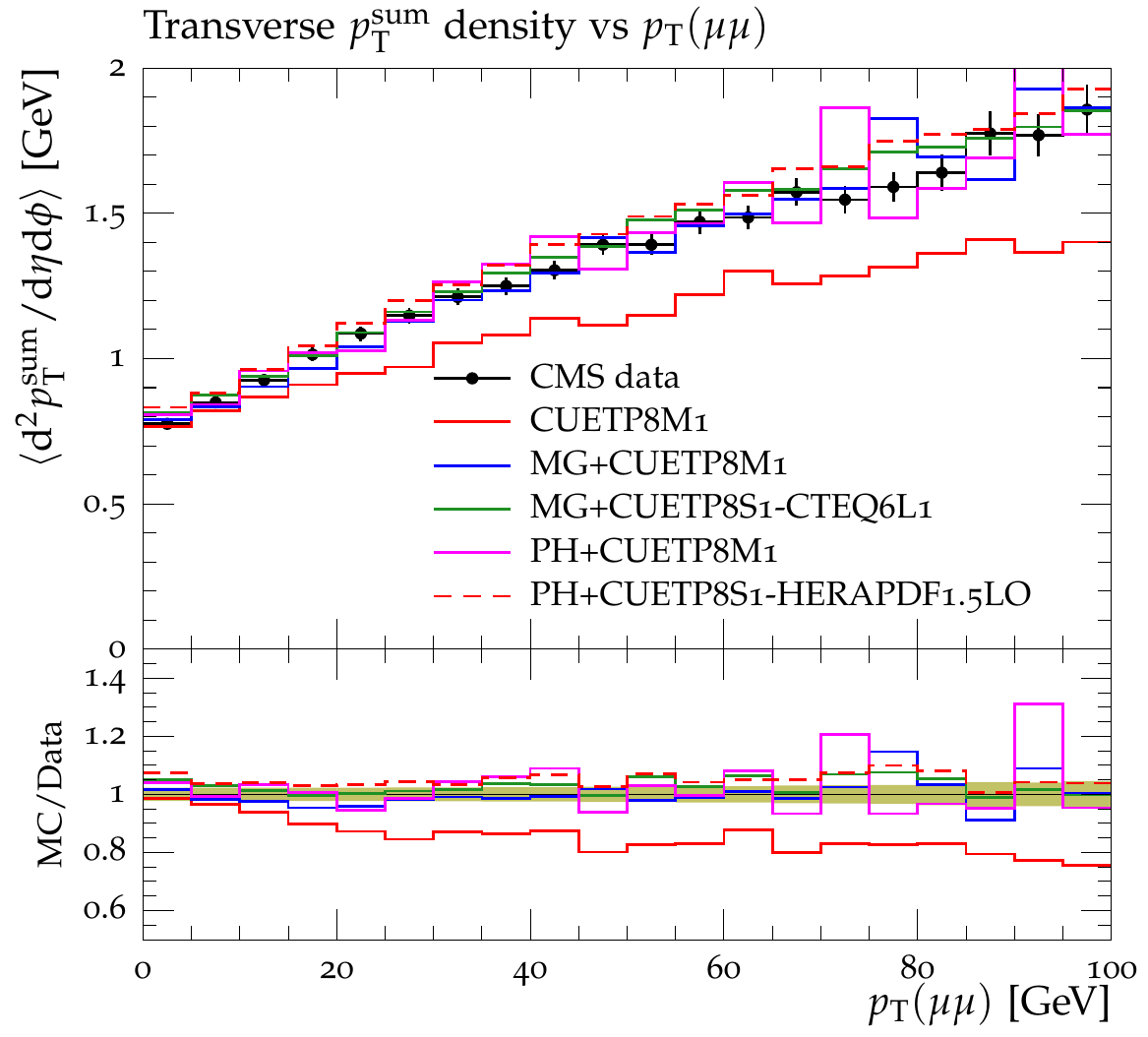}\\
\includegraphics[scale=0.6]{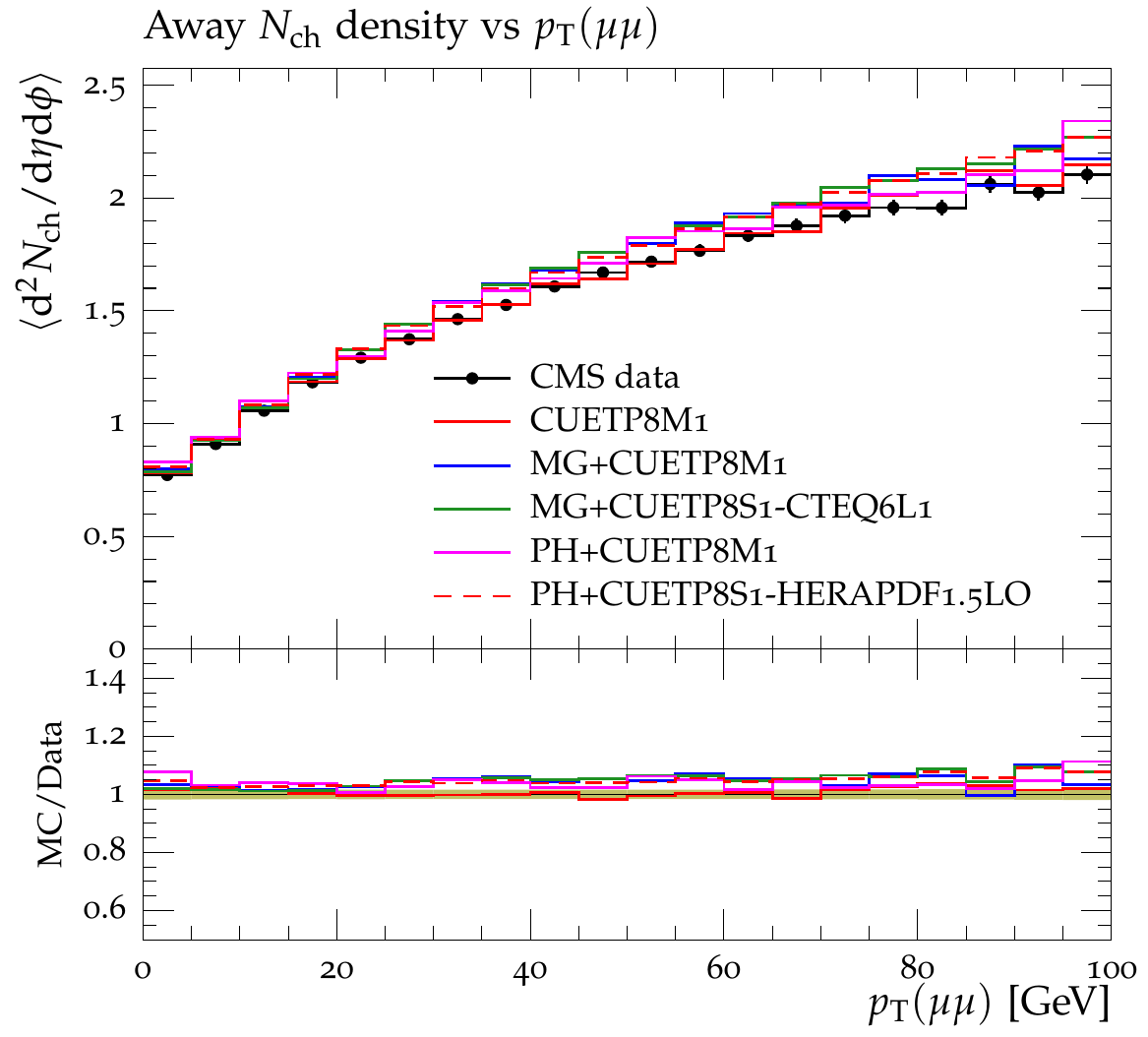}
\includegraphics[scale=0.6]{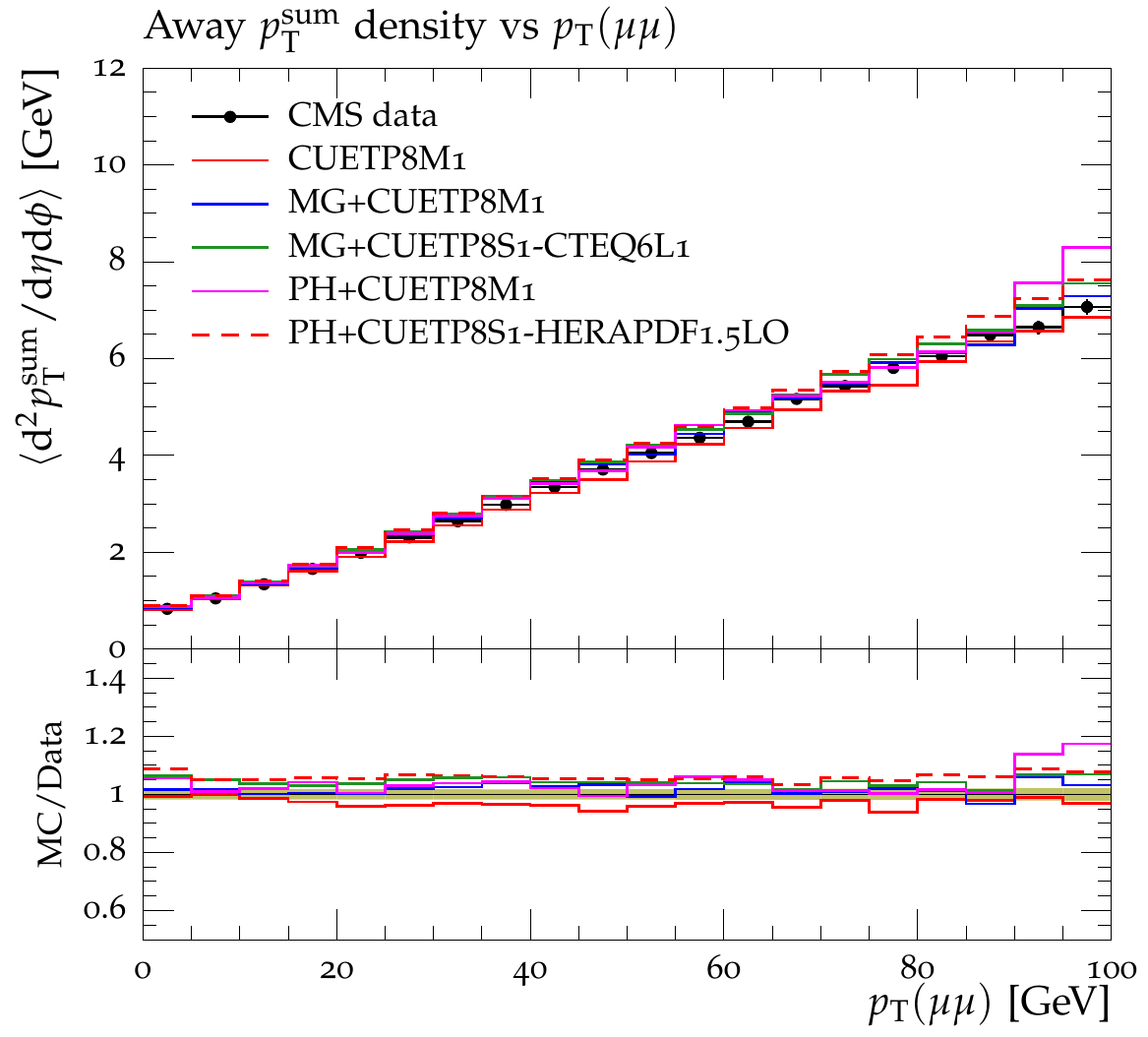}
\caption{Charged-particle (left)  and  \ptsum\  densities (right) in the toward (top), away (middle), and \TR\ (\tave) (bottom) regions, as defined by the Z-boson direction in Drell--Yan production at $\sqrt{s}=7\TeV$~\cite{Chatrchyan:2012tb}. The data are compared to \pynewhyphen\ using \cuePM, to \MADGRAPH (MG) interfaced to  \pynewhyphen\ using  \cuePB\ and \cuePM, and to \POWHEG (PH) interfaced to  \pynewhyphen\ using  \cuePH\ and \cuePM. The green bands in the ratios represent the total experimental uncertainty.}
\label{PUB_fig29}
\end{center}
\end{figure*}

{\tolerance=5000
In Fig.~\ref{PUB_fig29} the
charged-particle and \ptsum\ densities~\cite{Chatrchyan:2012tb}  in the toward, away, and transverse (\tave) regions as defined by the $\PZ$ boson in proton-proton collisions at $\sqrt{s}=7\TeV$ are compared to predictions of \pynewhyphen\ using \cuePM. Also shown are \MADGRAPH and \POWHEG results interfaced to  \pynewhyphen\ using \cuePH\ and \cuePM. The \MADGRAPH generator simulates Drell--Yan events with up to four partons, using the CTEQ6L1 PDF. The matching of ME partons and PS is performed at a scale of 20 GeV. The \POWHEG events are obtained using NLO inclusive Drell--Yan production, including up to one additional parton. The \POWHEG events are interfaced to  \pynewhyphen\ using \cuePM\ and \cuePH. The predictions based on \cuePM\ do not fit the $\PZ$ boson data unless they are interfaced to a higher-order ME generator. In \pynew\ only the Born term ($ \PQq \PAQq \rightarrow \PZ$), corrected for single-parton emission, is generated. This ME configuration agrees well with the observables in the away region in data, when the $\PZ$ boson recoils against one or more jets. In the transverse and toward regions, larger discrepancies between data and \pynew\ predictions appear at high \pt, where the occurrence of multijet emission has a large impact. To describe $\PZ$ boson production at $\sqrt{s}=7\TeV$ in all regions, higher-order contributions (starting with $\PZ$+2-jets), as used in interfacing \PYTHIA to \POWHEG or \MADGRAPH, must be included. 
\par}

\section{Extrapolation to 13 TeV}
\label{section-5}

\begin{figure*}[htbp]
\begin{center}
\includegraphics[scale=0.6]{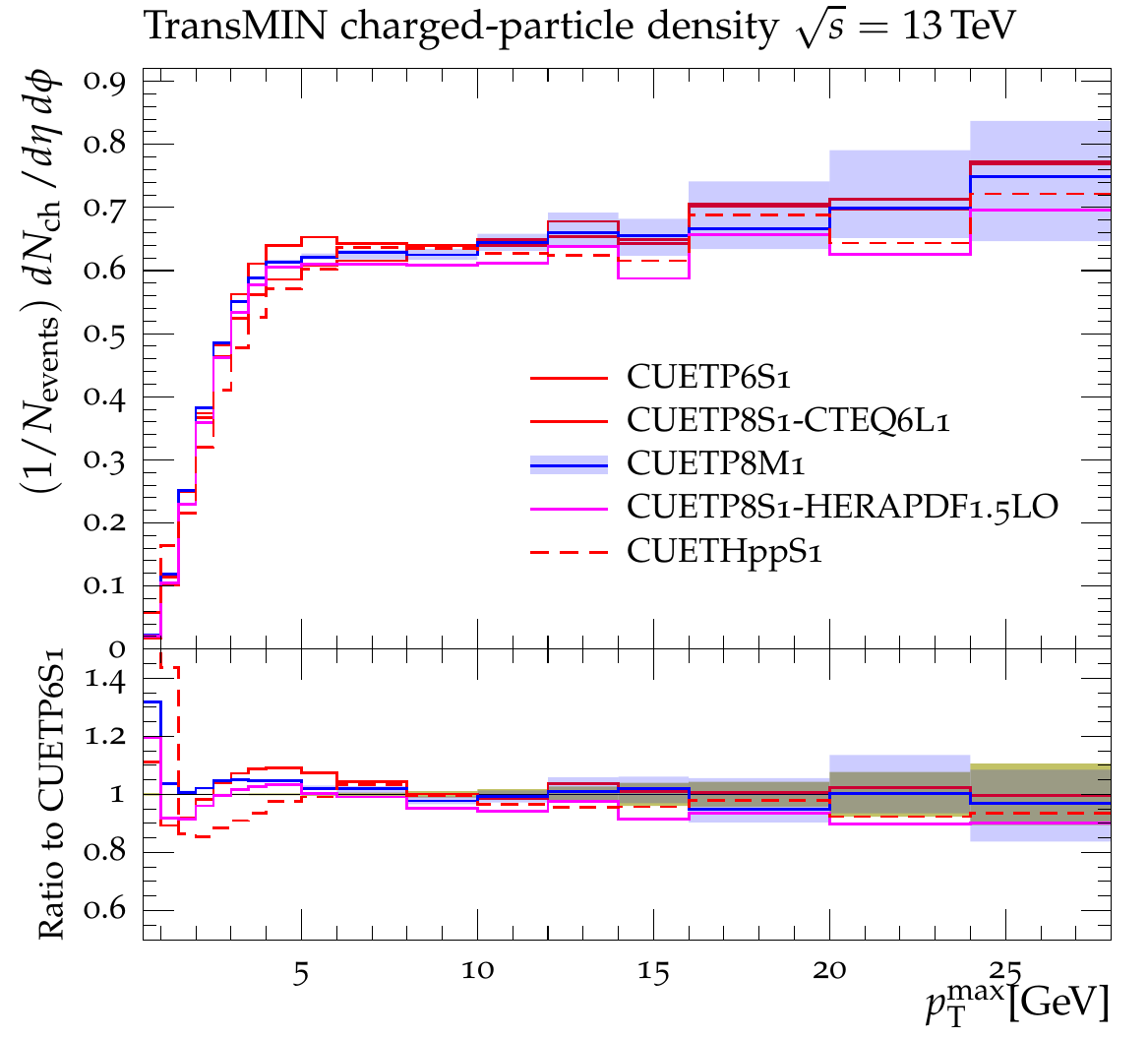}
\includegraphics[scale=0.6]{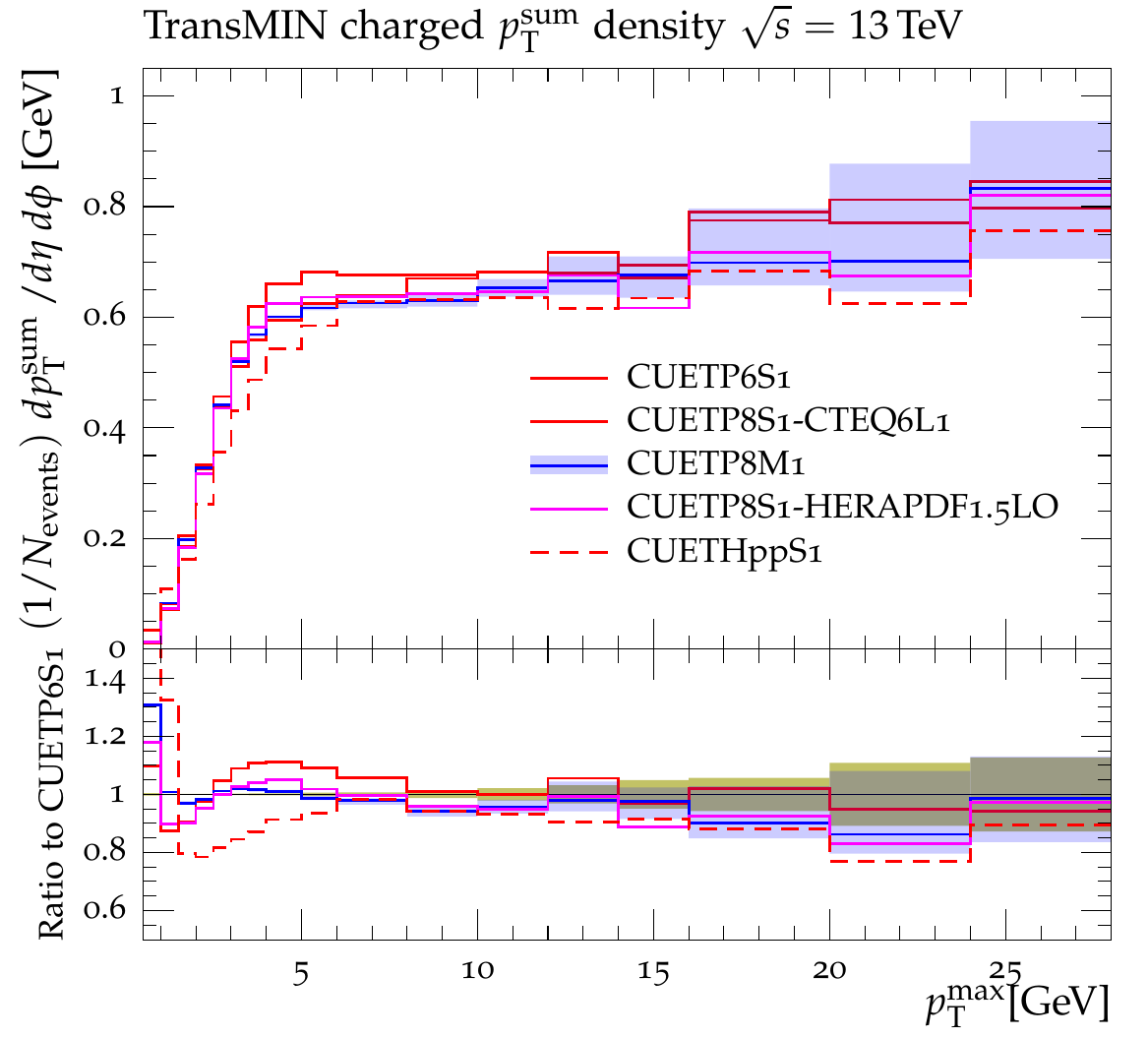}\\
\includegraphics[scale=0.6]{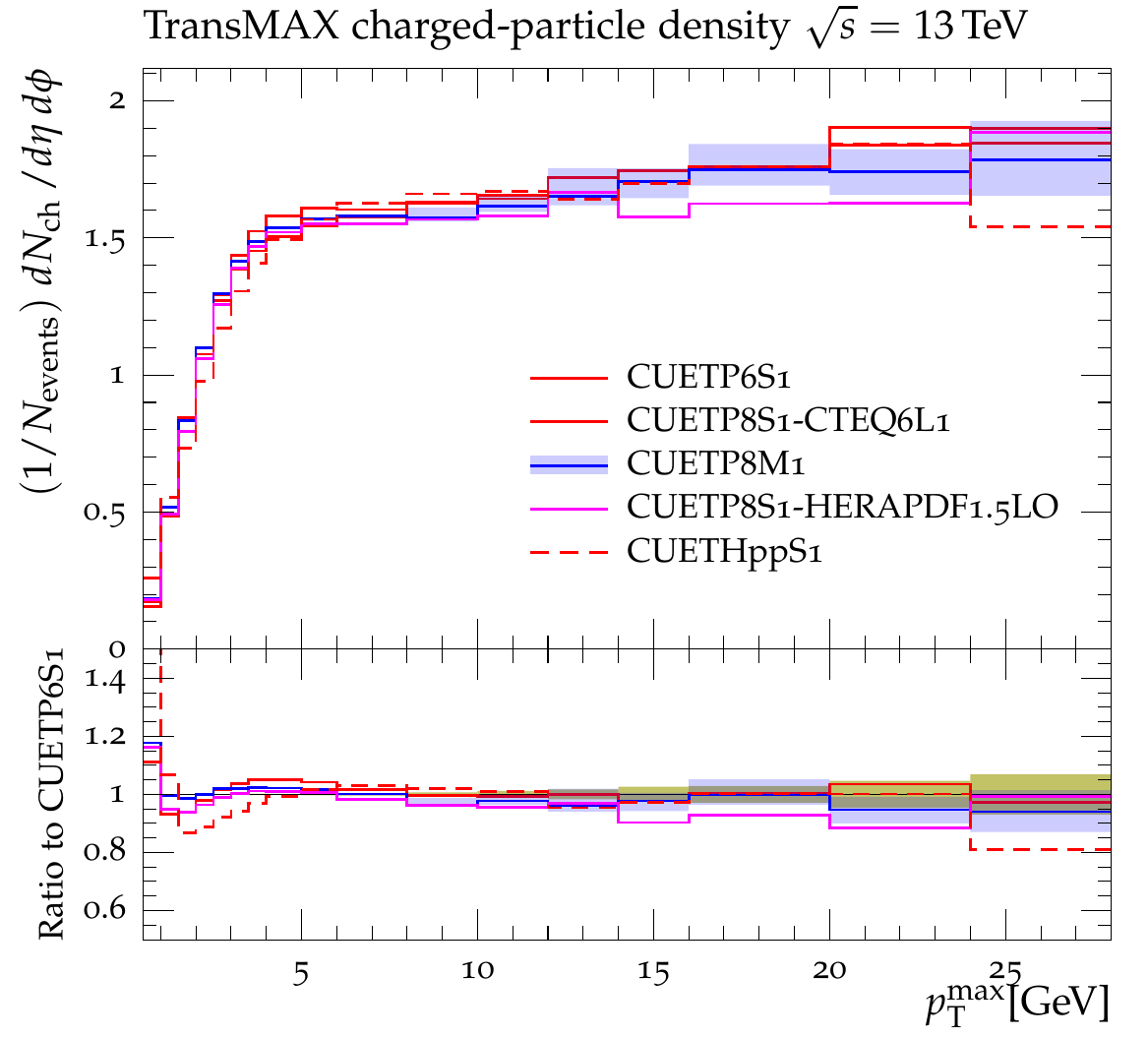}
\includegraphics[scale=0.6]{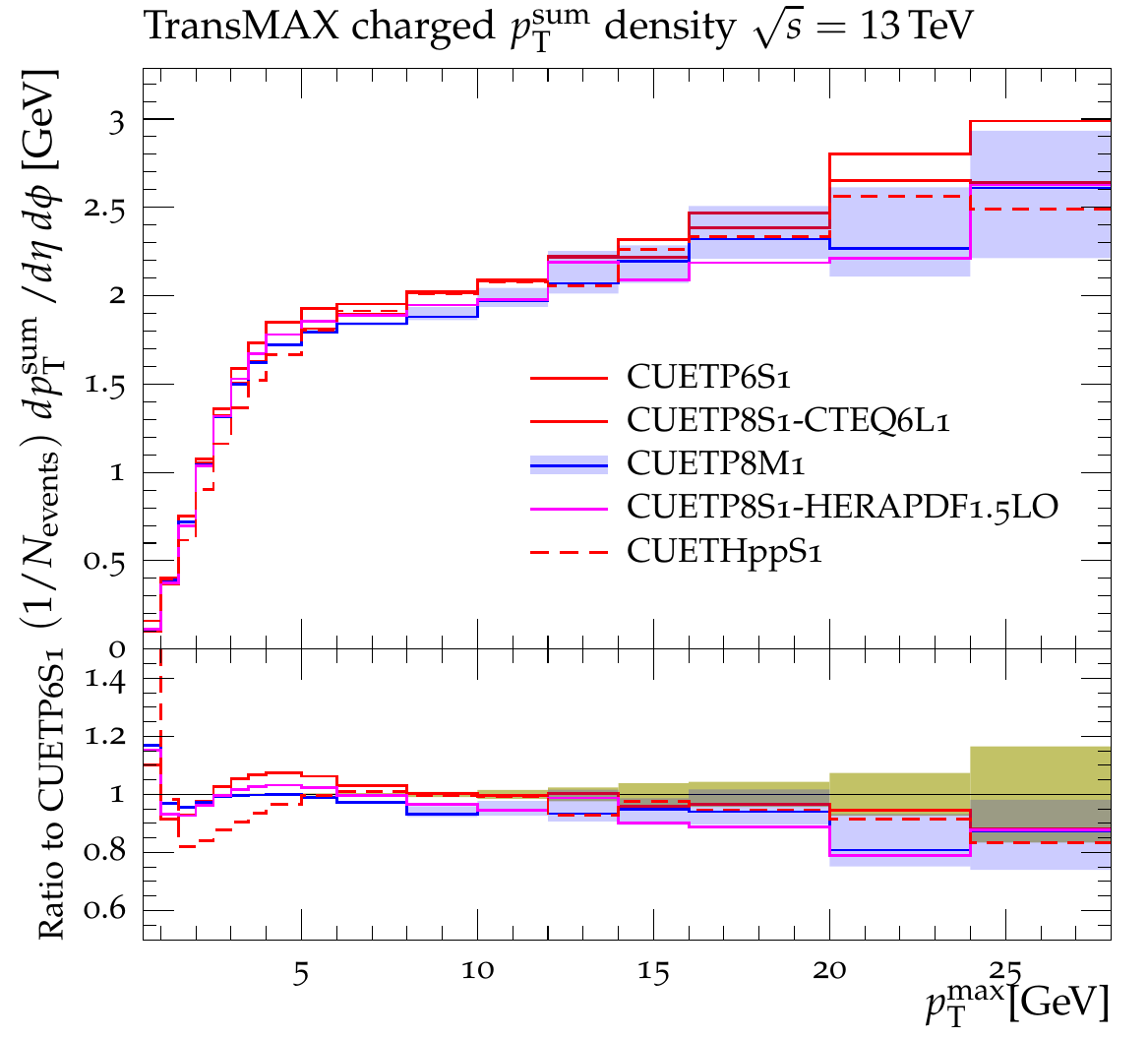}\\
\includegraphics[scale=0.6]{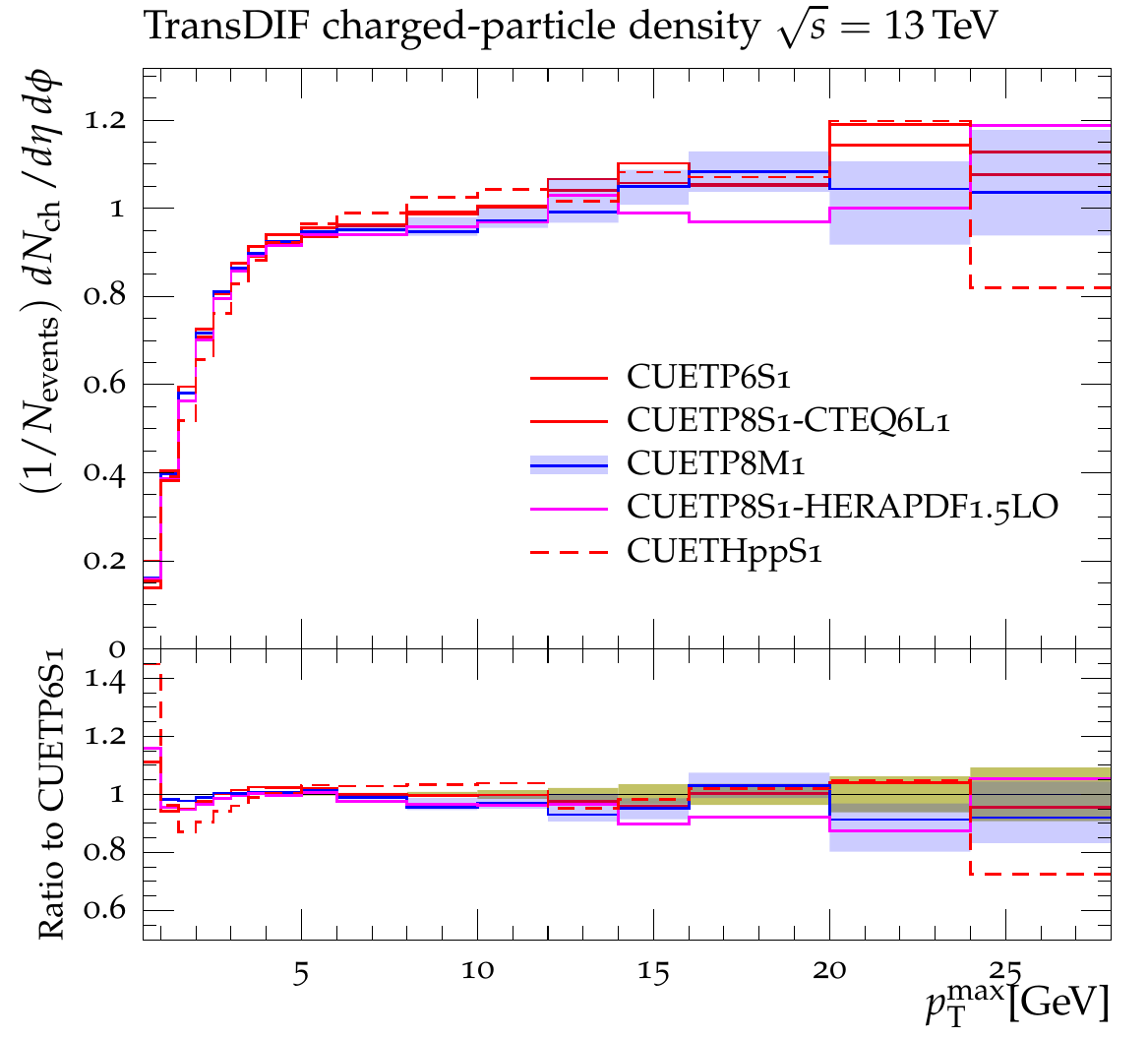}
\includegraphics[scale=0.6]{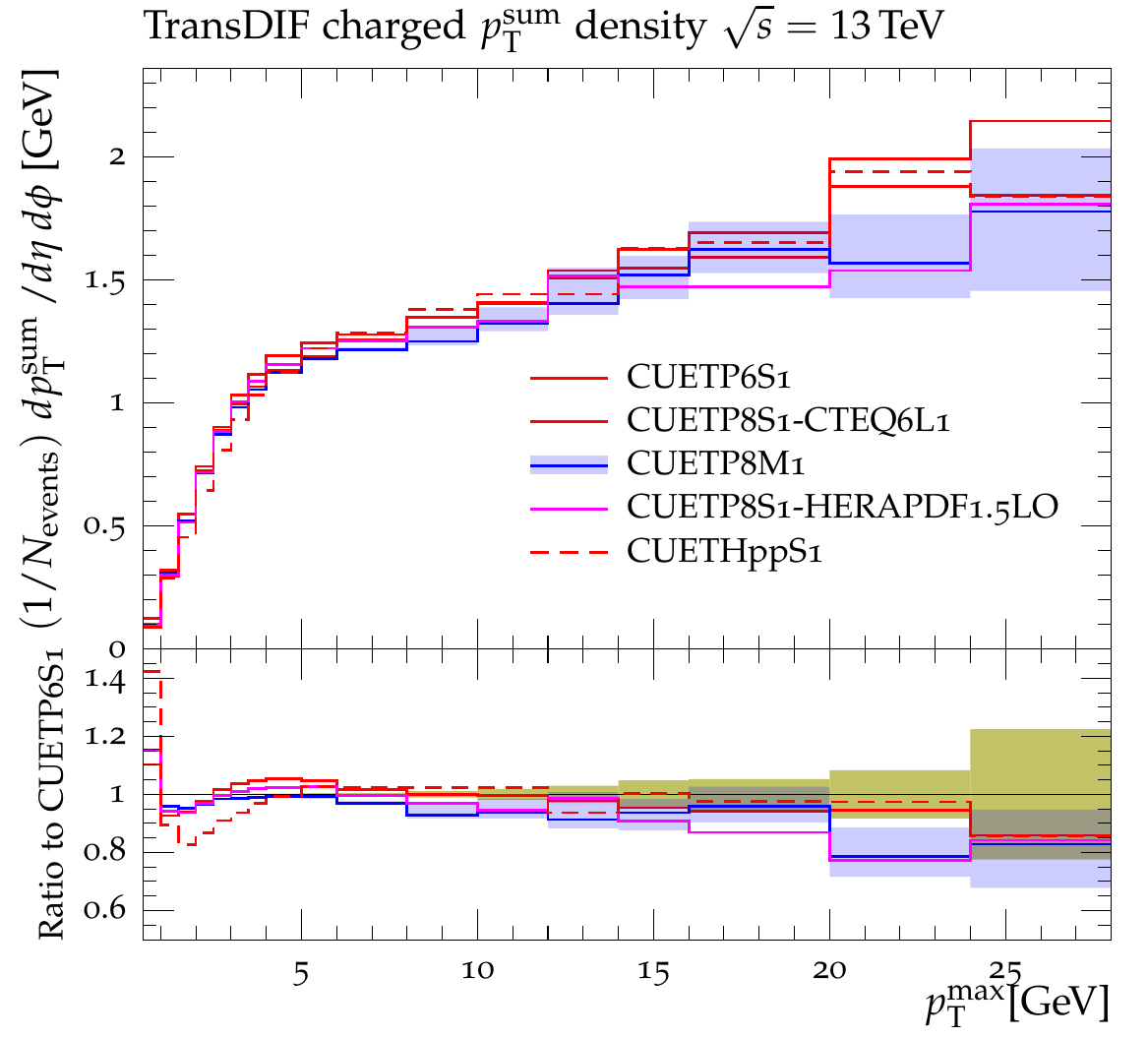}
\caption{Predictions at $\sqrt{s}=13\TeV$ for the particle (left) and the \ptsum\ densities (right) for charged particles with \ptcut\ and \etacut\ in the \tmin\ (top), \tmax\ (middle), and \tdif\ (bottom) regions, as defined by the leading charged particle, as a function of the leading charged-particle \ptmax\ for the five CMS UE tunes: \pyoldhyphen\ \cuePA, and \pynewhyphen\ \cuePB, \cuePH, and \cuePM, and \hwpp\ \cueHW. Also shown are the ratio of the tunes to predictions of \cuePB. Predictions for \cuePM\ are shown along with the envelope (green bands) of the corresponding eigentunes.}
\label{PUB_fig15}
\end{center}
\end{figure*}

In this section, predictions at $\sqrt{s}=13\TeV$, based on the new tunes, for observables sensitive to the UE  are presented. Figure~\ref{PUB_fig15} shows the predictions at 13 TeV for the charged-particle and the \ptsum\ densities in the \tmin, \tmax, and \tdif\ regions, as defined by the leading charged particle as a function of \ptmax\ based on the five new CMS UE tunes: \cuePA,  \cuePB, \cuePH, \cuePM, and \cueHW. In Fig.~\ref{PUB_fig15} the ratio of the predictions using the four CMS tunes to the one using \cuePM\  is shown. The predictions at $13\TeV$ of all these tunes are remarkably similar.  It does not seem to matter that the new CMS \pynewhyphen\ UE tunes do not fit very well to the $\sqrt{s}=300\GeV$ UE data.  The new \pynewhyphen\ tunes give results at $13\TeV$ similar to the new CMS \pyoldhyphen\ tune and the new CMS \hwpp\ tune. The uncertainties on the predictions based on the eigentunes do not exceed 10\% relative to the central value.

\begin{figure*}[htbp]
\begin{center}
\includegraphics[scale=0.6]{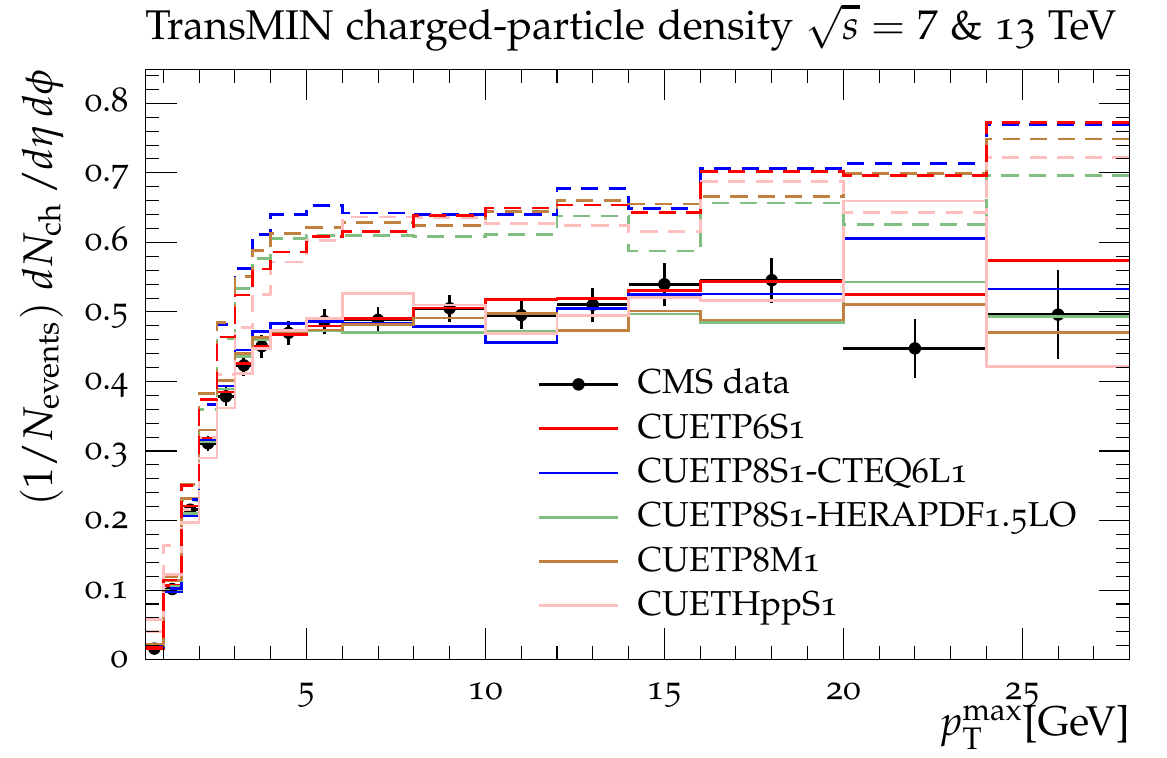}
\includegraphics[scale=0.6]{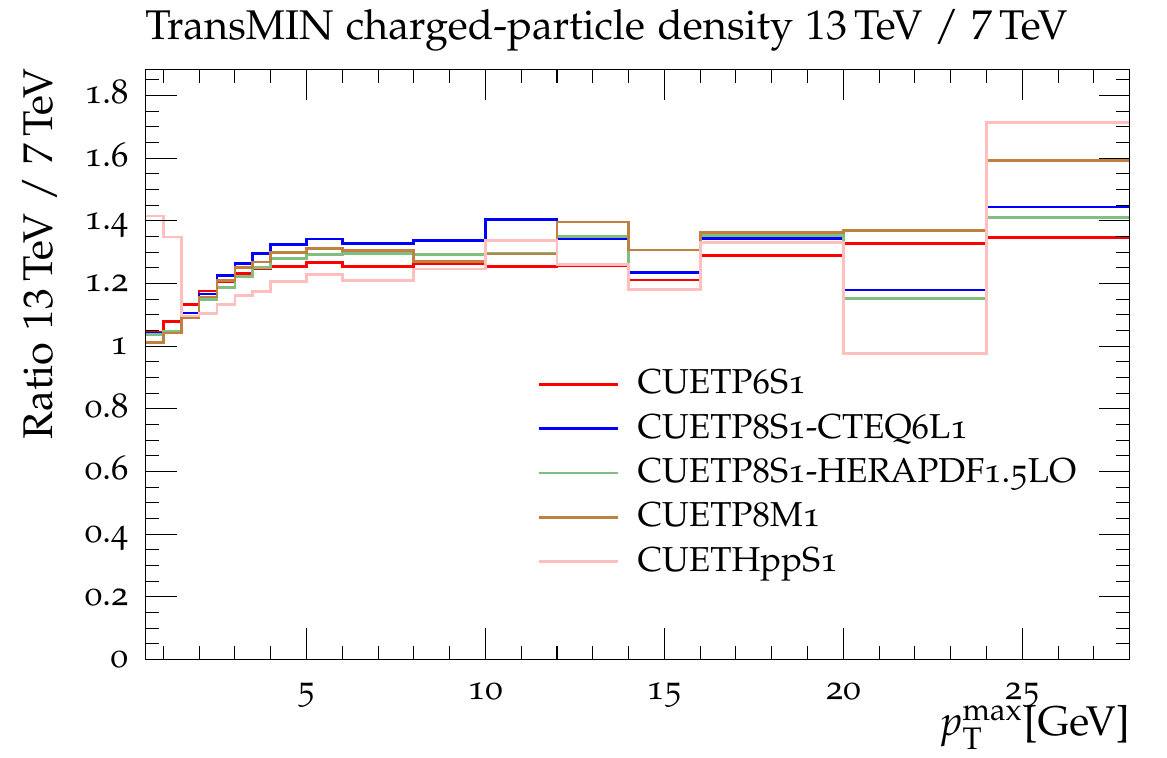}\\
\includegraphics[scale=0.6]{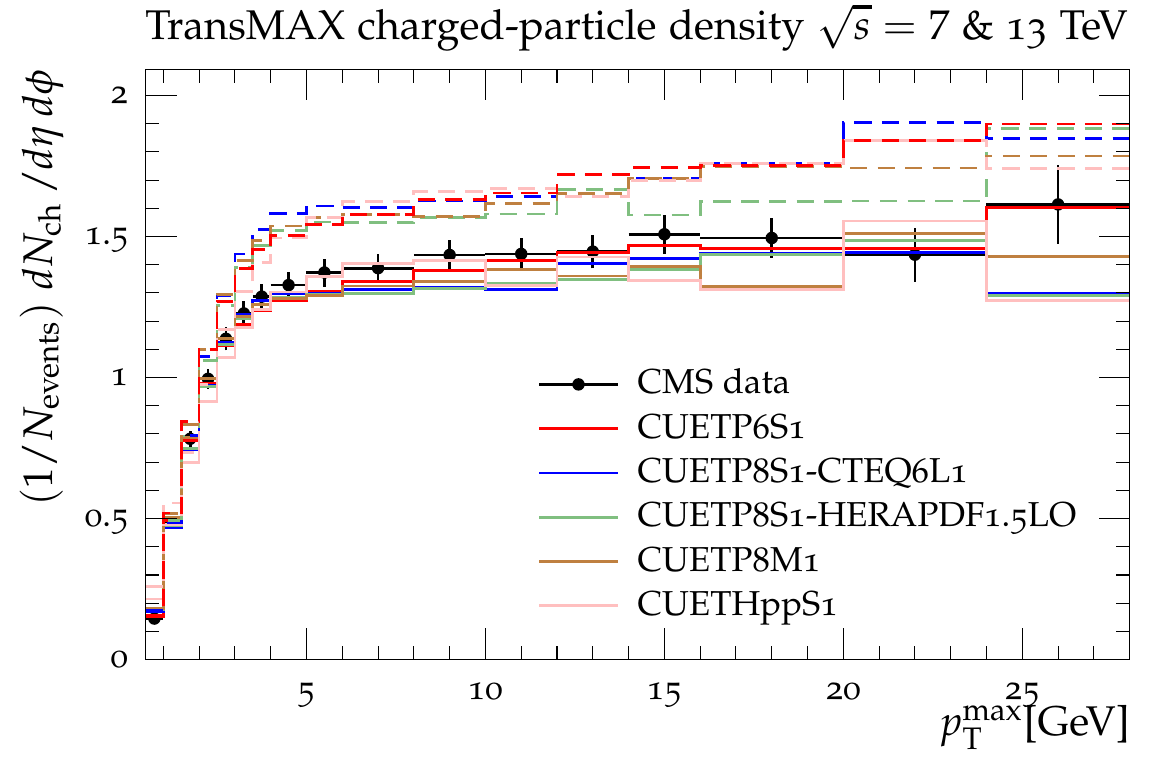}
\includegraphics[scale=0.6]{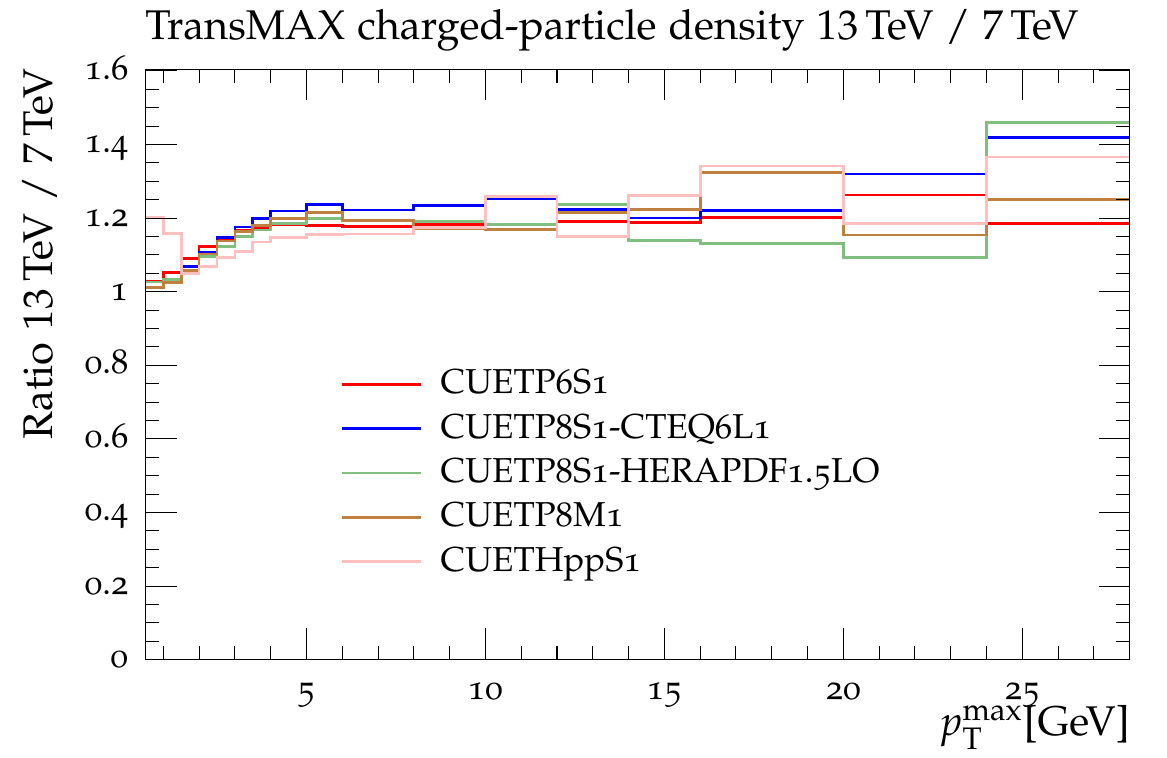}\\
\includegraphics[scale=0.6]{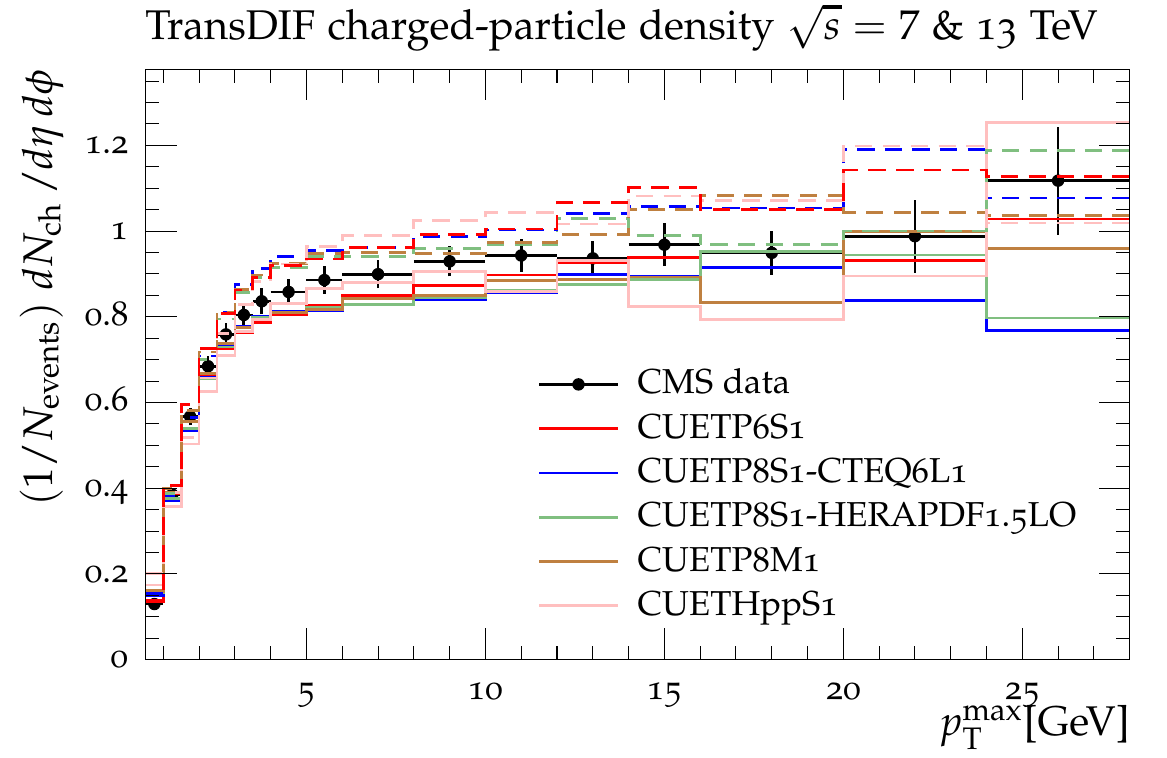}
\includegraphics[scale=0.6]{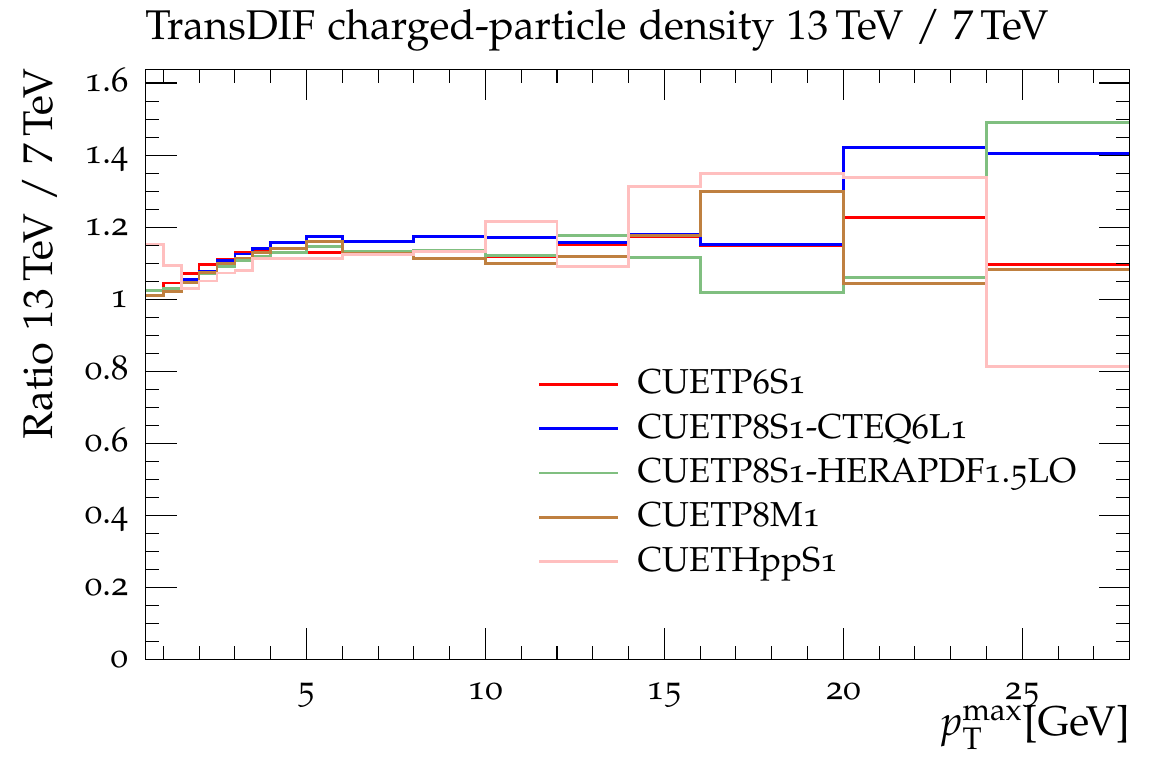}
\caption{Charged-particle density at $\sqrt{s}=7\TeV$ for particles with \ptcut\ and \etacut\ in the \tmin\ (top), \tmax\  (middle), and \tdif\ (bottom) regions, as defined by the leading charged particle, as a function of the leading charged-particle \ptmax. The data are compared to \pyoldhyphen\ using \cuePA, to \pynewhyphen\ using \cuePB, \cuePH, and \cuePM, and to \hwpp\ using \cueHW. Also shown are the predictions (left) based on the CMS UE tunes at $13\TeV$ (dashed lines), and the ratio of the $13\TeV$ to $7\TeV$ results for the five tunes (right).}
\label{PUB_fig16}
\end{center}
\end{figure*}

\begin{figure*}[htbp]
\begin{center}
\includegraphics[scale=0.6]{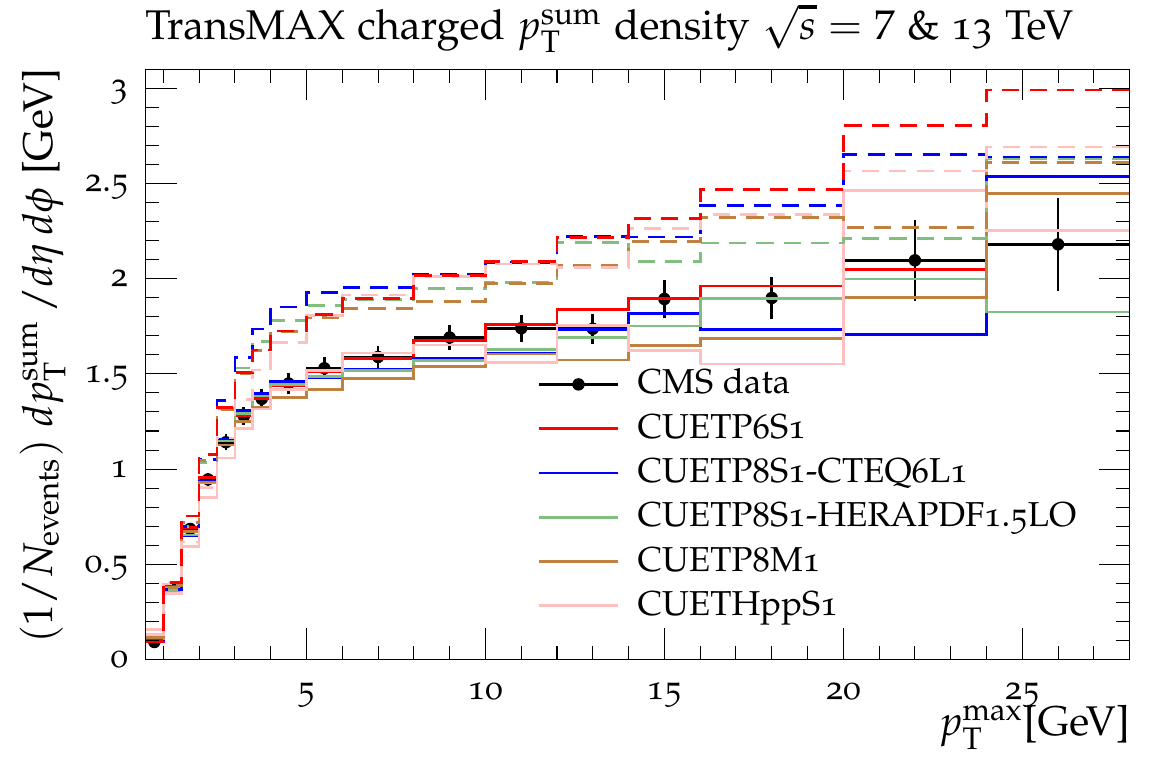}
\includegraphics[scale=0.6]{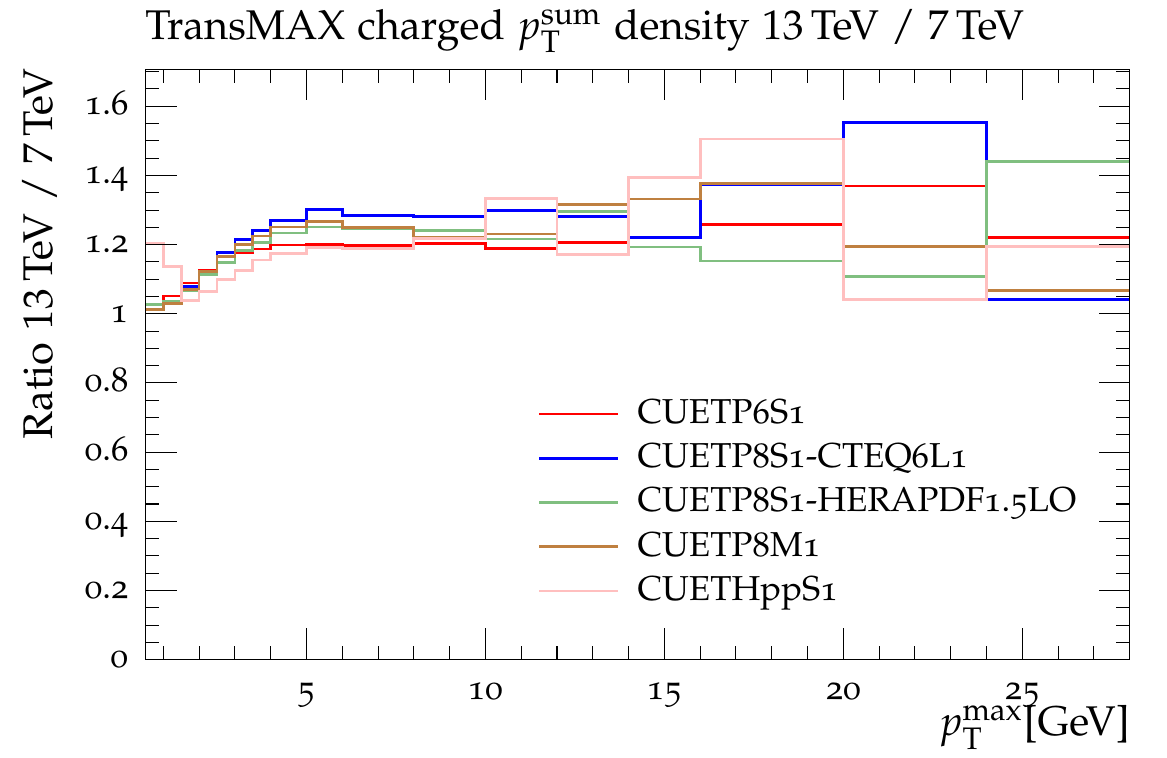}\\
\includegraphics[scale=0.6]{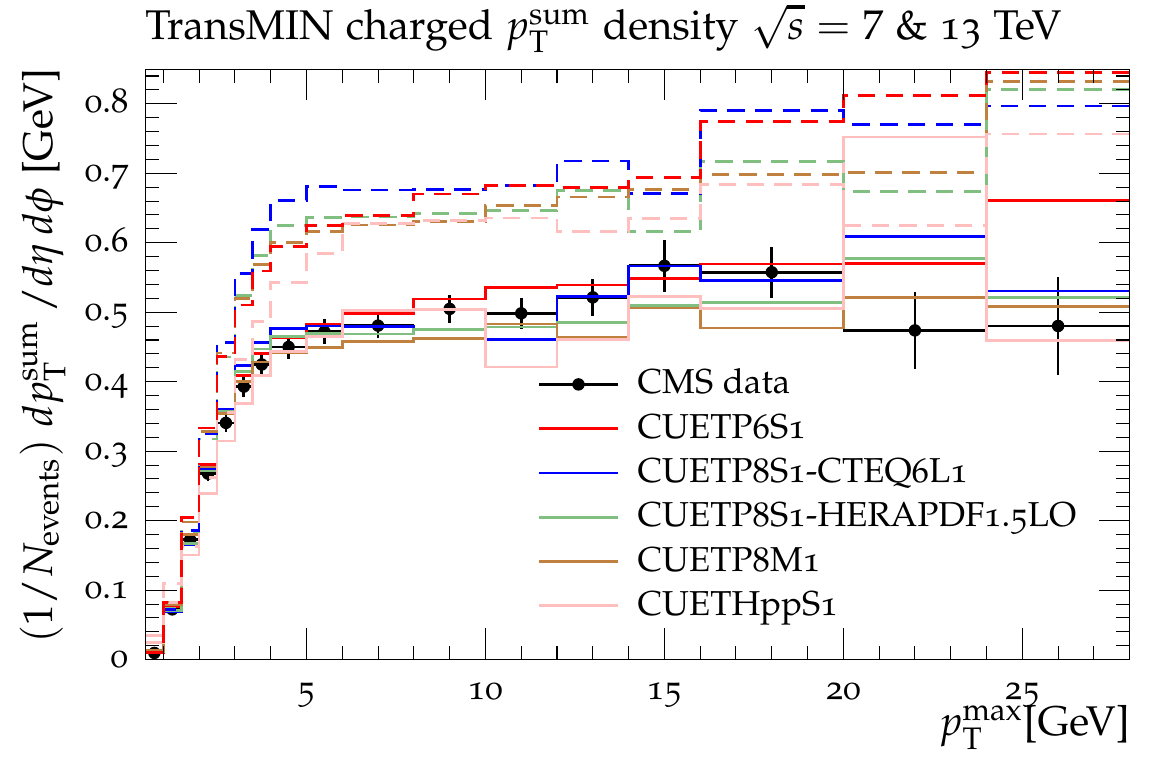}
\includegraphics[scale=0.6]{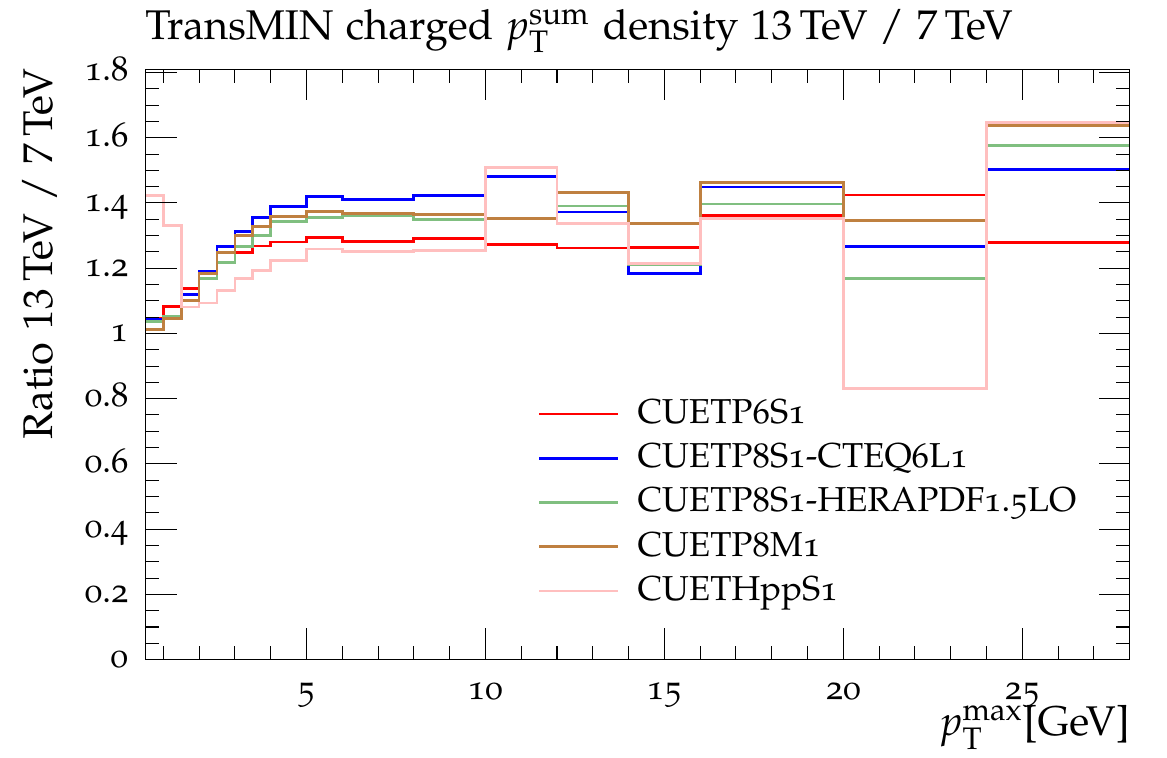}\\
\includegraphics[scale=0.6]{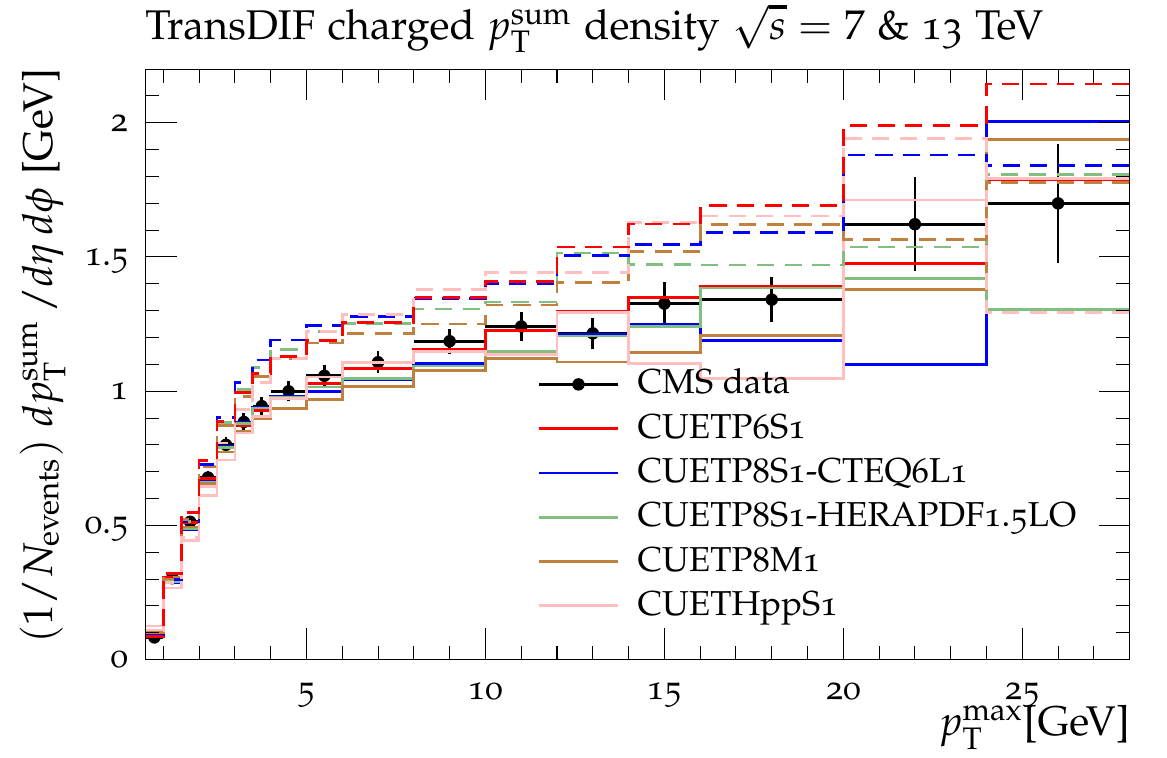}
\includegraphics[scale=0.6]{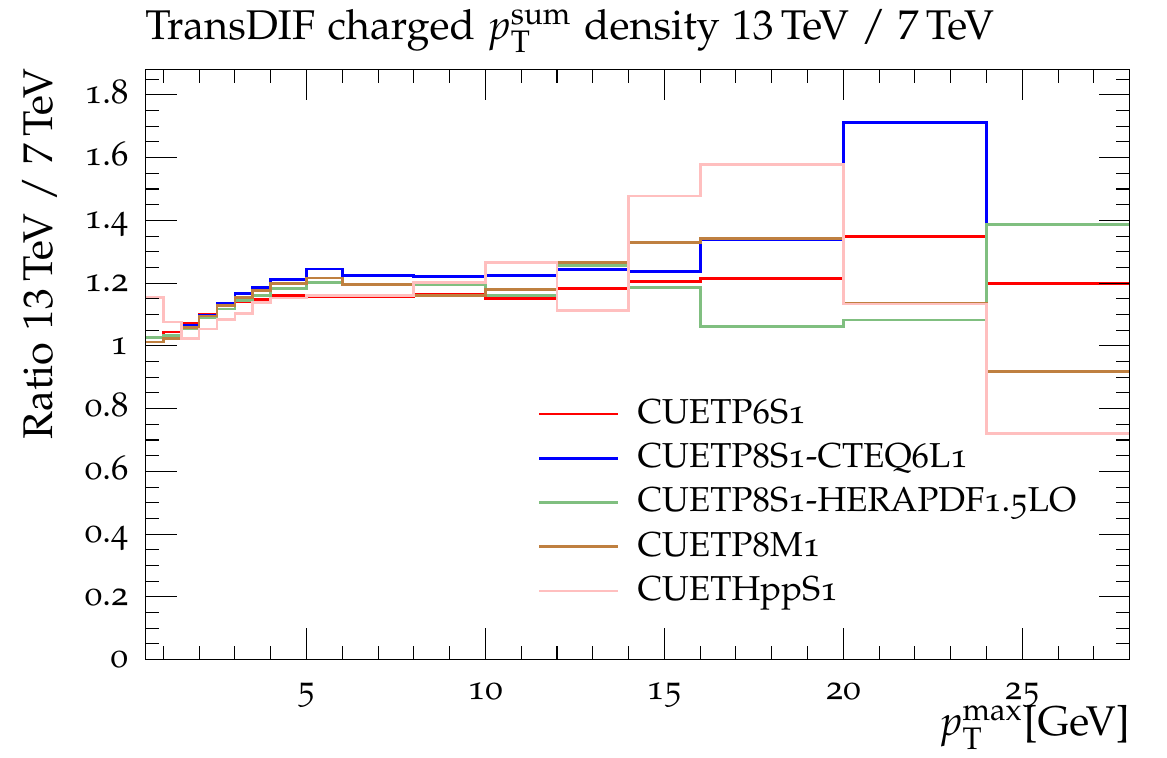}
\caption{Charged \ptsum\ density at $\sqrt{s}=7\TeV$  for particles with \ptcut\ and \etacut\ in the \tmin\ (top), \tmax\  (middle), and \tdif\ (bottom) regions, as defined by the leading charged particle, as a function of the leading charged-particle \ptmax. The data are compared to \pyoldhyphen\ using \cuePA, to \pynewhyphen\ using \cuePB, \cuePH, and \cuePM, and to \hwpp\ using \cueHW. Also shown are the predictions (left) based on the CMS UE tunes at $13\TeV$ (dashed lines), and the ratio of the $13\TeV$ to $7\TeV$ results for the five tunes (right).}
\label{PUB_fig17}
\end{center}
\end{figure*}

In Fig.~\ref{PUB_fig16} and \ref{PUB_fig17} the predictions at $\sqrt{s}=13\TeV$ obtained using the new tunes from $7\TeV$ are shown for the charged-particle and the \ptsum\ densities in the \tmin, \tmax, and \tdif\ regions, defined as a function of \ptmax.  Also shown is the ratio of $13\TeV$ to $7\TeV$ results for the five tunes.  The \tmin\ region increases much more rapidly with energy than the \tdif\ region. For example, when using \cuePM, the charged-particle and the \ptsum\ densities in the \tmin\ region for $5.0<$\ptmax$<6.0\GeV$ is predicted to increase by $28\%$ and $37\%$, respectively, while the \tdif\ region is predicted to increase by a factor of two less, \ie by $13\%$ and $18\%$ respectively. 

\begin{figure*}[htbp]
\begin{center}
\includegraphics[scale=0.7]{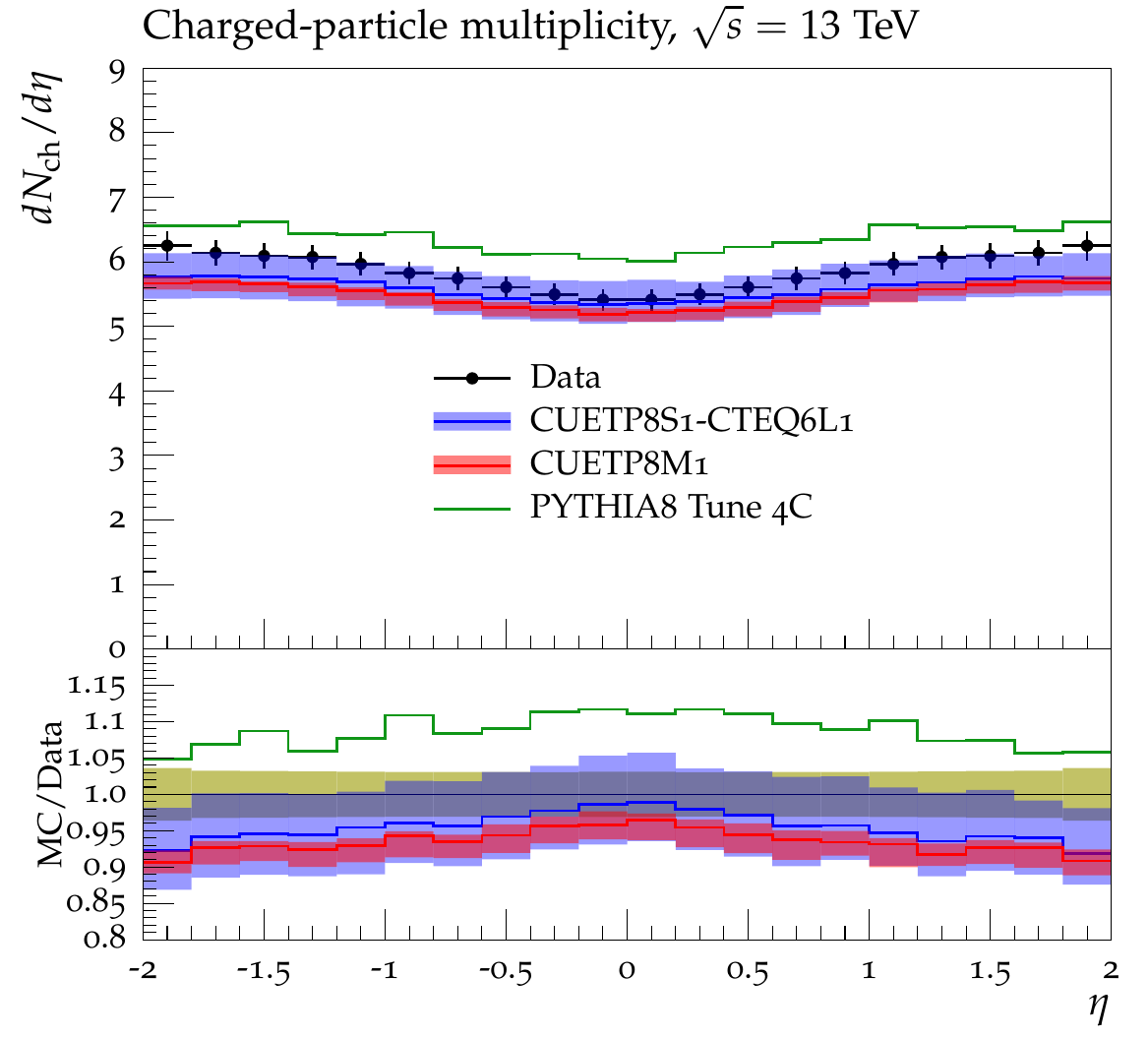}
\caption{CMS data at $\sqrt{s}=13\TeV$~\cite{Khachatryan:2015jna} for the charged-particle pseudorapidity distribution, $\rd \mathrm{N}_{\text{ch}}/\rd \eta$, in inelastic proton-proton collisions. The data are compared to predictions of \pynewhyphen\ using \cuePB, \cuePM, and \tunec. The predictions based on \cuePB\ and \cuePM\ are shown with an error band corresponding to the total uncertainty obtained from the eigentunes. Also shown are the ratios of these predictions to the data. The green band represents the total experimental uncertainty on the data.}
\label{PUB_fig291}
\end{center}
\end{figure*}

In Fig.~\ref{PUB_fig291}, predictions obtained with \pynewhyphen\ using \cuePB\ and \cuePM, and \tunec\ are compared to the recent CMS data measured at $\sqrt{s} = 13\TeV$~\cite{Khachatryan:2015jna} on charged-particle multiplicity as a function of pseudorapidity. Predictions from \cuePB\ and \cuePM\ are shown with the error bands corresponding to the uncertainties obtained from the eigentunes. These two new CMS tunes, although obtained from fits to UE data at 7\TeV, agree well with the MB measurements over the whole pseudorapidity range, while predictions from \pynew\ \tunec\ overestimate the data by about 10\%. This confirms that the collision-energy dependence of the CMS UE tunes parameters can be trusted for predictions of MB observables.

\newpage
\clearpage

\section{Summary and conclusions}
\label{section-6}

New tunes of the \PYTHIA\ event generator were constructed for different parton distribution functions using various sets of underlying-event (UE) data. By simultaneously fitting UE data at several center-of-mass energies, models for UE have been tested and their parameters constrained.
The improvement in the description of UE data provided by the new CMS tunes at different collision energies gives confidence that they can provide reliable predictions at $\sqrt{s} = 13\TeV$, where all the new UE tunes predict similar results for the UE observables.
   
The observables sensitive to double-parton scattering (DPS) were fitted directly by tuning the MPI parameters. Two $\PW$+dijet DPS tunes and two four-jet DPS tunes were constructed to study the dependence of the DPS-sensitive observables on the MPI parameters. The CMS UE tunes perform fairly well in the description of DPS observables, but they do not fit the DPS data as well as the DPS tunes do.  On the other hand, the CMS DPS tunes do not fit the UE data as well as the UE tunes. At present, it is not possible to accurately describe both soft and hard MPI within the current \PYTHIA and \textsc{herwig++} frameworks. Fitting DPS-sensitive observables has also provided the DPS effective cross section \eff\ associated to each model. This method can be applied to determine the \eff\ values associated with  different MPI models implemented in the current MC event generators for the production of any final-state with two hard particles.

Predictions of \pynewhyphen\ using the CMS UE tunes agree fairly well with the MB observables in the central region ($|\eta|<2$) and can be interfaced to higher-order and multileg matrix-element generators, such as \POWHEG\ and \MADGRAPH, while maintaining their good description of the UE.  It is not necessary to produce separate tunes for these generators. In addition, we have verified that the measured particle pseudorapidity density at 13\TeV is well reproduced by the new CMS UE Tunes. Furthermore, all of the new CMS tunes come with their eigentunes, which can be used to determine the uncertainties associated with the theoretical predictions. These new CMS tunes will play an important role in predicting and analyzing LHC data at $13$ and $14\TeV$. 

\appendix
\section{Tables of tune uncertainties}
This section provides the values of the parameters corresponding to the eigentunes of the new CMS \pynew\ and the \hwpp\ tunes. A change in the $\chi^2$ of the fit that equals the absolute $\chi^2$ value obtained in the tune defines the eigentunes listed in Tables \ref{table5Appendix}--\ref{table8Appendix} for the new \pynewhyphen\ and the new \hwpp\ tunes. The different parameter values indicated refer to the deviation tunes along each of the maximally independent directions in the parameter space, obtained by using the covariance matrix in the region of the best tune. The number of directions defined in the parameter space equals the number of free parameters $n$ used in the fit and results into 2$n$ parameter variations, \ie eigentunes. These variations represent a good set of systematic errors on the given  tune.

\begin{table*}[htbp]
\begin{center}
\topcaption{Eigentunes sets for CUETP8S1-CTEQ6L1.}
\label{table5Appendix}
\resizebox{\textwidth}{!}{
\begin{tabular}{l c c c c c c c c} \hline 
\pynew\ Parameter  &
 $1-$ &    
 $1+$ &   
 $2-$ &
 $2+$ & 
 $3-$ &   
 $3+$ & 
 $4-$ & 
 $4+$  \\ \hline 
  MultipartonInteractions:pT0Ref [GeV] & 
  2.101 & 2.101 & 2.068 & 2.135 & 2.100 & 2.102 &  2.079 & 2.123 \\
  MultipartonInteractions:ecmPow &
  0.191 & 0.231 & 0.210 & 0.211 & 0.231 & 0.191 & 0.191 & 0.231  \\
  MultipartonInteractions:expPow & 
  1.609 &  1.609 & 1.602 & 1.616 & 1.613 & 1.605 & 1.714 & 1.503  \\
  ColourReconnection:range &
  3.030  & 3.609 & 3.313 & 3.313 & 3.311 & 3.314 & 3.314 & 3.311\\ \hline
 \end{tabular}
}
\end{center}
\end{table*}

\begin{table*}[htbp]
\begin{center}
\topcaption{Eigentunes sets for CUETP8S1-HERAPDF.}
\label{table6Appendix}
\resizebox{\textwidth}{!}{
\begin{tabular}{l c c c c c c c c } \hline 
\pynew\ Parameter & $1-$ & $1+$ & $2-$ & $2+$ & $3-$ & $3+$ & $4-$ & $4+$  \\ \hline
MultipartonInteractions:pT0Ref [GeV] & 2.000 & 2.000 & 1.960 & 2.043 & 1.999 & 2.001 & 1.968 & 2.030 \\ 
MultipartonInteractions:ecmPow  & 0.275 & 0.226 & 0.250 & 0.250 & 0.226 & 0.275 & 0.274 & 0.227 \\
MultipartonInteractions:expPow &  1.691 & 1.690 & 1.681 & 1.700 & 1.695 & 1.686 & 1.831 & 1.559 \\
ColourReconnection:range & 6.224 & 5.972 & 6.096 & 6.096 & 6.101 & 6.091 & 6.091 & 6.101 \\ \hline
\end{tabular}
}
\end{center}
\end{table*}

\begin{table*}[htbp]
\begin{center}
\topcaption{Eigentunes sets for CUETP8M1.}
\label{table7Appendix}
\begin{tabular}{l c c c c} \hline 
\pynew\ Parameter  & $1-$ &$1+$ & $2-$ & $2+$  \\ \hline
MultipartonInteractions:pT0Ref [GeV] & 2.403 & 2.402 & 2.400 & 2.405 \\
MultipartonInteractions:ecmPow & 0.253 & 0.251 & 0.253 & 0.252 \\ \hline
\end{tabular}
\end{center}
\end{table*}

\begin{table*}[htbp]
\begin{center}
\topcaption{Eigentunes sets for CUETHppS1.}
\label{table8Appendix}
\resizebox{\textwidth}{!}{
\begin{tabular}{l c c c c c c c c } \hline 
\hwpp\ Parameter  & $1-$ & $1+$ & $2-$ & $2+$ & $3-$ & $3+$ & $4-$ & $4+$ \\ \hline
MPIHandler:InvRadius & 2.290 & 2.227 & 2.318 & 2.196 & 2.272 & 2.237 & 2.254 & 2.256 \\
RemnantDecayer:colourDisrupt &  0.396 & 0.811  & 0.634  & 0.623  &  0.632 & 0.625  & 0.596  &  0.666  \\
MPIHandler:Power & 0.396 &  0.351 & 0.331   &  0.408  &  0.399  &  0.342  & 0.361   &  0.381  \\
ColourReconnector:ReconnectionProbability  & 0.615  & 0.460  & 0.529  & 0.527  & 0.523  & 0.533  & 0.444  & 0.626\\ \hline
\end{tabular}
}
\end{center}
\end{table*}

\ifthenelse{\boolean{cms@external}}{\section{Comparisons of PYTHIA6 UE tunes to data}}{\section{Comparisons of \textsc{pythia6} UE tunes to data}}

\begin{figure*}[htbp]
\begin{center}
\includegraphics[scale=0.65]{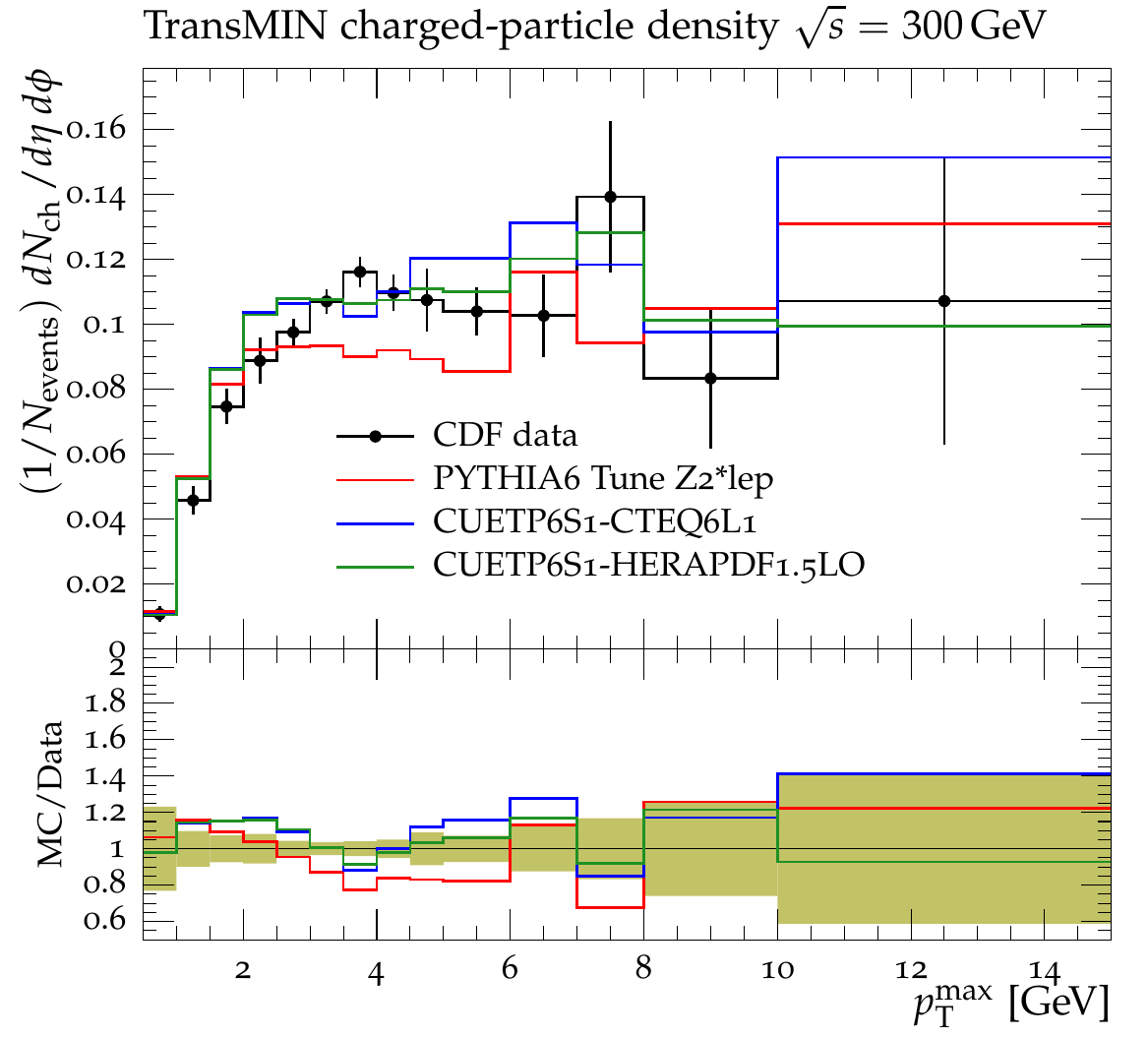}
\includegraphics[scale=0.65]{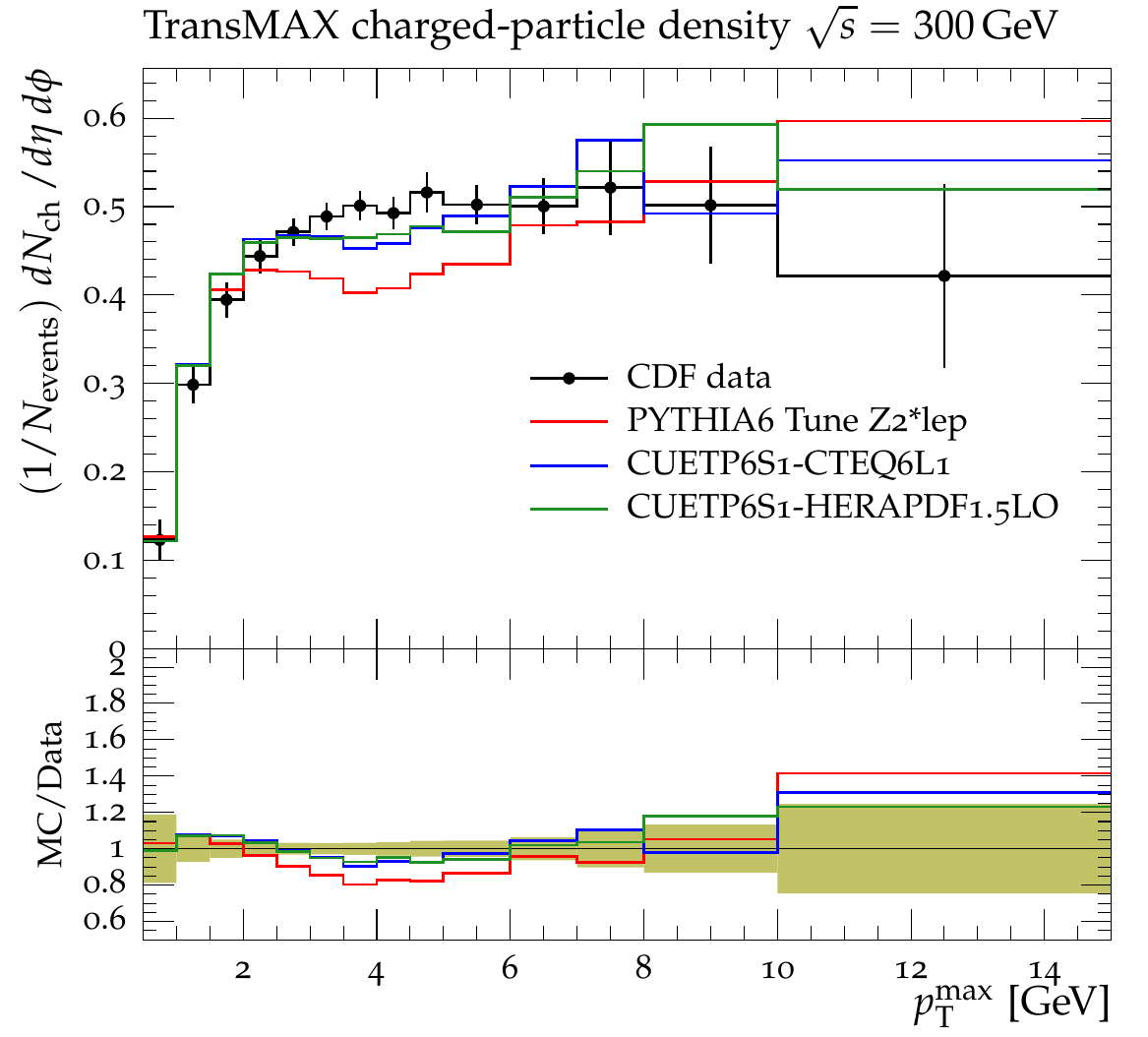}\\
\includegraphics[scale=0.65]{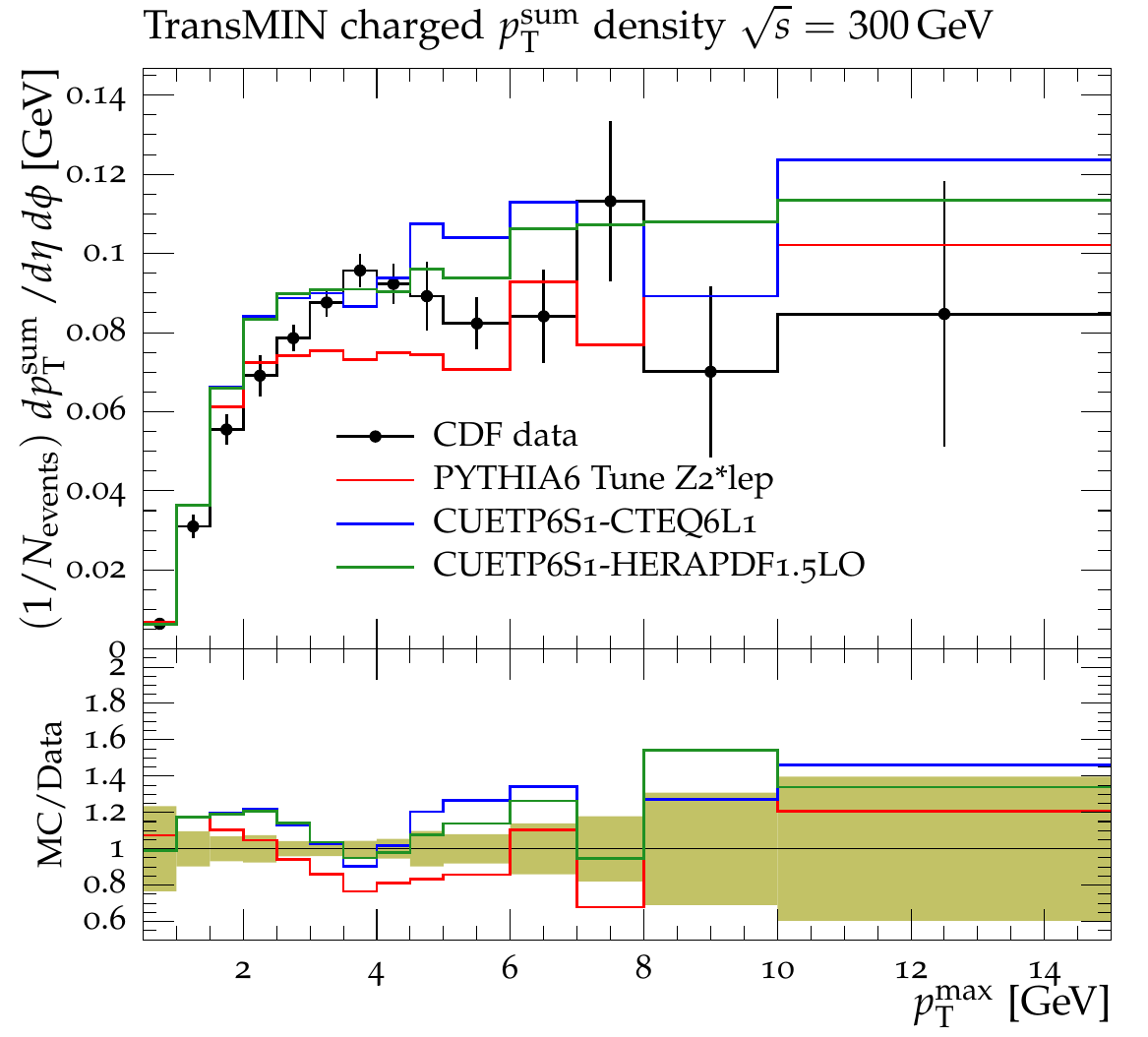}
\includegraphics[scale=0.65]{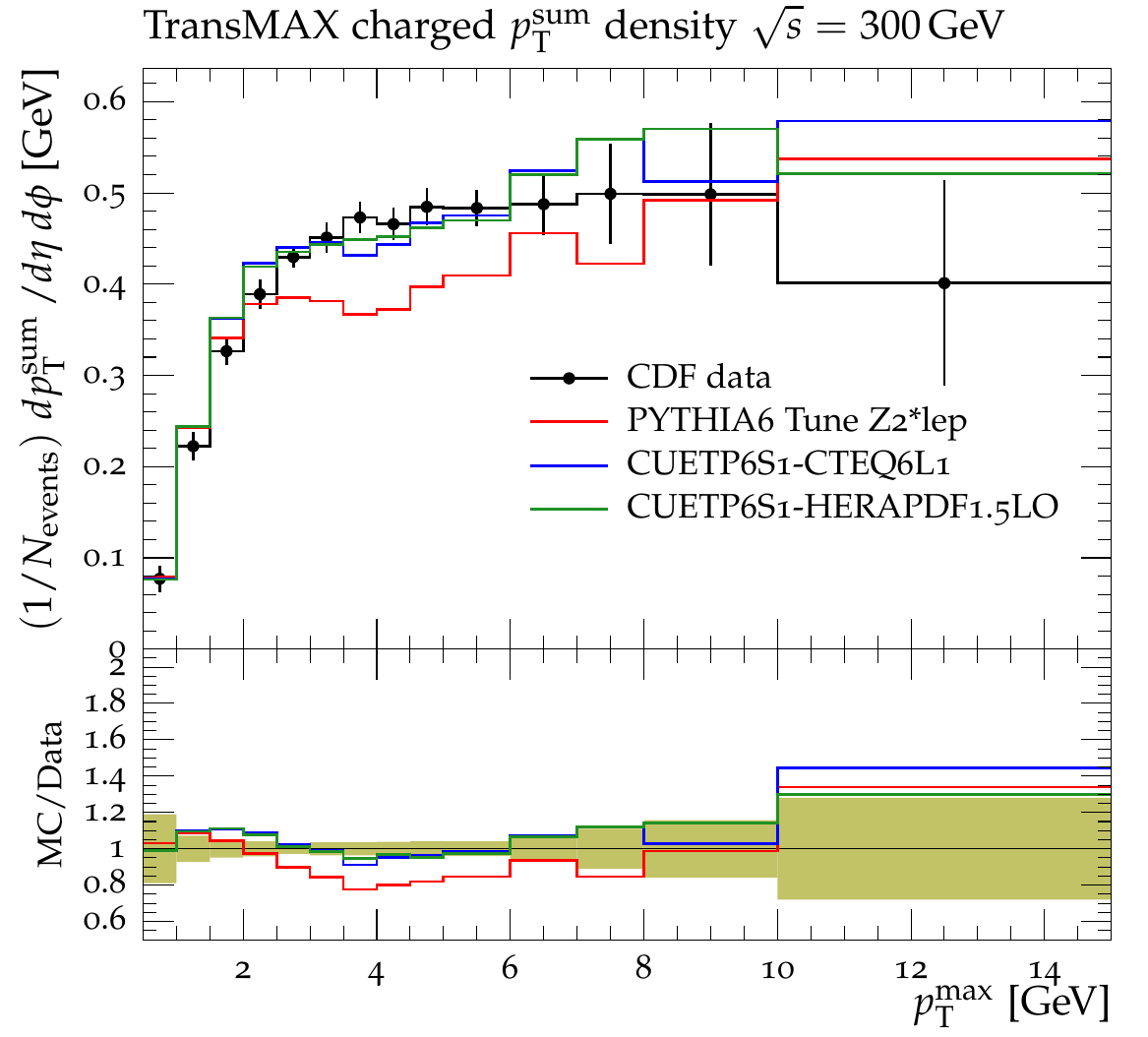}
\caption{CDF data at $\sqrt{s}=300\GeV$~\cite{Aaltonen:2015aoa} on the particle  (top) and \ptsum\ densities (bottom) for charged particles with \ptcut\ and \etacut\ in the \tmin\ (left) and \tmax\ (right) regions as defined by the leading charged particle, as a function of the transverse momentum of the leading charged-particle \ptmax. The data are compared to the \pyoldhyphen\ \tunelep, \cuePA\ and \cuePAH. The green bands in the ratios represent the total experimental uncertainties.}
\label{PUB_fig3}
\end{center}
\end{figure*}

\begin{figure*}[htbp]
\begin{center}
\includegraphics[scale=0.65]{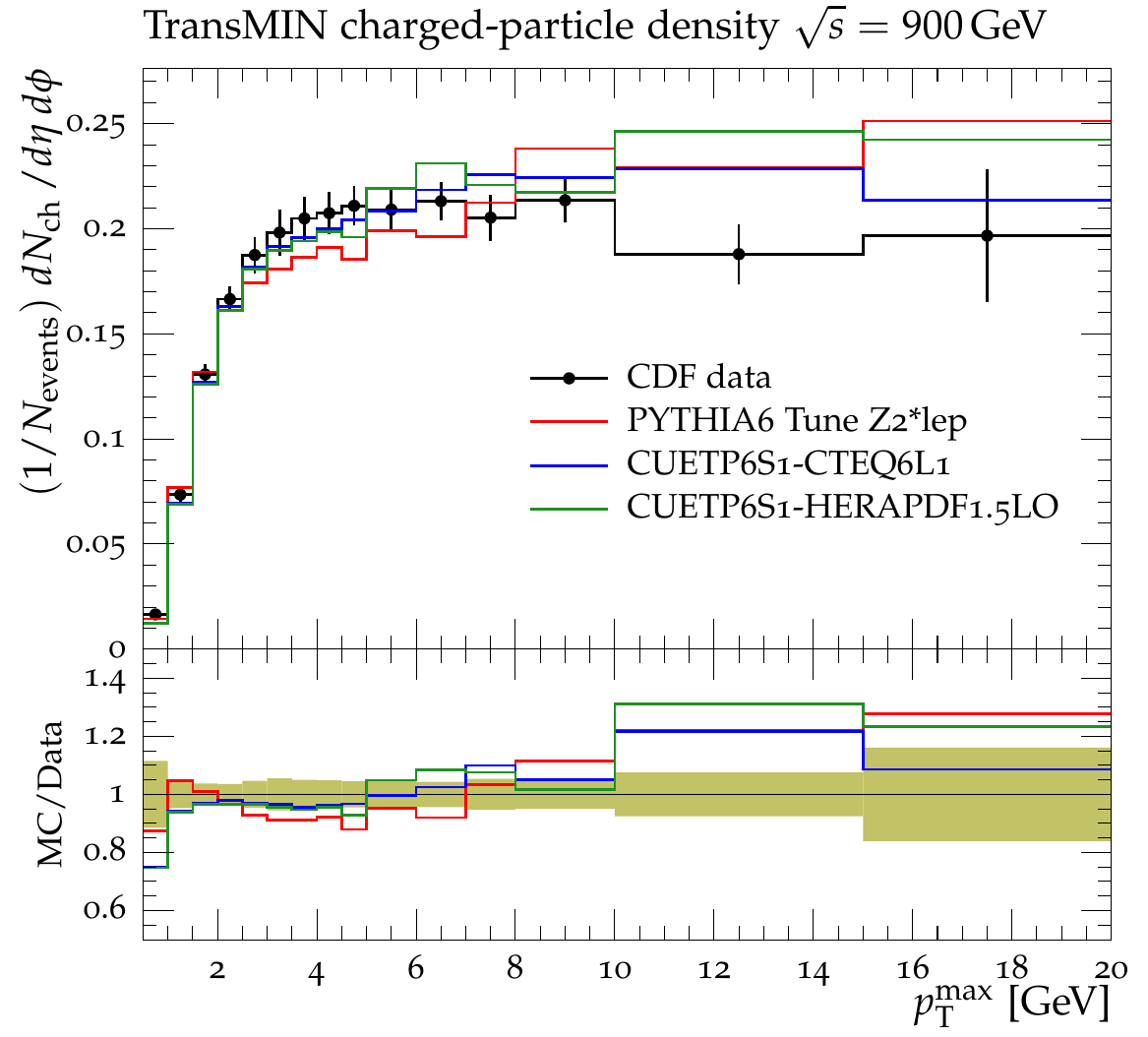}
\includegraphics[scale=0.65]{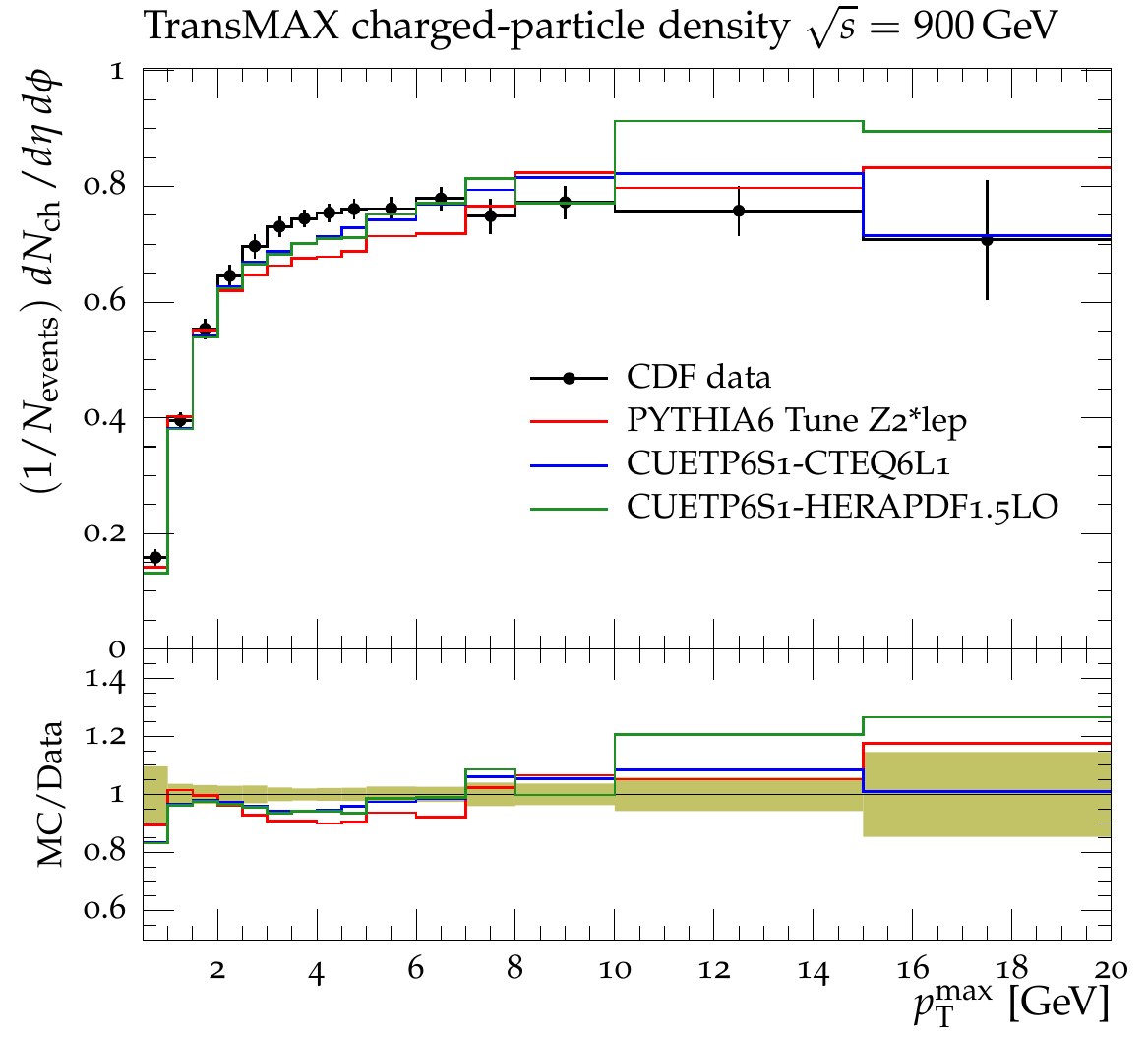}\\
\includegraphics[scale=0.65]{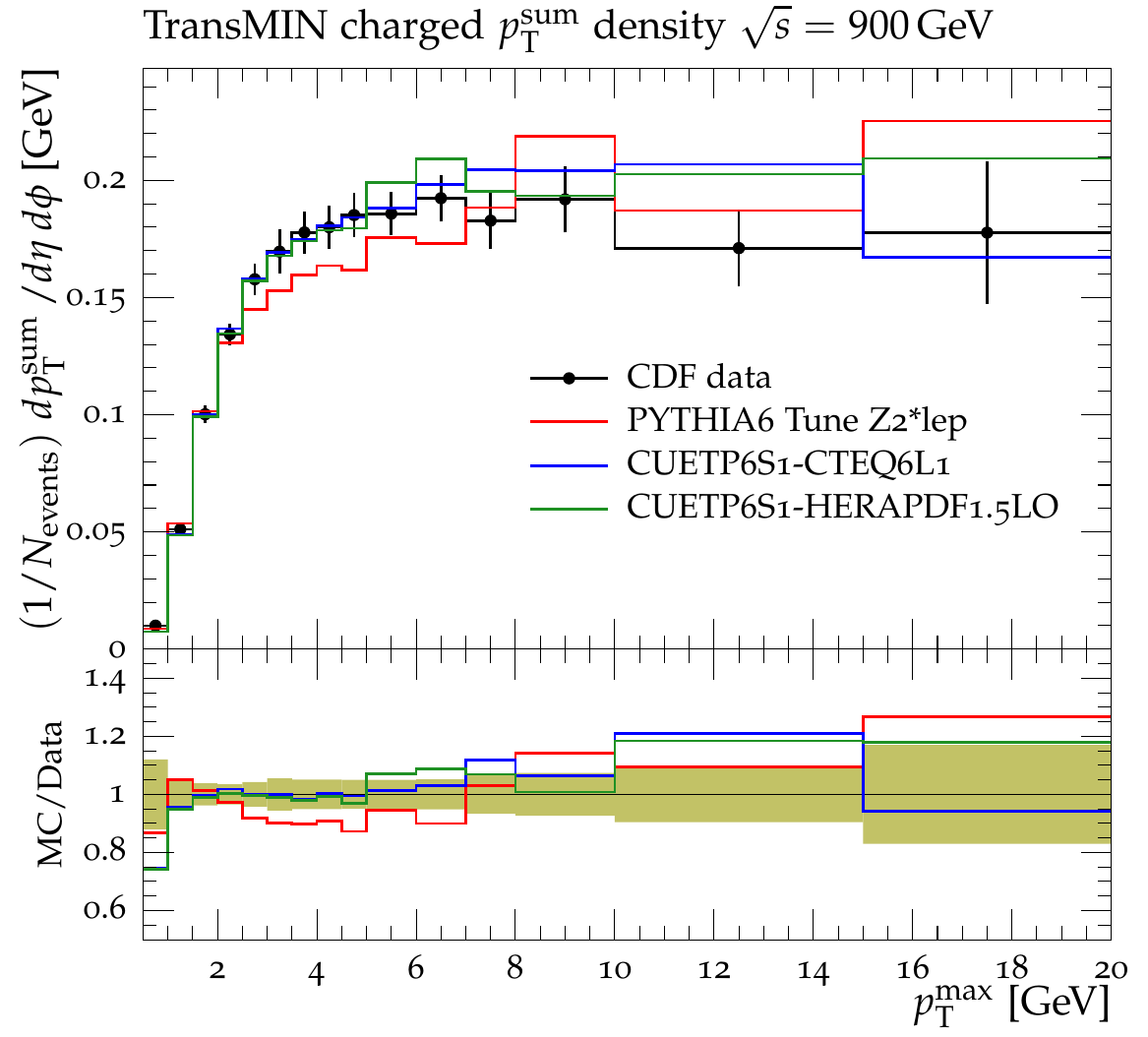}
\includegraphics[scale=0.65]{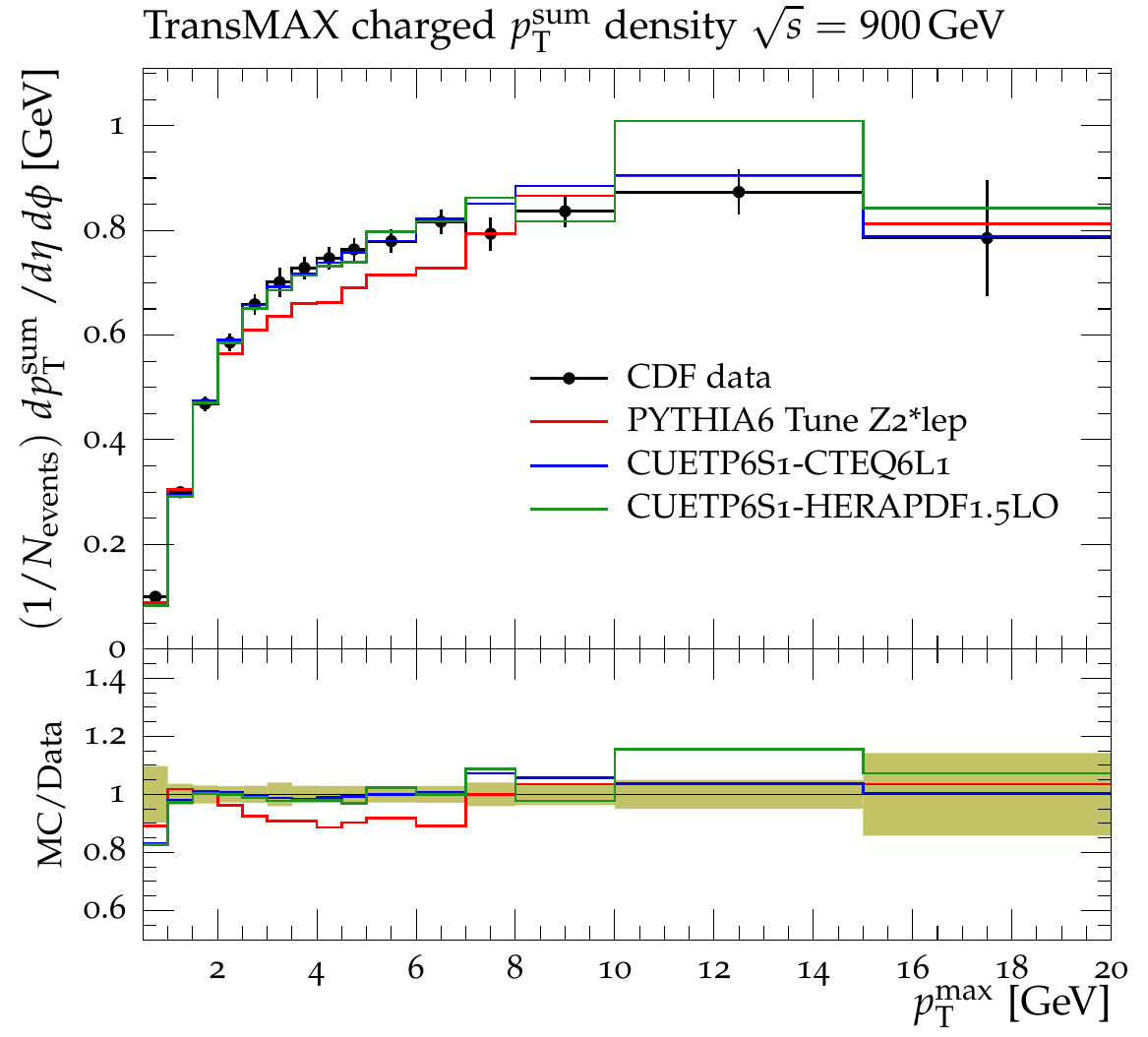}
\caption{CDF data at $\sqrt{s}=900\GeV$~\cite{Aaltonen:2015aoa} on the particle  (top) and \ptsum\ densities (bottom) for charged particles with \ptcut\ and \etacut\ in the \tmin\ (left) and \tmax\ (right) regions as defined by the leading charged particle, as a function of the transverse momentum of the leading-charged particle \ptmax. The data are compared to the \pyoldhyphen\ \tunelep, \cuePA\ and \cuePAH. The green bands in the ratios represent the total experimental uncertainties.}
\label{PUB_fig4}
\end{center}
\end{figure*}

\begin{figure*}[htbp]
\begin{center}
\includegraphics[scale=0.65]{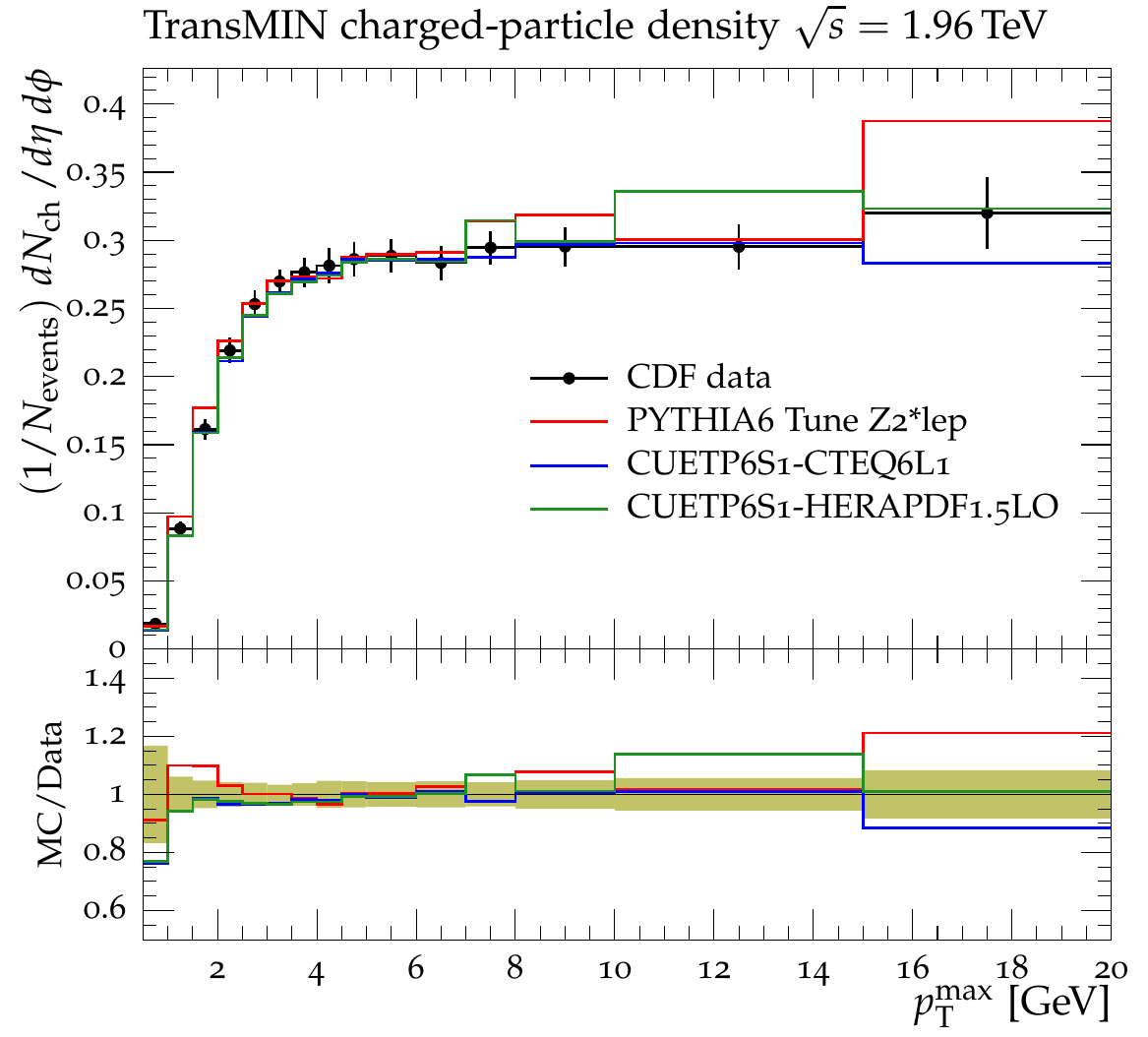}
\includegraphics[scale=0.65]{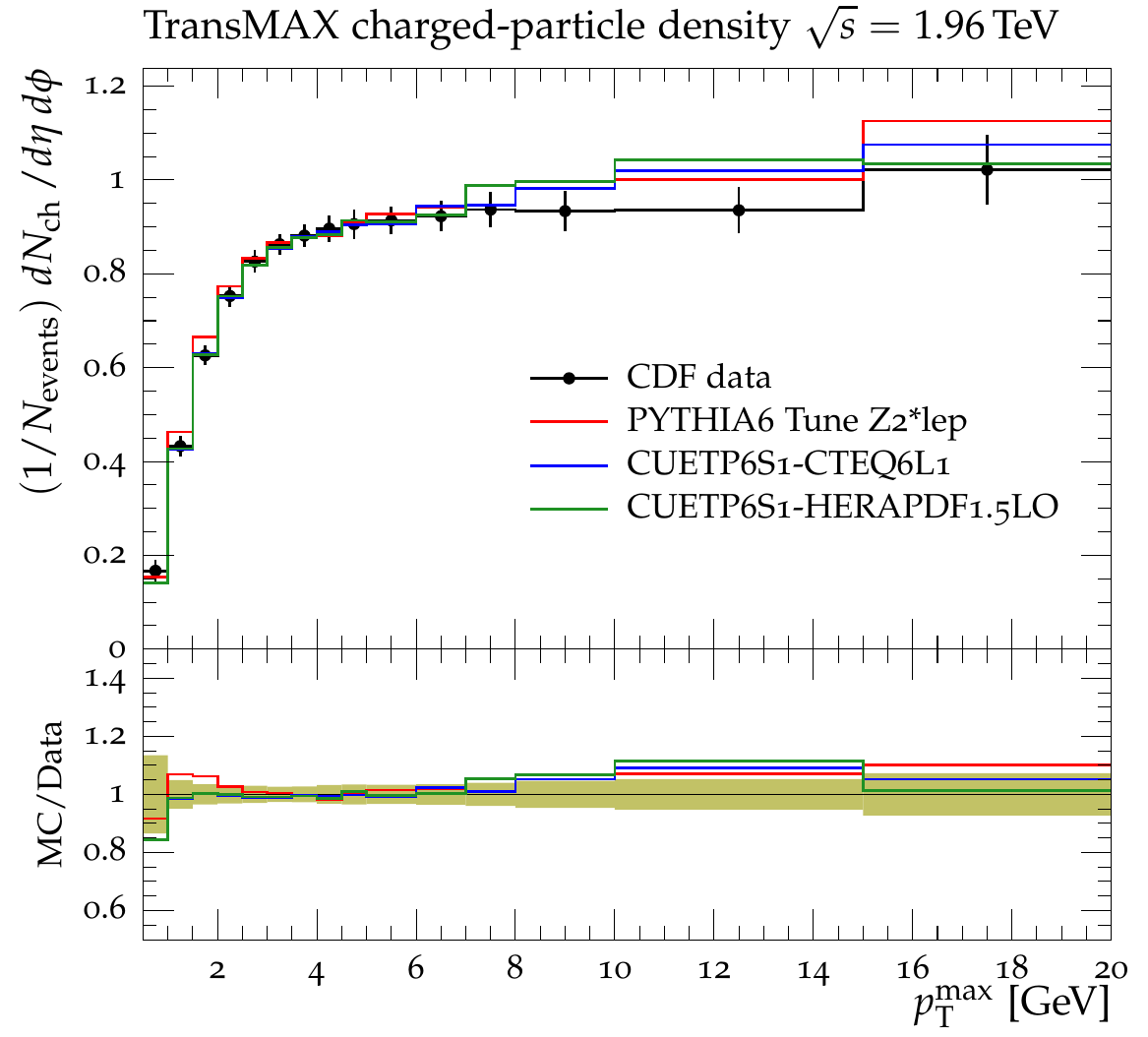}\\
\includegraphics[scale=0.65]{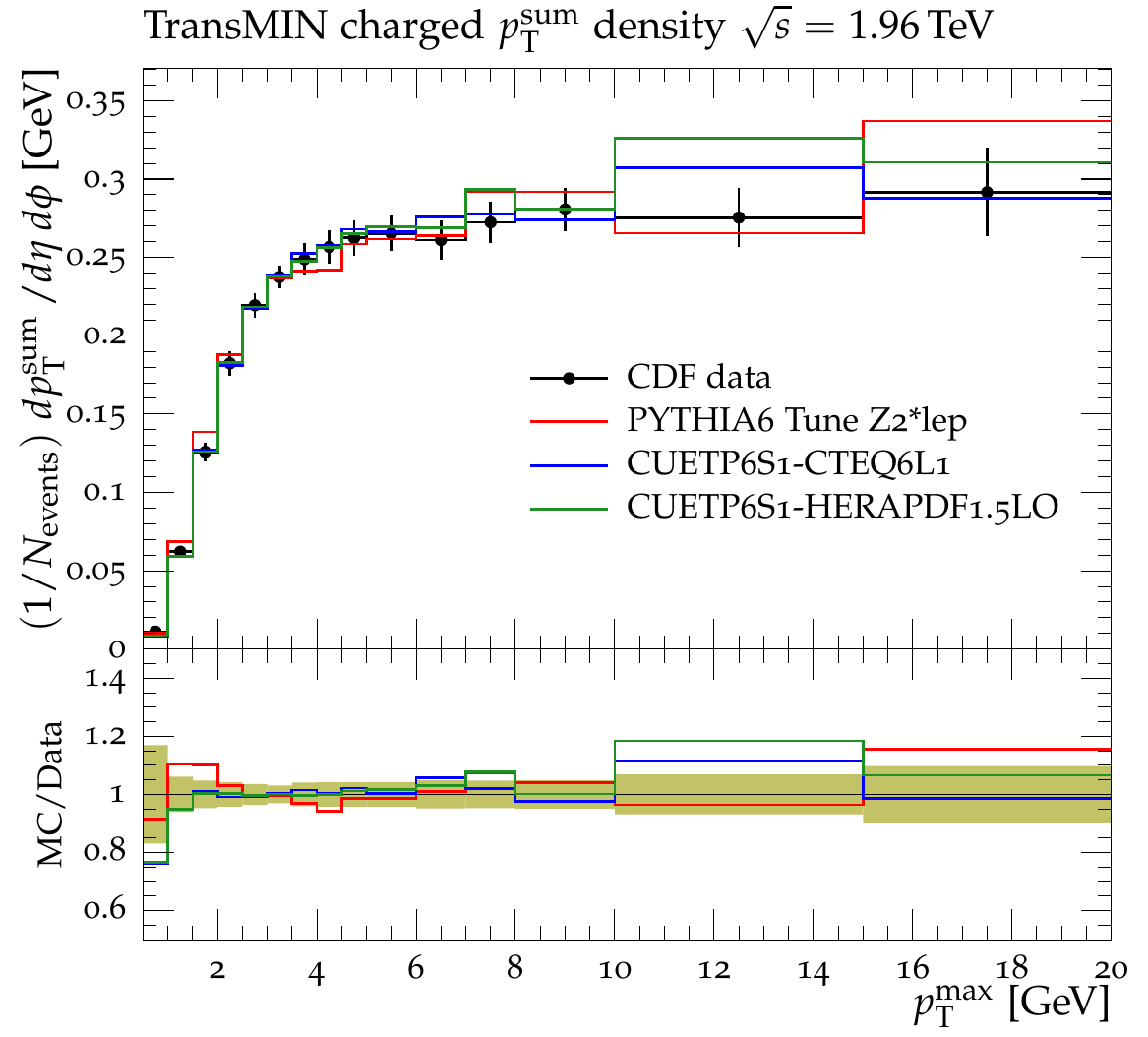}
\includegraphics[scale=0.65]{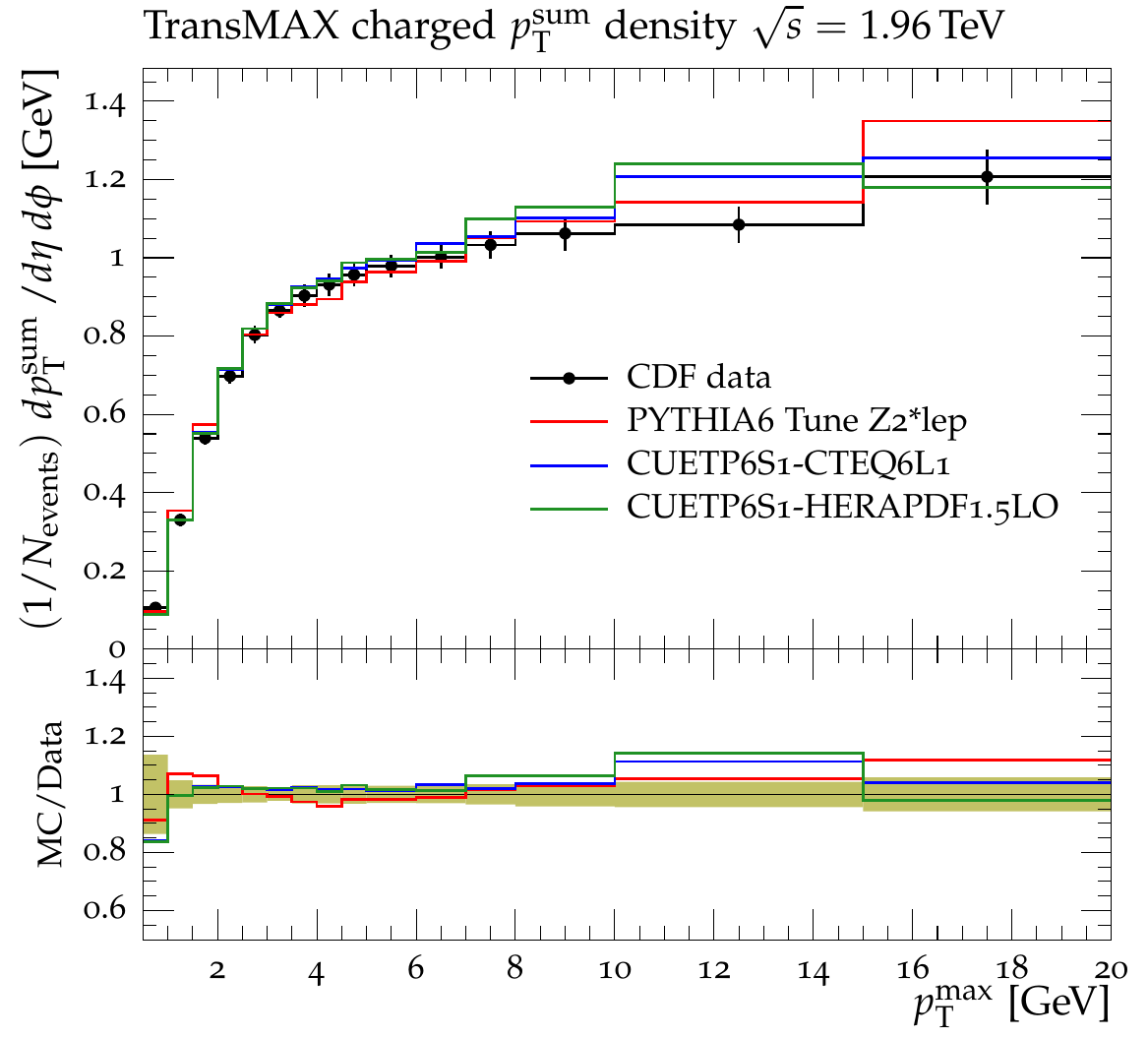}
\caption{CDF data at $\sqrt{s}=1.96\TeV$~\cite{Aaltonen:2015aoa} on the particle  (top) and \ptsum\ densities (bottom) for charged particles with \ptcut\ and \etacut\ in the \tmin\ (left) and \tmax\ (right) regions as defined by the leading charged particle, as a function of the transverse momentum of the leading charged-particle \ptmax. The data are compared to the \pyoldhyphen\ \tunelep, \cuePA\ and \cuePAH. The green bands in the ratios represent the total experimental uncertainties.}
\label{PUB_fig5}
\end{center}
\end{figure*}

\begin{figure*}[htbp]
\begin{center}
\includegraphics[scale=0.65]{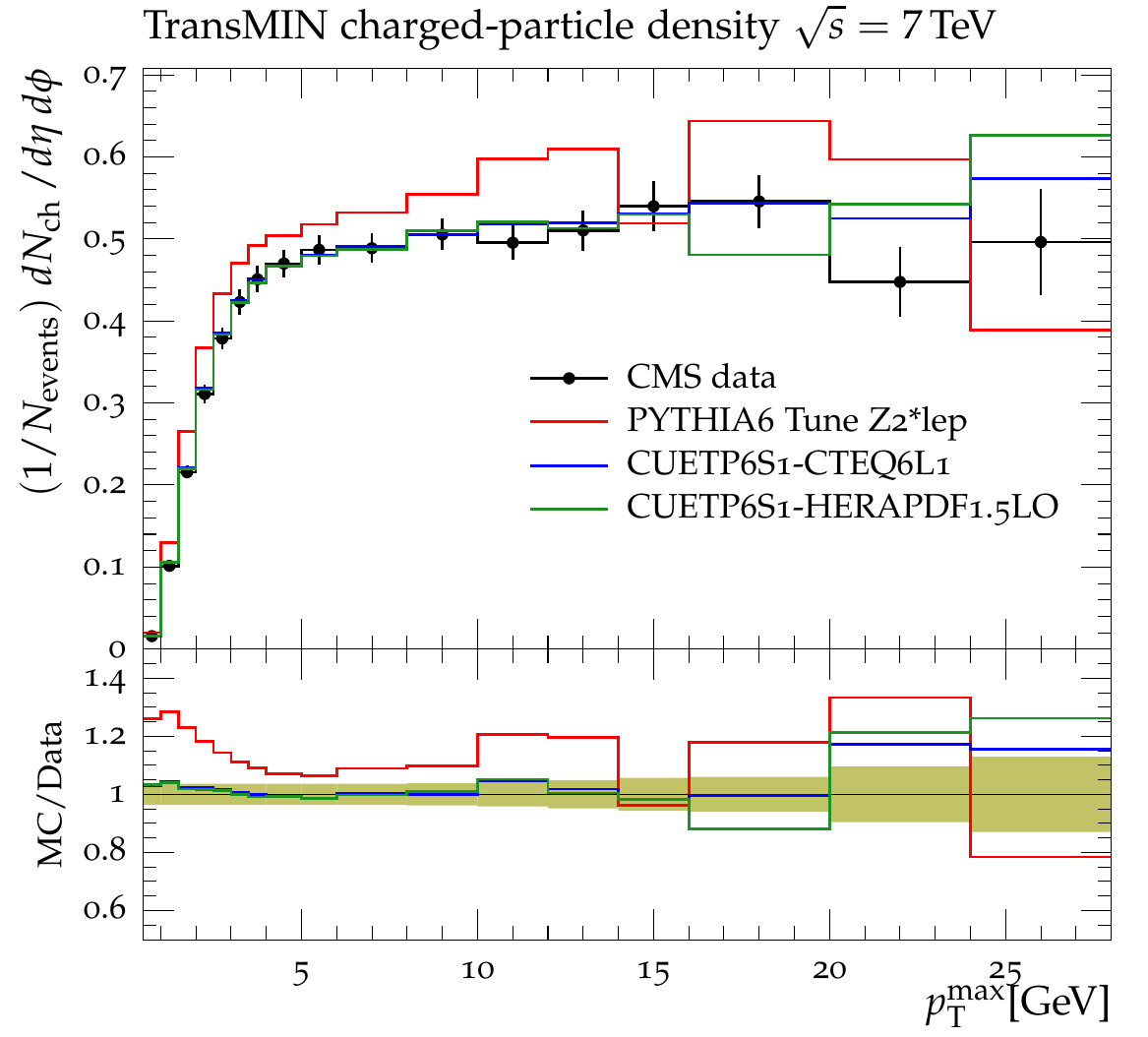}
\includegraphics[scale=0.65]{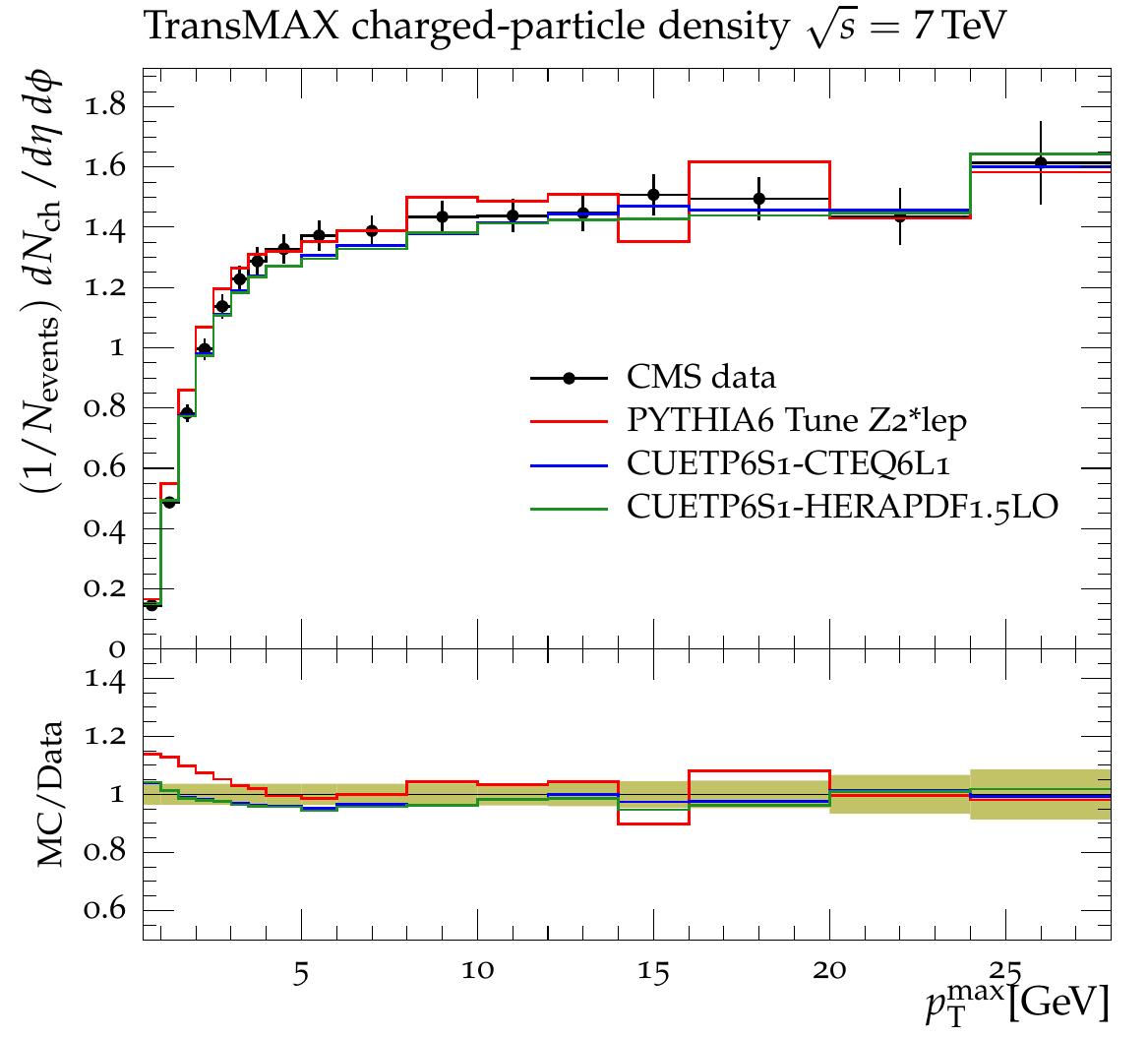}\\
\includegraphics[scale=0.65]{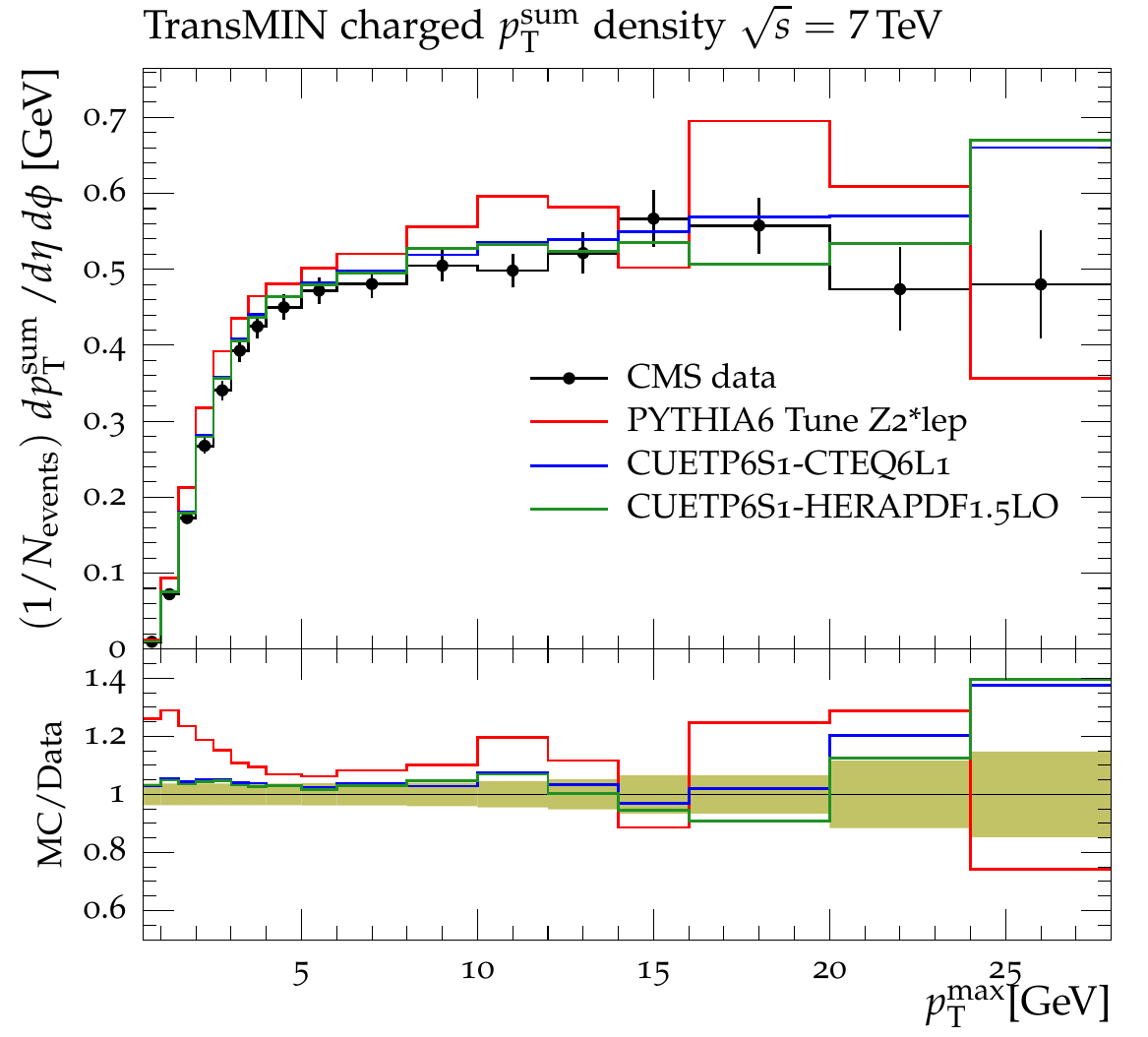}
\includegraphics[scale=0.65]{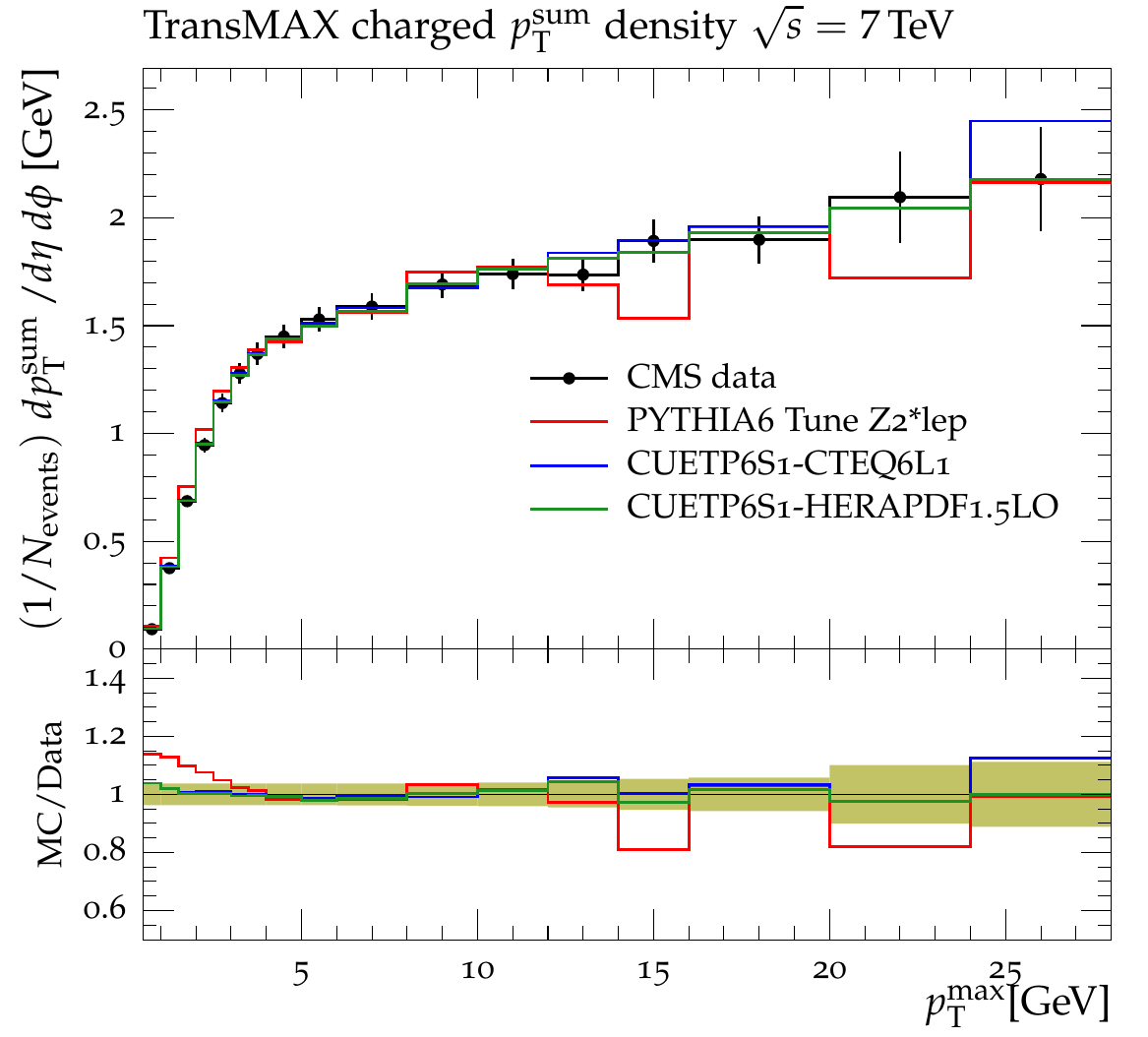}
\caption{CMS data at $\sqrt{s}=7\TeV$~\cite{CMS:2012kca} on the particle  (top) and \ptsum\ densities (bottom) for charged particles with \ptcut\ and \etacut\ in the \tmin\ (left) and \tmax\ (right) regions as defined by the leading charged particle, as a function of the transverse momentum of the leading charged-particle \ptmax. The data are compared to the \pyoldhyphen\ \tunelep, \cuePA\ and \cuePAH. The green bands in the ratios represent the total experimental uncertainties.}
\label{PUB_fig6}
\end{center}
\end{figure*}

Figures~\ref{PUB_fig3}--\ref{PUB_fig6} show the CDF data at $\sqrt{s} = 0.3$, $0.9$, and $1.96\TeV$, and the CMS data at $\sqrt{s} = 7\TeV$ on charged-particle and \ptsum\ densities in the \tmin\ and \tmax\ regions, as a function of the transverse momentum of the leading charged-particle \ptmax. The distributions are compared to predictions obtained with \pyoldhyphen\ \tunelep\ and the two new \cuePA\ and \cuePAH. The new CMS \pyold\ tunes are able to describe the measurements better than \tunelep, in both the rising and the plateau regions of the spectra.

\ifthenelse{\boolean{cms@external}}{\section{Comparisons to HERWIG++ UE tunes to data}}{\section{Comparisons to \textsc{herwig++} UE tunes to data}}

\begin{figure*}[htbp]
\begin{center}
\includegraphics[scale=0.65]{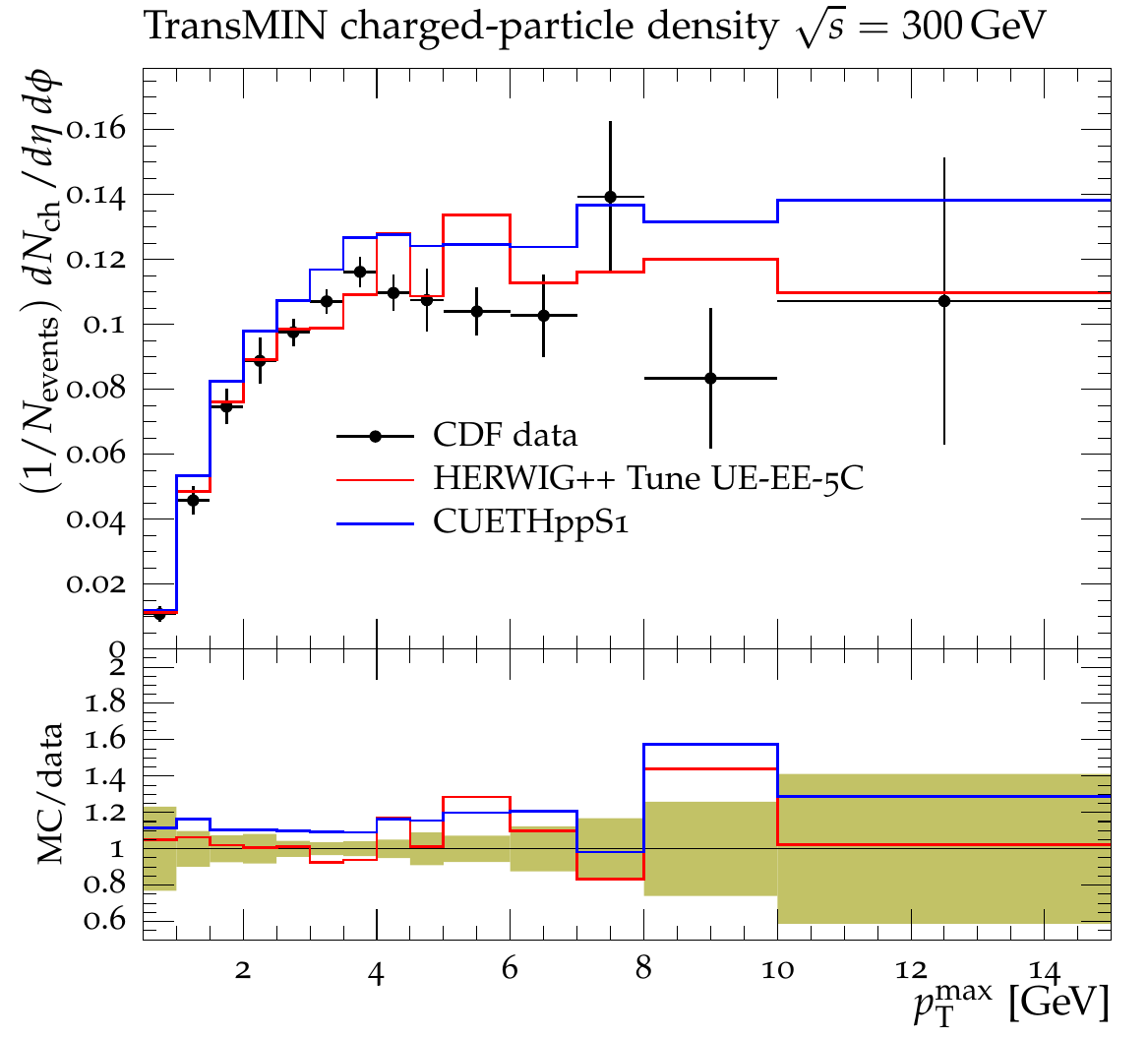}
\includegraphics[scale=0.65]{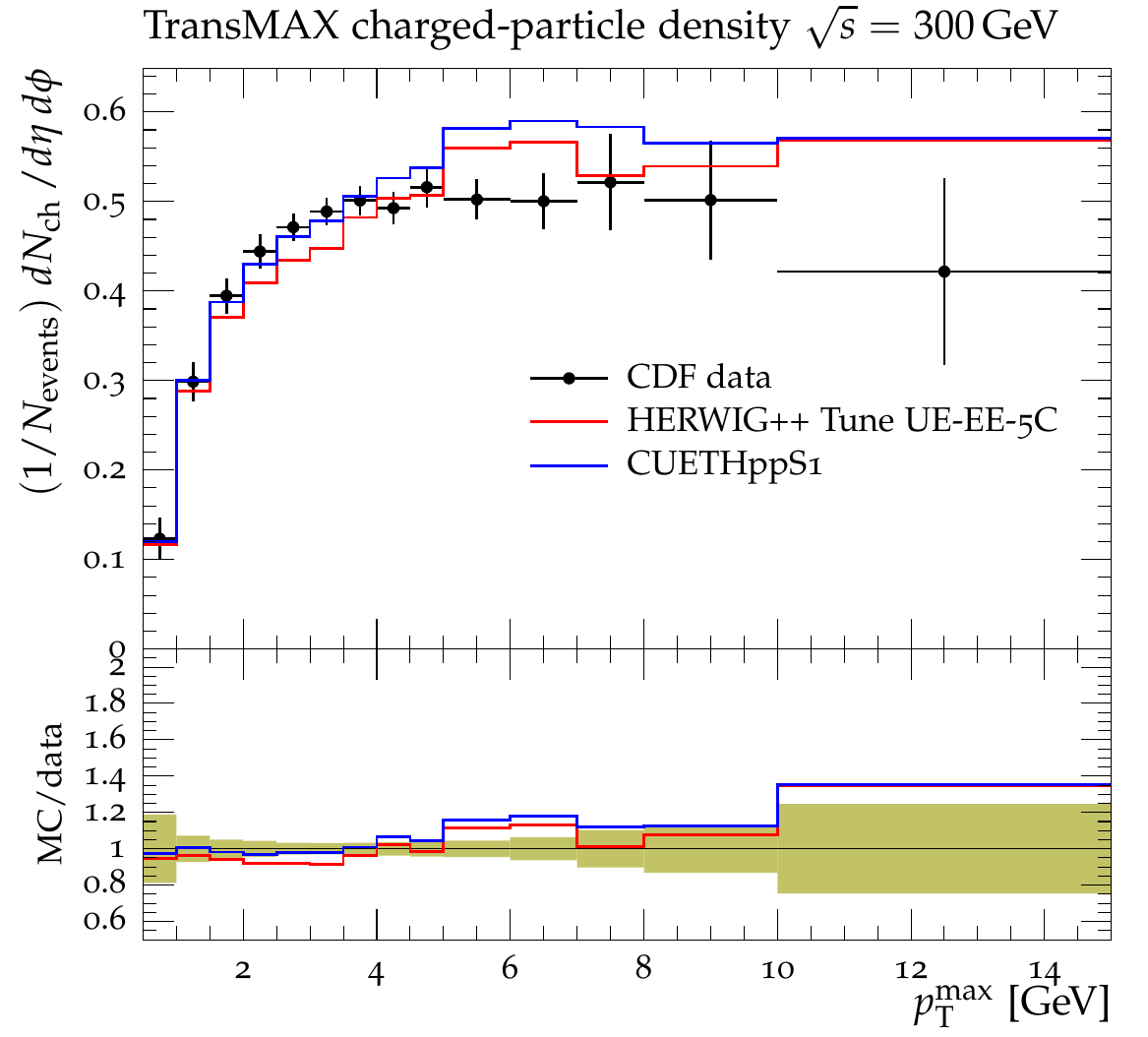}\\
\includegraphics[scale=0.65]{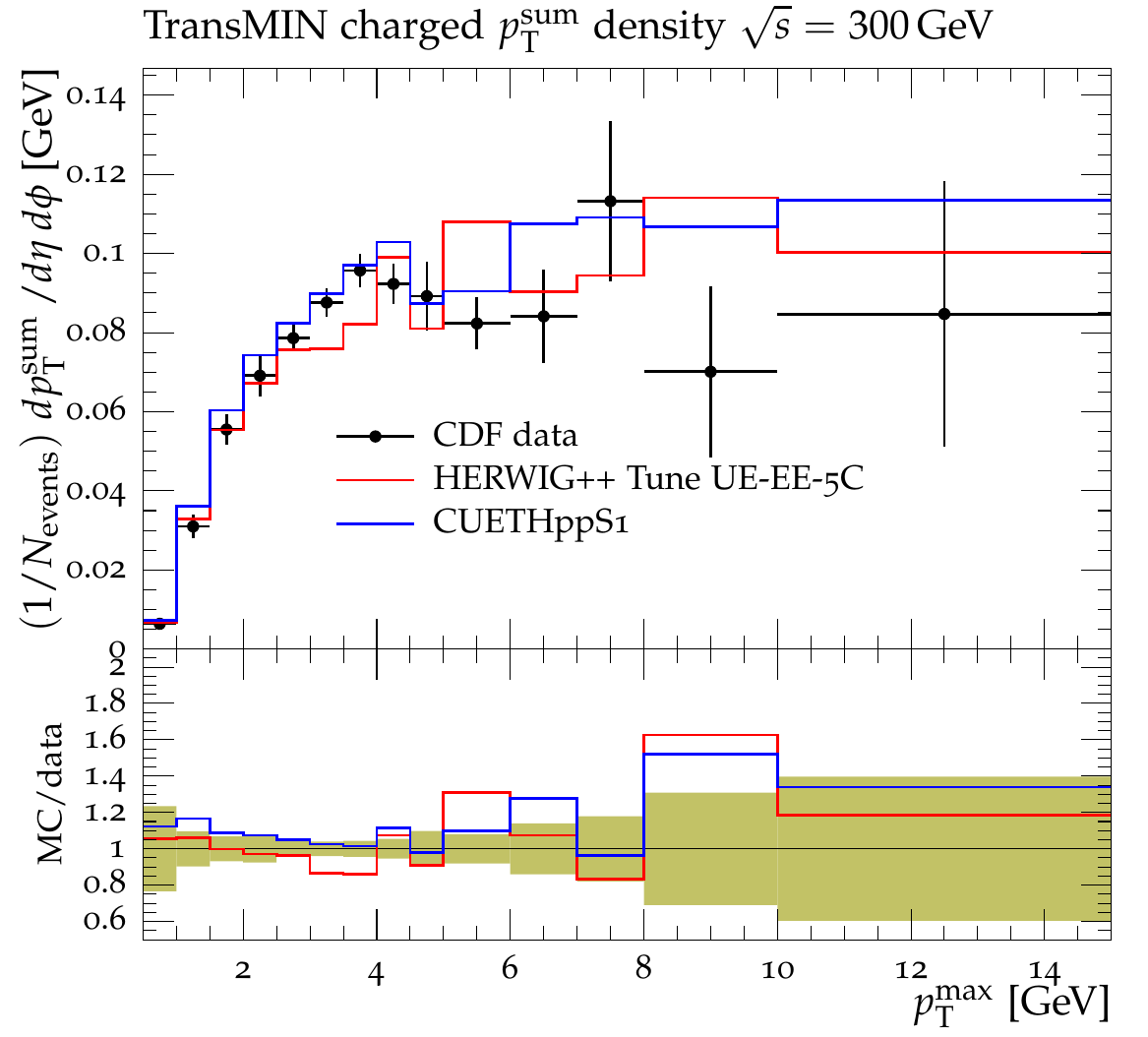}
\includegraphics[scale=0.65]{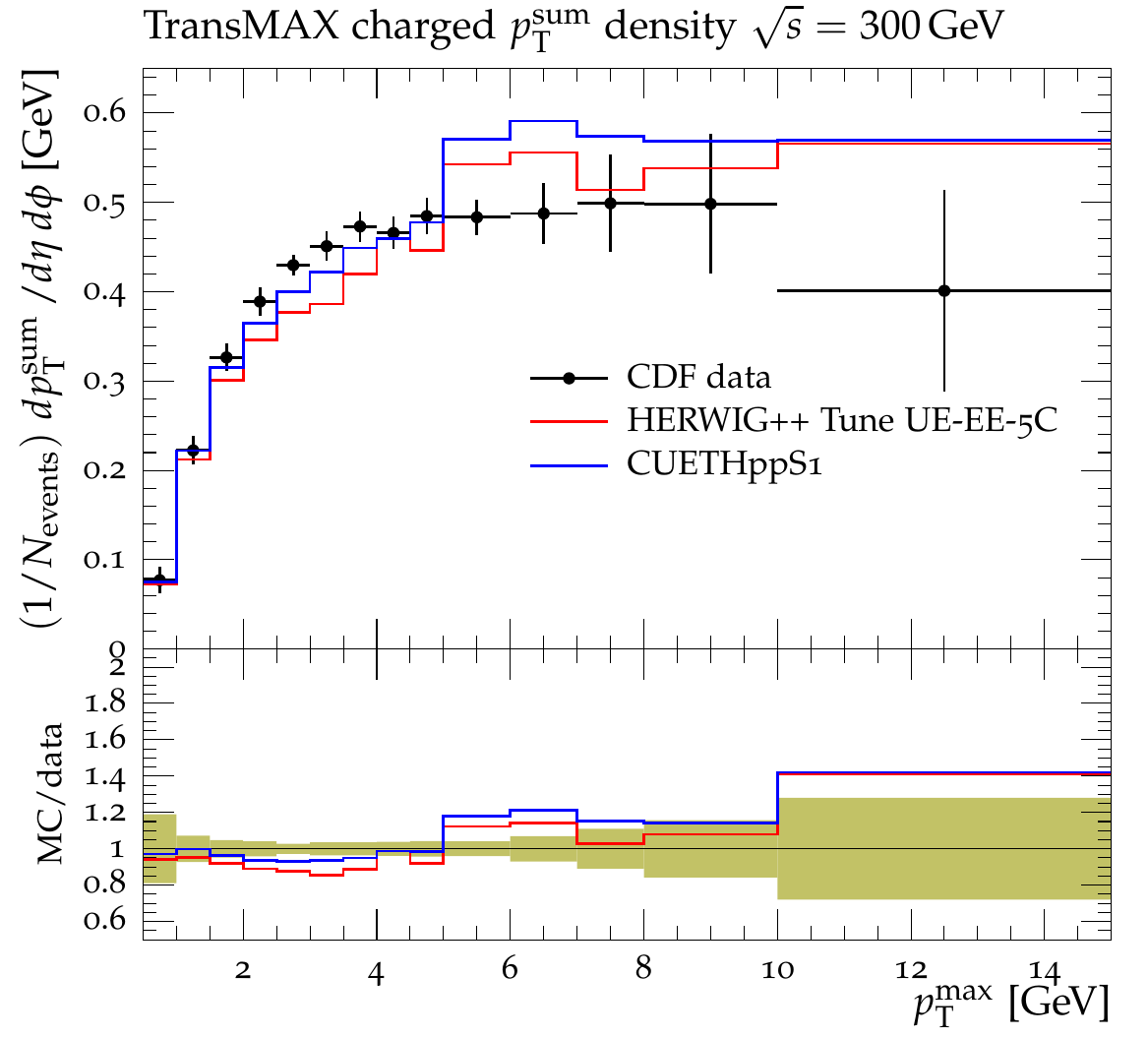}
\caption{CDF data at $\sqrt{s}=300\GeV$~\cite{Aaltonen:2015aoa} on particle  (top) and \ptsum\ densities (bottom) for charged particles with \ptcut\ and \etacut\ in the \tmin\ (left) and \tmax\ (right) regions as defined by the leading charged particle, as a function of the transverse momentum of the leading charged-particle \ptmax. The data are compared to the \hwpp\ Tune \tunee\ and \cueHW. The green bands in the ratios represent the total experimental uncertainties.}
\label{PUB_fig11}
\end{center}
\end{figure*}

\begin{figure*}[htbp]
\begin{center}
\includegraphics[scale=0.65]{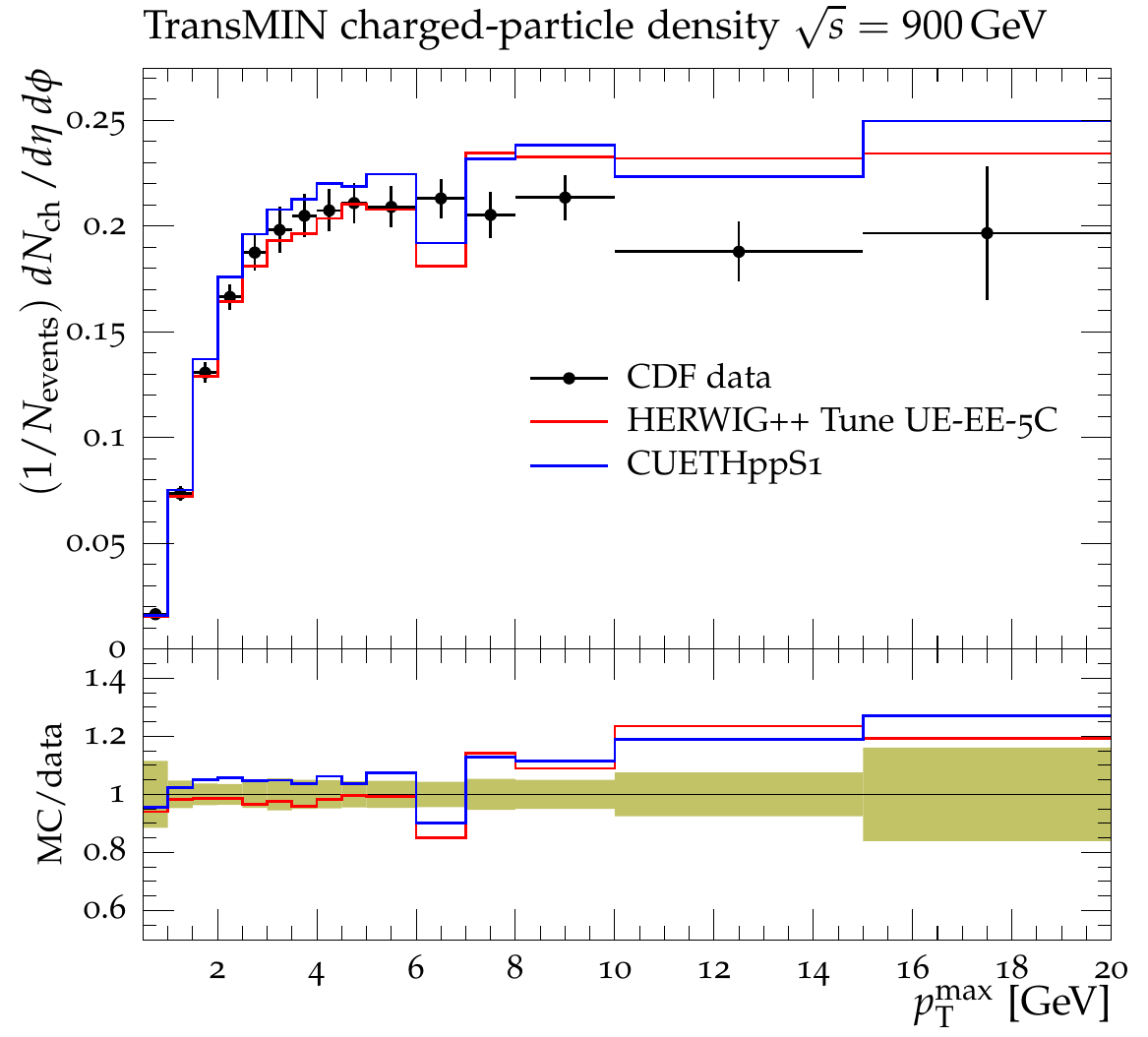}
\includegraphics[scale=0.65]{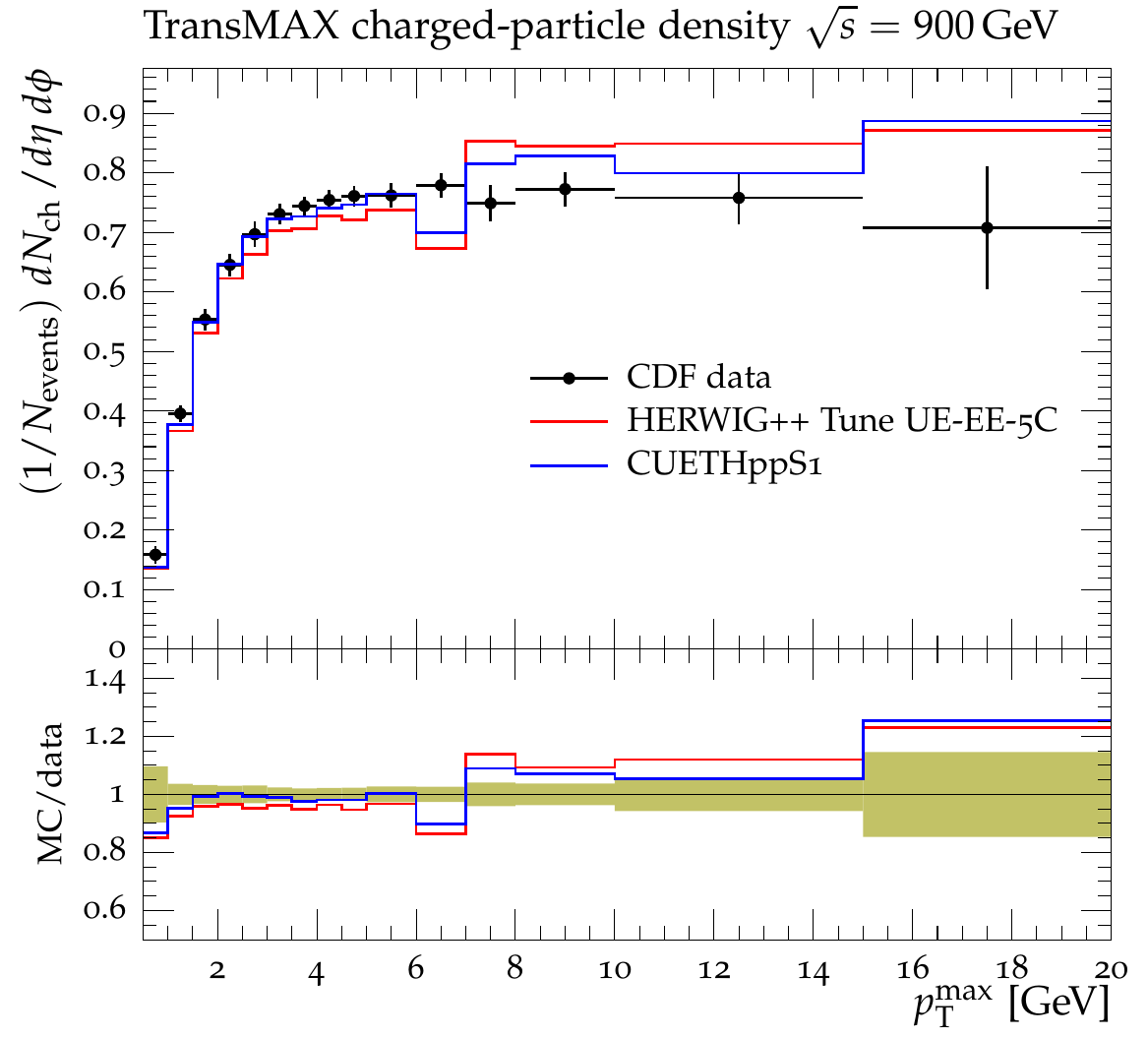}\\
\includegraphics[scale=0.65]{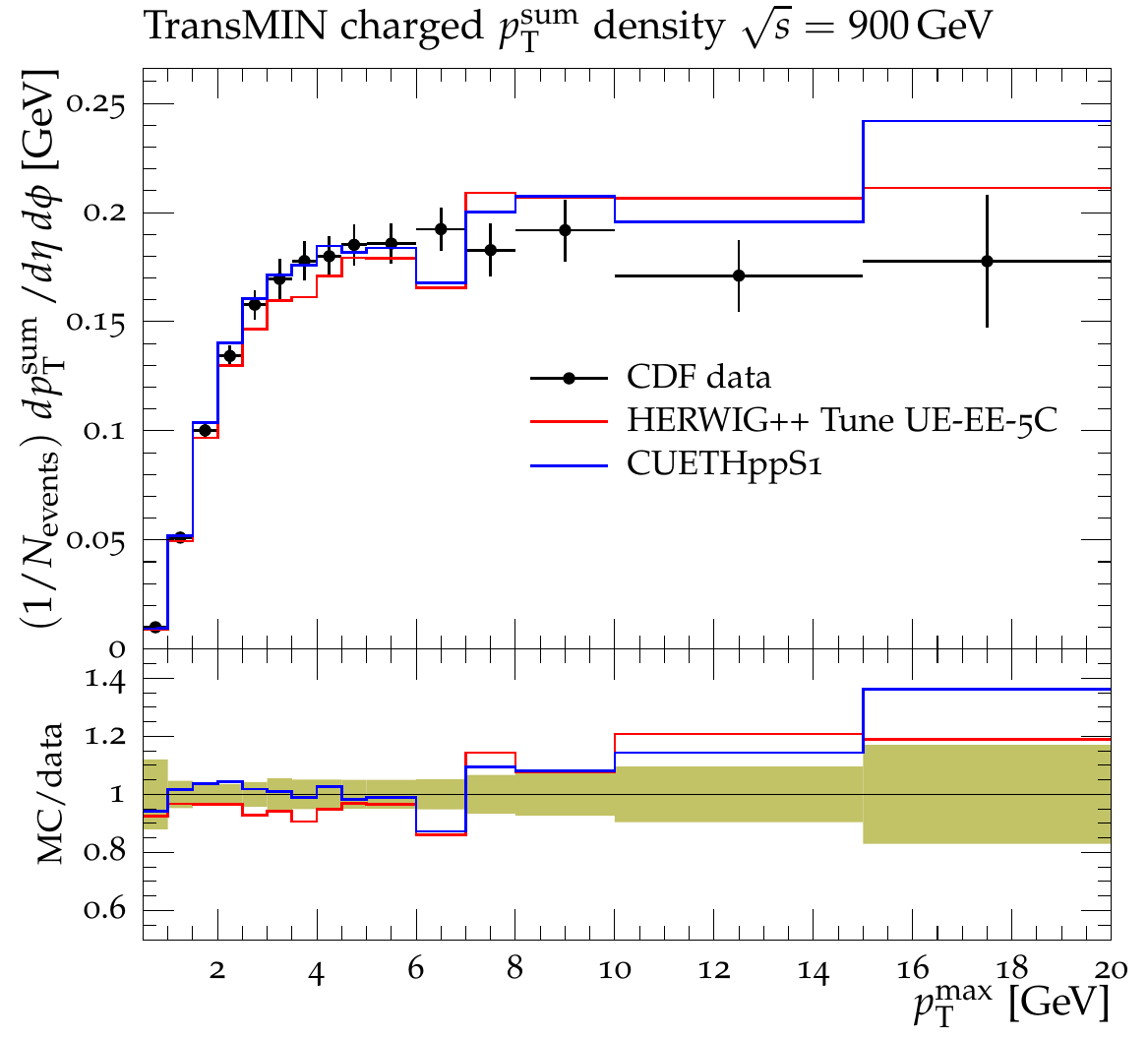}
\includegraphics[scale=0.65]{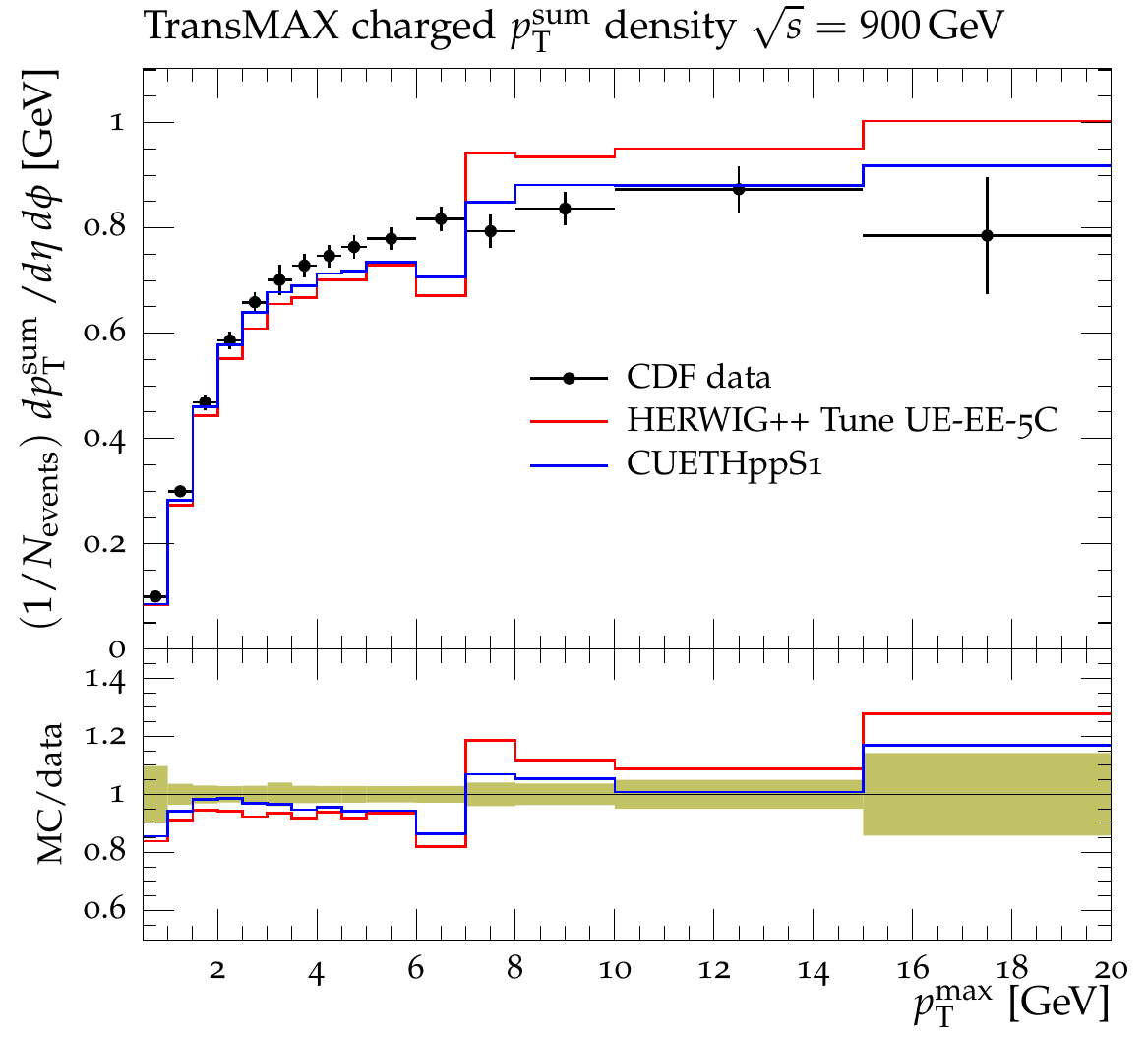}
\caption{CDF data at $\sqrt{s}=900\GeV$~\cite{Aaltonen:2015aoa} on particle  (top) and \ptsum\ densities (bottom) for charged particles with \ptcut\ and \etacut\ in the \tmin\ (left) and \tmax\ (right) regions as defined by the leading charged particle, as a function of the transverse momentum of the leading charged-particle \ptmax. The data are compared to the \hwpp\ Tune \tunee\ and \cueHW. The green bands in the ratios represent the total experimental uncertainties.}
\label{PUB_fig12}
\end{center}
\end{figure*}

\begin{figure*}[htbp]
\begin{center}
\includegraphics[scale=0.65]{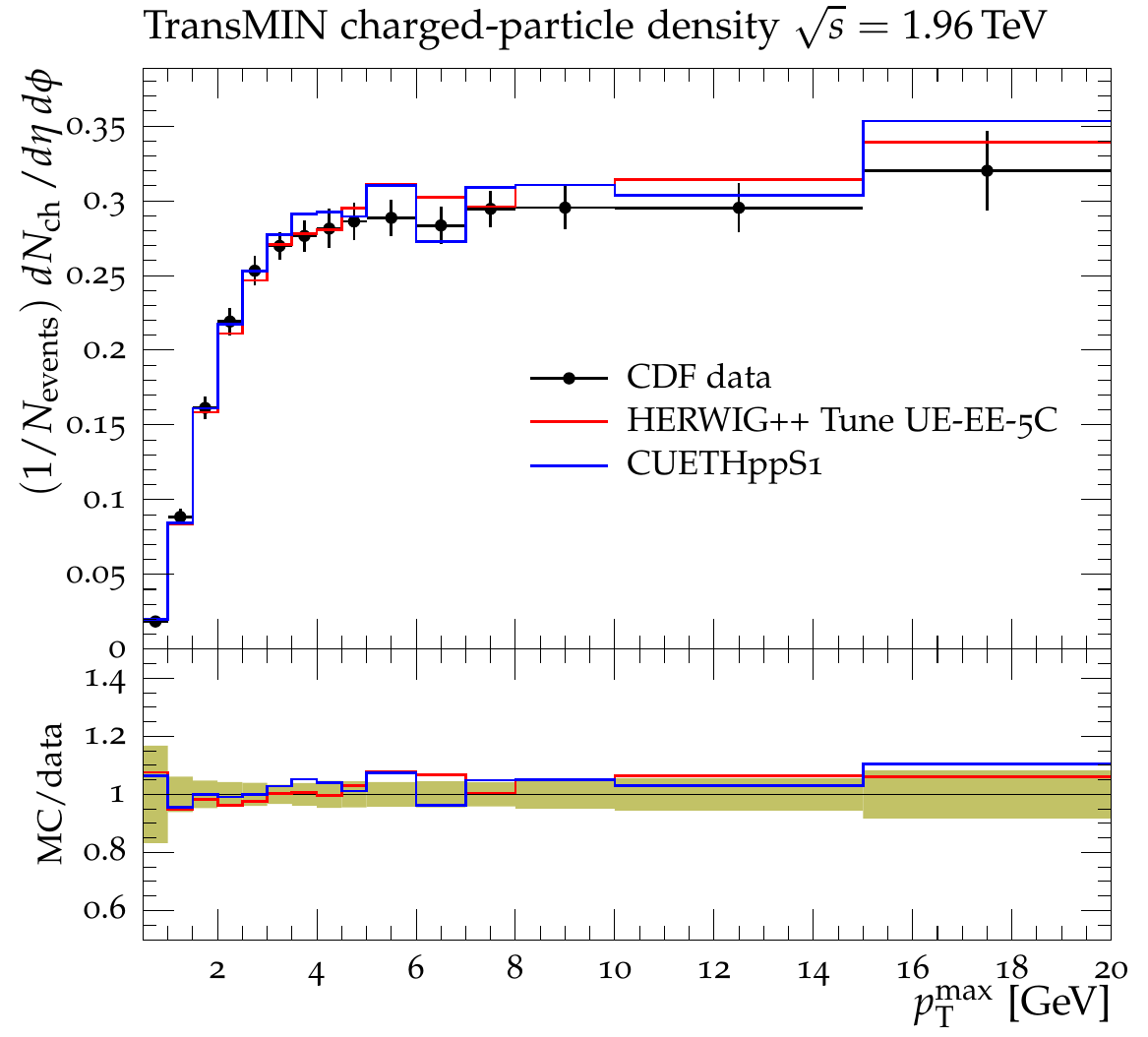}
\includegraphics[scale=0.65]{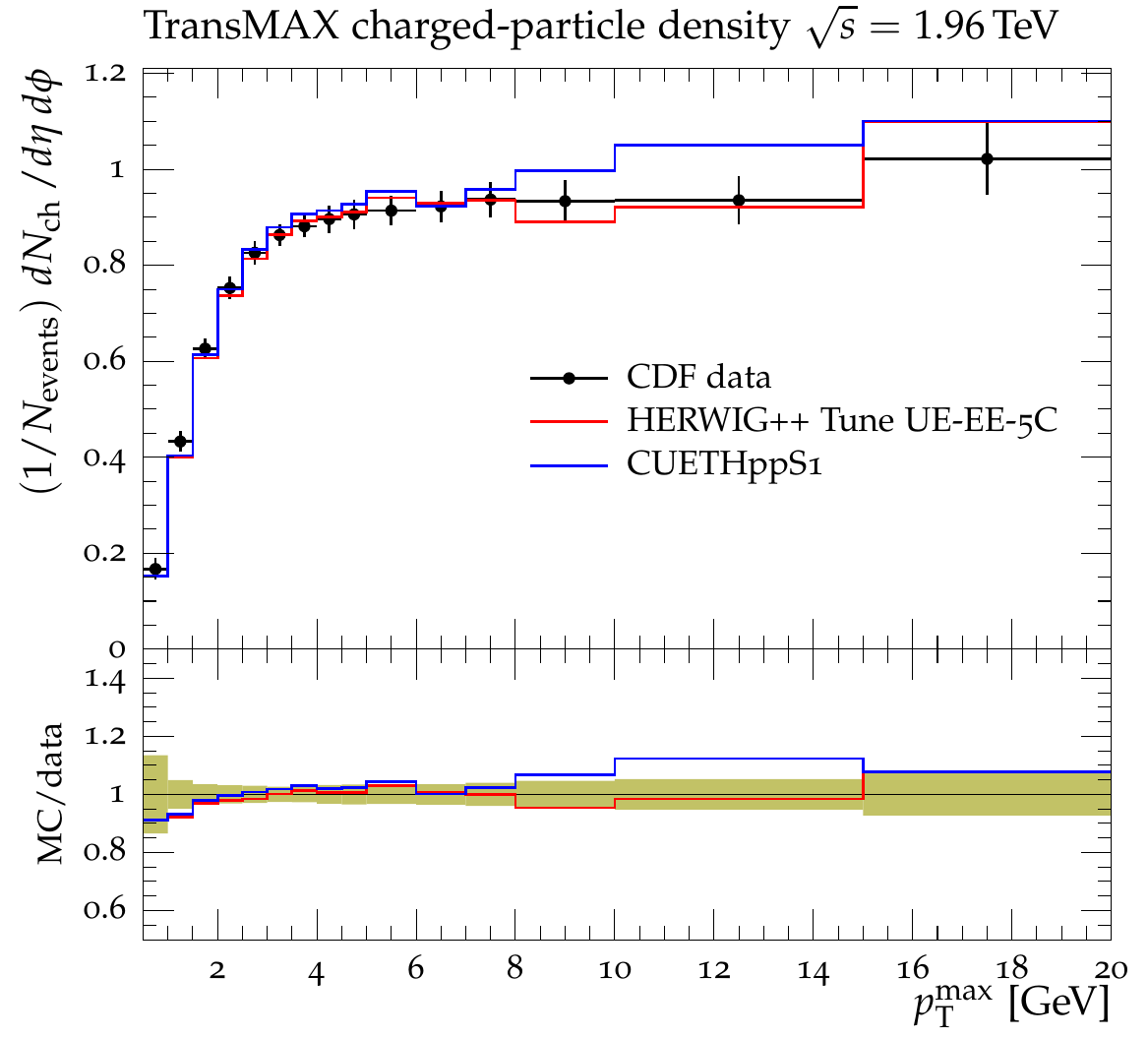}\\
\includegraphics[scale=0.65]{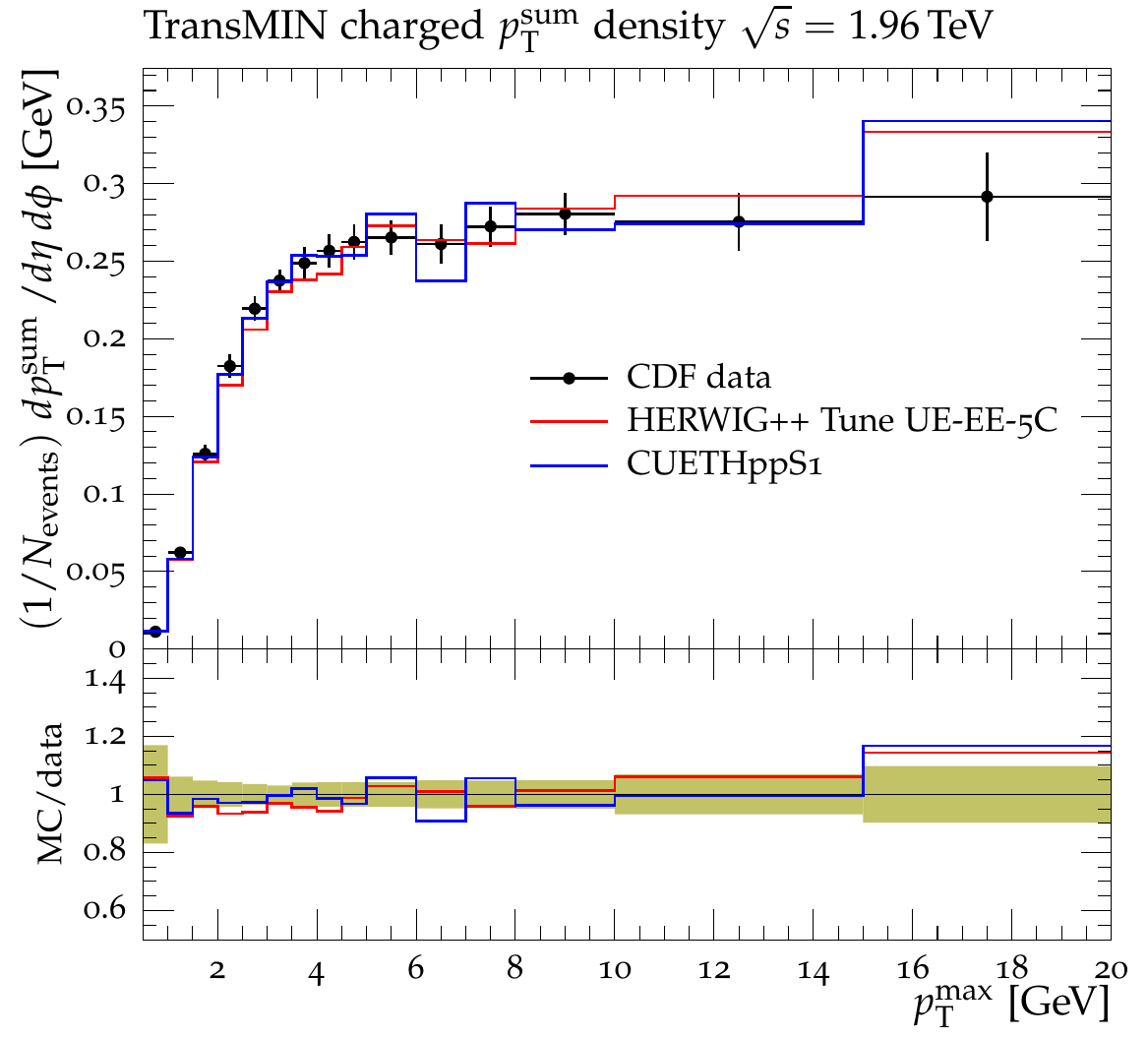}
\includegraphics[scale=0.65]{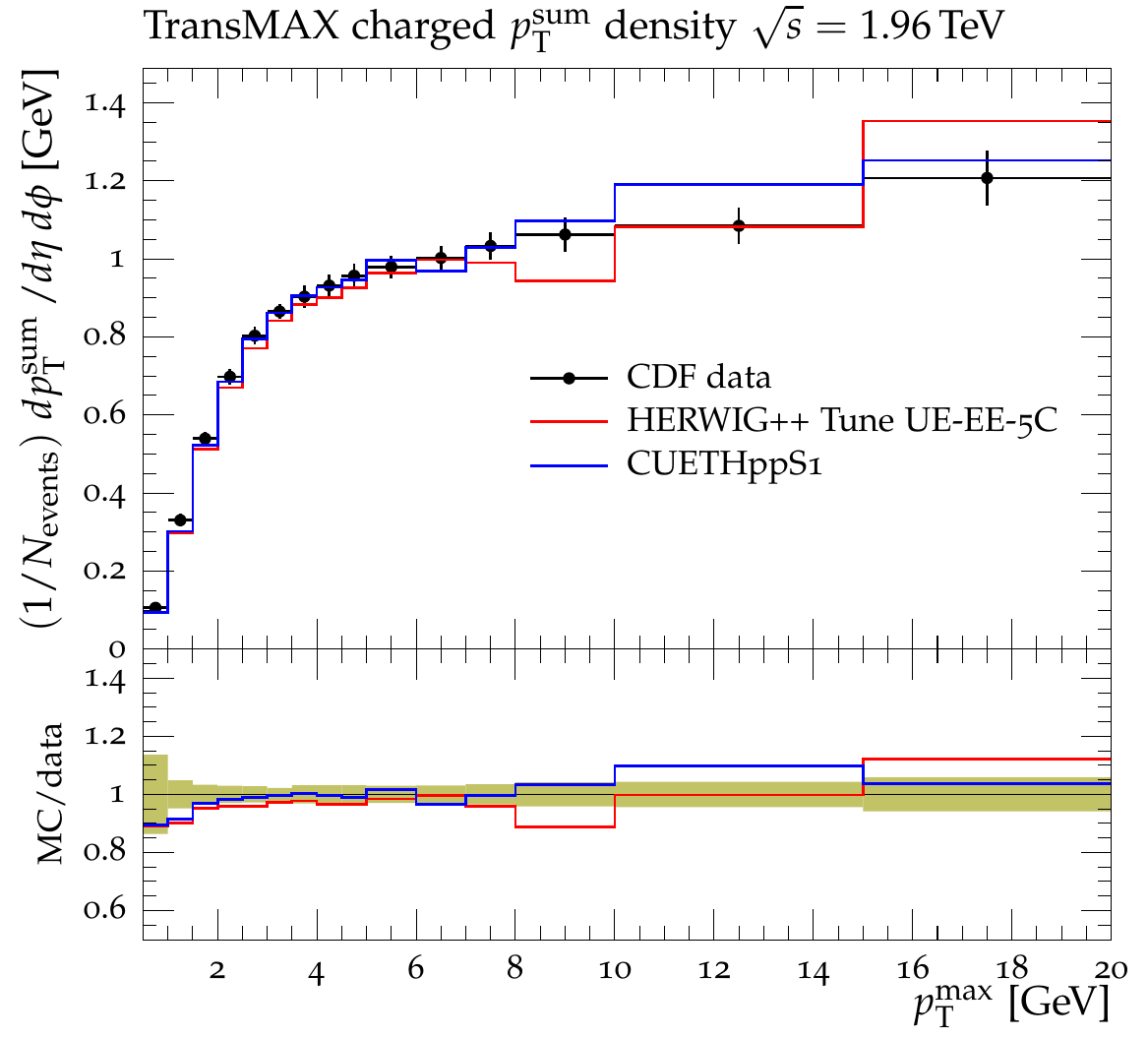}
\caption{CDF data at $\sqrt{s}=1.96\TeV$~\cite{Aaltonen:2015aoa} on particle  (top) and \ptsum\ densities (bottom) for charged particles with \ptcut\ and \etacut\ in the \tmin\ (left) and \tmax\ (right) regions as defined by the leading charged particle, as a function of the transverse momentum of the leading charged-particle \ptmax. The data are compared to the \hwpp\ Tune \tunee\ and \cueHW. The green bands in the ratios represent the total experimental uncertainties.}
\label{PUB_fig13}
\end{center}
\end{figure*}

\begin{figure*}[htbp]
\begin{center}
\includegraphics[scale=0.65]{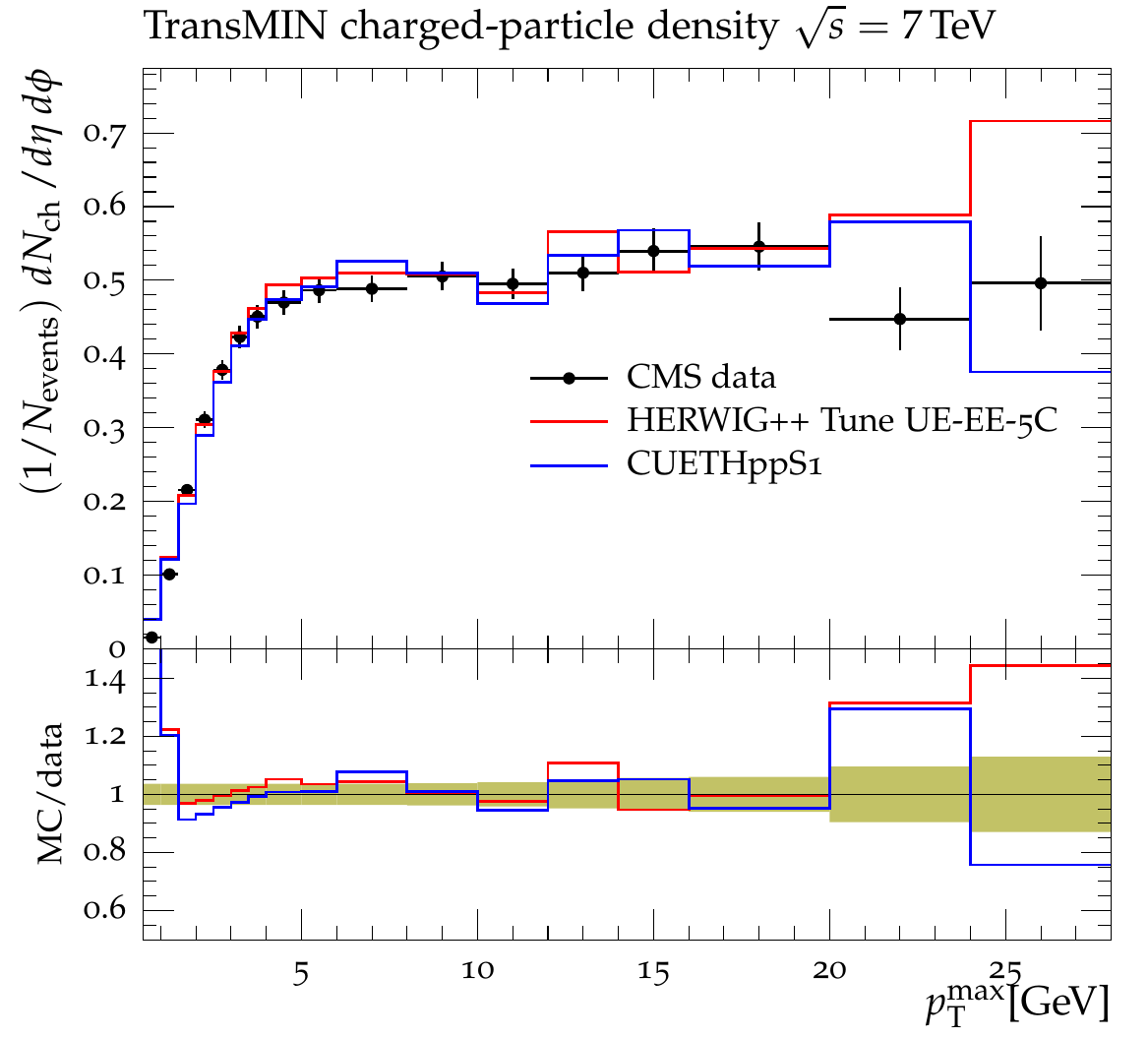}
\includegraphics[scale=0.65]{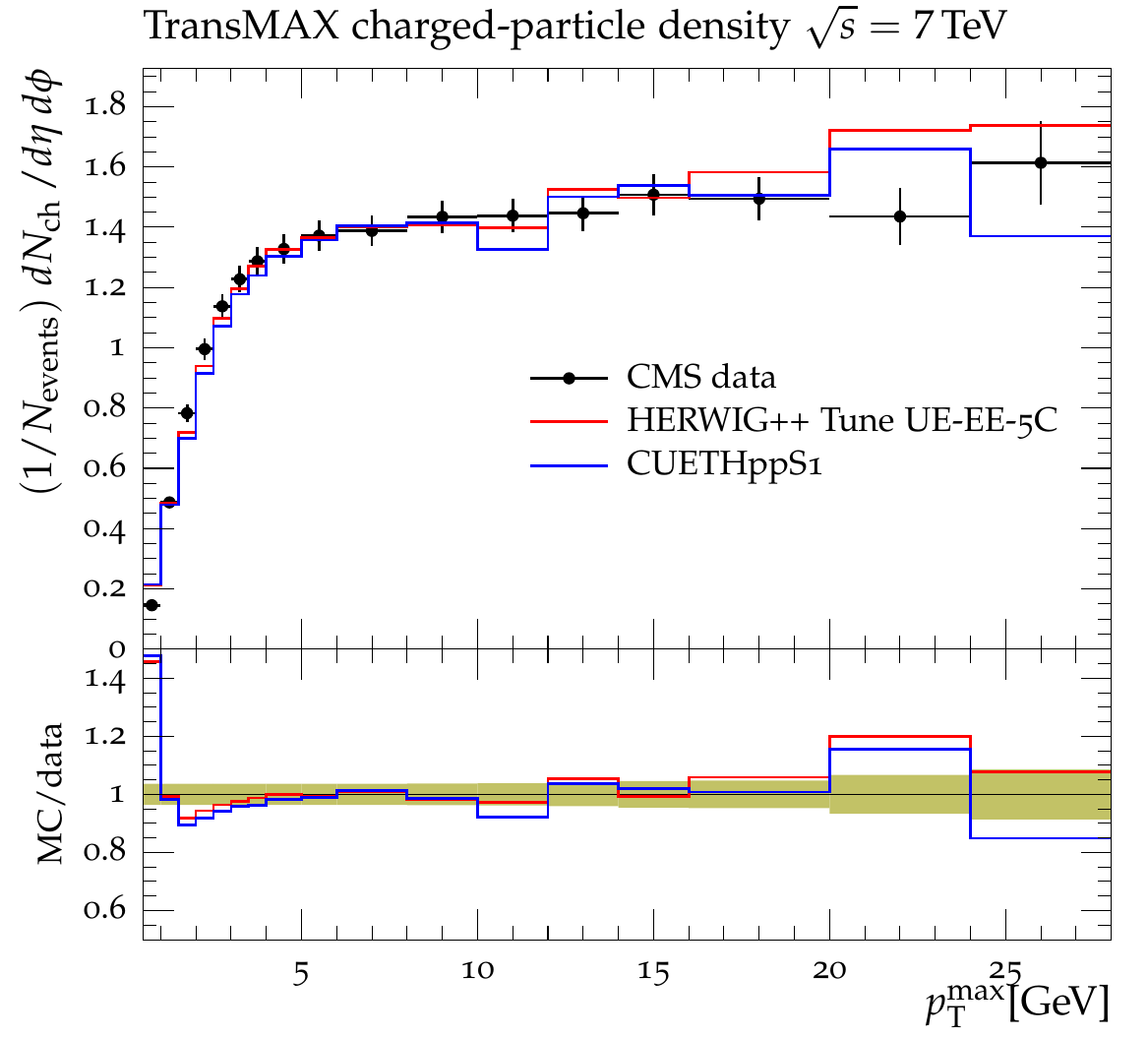}\\
\includegraphics[scale=0.65]{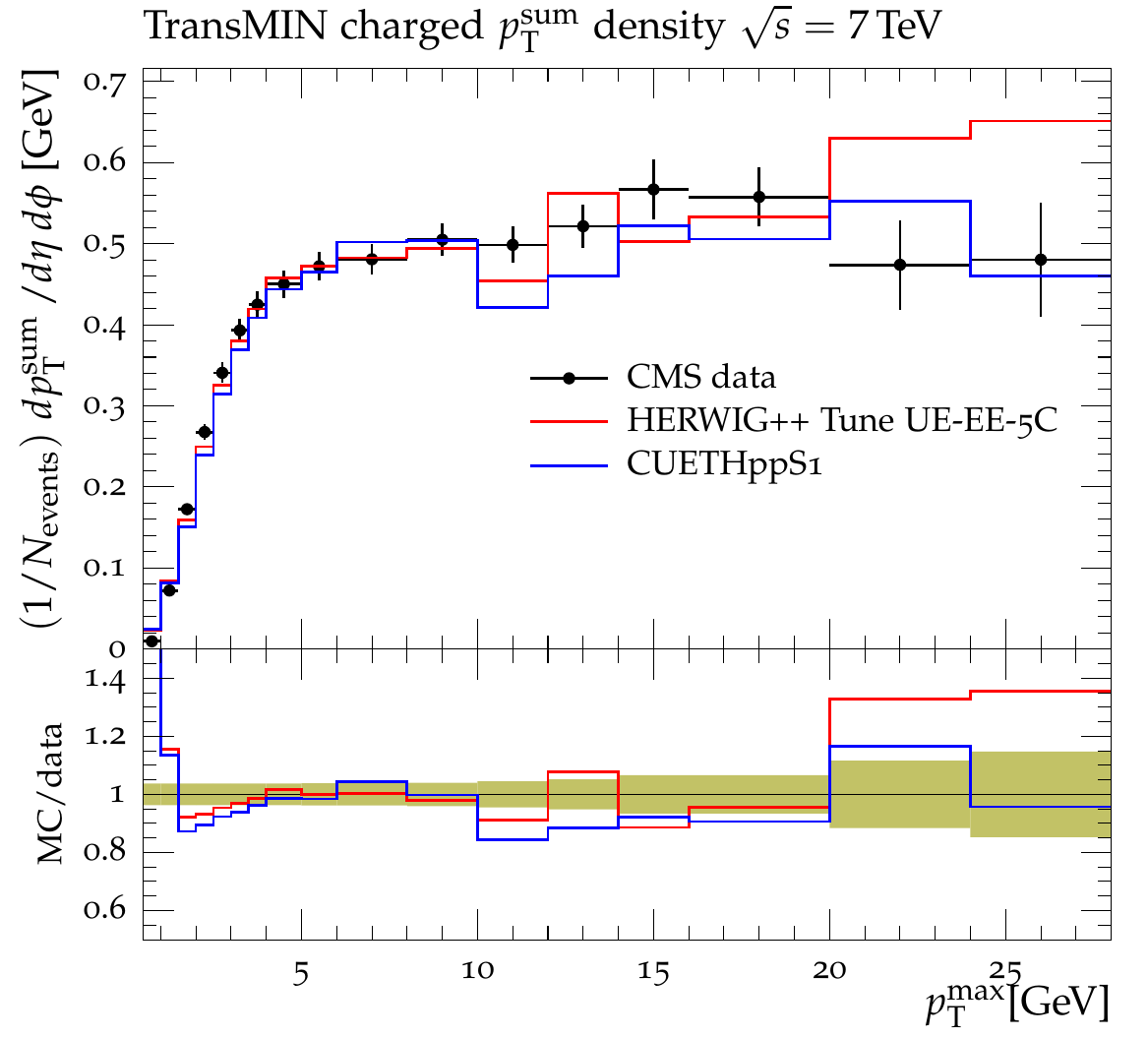}
\includegraphics[scale=0.65]{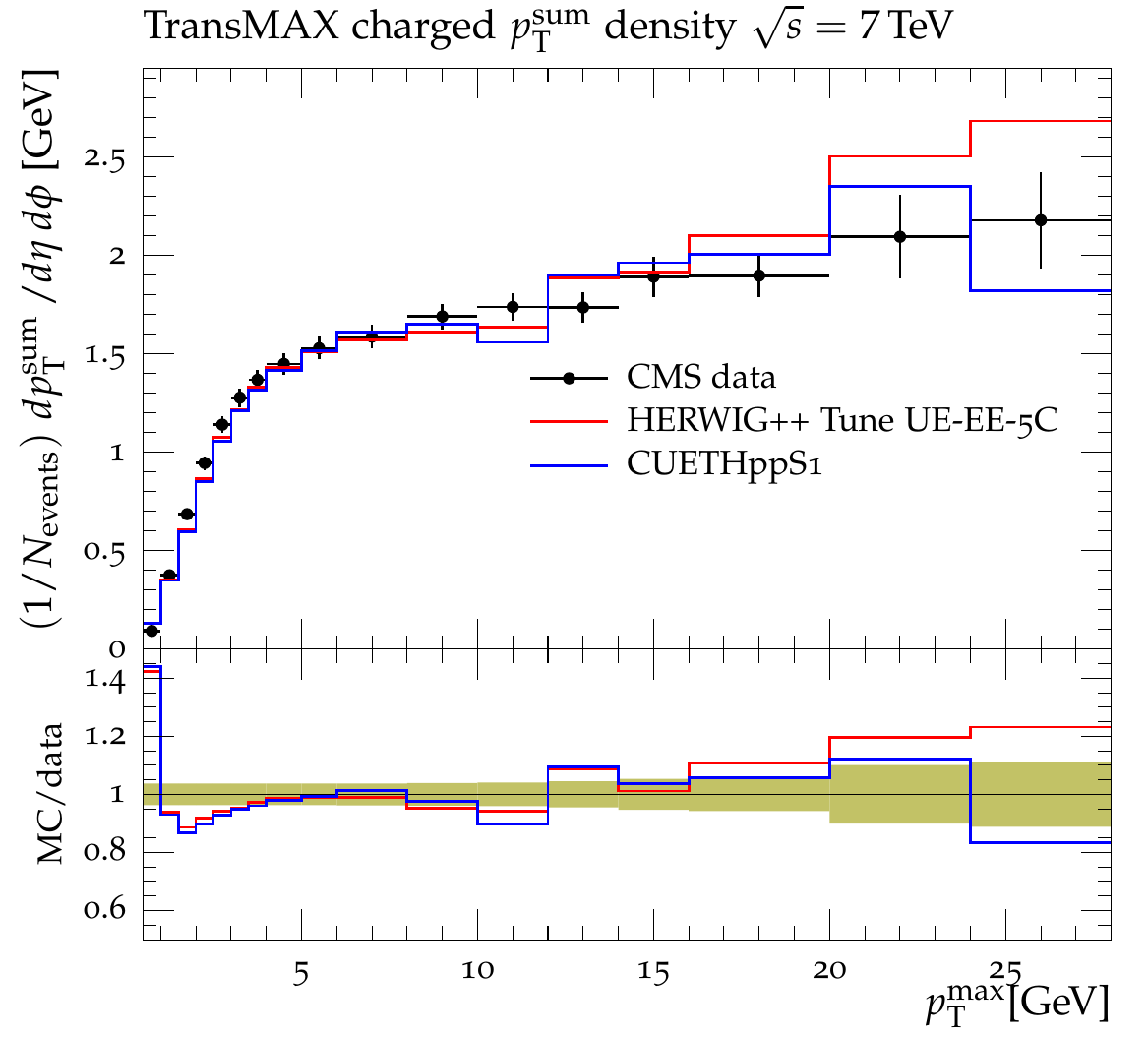}
\caption{CMS data at $\sqrt{s}=7\TeV$~\cite{CMS:2012kca} on particle  (top) and \ptsum\ densities (bottom) for charged particles with \ptcut\ and \etacut\ in the \tmin\ (left) and \tmax\ (right) regions as defined by the leading charged particle, as a function of the transverse momentum of the leading charged-particle \ptmax. The data are compared to the \hwpp\ Tune \tunee\ and \cueHW. The green bands in the ratios represent the total experimental uncertainties.}
\label{PUB_fig14}
\end{center}
\end{figure*}

Figures~\ref{PUB_fig11}--\ref{PUB_fig14} show the CDF data at $\sqrt{s} = 0.3$, $0.9$, and $1.96\TeV$, and the CMS data at $\sqrt{s} = 7\TeV$ on the charged-particle and \ptsum\ densities in the \tmin\ and \tmax\ regions as a function of \ptmax, and compared with predictions obtained with the \hwpp\ Tune UE-EE-5C and the new \cueHW.  These two \hwpp\ tunes are very similar and adequately describe the UE data at all four energies. 

\section{Additional comparisons at 13 TeV}
In this section, a supplementary collection of comparisons among predictions of the new tunes are shown for DPS and MB observables at 13\TeV. 

\subsection{DPS predictions at 13 TeV}

\begin{figure*}[htbp]
\begin{center}
\includegraphics[scale=0.65]{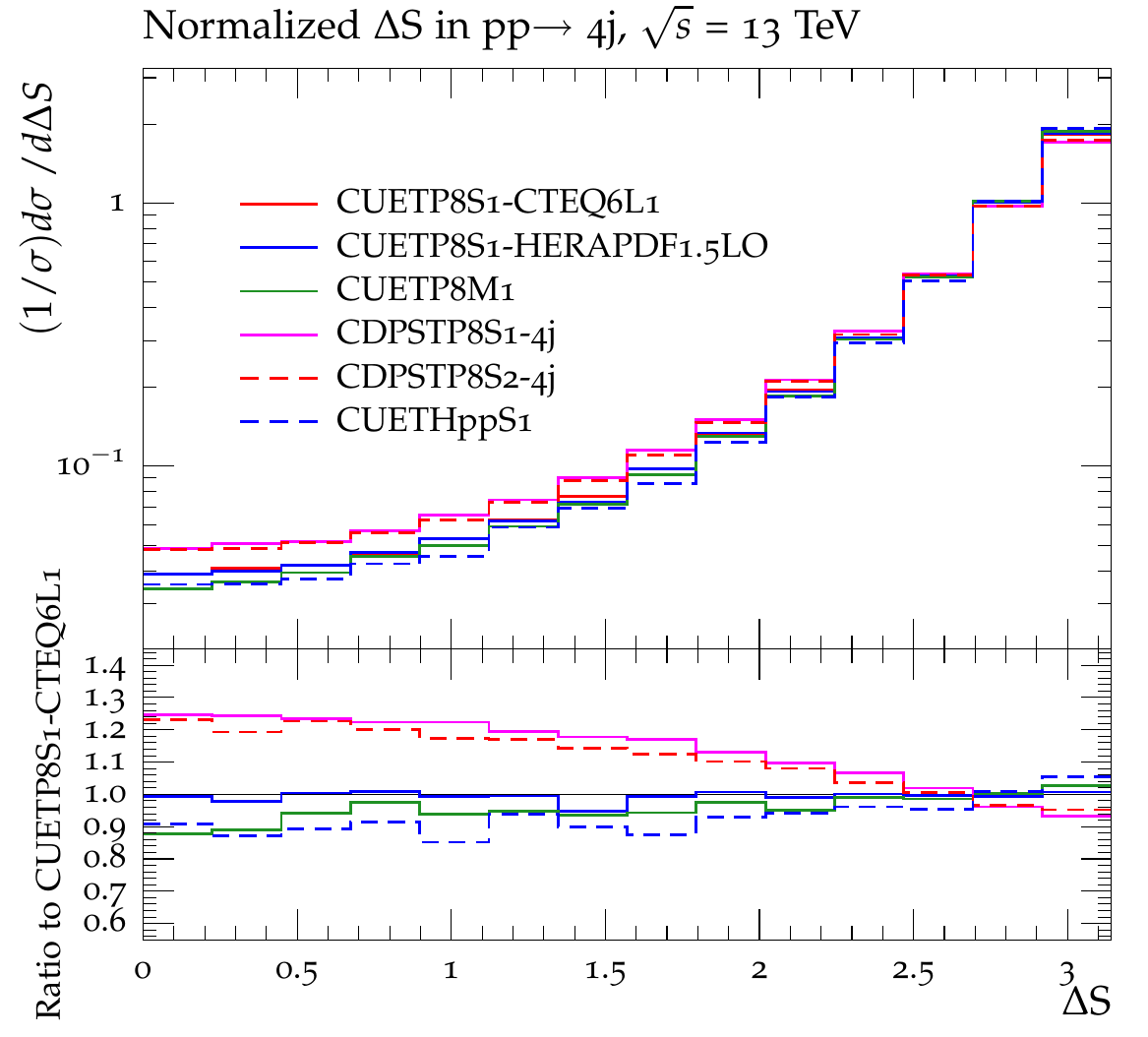}                                                                               
\includegraphics[scale=0.65]{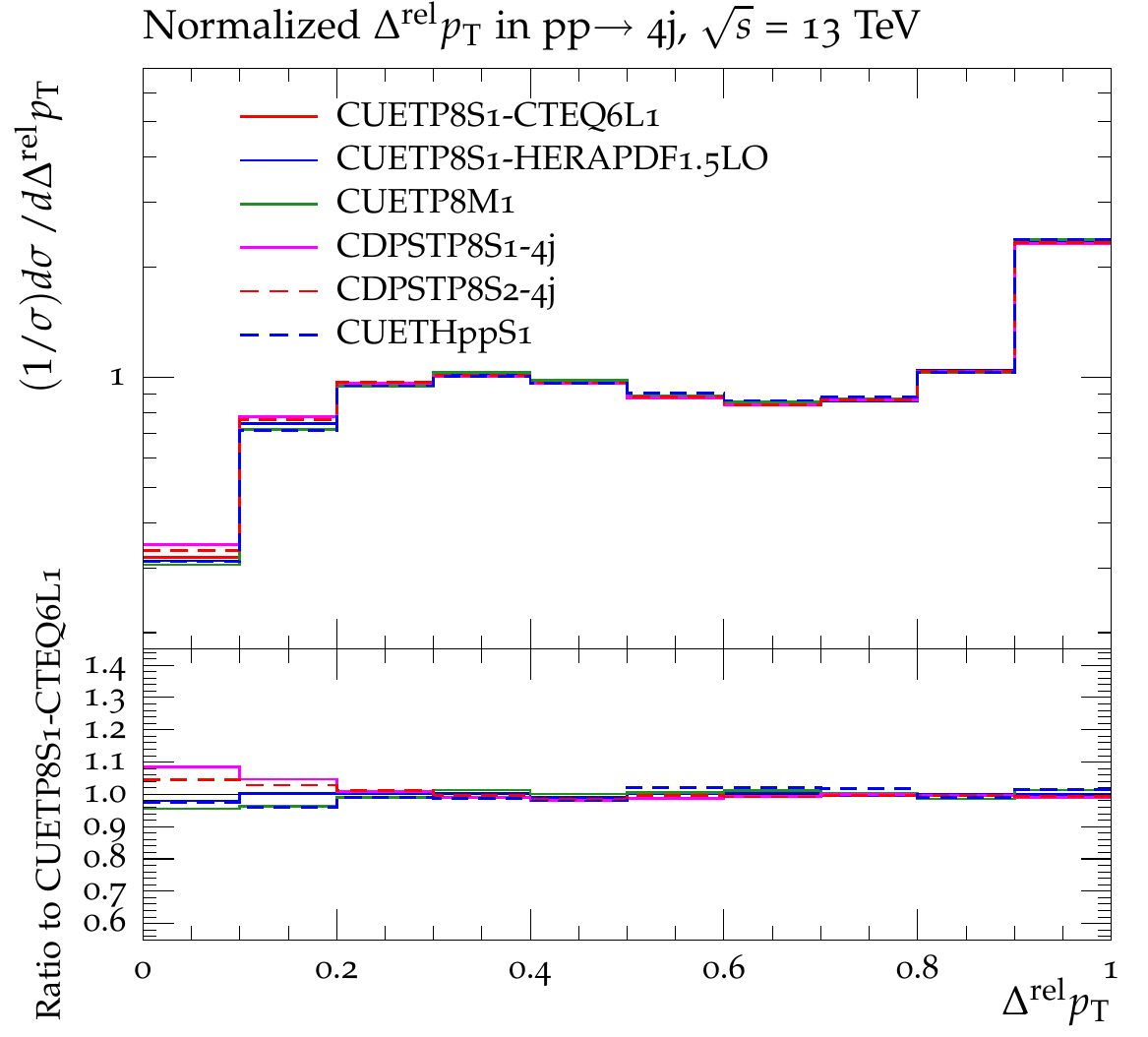}
\caption{Predictions at $\sqrt{s}=13\TeV$ for the normalized distributions of the correlation observables $\Delta$S (left), and \relpt\ (right) for four-jet production in pp collisions for the three CMS \pynewhyphen\ UE tunes \cuePB, \cuePH, and \cuePM, for \cueHW, and for \cdpJA\  and \cdpJB. Also shown are the ratios of the tunes to predictions of \cuePB.}
\label{PUB_fig21}
\end{center}
\end{figure*}

{\tolerance=5000
In Fig.~\ref{PUB_fig21}, the predictions for the DPS-sensitive observables at 13\TeV are shown for the three CMS \pynewhyphen\ UE tunes: \cuePB, \cuePH, and \cuePM, for \cueHW, and for the two CMS \pynewhyphen\ DPS tunes \cdpJA\ and \cdpJB. In \hwpp, \eff\ is independent of the center-of-mass energy, while \pynew\ gives a \eff\ that increases with energy. The \pynewhyphen\ UE tunes predict that \eff\ will increase by about $7\%$ between $7\TeV$ and $13\TeV$, while the \cdpJB\  predicts an increase of about $20\%$.  This results in slightly different predictions for the DPS-sensitive observables at $13\TeV$ for the CMS UE tunes and the CMS DPS tunes.
\par}

\subsection{MB predictions at 13 TeV}

\begin{figure*}[htbp]
\begin{center}
\includegraphics[scale=0.65]{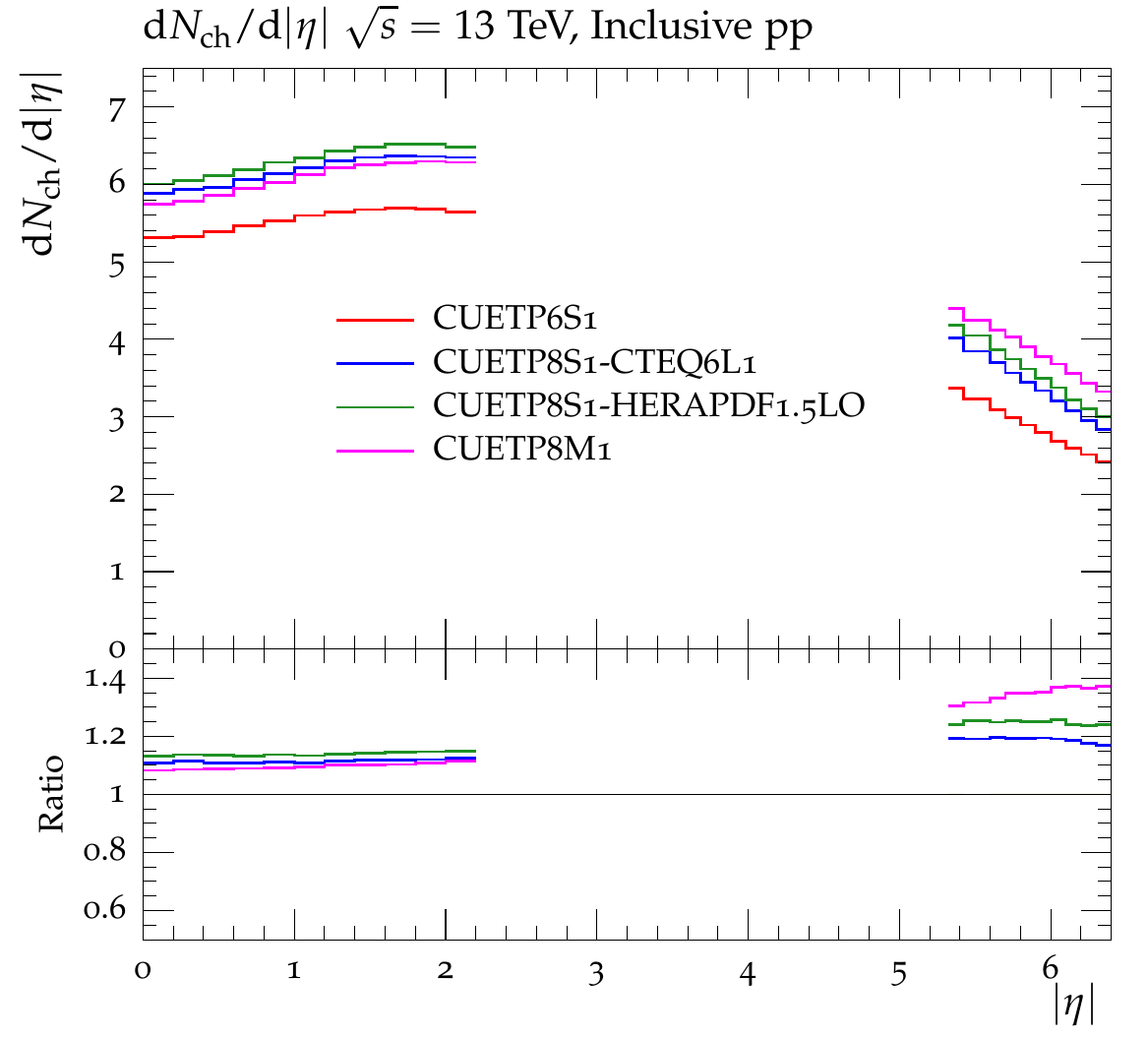}
\includegraphics[scale=0.58]{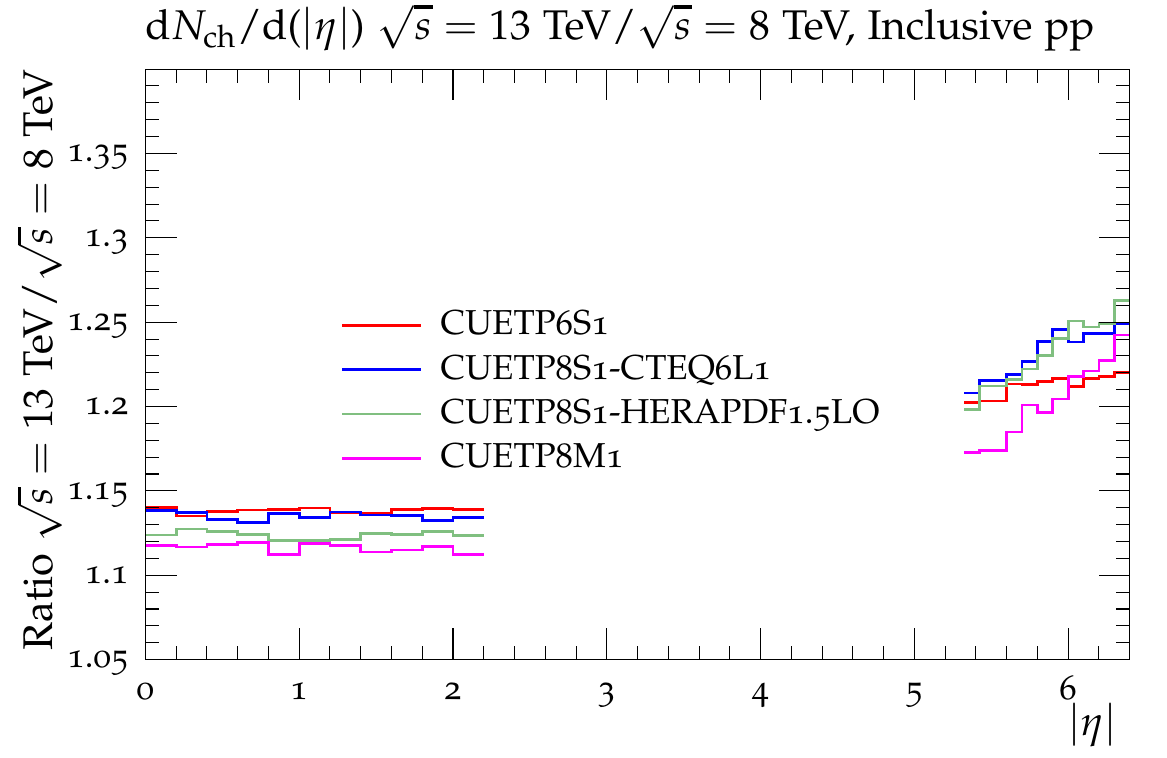}\\
\includegraphics[scale=0.65]{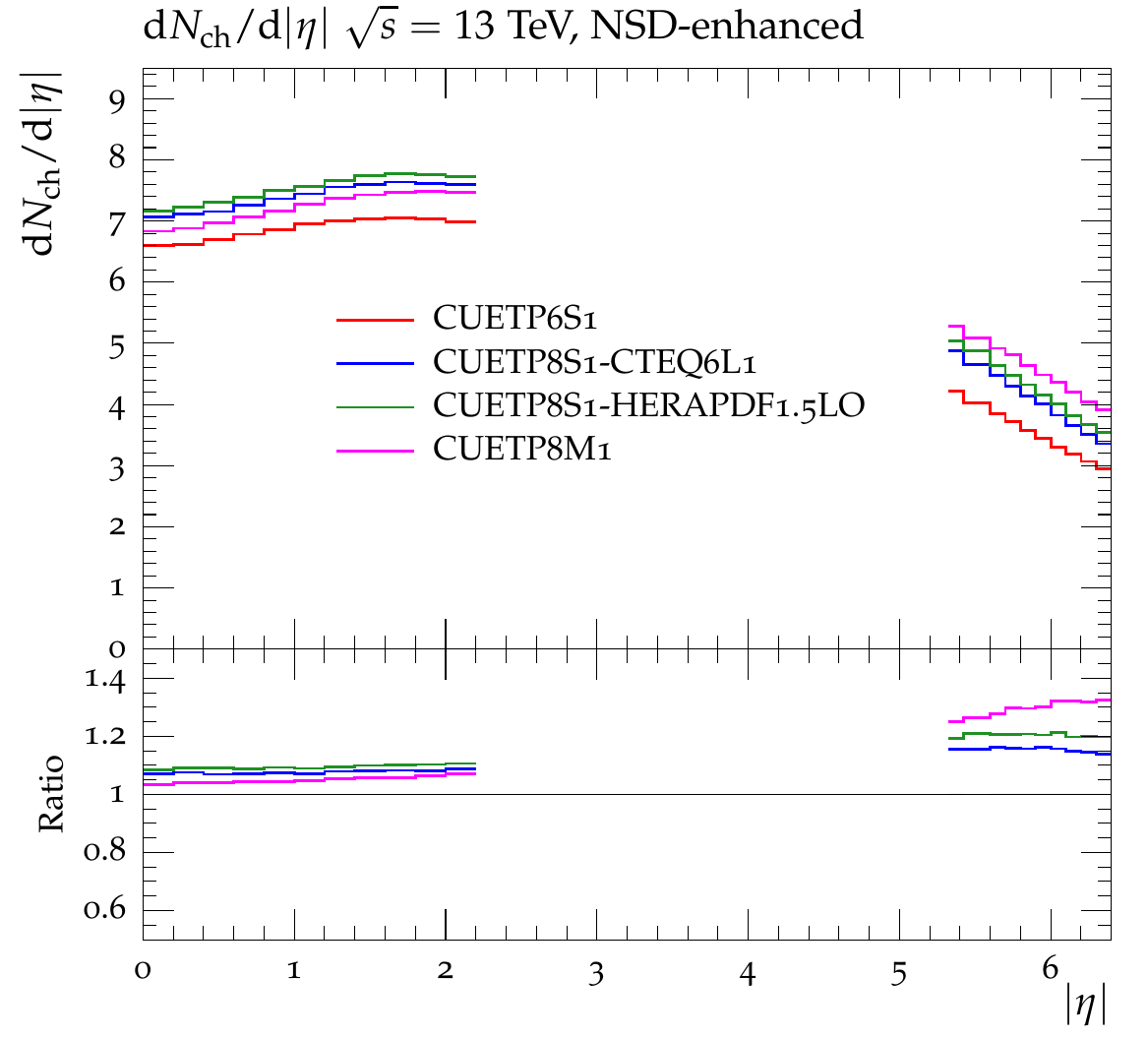}
\includegraphics[scale=0.58]{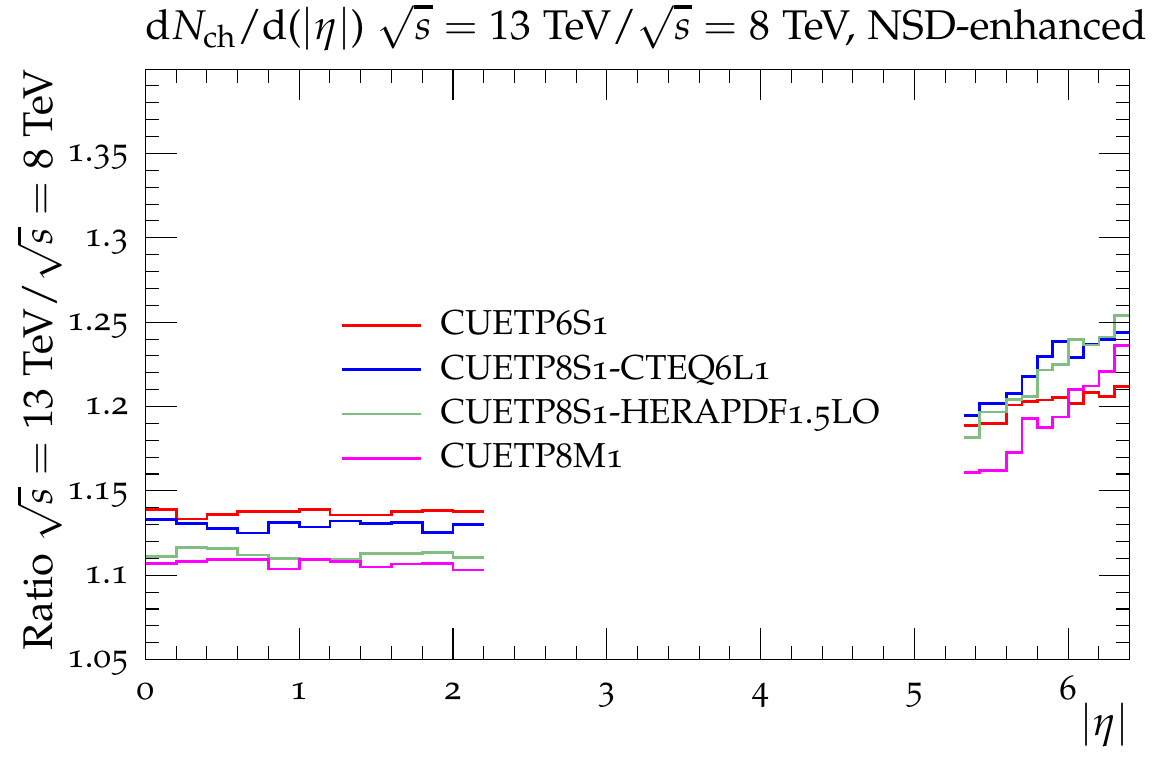}\\
\includegraphics[scale=0.65]{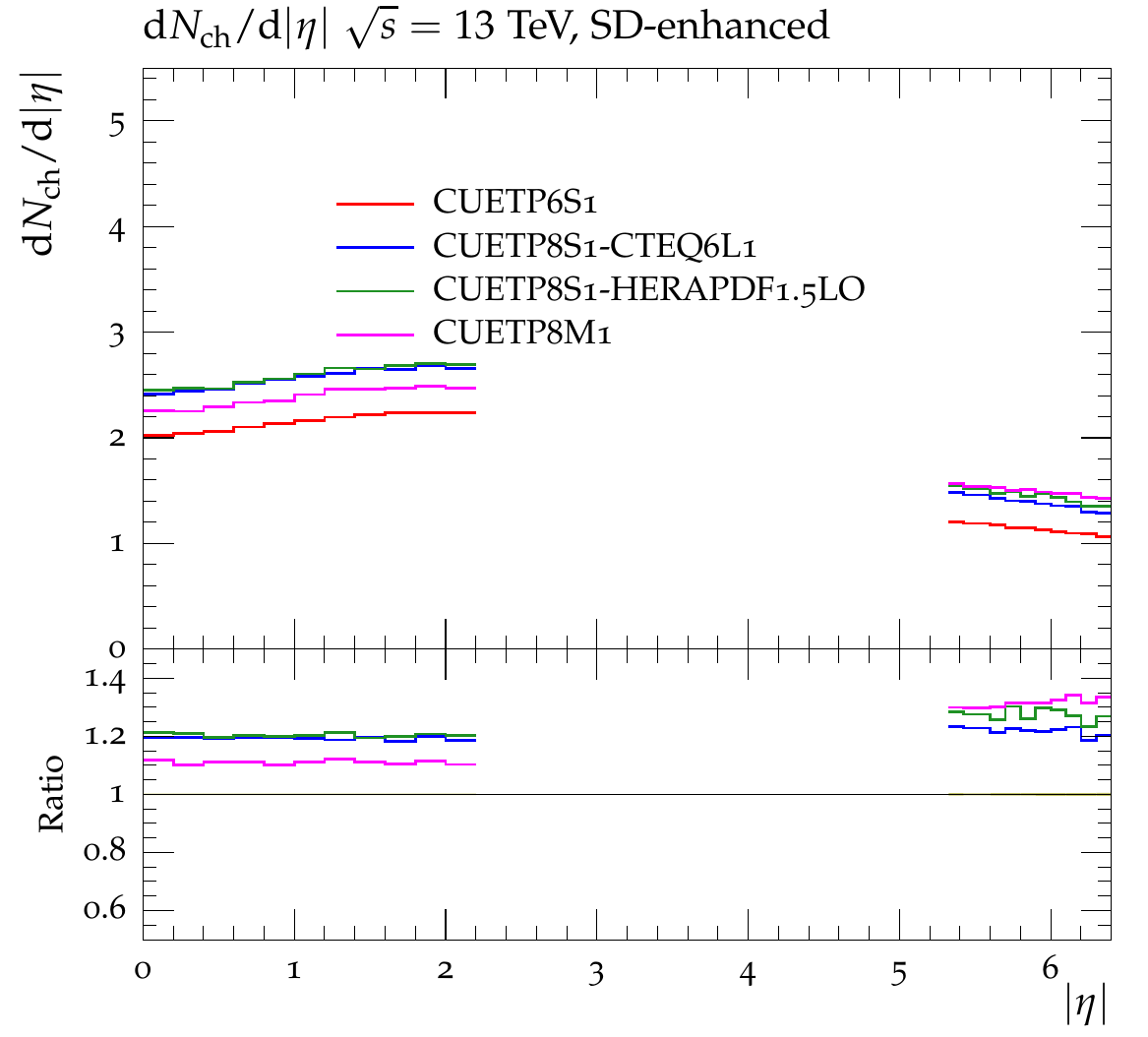}
\includegraphics[scale=0.58]{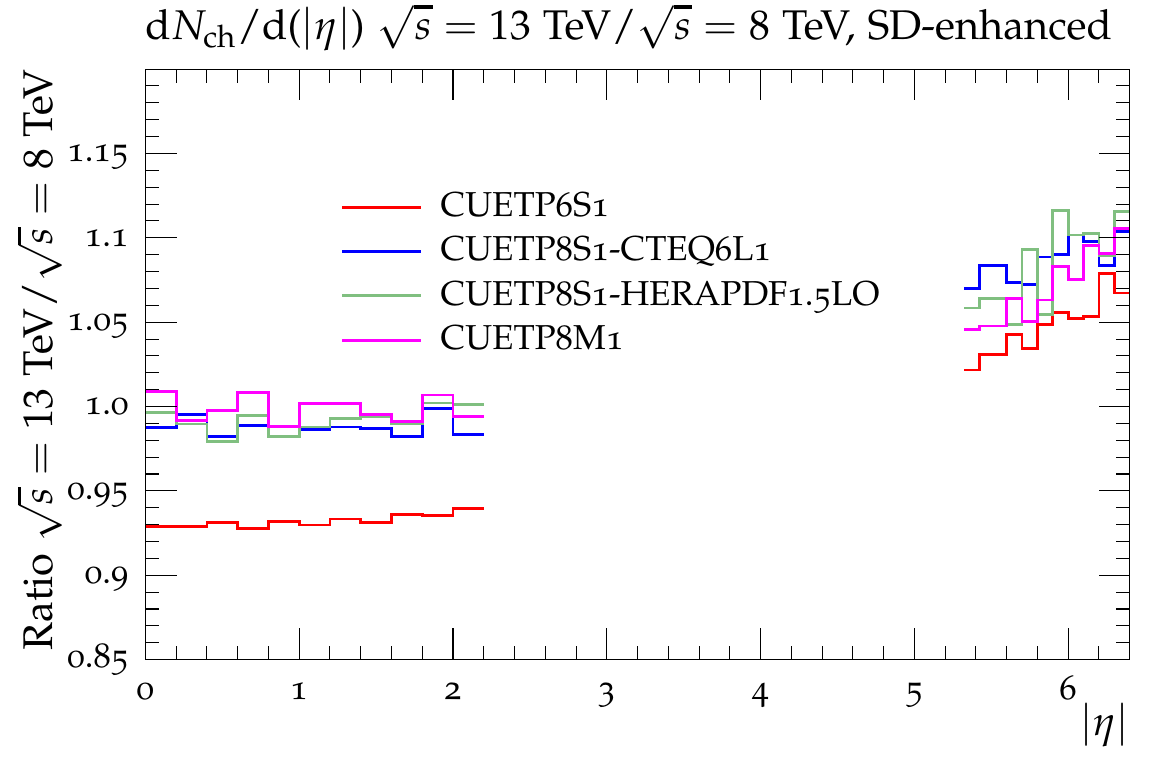}
\caption{Predictions at $\sqrt{s}=13\TeV$ for the charged-particle pseudorapidity distribution $\rd \mathrm{N}_{\text{ch}}/\rd \eta$, for (top) inelastic, (middle) NSD-enhanced, and (bottom) SD-enhanced pp collisions from \cuePA, \cuePB, \cuePH, and \cuePM.  Also shown are the ratios of the tunes to predictions of \cuePM, and the ratio of $13\TeV$ to $8\TeV$ results for each of the tunes (right).}
\label{PUB_fig26}
\end{center}
\end{figure*}

Predictions of the CMS UE tunes at $\sqrt{s}=13\TeV$ are shown in Fig.~\ref{PUB_fig26} for the charged-particle pseudorapidity distribution, $\rd \mathrm{N}_{\text{ch}}/\rd \eta$, for inelastic, non single-diffraction-enhanced, and single-diffraction-enhanced proton-proton collisions. In Fig.~\ref{PUB_fig26}, the ratio of $13\TeV$ to $8\TeV$ results is shown for each of the tunes. The densities in the forward region are predicted to increase more rapidly than the central region between $8\TeV$ and $13\TeV$.  However, the UE observables in Figs.~\ref{PUB_fig16} and \ref{PUB_fig17} increase much faster with center-of-mass energy than do these MB observables.

\clearpage

\begin{acknowledgments}

 \hyphenation{Bundes-ministerium Forschungs-gemeinschaft Forschungs-zentren} We congratulate our colleagues in the CERN accelerator departments for the excellent performance of the LHC and thank the technical and administrative staffs at CERN and at other CMS institutes for their contributions to the success of the CMS effort. In addition, we gratefully acknowledge the computing centres and personnel of the Worldwide LHC Computing Grid for delivering so effectively the computing infrastructure essential to our analyses. Finally, we acknowledge the enduring support for the construction and operation of the LHC and the CMS detector provided by the following funding agencies: the Austrian Federal Ministry of Science, Research and Economy and the Austrian Science Fund; the Belgian Fonds de la Recherche Scientifique, and Fonds voor Wetenschappelijk Onderzoek; the Brazilian Funding Agencies (CNPq, CAPES, FAPERJ, and FAPESP); the Bulgarian Ministry of Education and Science; CERN; the Chinese Academy of Sciences, Ministry of Science and Technology, and National Natural Science Foundation of China; the Colombian Funding Agency (COLCIENCIAS); the Croatian Ministry of Science, Education and Sport, and the Croatian Science Foundation; the Research Promotion Foundation, Cyprus; the Ministry of Education and Research, Estonian Research Council via IUT23-4 and IUT23-6 and European Regional Development Fund, Estonia; the Academy of Finland, Finnish Ministry of Education and Culture, and Helsinki Institute of Physics; the Institut National de Physique Nucl\'eaire et de Physique des Particules~/~CNRS, and Commissariat \`a l'\'Energie Atomique et aux \'Energies Alternatives~/~CEA, France; the Bundesministerium f\"ur Bildung und Forschung, Deutsche Forschungsgemeinschaft, and Helmholtz-Gemeinschaft Deutscher Forschungszentren, Germany; the General Secretariat for Research and Technology, Greece; the National Scientific Research Foundation, and National Innovation Office, Hungary; the Department of Atomic Energy and the Department of Science and Technology, India; the Institute for Studies in Theoretical Physics and Mathematics, Iran; the Science Foundation, Ireland; the Istituto Nazionale di Fisica Nucleare, Italy; the Ministry of Science, ICT and Future Planning, and National Research Foundation (NRF), Republic of Korea; the Lithuanian Academy of Sciences; the Ministry of Education, and University of Malaya (Malaysia); the Mexican Funding Agencies (CINVESTAV, CONACYT, SEP, and UASLP-FAI); the Ministry of Business, Innovation and Employment, New Zealand; the Pakistan Atomic Energy Commission; the Ministry of Science and Higher Education and the National Science Centre, Poland; the Funda\c{c}\~ao para a Ci\^encia e a Tecnologia, Portugal; JINR, Dubna; the Ministry of Education and Science of the Russian Federation, the Federal Agency of Atomic Energy of the Russian Federation, Russian Academy of Sciences, and the Russian Foundation for Basic Research; the Ministry of Education, Science and Technological Development of Serbia; the Secretar\'{\i}a de Estado de Investigaci\'on, Desarrollo e Innovaci\'on and Programa Consolider-Ingenio 2010, Spain; the Swiss Funding Agencies (ETH Board, ETH Zurich, PSI, SNF, UniZH, Canton Zurich, and SER); the Ministry of Science and Technology, Taipei; the Thailand Center of Excellence in Physics, the Institute for the Promotion of Teaching Science and Technology of Thailand, Special Task Force for Activating Research and the National Science and Technology Development Agency of Thailand; the Scientific and Technical Research Council of Turkey, and Turkish Atomic Energy Authority; the National Academy of Sciences of Ukraine, and State Fund for Fundamental Researches, Ukraine; the Science and Technology Facilities Council, UK; the US Department of Energy, and the US National Science Foundation.

Individuals have received support from the Marie-Curie programme and the European Research Council and EPLANET (European Union); the Leventis Foundation; the A. P. Sloan Foundation; the Alexander von Humboldt Foundation; the Belgian Federal Science Policy Office; the Fonds pour la Formation \`a la Recherche dans l'Industrie et dans l'Agriculture (FRIA-Belgium); the Agentschap voor Innovatie door Wetenschap en Technologie (IWT-Belgium); the Ministry of Education, Youth and Sports (MEYS) of the Czech Republic; the Council of Science and Industrial Research, India; the HOMING PLUS programme of the Foundation for Polish Science, cofinanced from European Union, Regional Development Fund; the OPUS programme of the National Science Center (Poland); the Compagnia di San Paolo (Torino); MIUR project 20108T4XTM (Italy); the Thalis and Aristeia programmes cofinanced by EU-ESF and the Greek NSRF; the National Priorities Research Program by Qatar National Research Fund; the Rachadapisek Sompot Fund for Postdoctoral Fellowship, Chulalongkorn University (Thailand); and the Welch Foundation, contract C-1845. 

\end{acknowledgments}

\bibliography{auto_generated}

\providecommand{\href}[2]{#2}\begingroup\raggedright\begin{thebibliography}{10}%
\makeatletter
\providecommand{\hrefCMSnoop }[0]{\@secondoftwo}%
\makeatother
\providecommand{\doi}{\texttt{doi:}\begingroup \urlstyle{tt}\Url}

\bibitem{Sjostrand:2006za}
\hrefCMSnoop {}{T.~Sj{\"o}strand, S.~Mrenna, and P.~Skands, ``{PYTHIA} 6.4
  physics and manual'',} \textit{ JHEP} \textbf{ 05} (2006) 026,
  \href{http://dx.doi.org/10.1088/1126-6708/2006/05/026}{\doi{10.1088/1126-6708/2006/05/026}},
\href{http://www.arXiv.org/abs/hep-ph/0603175}{\texttt{arXiv:hep-ph/0603175}}.

\bibitem{Sjostrand:2001yu}
\hrefCMSnoop {}{T.~Sj{\"o}strand, L.~Lonnblad, and S.~Mrenna, ``{PYTHIA 6.2:
  physics and manual}'',} (2001).
\href{http://www.arXiv.org/abs/hep-ph/0108264}{\texttt{arXiv:hep-ph/0108264}}.

\bibitem{Sjostrand:1987su}
\hrefCMSnoop {}{T.~Sj{\"o}strand and M.~van Zijl, ``{A multiple interaction
  model for the event structure in hadron collisions}'',} \textit{ Phys. Rev.
  D} \textbf{ 36} (1987) 2019,
\href{http://dx.doi.org/10.1103/PhysRevD.36.2019}{\doi{10.1103/PhysRevD.36.2019}}.

\bibitem{Bengtsson:1986gz}
\hrefCMSnoop {}{M.~Bengtsson, T.~Sj{\"o}strand, and M.~van Zijl, ``{Initial
  state radiation effects on $W$ and jet production}'',} \textit{ Z. Phys. C}
  \textbf{ 32} (1986) 67,
\href{http://dx.doi.org/10.1007/BF01441353}{\doi{10.1007/BF01441353}}.

\bibitem{oai:arXiv.org:0710.3820}
\hrefCMSnoop {}{T.~Sj{\"o}strand, S.~Mrenna, and P.~Z. Skands, ``{A brief
  introduction to PYTHIA 8.1}'',} \textit{ Comput. Phys. Commun.} \textbf{ 178}
  (2008) 852,
  \href{http://dx.doi.org/10.1016/j.cpc.2008.01.036}{\doi{10.1016/j.cpc.2008.01.036}},
\href{http://www.arXiv.org/abs/0710.3820}{\texttt{arXiv:0710.3820}}.

\bibitem{Bahr:2008pv}
M.~Bahr\hrefCMSnoop {}{ {et~al.}, ``{Herwig++ physics and manual}'',} \textit{
  Eur. Phys. J. C} \textbf{ 58} (2008) 639,
  \href{http://dx.doi.org/10.1140/epjc/s10052-008-0798-9}{\doi{10.1140/epjc/s10052-008-0798-9}},
\href{http://www.arXiv.org/abs/0803.0883}{\texttt{arXiv:0803.0883}}.

\bibitem{Bellm:2013lba}
J.~Bellm\hrefCMSnoop {}{ {et~al.}, ``{Herwig++ 2.7 Release Note}'',} (2013).
\href{http://www.arXiv.org/abs/1310.6877}{\texttt{arXiv:1310.6877}}.

\bibitem{Albrow:2006rt}
\hrefCMSnoop {}{M.~G. Albrow {et~al.}, ``{Tevatron-for-LHC report of the QCD
  working group}'',} (2006).
\href{http://www.arXiv.org/abs/hep-ph/0610012}{\texttt{arXiv:hep-ph/0610012}}.

\bibitem{Field:2009zz}
\hrefCMSnoop {}{{CDF} Collaboration, ``{Studying the 'underlying event' at CDF
  and the LHC}'',} (2009).
FERMILAB-CONF-09-792-E.

\bibitem{Skands:2009zm}
\href {http://lss.fnal.gov/cgi-bin/find_paper.pl?conf-09-113}{P.~Z. Skands,
  ``{The Perugia tunes}'',} in \textit{ {Proceedings, 1st International
  Workshop on Multiple Partonic Interactions at the LHC (MPI08)}}.
\newblock 2009.
\newblock
\href{http://www.arXiv.org/abs/0905.3418}{\texttt{arXiv:0905.3418}}.
\newblock

\bibitem{Aaltonen:2015aoa}
\hrefCMSnoop {}{{CDF} Collaboration, ``{A study of the energy dependence of the
  underlying event in proton-antiproton collisions}'',} \textit{ Phys. Rev. D}
  \textbf{ 92} (2015), no.~9, 092009,
  \href{http://dx.doi.org/10.1103/PhysRevD.92.092009}{\doi{10.1103/PhysRevD.92.092009}},
\href{http://www.arXiv.org/abs/1508.05340}{\texttt{arXiv:1508.05340}}.

\bibitem{Schuler:1993wr}
\hrefCMSnoop {}{G.~A. Schuler and T.~Sj{\"o}strand, ``{Hadronic diffractive
  cross-sections and the rise of the total cross-section}'',} \textit{ Phys.
  Rev. D} \textbf{ 49} (1994) 2257,
\href{http://dx.doi.org/10.1103/PhysRevD.49.2257}{\doi{10.1103/PhysRevD.49.2257}}.

\bibitem{Manohar:2012pe}
\hrefCMSnoop {}{A.~V. Manohar and W.~J. Waalewijn, ``What is double parton
  scattering?'',} \textit{ Phys. Lett. B} \textbf{ 713} (2012) 196,
  \href{http://dx.doi.org/10.1016/j.physletb.2012.05.044}{\doi{10.1016/j.physletb.2012.05.044}},
\href{http://www.arXiv.org/abs/1202.5034}{\texttt{arXiv:1202.5034}}.

\bibitem{DelFabbro:2001rs}
\hrefCMSnoop {}{A.~Del~Fabbro and D.~Treleani, ``{Double parton scatterings at
  the CERN LHC}'',} in \textit{ {Multiparticle production: New frontiers in
  soft physics and correlations on the threshold of the third millennium.
  Proceedings, 9th International Workshop, Torino, Italy, June 12-17, 2000}},
  volume~92, p.~130.
\newblock 2001.
\newblock
\href{http://dx.doi.org/10.1016/S0920-5632(00)01027-6}{\doi{10.1016/S0920-5632(00)01027-6}}.

\bibitem{Blok:2013bpa}
\hrefCMSnoop {}{B.~Blok, Y.~Dokshitzer, L.~Frankfurt, and M.~Strikman,
  ``{Perturbative QCD correlations in multi-parton collisions}'',} \textit{
  Eur. Phys. J. C} \textbf{ 74} (2014) 2926,
  \href{http://dx.doi.org/10.1140/epjc/s10052-014-2926-z}{\doi{10.1140/epjc/s10052-014-2926-z}},
\href{http://www.arXiv.org/abs/1306.3763}{\texttt{arXiv:1306.3763}}.

\bibitem{Bartalini:2011jp}
\href
  {http://inspirehep.net/record/944170/files/arXiv:1111.0469.pdf}{P.~Bartalini
  {et~al.}, ``{Multi-parton interactions at the LHC}'',} (2011).
\href{http://www.arXiv.org/abs/1111.0469}{\texttt{arXiv:1111.0469}}.

\bibitem{CMS:2012kca}
\href {http://cdsweb.cern.ch/record/1478982}{{CMS Collaboration},
  ``{Measurement of the underlying event activity at the LHC at 7\TeV and
  comparison with 0.9\TeV}'',} CMS Physics Analysis Summary CMS-PAS-FSQ-12-020,
  2012.

\bibitem{Rivet}
A.~Buckley\hrefCMSnoop {}{ {et~al.}, ``{Rivet user manual}'',} \textit{ Comput.
  Phys. Commun.} \textbf{ 184} (2013) 2803,
  \href{http://dx.doi.org/10.1016/j.cpc.2013.05.021}{\doi{10.1016/j.cpc.2013.05.021}},
\href{http://www.arXiv.org/abs/1003.0694}{\texttt{arXiv:1003.0694}}.

\bibitem{Buckley:2009bj}
A.~Buckley\hrefCMSnoop {}{ {et~al.}, ``{Systematic event generator tuning for
  the LHC}'',} \textit{ Eur. Phys. J. C} \textbf{ 65} (2010) 331,
  \href{http://dx.doi.org/10.1140/epjc/s10052-009-1196-7}{\doi{10.1140/epjc/s10052-009-1196-7}},
\href{http://www.arXiv.org/abs/0907.2973}{\texttt{arXiv:0907.2973}}.

\bibitem{Pumplin:2002vw}
J.~Pumplin\hrefCMSnoop {}{ {et~al.}, ``{New generation of parton distributions
  with uncertainties from global QCD analysis}'',} \textit{ JHEP} \textbf{ 07}
  (2002) 012,
  \href{http://dx.doi.org/10.1088/1126-6708/2002/07/012}{\doi{10.1088/1126-6708/2002/07/012}},
\href{http://www.arXiv.org/abs/hep-ph/0201195}{\texttt{arXiv:hep-ph/0201195}}.

\bibitem{Sarkar:2014zua}
\hrefCMSnoop {}{A.~M. Cooper-Sarkar, ``{HERAPDF1.5LO PDF set with experimental
  uncertainties}'',} in \textit{ {Proceedings, 22nd International Workshop on
  Deep-Inelastic Scattering and Related Subjects (DIS 2014)}}, volume DIS2014,
  p.~032.
\newblock
2014.
\newblock

\bibitem{Ball:2013hta}
\hrefCMSnoop {}{{NNPDF} Collaboration, ``{Parton distributions with QED
  corrections}'',} \textit{ Nucl. Phys. B} \textbf{ 877} (2013) 290,
  \href{http://dx.doi.org/10.1016/j.nuclphysb.2013.10.010}{\doi{10.1016/j.nuclphysb.2013.10.010}},
\href{http://www.arXiv.org/abs/1308.0598}{\texttt{arXiv:1308.0598}}.

\bibitem{Ball:2011uy}
\hrefCMSnoop {}{{NNPDF} Collaboration, ``{Unbiased global determination of
  parton distributions and their uncertainties at NNLO and at LO}'',} \textit{
  Nucl. Phys. B} \textbf{ 855} (2012) 153,
  \href{http://dx.doi.org/10.1016/j.nuclphysb.2011.09.024}{\doi{10.1016/j.nuclphysb.2011.09.024}},
\href{http://www.arXiv.org/abs/1107.2652}{\texttt{arXiv:1107.2652}}.

\bibitem{Chatrchyan:2011id}
\hrefCMSnoop {}{{CMS Collaboration}, ``{Measurement of the underlying event
  activity at the LHC with $\sqrt{s} = 7\TeV$ and comparison with $\sqrt{s} =
  0.9\TeV$}'',} \textit{ JHEP} \textbf{ 09} (2011) 109,
  \href{http://dx.doi.org/10.1007/JHEP09(2011)109}{\doi{10.1007/JHEP09(2011)109}},
\href{http://www.arXiv.org/abs/1107.0330}{\texttt{arXiv:1107.0330}}.

\bibitem{Chatrchyan:2013gfi}
\hrefCMSnoop {}{{CMS Collaboration}, ``{Study of the underlying event at
  forward rapidity in pp collisions at $\sqrt{s} = 0.9$, $2.76$, and
  $7\TeV$}'',} \textit{ JHEP} \textbf{ 04} (2013) 072,
  \href{http://dx.doi.org/10.1007/JHEP04(2013)072}{\doi{10.1007/JHEP04(2013)072}},
\href{http://www.arXiv.org/abs/1302.2394}{\texttt{arXiv:1302.2394}}.

\bibitem{Chatrchyan:2012tb}
\hrefCMSnoop {}{{CMS Collaboration}, ``{Measurement of the underlying event in
  the Drell-Yan process in proton-proton collisions at $\sqrt{s}$ = 7 TeV}'',}
  \textit{ Eur. Phys. J. C} \textbf{ 72} (2012) 2080,
  \href{http://dx.doi.org/10.1140/epjc/s10052-012-2080-4}{\doi{10.1140/epjc/s10052-012-2080-4}},
\href{http://www.arXiv.org/abs/1204.1411}{\texttt{arXiv:1204.1411}}.

\bibitem{Pumplin:1997ix}
\hrefCMSnoop {}{J.~Pumplin, ``{Hard underlying event correction to inclusive
  jet cross-sections}'',} \textit{ Phys. Rev. D} \textbf{ 57} (1998) 5787,
  \href{http://dx.doi.org/10.1103/PhysRevD.57.5787}{\doi{10.1103/PhysRevD.57.5787}},
\href{http://www.arXiv.org/abs/hep-ph/9708464}{\texttt{arXiv:hep-ph/9708464}}.

\bibitem{Corke:2010yf}
\hrefCMSnoop {}{R.~Corke and T.~Sj{\"o}strand, ``{Interleaved parton showers
  and tuning prospects}'',} \textit{ JHEP} \textbf{ 03} (2011) 032,
  \href{http://dx.doi.org/10.1007/JHEP03(2011)032}{\doi{10.1007/JHEP03(2011)032}},
\href{http://www.arXiv.org/abs/1011.1759}{\texttt{arXiv:1011.1759}}.

\bibitem{Skands:2014pea}
\hrefCMSnoop {}{P.~Skands, S.~Carrazza, and J.~Rojo, ``{Tuning PYTHIA 8.1: the
  Monash 2013 tune}'',} \textit{ Eur. Phys. J. C} \textbf{ 74} (2014) 3024,
  \href{http://dx.doi.org/10.1140/epjc/s10052-014-3024-y}{\doi{10.1140/epjc/s10052-014-3024-y}},
\href{http://www.arXiv.org/abs/1404.5630}{\texttt{arXiv:1404.5630}}.

\bibitem{Seymour:2013qka}
\hrefCMSnoop {}{M.~H. Seymour and A.~Siodmok, ``{Constraining MPI models using
  $\sigma_{\text{eff}}$ and recent Tevatron and LHC underlying event data}'',}
  \textit{ JHEP} \textbf{ 10} (2013) 113,
  \href{http://dx.doi.org/10.1007/JHEP10(2013)113}{\doi{10.1007/JHEP10(2013)113}},
\href{http://www.arXiv.org/abs/1307.5015}{\texttt{arXiv:1307.5015}}.

\bibitem{Skands:2010ak}
\hrefCMSnoop {}{P.~Z. Skands, ``{Tuning Monte Carlo generators: the Perugia
  tunes}'',} \textit{ Phys. Rev. D} \textbf{ 82} (2010) 074018,
  \href{http://dx.doi.org/10.1103/PhysRevD.82.074018}{\doi{10.1103/PhysRevD.82.074018}},
\href{http://www.arXiv.org/abs/1005.3457}{\texttt{arXiv:1005.3457}}.

\bibitem{afsdps}
\hrefCMSnoop {}{{AFS} Collaboration, ``{Double parton scattering in pp
  collisions at $\sqrt{s} = 63\GeV$}'',} \textit{ Z. Phys. C} \textbf{ 34}
  (1987)
\href{http://dx.doi.org/10.1007/BF01566757}{\doi{10.1007/BF01566757}}.

\bibitem{cdfdps}
\hrefCMSnoop {}{{CDF} Collaboration, ``{Double parton scattering in $\bar{p} p$
  collisions at $\sqrt{s} = 1.8\TeV$}'',} \textit{ Z. Phys. D} \textbf{ 56}
  (1997)
\href{http://dx.doi.org/10.1103/PhysRevD.56.3811}{\doi{10.1103/PhysRevD.56.3811}}.

\bibitem{uadps}
\hrefCMSnoop {}{{UA2} Collaboration, ``{A study of multi-jet events at the CERN
  $\bar{p} p$ collider and a search for double parton scattering}'',} \textit{
  Phys. Lett. B} \textbf{ 268} (1991) 145,
\href{http://dx.doi.org/10.1016/0370-2693(91)90937-L}{\doi{10.1016/0370-2693(91)90937-L}}.

\bibitem{Aad:2013bjm}
\hrefCMSnoop {}{{ATLAS Collaboration}, ``{Measurement of hard double-parton
  interactions in $W(\to \ell\nu)$ + 2-jet events at $\sqrt{s} = 7\TeV$ with
  the ATLAS detector}'',} \textit{ New J. Phys.} \textbf{ 15} (2013) 033038,
  \href{http://dx.doi.org/10.1088/1367-2630/15/3/033038}{\doi{10.1088/1367-2630/15/3/033038}},
\href{http://www.arXiv.org/abs/1301.6872}{\texttt{arXiv:1301.6872}}.

\bibitem{Chatrchyan:2013xxa}
\hrefCMSnoop {}{{CMS Collaboration}, ``{Study of double parton scattering using
  W + 2-jet events in proton-proton collisions at $\sqrt{s} = 7\TeV$}'',}
  \textit{ JHEP} \textbf{ 03} (2014) 032,
  \href{http://dx.doi.org/10.1007/JHEP03(2014)032}{\doi{10.1007/JHEP03(2014)032}},
\href{http://www.arXiv.org/abs/1312.5729}{\texttt{arXiv:1312.5729}}.

\bibitem{Chatrchyan:2013qza}
\hrefCMSnoop {}{{CMS Collaboration}, ``{Measurement of four-jet production in
  proton-proton collisions at $\sqrt{s} = 7\TeV$}'',} \textit{ Phys. Rev. D}
  \textbf{ 89} (2014) 092010,
  \href{http://dx.doi.org/10.1103/PhysRevD.89.092010}{\doi{10.1103/PhysRevD.89.092010}},
\href{http://www.arXiv.org/abs/1312.6440}{\texttt{arXiv:1312.6440}}.

\bibitem{Alwall:2011uj}
J.~Alwall\hrefCMSnoop {}{ {et~al.}, ``{MadGraph 5 : going beyond}'',} \textit{
  JHEP} \textbf{ 06} (2011) 128,
  \href{http://dx.doi.org/10.1007/JHEP06(2011)128}{\doi{10.1007/JHEP06(2011)128}},
\href{http://www.arXiv.org/abs/1106.0522}{\texttt{arXiv:1106.0522}}.

\bibitem{Aad:2010fh}
\hrefCMSnoop {}{{ATLAS Collaboration}, ``{Measurement of underlying event
  characteristics using charged particles in $pp$ collisions at $\sqrt{s} =
  900\GeV$ and $7\TeV$ with the ATLAS detector}'',} \textit{ Phys. Rev. D}
  \textbf{ 83} (2011) 112001,
  \href{http://dx.doi.org/10.1103/PhysRevD.83.112001}{\doi{10.1103/PhysRevD.83.112001}},
\href{http://www.arXiv.org/abs/1012.0791}{\texttt{arXiv:1012.0791}}.

\bibitem{Blok:2015rka}
\hrefCMSnoop {}{B.~Blok and P.~Gunnellini, ``{Dynamical approach to MPI
  four-jet production in Pythia}'',} \textit{ Eur. Phys. J. C} \textbf{ 75}
  (2015) 282,
  \href{http://dx.doi.org/10.1140/epjc/s10052-015-3520-8}{\doi{10.1140/epjc/s10052-015-3520-8}},
\href{http://www.arXiv.org/abs/1503.08246}{\texttt{arXiv:1503.08246}}.

\bibitem{Diehl:2014vaa}
\hrefCMSnoop {}{M.~Diehl, T.~Kasemets, and S.~Keane, ``Correlations in double
  parton distributions: effects of evolution'',} \textit{ JHEP} \textbf{ 05}
  (2014) 118,
  \href{http://dx.doi.org/10.1007/JHEP05(2014)118}{\doi{10.1007/JHEP05(2014)118}},
\href{http://www.arXiv.org/abs/1401.1233}{\texttt{arXiv:1401.1233}}.

\bibitem{Khachatryan:2015jza}
\hrefCMSnoop {}{{CMS Collaboration}, ``{Measurement of the underlying event
  activity using charged-particle jets in proton-proton collisions at $\sqrt(s)
  = 2.76\TeV$}'',} \textit{ JHEP} \textbf{ 09} (2015) 137,
  \href{http://dx.doi.org/10.1007/JHEP09(2015)137}{\doi{10.1007/JHEP09(2015)137}},
\href{http://www.arXiv.org/abs/1507.07229}{\texttt{arXiv:1507.07229}}.

\bibitem{Cooper:2011gk}
B.~Cooper\hrefCMSnoop {}{ {et~al.}, ``{Importance of a consistent choice of
  $\alpha_s$ in the matching of AlpGen and Pythia}'',} \textit{ Eur. Phys. J.C}
  \textbf{ 72} (2012) 2078,
  \href{http://dx.doi.org/10.1140/epjc/s10052-012-2078-y}{\doi{10.1140/epjc/s10052-012-2078-y}},
\href{http://www.arXiv.org/abs/1109.5295}{\texttt{arXiv:1109.5295}}.

\bibitem{Nason:2010ap}
\hrefCMSnoop {}{P.~A. Nason, ``{Recent developments in POWHEG}'',} in \textit{
  {Proceedings, 9th International Symposium on Radiative Corrections:
  Applications of quantum field theory to phenomenology. (RADCOR 2009)}},
  volume RADCOR2009, p.~018.
\newblock 2010.
\newblock
\href{http://www.arXiv.org/abs/1001.2747}{\texttt{arXiv:1001.2747}}.
\newblock

\bibitem{Alioli:2010xd}
\hrefCMSnoop {}{S.~Alioli, P.~Nason, C.~Oleari, and E.~Re, ``{A general
  framework for implementing NLO calculations in shower Monte Carlo programs:
  the POWHEG BOX}'',} \textit{ JHEP} \textbf{ 06} (2010) 043,
  \href{http://dx.doi.org/10.1007/JHEP06(2010)043}{\doi{10.1007/JHEP06(2010)043}},
\href{http://www.arXiv.org/abs/1002.2581}{\texttt{arXiv:1002.2581}}.

\bibitem{Lai:2010vv}
H.-L. Lai\hrefCMSnoop {}{ {et~al.}, ``{New parton distributions for collider
  physics}'',} \textit{ Phys. Rev. D} \textbf{ 82} (2010) 074024,
  \href{http://dx.doi.org/10.1103/PhysRevD.82.074024}{\doi{10.1103/PhysRevD.82.074024}},
\href{http://www.arXiv.org/abs/1007.2241}{\texttt{arXiv:1007.2241}}.

\bibitem{Aamodt:2010pp}
\hrefCMSnoop {}{{ALICE Collaboration}, ``{Charged-particle multiplicity
  measurement in proton-proton collisions at $\sqrt{s} = 7\TeV$ with ALICE at
  LHC}'',} \textit{ Eur. Phys. J. C} \textbf{ 68} (2010) 345,
  \href{http://dx.doi.org/10.1140/epjc/s10052-010-1350-2}{\doi{10.1140/epjc/s10052-010-1350-2}},
\href{http://www.arXiv.org/abs/1004.3514}{\texttt{arXiv:1004.3514}}.

\bibitem{Antchev:2011vs}
\hrefCMSnoop {}{{{TOTEM}} Collaboration, ``{First measurement of the total
  proton-proton cross section at the LHC energy of $\sqrt{s} = 7\TeV$}'',}
  \textit{ Europhys. Lett.} \textbf{ 96} (2011) 21002,
  \href{http://dx.doi.org/10.1209/0295-5075/96/21002}{\doi{10.1209/0295-5075/96/21002}},
\href{http://www.arXiv.org/abs/1110.1395}{\texttt{arXiv:1110.1395}}.

\bibitem{Chatrchyan:2011wm}
\hrefCMSnoop {}{{CMS Collaboration}, ``{Measurement of energy flow at large
  pseudorapidities in pp collisions at $\sqrt{s} = 0.9$ and $7\TeV$}'',}
  \textit{ JHEP} \textbf{ 11} (2011) 148,
  \href{http://dx.doi.org/10.1007/JHEP02(2012)055}{\doi{10.1007/JHEP02(2012)055}},
\href{http://www.arXiv.org/abs/1110.0211}{\texttt{arXiv:1110.0211}}.

\bibitem{Chatrchyan:2014qka}
\hrefCMSnoop {}{{CMS and TOTEM Collaborations}, ``{Measurement of
  pseudorapidity distributions of charged particles in proton-proton collisions
  at $\sqrt{s} = 8\TeV$ by the CMS and TOTEM experiments}'',} \textit{ Eur.
  Phys. J. C} \textbf{ 74} (2014) 3053,
  \href{http://dx.doi.org/10.1140/epjc/s10052-014-3053-6}{\doi{10.1140/epjc/s10052-014-3053-6}},
\href{http://www.arXiv.org/abs/1405.0722}{\texttt{arXiv:1405.0722}}.

\bibitem{CMS:2011ab}
\hrefCMSnoop {}{{CMS Collaboration}, ``{Measurement of the inclusive jet cross
  section in pp collisions at $\sqrt{s} = 7\TeV$}'',} \textit{ Phys. Rev.
  Lett.} \textbf{ 107} (2011) 132001,
  \href{http://dx.doi.org/10.1103/PhysRevLett.107.132001}{\doi{10.1103/PhysRevLett.107.132001}},
\href{http://www.arXiv.org/abs/1106.0208}{\texttt{arXiv:1106.0208}}.

\bibitem{Chatrchyan:2011wt}
\hrefCMSnoop {}{{CMS Collaboration}, ``{Measurement of the rapidity and
  transverse momentum distributions of Z bosons in pp collisions at $\sqrt{s} =
  7\TeV$}'',} \textit{ Phys. Rev. D} \textbf{ 85} (2012) 032002,
  \href{http://dx.doi.org/10.1103/PhysRevD.85.032002}{\doi{10.1103/PhysRevD.85.032002}},
\href{http://www.arXiv.org/abs/1110.4973}{\texttt{arXiv:1110.4973}}.

\bibitem{Khachatryan:2015jna}
\hrefCMSnoop {}{{CMS Collaboration}, ``{Pseudorapidity distribution of charged
  hadrons in proton-proton collisions at $\sqrt{s} = 13\TeV$}'',} \textit{
  Phys. Lett. B} \textbf{ 751} (2015) 143,
  \href{http://dx.doi.org/10.1016/j.physletb.2015.10.004}{\doi{10.1016/j.physletb.2015.10.004}},
\href{http://www.arXiv.org/abs/1507.05915}{\texttt{arXiv:1507.05915}}.

\end{thebibliography}\endgroup

\cleardoublepage \section{The CMS Collaboration \label{app:collab}}\begin{sloppypar}\hyphenpenalty=5000\widowpenalty=500\clubpenalty=5000\textbf{Yerevan Physics Institute,  Yerevan,  Armenia}\\*[0pt]
V.~Khachatryan, A.M.~Sirunyan, A.~Tumasyan
\vskip\cmsinstskip
\textbf{Institut f\"{u}r Hochenergiephysik der OeAW,  Wien,  Austria}\\*[0pt]
W.~Adam, E.~Asilar, T.~Bergauer, J.~Brandstetter, E.~Brondolin, M.~Dragicevic, J.~Er\"{o}, M.~Flechl, M.~Friedl, R.~Fr\"{u}hwirth\cmsAuthorMark{1}, V.M.~Ghete, C.~Hartl, N.~H\"{o}rmann, J.~Hrubec, M.~Jeitler\cmsAuthorMark{1}, V.~Kn\"{u}nz, A.~K\"{o}nig, M.~Krammer\cmsAuthorMark{1}, I.~Kr\"{a}tschmer, D.~Liko, T.~Matsushita, I.~Mikulec, D.~Rabady\cmsAuthorMark{2}, B.~Rahbaran, H.~Rohringer, J.~Schieck\cmsAuthorMark{1}, R.~Sch\"{o}fbeck, J.~Strauss, W.~Treberer-Treberspurg, W.~Waltenberger, C.-E.~Wulz\cmsAuthorMark{1}
% ...
\vskip\cmsinstskip
% ...
1:~~Also at Vienna University of Technology, Vienna, Austria\\
2:~~Also at CERN, European Organization for Nuclear Research, Geneva, Switzerland\\
% ...

\end{sloppypar}
\end{document}